# A statistical mechanical model of closed loop plectoneme supercoiling and its variational approximation

Dominic J. (0') Lee[1,a.)],

[1]Department of Chemistry, Imperial College London, SW7 2AZ, London, UK

## Abstract

Presented here, is a technical manuscript that may form the basis of later published work. In it, we develop a statistical mechanical model to describe a closed loop plectoneme, applicable for when the closed loop is sufficiently supercoiled. The model divides the system up into end loops and a braided section; the end loops are assumed to contribute little to the super-coil writhe. Within the braided section, the model incorporates interactions that depend on the structure of the molecule; in particular, we consider those that depend on helical structure. A method for approximating the steric interactions is utilized that we had previously used in other publications. We go on to construct variational approximations for our closed loop plectoneme model in two cases. The first case is where helix dependent interactions are strong, and in the second case they are considered weak. In developing these approximations, we approximate the Fuller-White condition by replacing, in all expressions that depend on twist, writhe with average writhe, valid when the braided section is sufficiently long. How this approximation is made and the conditions when this approximation is valid are also discussed. The approximation allows for a Legendre transformation of the free energy, which with the introduction of moment (or torque), effectively allowing for twist and average writhe to be treated independently in the transformed (Gibbs like) free energy. Next, we then show how one may compute the average writhe of the braided section. Lastly, we discuss how some of the approximations considered may be relaxed, and discuss how the resulting model free energy might be computed by MC simulation.

## 0. Introduction

This is technical archival document that formulates a statistical mechanical model for closed loop supercoiling, which allows for the inclusion of interactions that depend on the helix structure of the molecule. This work may form a basis for later publications. This study is mainly directed plectonemes formed by DNA molecules, but might be applied to other helical molecules. For instance, recently closed loop plectonemes of super-coiled actin have been constructed [1]. For helical molecules, the helix structure is distorted by thermal fluctuations, as well as, in the case of DNA, sequence dependent, imperfect base pair stacking. Modes in intersegment interactions that depend on helix structure depend on the two segment helices being able to maintain register with each other [2,3]; their helices being commensurate with each other. If thermal fluctuations or other forms of helix disorder are too strong, the helix dependent interactions can be significantly weakened.

---

[a.)] Electronic Mail: domolee@hotmail.com

To begin, let us discuss the interplay between helix structure dependent force and helix distortions in a little more detail. A distorted helix $\mu$, with a straight axis, can be described by the position vector equation

$$\mathbf{r}_{helix,\mu}(z) = a\cos\left(gz + \delta\xi_\mu(z)\right)\hat{\mathbf{i}} + a\sin\left(gz + \delta\xi_\mu(z)\right)\hat{\mathbf{j}} + z\hat{\mathbf{k}}, \qquad (0.1)$$

where $z$ is the position along the central axis of the helix and $a$ is the helix radius. Any $z$ variation in $\delta\xi_\mu(z)$ represents the distortions of the helix away from an ideal helix structure with constant pitch $H = 2\pi/g$. Between two straight distorted helices (labelled 1 and 2), parallel to each other, we may define an angle that describes the local relative orientation of the two helices $\Delta\Phi(z) = \delta\xi_1(z) - \delta\xi_2(z)$, as we change $z$. The accumulation of mismatch between the two helices can then be characterized by the correlation function

$$\mathcal{G}(z - z') = \left\langle \left(\Delta\Phi(z) - \Delta\Phi(z')\right)^2 \right\rangle. \qquad (0.2)$$

A perfect alignment of two helices is characterized by $\mathcal{G}(z - z') = 0$; this corresponds either to perfect helices $\delta\xi_1(z) = \delta\xi_2(z) = 0$, or helices with the same pattern of distortions $\delta\xi_1(z) = \delta\xi_2(z) \neq 0$. It is in either of these two situations where helix structure dependent interactions are strongest.

Now, we suppose now that the helices lie on elastic rods with a twisting rigidity $C$ and helices move with the twisting of the rods. Then thermal twisting fluctuations distort the helix geometries so that $\delta\xi_1(z) \neq \delta\xi_2(z)$. If we consider only thermal fluctuations, with no helix structure specific interactions, we have that for the correlation function

$$\mathcal{G}(z - z') = \frac{2k_B T}{C}|z - z'| \qquad (0.3)$$

If one includes also base-pair specific distortions for DNA, still we have $\mathcal{G}(z - z') \propto |z - z'|$, in the absence of any helix structure dependent interactions [2,3]. The tendency of helical interactions is to reduce $\mathcal{G}(z - z')$ at the cost of entropy and twisting elastic energy to increase the degree of attraction, or reduce repulsion.

For plectonemes, we will examine two situations for helix dependent forces. In the first regime, helix dependent forces are sufficiently strong enough to significantly change the form of $\mathcal{G}(z - z')$, and introduce a new length scale into the correlation function, $\lambda_h^*$ the helical adaptation length. For $|z - z'| < \lambda_h^*$ we still have that $\mathcal{G}(z - z') \propto |z - z'|$; however, now when $|z - z'| > \lambda_h^*$, $\mathcal{G}(z - z') \propto \lambda_h^*$. Over large length scales an imperfect alignment of helices exists and a preferred thermal average azimuthal orientation $\langle\Delta\Phi(z)\rangle$ is present. How imperfect this alignment is characterized by the size of the length $\lambda_h^*$; and helix dependent interactions are weakened as $\lambda_h^*$ is increased. On the other hand, the stronger the helix structure dependent interaction is, the smaller $\lambda_h^*$ becomes. We call this regime where there exists a finite $\lambda_h^*$ the strong helix dependent force regime.

In the second case we consider, the helix dependent forces are too weak for $\lambda_h^*$ to remain finite. However, this does not mean that these forces are completely washed out. Instead, a different picture arises, that of correlation forces; but rather between local helix structures than the more conventional ionic correlation forces. In this regime, for the two helices (1 and 2) the phase $\delta\xi_1(z)$ is still able to correlate with $\delta\xi_2(z)$, but only over small fluctuating regions along the two segments. However, there is no thermal average preferred alignment between the two helices $\langle\Delta\Phi(z)\rangle$. We call this situation the weak helix dependent force regime.

The purpose of this study is to develop a statistical mechanical model of plectonemes that takes account of helix dependent forces in both regimes. This extends beyond the ground state model considered in Ref. [4], and is based on elements of previous work [5,6,7]. From the model considered here, we develop equations that characterize both the average geometry of a plectoneme and the fluctuations about this average structure, where it is supposed that the length of the braided section is much larger than the bending persistence length of the molecule. Numerical solutions of the equations developed here, to be performed in later work, should give insight into the possible role of helix dependent forces in plectonemes. Already, there may be indications of a slight role for helix dependent forces in DNA braiding experiments [8]. This is marked by a slight asymmetry between left and right braids. In Ref [7] it was postulated that the weak helix force regime may indeed be present here. In this work, we present a first time study of the weak helix dependent force regime for braids; although such a regime has been studied in the past for assemblies [9,10].

The work is structured in the following way. In the next section we start by looking at some initial considerations of plectoneme geometry and topology. In describing our model, we start by dividing the plectoneme into constituent end loops and a braided section. In section 2, we discuss each of the various contributions to the energy functional that describes the energy of a particular configuration of the plectoneme. In section 3, we discuss how the steric interaction may be approximated using a procedure first developed in Ref. [11] for assemblies, based on the work of Ref. [12]. This prescription allows us, in section 4, to construct a variational approximation for the strong helix specific force regime; and in section 5, a variational approximation for the weak regime. Both calculations rely on an approximation where the writhe, appearing in various expressions in the energy functional, is replaced by its average value. This should be valid when the length of the braid section is sufficiently long, as is discussed in the text. In both sections, through minimizing the resulting free energy, we obtain a system of equations describing the average plectoneme geometry and the fluctuations about it, for each case. However, what still remains is to compute the average writhe. In section 6 we utilize Gauss's integral evaluation of writhe, considering contributions from only the braided section. Unlike the single integral in Fuller's approximation, this does suffer from any divergence due to the orientation of the tangent vector [13], when we consider thermal fluctuations. In section 7, we obtain expressions for the derivatives of the average writhe that can be substituted back into the system of equations, fully formulating them for later work. In section 8, we examine the limit of our expressions for the average writhe when thermal fluctuations are infinitesimally small, the ground state. When the tilt angle (the angle between the two segments making the braid) is small we recover the expression for writhe used in Ref. [4]. In the last section, we relax a large number of the approximations to formulate a model partition function that could be evaluated by Monte-Carlo simulation.

## 1. Initial Considerations

To describe the molecular centre line of the complete plectoneme we will start by defining the arc-length coordinate $s$. The value of $s$ runs from $0$ to $2L$, where $2L$ is the total molecular contour length of one cycle around the plectoneme. To describe the plectoneme, as well as the molecular centre line, one needs a second trajectory that processes around it; this second trajectory is traced out by position vectors of unit length. We will term this second trajectory the helix, which is sometimes termed as a ribbon. These unit vectors have origins that lie on the molecular centre line and lie perpendicular to the tangent vector of the molecular centre line.

Associated with molecular centre line is quantity called writhe. This is computed through Guass' Integral

$$Wr = \frac{1}{4\pi} \oint \oint ds ds' \frac{(\mathbf{r}(s)-\mathbf{r}(s')).\hat{\mathbf{t}}(s) \times \hat{\mathbf{t}}(s')}{|\mathbf{r}(s)-\mathbf{r}(s')|^{3/2}}, \qquad (1.1)$$

where $\mathbf{r}(s)$ is the position vector of the molecular centreline and $\hat{\mathbf{t}}(s) = \mathbf{r}'(s)$ is the tangent vector (here the prime refers to differentiation with respect to argument). Essentially, the writhe is an average over all possible 2-D projections of the number of times within each projection the molecular centre line crosses its self. The helix has associated with it a quantity called twist, which is essentially the number of times the unit position vector describing the helix processes about the tangent vector in the lab frame. The braid twist is given by

$$Tw = \frac{1}{2\pi} \oint g(s) ds, \qquad (1.2)$$

where $g(s)$ is the local twist density. The sum of these two quantities define the linking number, which is given through the Fuller-White formula

$$Lk = Wr + Tw. \qquad (1.3)$$

As the plectoneme is in a closed loop the total linking number $Lk$ is a number that must be constrained to a fixed value that cannot change without cutting the loop; a topological invariant.

Now, it is useful to define the twist of a torsionally relaxed molecule as

$$Tw^0 = \frac{1}{2\pi} \oint g^0(s) ds, \qquad (1.4)$$

where $g^0(s)$ is the twist density of the torsionally relaxed state. For an ideal helix $g^0(s)$ is constant with respect to $s$. This definition allows us to consider torsional elastic strain, which is the difference between the actual twist $Tw$ and $Tw^0$.

Let us now consider $g^0(s)$ should be for DNA. We define the relaxed helix as the trajectory traced out by the minor groove of the DNA molecule around the plectoneme, in the absence of torsional stress. The function $g^0(s)$ is determined by the DNA base pair sequence [14]. Indeed, for DNA, the helix is actually a distorted one due to the imperfect stacking of base pairs (see Refs. [2,15]). For the relaxed DNA twist density, we write

$$g^0(s) = \bar{g}_0 + \Delta g^0(s). \tag{1.5}$$

Here, $\bar{g}_0$ is an average of $g^0(s)$ over all possible base pair sequences. We have that $\bar{g}_0 = 2\pi/H$, where $H$ is the average helical pitch of relaxed molecule; for DNA it is $H \approx 33.8\text{Å}$. The base pair sequence dependent function $\Delta g^0(s)$ is assumed to be a random field- the randomness is to do with the difference in choices of base pairs along the sequence- with an uncorrelated Gaussian distribution (for a justification see Ref. [14]). This means that

$$\left\langle \Delta g^0(s) \Delta g^0(s') \right\rangle_g = \frac{1}{\lambda_c^{(0)}} \delta(s-s'), \tag{1.6}$$

$$\left\langle \Delta g^0(s) \right\rangle_g = \left\langle \Delta g^0(s) \right\rangle_g = 0. \tag{1.7}$$

Here the subscript $g$ refers to ensemble averaging over all possible realizations of $\Delta g^0(s)$. We call $\lambda_c^{(0)}$ the intrinsic helical coherence length, which is a measure how distorted the DNA helix is in the relaxed state (in the absence of thermal fluctuations). For DNA, $\lambda_c^{(0)}$ is estimated to be $\lambda_c^{(0)} \approx 150\text{Å}$ [14].

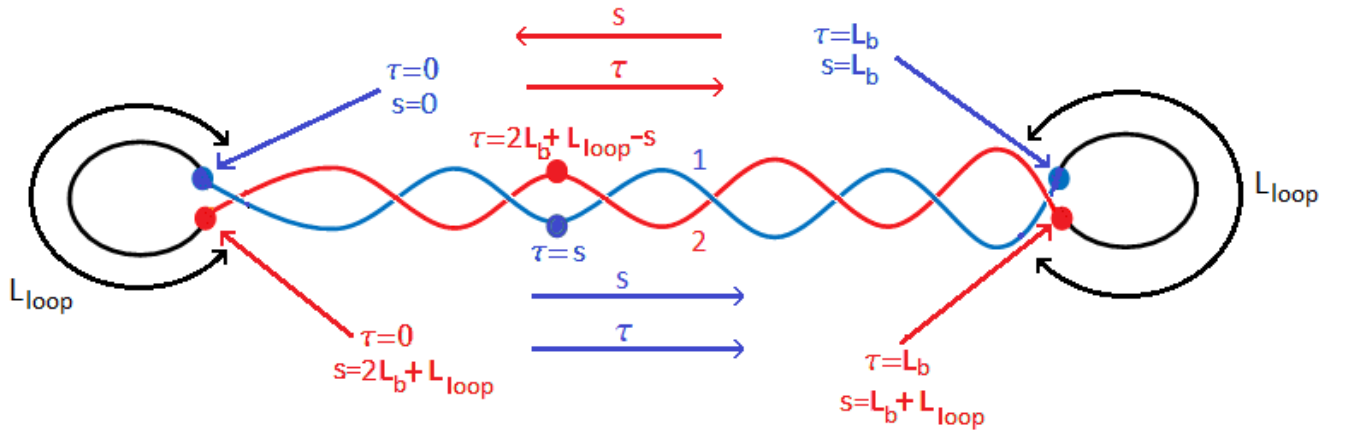

Fig.1. Schematic illustration of how the plectoneme is divided up in the model. The black parts of the plectoneme correspond to the end loop sections of length $L_{loop}$, the blue corresponds to segment 1 of braided section, and the red corresponds to segment 2. There are two sets of coordinates $s$ and $\tau$, $0 \leq s \leq 2L$ corresponds to any position around the entire closed loop, and $0 \leq \tau \leq L_b$ is defined only for the braided section. Shown as blue and red blobs are the ends of the braided section, corresponding to the sets of values $s=0$, $\tau=0$; $s=2L_b+L_{loop}$, $\tau=0$; $s=L_b$, $\tau=L_b$; and $s=L_b+L_{loop}$, $\tau=L_b$. Also shown (within the braided

section) are coloured blobs corresponding to arbitrary points along the braid, where relationships between $\tau$ and $s$ are given for both segments 1 and 2. Also, the flat arrows show the directions of increasing $\tau$ and $s$ for both segments 1 and 2, along the braided section.

Now, in our model of closed loop supercoiling, we will also divide the plectoneme into a braided section, with two attached end loops of length $L_{loop}$. The two sections that intertwine in the braided part we will number 1 and 2. Next, we define a new arc-length coordinate $\tau$ to describe the braid, which runs from one end of the braided section $\tau = 0$ to the other end $\tau = L_b$ (see Fig.1), positions along both segments 1 and 2 are described by $\tau$. Thus, $L_b$ is the contour length of each of the two sections 1 and 2, and we must have the fixed length constraint $L = L_b + L_{loop}$.

In this present study we will assume that the braid axis is straight, so that we may write for the position vectors of the segments

$$\mathbf{r}_1(\tau) = \frac{1}{2}\left(R(\tau)\cos\theta(\tau)\hat{\mathbf{i}} + R(\tau)\sin\theta(\tau)\hat{\mathbf{j}}\right) + Z(\tau)\hat{\mathbf{k}}, \qquad (1.8)$$

$$\mathbf{r}_2(\tau) = -\frac{1}{2}\left(R(\tau)\cos\theta(\tau)\hat{\mathbf{i}} + R(\tau)\sin\theta(\tau)\hat{\mathbf{j}}\right) + Z(\tau)\hat{\mathbf{k}}. \qquad (1.9)$$

Here, $R(\tau)$ is the distance between the two molecular centre lines of segments 1 and 2 making up the braid and $\theta(\tau)$ is their angle of rotation around the braid axis in the lab frame. In writing Eqs. (1.8) and (1.9), the braid axis is supposed to lie along the $z$-direction. For a regular braided structure we have that

$$R(\tau) = R_0 \quad \text{and} \quad \theta(\tau) = \theta_0 - Q\tau, \qquad (1.10)$$

where $R_0$, $\theta_0$ and $Q$ are all constants. By writing down the general forms given by Eqs. (1.8) and (1.9), we can allow for thermal fluctuations in the braid structure; only the average braid structure will be assumed to be regular. Both $\mathbf{r}_1(\tau)$ and $\mathbf{r}_2(\tau)$ are related back to the full centre line trajectory $\mathbf{r}(s)$ around the plectoneme through the relationships

$$\mathbf{r}_1(\tau) = \mathbf{r}(s), \qquad 0 < s \leq L_b, \qquad (1.11)$$

$$\mathbf{r}_2(\tau) = \mathbf{r}(2L_b + L_{loop} - s), \qquad L_b + L_{loop} < s \leq 2L_b + L_{loop}. \qquad (1.12)$$

Simply by differentiating $\mathbf{r}_1(\tau)$ and $\mathbf{r}_2(\tau)$ with respect to $\tau$ we obtain the tangent vectors, which can be written in the form (where we have used the requirement that both $\left|\hat{\mathbf{t}}_1(\tau)\right| = \left|\hat{\mathbf{t}}_2(\tau)\right| = 1$) as

$$\hat{\mathbf{t}}_1(\tau) = \sin\left(\frac{\eta(\tau)}{2}\right)\sin\left(\theta(\tau) + \gamma(\tau)\right)\hat{\mathbf{i}} - \sin\left(\frac{\eta(\tau)}{2}\right)\cos\left(\theta(\tau) + \gamma(\tau)\right)\hat{\mathbf{j}} + \cos\left(\frac{\eta(\tau)}{2}\right)\hat{\mathbf{k}},$$

(1.13)

$$\hat{\mathbf{t}}_2(\tau) = -\sin\left(\frac{\eta(\tau)}{2}\right)\sin(\theta(\tau)+\gamma(\tau))\hat{\mathbf{i}} + \sin\left(\frac{\eta(\tau)}{2}\right)\cos(\theta(\tau)+\gamma(\tau))\hat{\mathbf{j}} + \cos\left(\frac{\eta(\tau)}{2}\right)\hat{\mathbf{k}}.$$

(1.14)

Thus, we require the following interrelationships

$$Z(\tau) = \int_0^\tau d\tau' \cos\left(\frac{\eta(\tau')}{2}\right), \quad \theta(\tau) - \theta_0 = -\int_0^\tau d\tau' \frac{1}{R(\tau')}\sqrt{4\sin^2\left(\frac{\eta(\tau')}{2}\right) - \left(\frac{dR(\tau')}{d\tau'}\right)^2}, \quad (1.15)$$

and

$$\cos(\gamma(\tau)) = \sqrt{1 - \frac{1}{4\sin(\eta(\tau)/2)^2}\left(\frac{dR(\tau)}{d\tau}\right)^2}, \quad \sin(\gamma(\tau)) = \frac{1}{2\sin(\eta(\tau)/2)}\left(\frac{dR(\tau)}{d\tau}\right). \quad (1.16)$$

Also, from Eqs. (1.13) and (1.14) one should note that

$$\hat{\mathbf{t}}_1(\tau).\hat{\mathbf{t}}_2(\tau) = \cos\eta(s). \tag{1.17}$$

For a regular helix, with $\eta(s) = \eta_0$, Eqs. (1.15) and (1.16) simply reduce to

$$Z(\tau) = \tau \cos\left(\frac{\eta_0}{2}\right), \quad \theta(\tau) - \theta_0 = -\frac{2\tau}{R_0}\sin\left(\frac{\eta_0}{2}\right) \equiv -Q\tau. \tag{1.18}$$

The tangent vectors $\hat{\mathbf{t}}_1(\tau)$ and $\hat{\mathbf{t}}_2(\tau)$ are related back to $\hat{\mathbf{t}}(s)$ through the relationships

$$\hat{\mathbf{t}}_1(\tau) = \hat{\mathbf{t}}(s), \qquad 0 < s \leq L_b, \tag{1.19}$$

$$\hat{\mathbf{t}}_2(\tau) = -\hat{\mathbf{t}}(2L_b + L_{loop} - s), \qquad L_b + L_{loop} < s \leq 2L_b + L_{loop}. \tag{1.20}$$

Now, for the braided section, we will need to construct a way of defining the relative orientation of the helices of the two segments in the braid in relation to the local braid geometry. We do this by constructing two braid frames [16], spanned by vector set $\{\hat{\mathbf{d}}_\mu(\tau), \hat{\mathbf{n}}_\mu(\tau), \hat{\mathbf{t}}_\mu(\tau)\}$ where $\mu = 1,2$. These are computed through

$$\hat{\mathbf{n}}_\mu(\tau) = \frac{\hat{\mathbf{t}}_\mu(\tau) \times \hat{\mathbf{d}}(\tau)}{|\hat{\mathbf{t}}_\mu(\tau) \times \hat{\mathbf{d}}(\tau)|}, \qquad \hat{\mathbf{d}}_\mu(\tau) = \hat{\mathbf{n}}_\mu(\tau) \times \hat{\mathbf{t}}_\mu(\tau), \tag{1.21}$$

where the vector $\hat{\mathbf{d}}(\tau)$ is given by $\hat{\mathbf{d}}(\tau) = (\mathbf{r}_1(\tau) - \mathbf{r}_2(\tau))/R(\tau)$. Using the vectors $\hat{\mathbf{n}}_\mu(s)$ and $\hat{\mathbf{d}}_\mu(s)$, expressions for vectors (perpendicular $\hat{\mathbf{t}}_\mu(s)$), charactering the local azimuthal orientations of the two helices of the two braided segments, can now be written in the braid frame, namely

$$\hat{\mathbf{v}}_\mu(\tau) = \cos\xi_\mu(\tau)\hat{\mathbf{d}}_\mu(\tau) + \sin\xi_\mu(\tau)\hat{\mathbf{n}}_\mu(\tau). \tag{1.22}$$

From these vectors we can compute the twist densities through the relation

$$g_\mu(\tau) = \left(\hat{\mathbf{t}}_\mu(\tau) \times \hat{\mathbf{v}}_\mu(\tau)\right) \cdot \frac{d\hat{\mathbf{v}}_\mu(\tau)}{d\tau}. \tag{1.23}$$

The twist densities $g_1(\tau)$ and $g_2(\tau)$ are related back to $g(s)$ through the relationships

$$g_1(\tau) = g(s), \qquad 0 < s \leq L_b, \tag{1.24}$$

$$g_2(\tau) = g(2L_b + L_{loop} - s), \qquad L_b + L_{loop} < s \leq 2L_b + L_{loop}. \tag{1.25}$$

We also define the twist densities for segments 1 and 2 in the torsionally relaxed state

$$g_1^0(\tau) = g^0(s), \qquad 0 < s \leq L_b, \tag{1.26}$$

$$g_2^0(\tau) = g^0(2L_b + L_{loop} - s), \qquad L_b + L_{loop} < s \leq 2L_b + L_{loop}. \tag{1.27}$$

Next we can divide the twist into contributions from the end loops and the braid. Thus, we can write Eq. (1.2) as

$$\begin{aligned}
2\pi Tw &= \int_0^{L_b} g(s)ds + \int_{L_b}^{L_b+L_{loop}} g(s)ds + \int_{L_b+L_{loop}}^{2L_b+L_{loop}} g(s)ds + \int_{2L_b+L_{loop}}^{2L_b+2L_{loop}} g(s)ds \\
&= \int_0^{L_b} \left(g_1(\tau) + g_2(\tau)\right) d\tau + \int_{L_b}^{L_b+L_{loop}} g(s)ds + \int_{2L_b+L_{loop}}^{2L_b+2L_{loop}} g(s)ds.
\end{aligned} \tag{1.28}$$

A similar expression for $Tw^0$ can be written. In what follows it will be useful to define a spatial average of the twist density $g_{av} = \pi Tw/L$; in addition, for the relaxed state, the spatial average $g_{av}^0 = \pi Tw^0/L \approx \bar{g}_0 = 2\pi/H$. Thus we may write $g(s) = g_{av} + \delta g(s)$, $g_1(\tau) = g_{av} + \delta g_1(\tau)$, $g_2(\tau) = g_{av} + \delta g_2(\tau)$, $g^0(s) = g_{av}^0 + \delta g(s)$, $g_1^0(\tau) = g_{av}^0 + \delta g_1^0(\tau)$, and $g_2^0(\tau) = g_{av}^0 + \delta g_2^0(\tau)$ where have the following relationship

$$\int_0^{L_b} \left(\delta g_1(\tau) + \delta g_2(\tau)\right) d\tau + \int_{L_b}^{L_b+L_{loop}} \delta g(s)ds + \int_{2L_b+L_{loop}}^{2L_b+2L_{loop}} \delta g(s)ds = 0. \tag{1.29}$$

A similar relationship for $\delta g^0(s)$ can be written.

For the writhe, we will neglect the writhe contributions from the end loops. This allows us to write Eq. (1.1) approximately as

$$4\pi Wr \approx \int_0^{L_b} d\tau \int_0^{L_b} d\tau' \frac{(\mathbf{r}_1(\tau)-\mathbf{r}_1(\tau')).\hat{\mathbf{t}}_1(\tau)\times\hat{\mathbf{t}}_1(\tau')}{|\mathbf{r}_1(\tau)-\mathbf{r}_1(\tau')|^{3/2}} + \int_0^{L_b} d\tau \int_0^{L_b} d\tau' \frac{(\mathbf{r}_2(\tau)-\mathbf{r}_2(\tau')).\hat{\mathbf{t}}_2(\tau)\times\hat{\mathbf{t}}_2(\tau')}{|\mathbf{r}_2(\tau)-\mathbf{r}_2(\tau')|^{3/2}}$$
$$-2\int_0^{L_b} d\tau \int_0^{L_b} d\tau' \frac{(\mathbf{r}_1(\tau)-\mathbf{r}_2(\tau')).\hat{\mathbf{t}}_1(\tau)\times\hat{\mathbf{t}}_2(\tau')}{|\mathbf{r}_1(\tau)-\mathbf{r}_2(\tau')|^{3/2}}$$
(1.30)

In the next section we will construct the model energy functional.

## 2. Model Energy functional

The total energy functional can be written as a sum of contributions, such that

$$E_T = E_B + E_{Tw} + E_{int} + E_{St}. \tag{2.1}$$

For the total bending elastic energy we may approximate with (combining the results of Refs. [17] and [4])

$$\frac{E_B}{k_B T} \approx \frac{E_{braid,B}}{k_B T} + \frac{E_{loop1,B}}{k_B T} + \frac{E_{loop2,B}}{k_B T}, \tag{2.2}$$

where $E_{braid,B}$ is the bending energy of the braid; both $E_{loop1,B}$ and $E_{loop2,B}$ are bending energies of the end loops. The braid bending energy contribution is given as

$$\frac{E_{R,B}}{k_B T} = \int_0^{L_b} d\tau \mathcal{E}_R(R''(\tau), R'(\tau), R(\tau), \delta\eta'(\tau), \delta\eta(\tau)), \tag{2.3}$$

where

$$\mathcal{E}_R(R''(\tau), R'(\tau), R(\tau), \delta\eta'(\tau), \delta\eta(\tau)) =$$
$$\frac{l_p}{4}\left(\frac{d^2 R(\tau)}{d\tau^2}\right)^2 + \frac{l_p}{4}\left(\frac{d\delta\eta(\tau)}{d\tau}\right)^2 - \left(\frac{dR(\tau)}{d\tau}\right)^2 \frac{l_p}{R(\tau)^2}\sin^2\left(\frac{\eta_0}{2}\right)$$
$$+\frac{4l_p}{R(\tau)^2}\left[\sin^4\left(\frac{\eta_0}{2}\right) + \frac{\delta\eta(\tau)^2}{2}\left(3\cos^2\left(\frac{\eta_0}{2}\right)\sin^2\left(\frac{\eta_0}{2}\right) - \sin^4\left(\frac{\eta_0}{2}\right)\right)\right] \tag{2.4}$$
$$+\left(\frac{dR(\tau)}{d\tau}\right)\left(\frac{d\delta\eta(\tau)}{d\tau}\right)\frac{3}{R(\tau)}\sin\left(\frac{\eta_0}{2}\right)\cos\left(\frac{\eta_0}{2}\right).$$

where $\eta(\tau) = \eta_0 + \delta\eta(\tau)$, and we have that the thermal average $\langle\delta\eta(\tau)\rangle = 0$, the prime and double prime, here refer to first and second derivatives with respect to argument. For bending energies of the loops we shall use approximate forms based on those contained in Ref. [4]. For the end loops we write

$$\frac{E_{loop1,B}}{k_B T} = \frac{l_p}{L_{loop}}\left(2\pi - \frac{\pi^2 R(0)}{2L_{loop}}\right)^2 \approx \frac{l_p}{L_{loop}}\left(2\pi - \frac{\pi^2\langle R(0)\rangle}{2L_{loop}}\right)^2 + \frac{l_p\pi^4\delta R(0)^2}{4L_{loop}^3}, \tag{2.5}$$

$$\frac{E_{loop2,B}}{k_B T} = \frac{l_p}{L_{loop}} \left(2\pi - \frac{\pi^2 R(L_b)}{2L_{loop}}\right)^2 \approx \frac{l_p}{L_{loop}} \left(2\pi - \frac{\pi^2 \langle R(L_b)\rangle}{2L_{loop}}\right)^2 + \frac{l_p \pi^4 \delta R(L_b)^2}{4L_{loop}^3}, \quad (2.6)$$

where $\delta R(0) = R(0) - \langle R(0)\rangle$ and $\delta R(L_b) = R(L_b) - \langle R(L_b)\rangle$ with also

$$\begin{aligned}L_{loop} &= \bar{L}_{loop}\theta\left(2\bar{L}_{loop} - \pi\langle R(L_b)\rangle\right) + \pi\langle R(L_b)\rangle \theta\left(\pi\langle R(L_b)\rangle - 2\bar{L}_{loop}\right)/2 \\ &= \bar{L}_{loop}\theta\left(2\bar{L}_{loop} - \pi\langle R(0)\rangle\right) + \pi\langle R(0)\rangle \theta\left(\pi\langle R(0)\rangle - 2\bar{L}_{loop}\right)/2,\end{aligned} \quad (2.7)$$

where $\bar{L}_{loop}$ is a evaluated at a turning point that minimizes the free energy. The theta functions in Eq. (2.7) constrains the minimum value of $L_{loop}$ to be $\pi\langle R(L_b)\rangle/2 = \pi\langle R(0)\rangle/2$, where both average values are supposed to be the same. This restriction supposes that the smallest length configuration of end loops, on average, is a semi-circle. This is an adaptation of what was considered in Ref. [4].

The only fluctuations that will be considered are those of $R(0)$ and $R(L_b)$, thus we will assume that the major contribution from the end loops to the free energy is elastic and not entropic. This should indeed be valid if the length of the loop does not exceed the order of magnitude of the bending persistence length $l_p$. Here, in this approximation, we have not matched tangent vectors between the end loops and braided sections. A more accurate and sophisticated end loop function that does this, based on what is considered in Ref. [18], could be a refinement to this. Here, bending fluctuations might be handed in a more careful manner. We also discuss in the last section how a better approximation of the end loops could be handled using MC simulation. The simple approximation given by Eqs. (2.5), (2.6) and (2.7) for the end loops is probably sufficient when the length of the braided section is such that $L_b \gg L_{loop}$, as the end loops are likely to be a second order effect.

The twisting (stretching) elastic energy expression comes from

$$\frac{E_{Tw}}{k_B T} = \frac{l_c}{2} \oint ds \left(\delta g(s) - \delta g^0(s)\right)^2 + l_{tw} L \left(g_{av} - \bar{g}_0\right)^2, \quad (2.8)$$

where the combined persistence length $l_c \approx 400\text{Å}$ (which combines fluctuations from both twisting and stretching; to see how they combine see Ref. [11]) and the twisting persistence length $l_{tw} = 1000\text{Å}$. Note that the second term only depends twisting rigidity, as the molecule is considered not to be under any stretching strain, however stretching fluctuations affect $\delta g(s)$, deviations of the twist density away from its average value, as well as affected by the intrinsic base-pair dependent pattern of distortions $\delta g^0(s)$. Note that $\delta g(s)$ and $g_{av}$ are decoupled through Eq. (1.29). The twisting elastic energy can be rewritten as

$$E_{Tw} = E_{Tw,braid,1} + E_{Tw,braid,2} + E_{Tw,loop}. \quad (2.9)$$

The first contribution to the twisting elastic energy is the contribution from the phase difference $\Delta\Phi(\tau) = \xi_1(\tau) - \xi_2(\tau)$

$$\frac{E_{Tw,braid,1}}{k_B T} = \frac{l_c}{4} \int_0^{L_b} d\tau \left( \frac{d\Delta\Phi(\tau)}{d\tau} - \Delta g^0(\tau) \right), \tag{2.10}$$

where $\Delta g^0(\tau) = g_1^0(\tau) - g_2^0(\tau)$.

We may write for the braid, from Eq. (2.8),

$$\frac{E_{Tw,braid,2}}{k_B T} = l_{tw} L_b \left( g_{av} - \bar{g}_0 \right)^2 + \frac{l_c}{4} \int_0^{L_b} d\tau \bar{\omega}(\tau)^2, \tag{2.11}$$

and for the end loops

$$\frac{E_{Tw,loop}}{k_B T} \simeq l_{tw} L_{loop} \left( g_{av} - \bar{g}_0 \right)^2 + \frac{l_c}{2} \int_{L_b}^{L_b + L_{loop}} \omega(s)^2 ds + \frac{l_c}{2} \int_{2L_b + L_{loop}}^{2L_b + 2L_{loop}} \omega(s)^2 ds, \tag{2.12}$$

where have defined $\bar{\omega}(\tau) = \delta g_1(\tau) + \delta g_2(\tau) - \delta g_1^0(\tau) - \delta g_2^0(\tau)$ and $\omega(s) = \delta g(s) - \delta g^0(s)$.

Next, we consider interaction energy for the two segments in the braid, where we will assume that it takes the form

$$\frac{E_{int}}{k_B T} = \int_0^{L_b} d\tau \varepsilon_{int}(\eta(\tau), R(\tau), \Delta\Phi(\tau), \bar{g}(\tau)/2) = \int_0^{L_b} d\tau \varepsilon_{int}(\eta(\tau), R(\tau), \Delta\Phi(\tau), g_{av} + (\delta \bar{g}^0(\tau) + \Delta g(\tau))/2). \tag{2.13}$$

where $\delta \bar{g}^0(\tau) = \delta g_1^0(\tau) + \delta g_2^0(\tau)$. It is relatively straightforward to extend Eq. (2.13) to the more general form given in Ref. [19], and generalize the analysis. However in the interests of simplicity we stick with Eq. (2.13), as so far we not yet calculated a $R'(s)$ correction to the electrostatic energy within the mean field framework of Ref. [20] (see Ref. [21] to see what has been calculated, in this framework, and how to extend the calculation). Note fluctuations in $\delta \bar{g}^0(\tau)$ and $\Delta g(\tau)$ are independent of bending, as they do not contribute to the overall twist $Tw$; thus, they can be considered as unconstrained. However $g_{av}$ is indeed constrained such that $g_{av} = \pi(Lk - Wr)/L$; the total twist around the loop depends on bending fluctuations through the Fuller-White formula (Eq. (1.3)). Now, if we suppose that $g_{av} l_c \gg 1$ and $g_{av} \lambda_c^0 \gg 1$, we can neglect both $\Delta g(\tau)$ and $\delta \bar{g}^0(\tau)$ from Eq. (2.13). Thus, we can approximate Eq. (2.13) with

$$\frac{E_{int}}{k_B T} \approx \int_0^{L_b} d\tau \varepsilon_{int}(\eta(\tau), R(\tau), \Delta\Phi(\tau), g_{av}). \tag{2.14}$$

Corrections can be considered by expanding out Eq. (2.13) in powers of both $\Delta g(\tau)$ and $\delta \bar{g}^0(\tau)$.

Last of all, we have the steric contribution for the molecules in the braid, for which

$$E_{st} = 0 \quad R(s) > 2a, \tag{2.15}$$

$$E_{st} = \infty \quad R(s) \leq 2a, \tag{2.16}$$

where we have assumed the molecules to be smooth rods of radius $a$.

In the partition function, as well as ensemble average, we can now easily integrate out the $\bar{\omega}(\tau)$ and $\delta \bar{g}^0(\tau)$ degrees of freedom in the braid and $\omega(s)$ and $\delta g^0(s)$ for the loops, not worrying too much about boundary conditions (provided that we consider large supercoils). Thus terms that depend on $\bar{\omega}(\tau)$, $\omega(s)$, $\delta \bar{g}^0(\tau)$, and $\delta g^0(s)$ can simply be discarded.

## 3. Approximating Steric Interactions

We approximate the effects of steric interactions by using the approach of Ref. [12], where we introduce a pseudo-potential of the form

$$\tilde{E}_{st} = k_B T \int_0^{L_p} d\tau \frac{\alpha_H}{2} (R(\tau) - R_0)^2 = k_B T \int_0^{L_p} d\tau \frac{\alpha_H}{2} \delta R(\tau)^2, \tag{3.1}$$

where $R_0 = \langle R(\tau) \rangle$ and $\delta R(\tau) = R(\tau) - R_0$. As was done in previous works (Refs. [7], [11], [17], [19]), we now modify the terms in the energy functional. For $\delta R(\tau) > d_{max}$ we replace $\delta R(\tau)$ with $d_{max}$, and for $\delta R(\tau) < d_{min}$ we replace $\delta R(\tau)$ with $d_{min}$. Where $d_{min}$ and $d_{max}$ are the minimum and maximum displacements away from $R_0$ allowed by steric interactions, respectively. This replacement prevents an unphysical overestimation (or underestimation) of all the other terms in the energy when we replace the steric interaction term with the harmonic potential. The parameter $\alpha_H$ is chosen in the absence of ranged interactions between the braided segments and $l_c \to 0$ (this limit needs to be considered due to the topological coupling of bending to twisting), so that we can write

$$\langle \delta R(s)^2 \rangle \approx \frac{(d_{min} - d_{max})^2}{4}. \tag{3.2}$$

If we suppose that Eq. (3.2), and that the first term in Eq. (2.4), are the dominant terms in determining $\delta R(s)$ fluctuations when there are no-ranged interactions (the effects of the other terms in Eq. (2.4) on $\langle \delta R(s)^2 \rangle$ will determine the actual extent of the $\delta R(s)$ fluctuations in the variational approximations given below), Eq. (3.2) yields an expression for $\alpha_H$

$$\alpha_H \approx \frac{2}{(d_{max} - d_{min})^{8/3} (l_p)^{1/3}}. \tag{3.3}$$

Using the cut-off prescription above, we now have the modified total energy

$$\tilde{E}_T = \tilde{E}_B + E_{Tw,braid,1} + l_c L\left(g_{av} - \overline{g}_0\right)^2 + \tilde{E}_{int} + \tilde{E}_{St}. \tag{3.4}$$

First, the modified bending energy functional is given by

$$\frac{\tilde{E}_B}{k_B T} \approx \frac{\tilde{E}_{R,B}}{k_B T} + \frac{\tilde{E}_{loop1,B}}{k_B T} + \frac{\tilde{E}_{loop2,B}}{k_B T}, \tag{3.5}$$

where

$$\frac{\tilde{E}_{braid,B}}{k_B T} = \int_0^{L_b} d\tau \big[ \mathcal{E}_R(R''(\tau), R'(\tau), R(\tau), \delta\eta'(\tau), \delta\eta(\tau))\theta(\delta R(\tau) - d_{min})\theta(d_{max} - \delta R(\tau))$$
$$+ \mathcal{E}_R(R''(\tau), R'(\tau), R_0 + d_{max}, \delta\eta'(\tau), \delta\eta(\tau))\theta(\delta R(\tau) - d_{max}) \tag{3.6}$$
$$+ \mathcal{E}_R(R''(\tau), R'(\tau), R_0 + d_{min}, \delta\eta'(\tau), \delta\eta(\tau))\theta(d_{min} - \delta R(\tau))\big],$$

$$\frac{\tilde{E}_{loop1,B}}{k_B T} = \left[ \frac{l_p}{L_{loop}} \left( 2\pi - \frac{\pi^2 R_0}{2L_{loop}} \right)^2 + \frac{l_p \pi^4 \delta R(0)^2}{4L_{loop}^3}\theta(\delta R(0) - d_{min})\theta(d_{max} - \delta R(0)) \right.$$
$$\left. + \frac{l_p \pi^4 d_{max}^2}{4L_{loop}^3}\theta(\delta R(0) - d_{max}) + \frac{l_p \pi^4 d_{min}^2}{4L_{loop}^3}\theta(d_{min} - \delta R(0)) \right], \tag{3.7}$$

$$\frac{\tilde{E}_{loop2,B}}{k_B T} = \left[ \frac{l_p}{L_{loop}} \left( 2\pi - \frac{\pi^2 R_0}{2L_{loop}} \right)^2 + \frac{l_p \pi^4 \delta R(L_b)^2}{4L_{loop}^3}\theta(\delta R(L_b) - d_{min})\theta(d_{max} - \delta R(L_b)) \right.$$
$$\left. + \frac{l_p \pi^4 d_{max}^2}{4L_{loop}^3}\theta(\delta R(L_b) - d_{max}) + \frac{l_p \pi^4 d_{min}^2}{4L_{loop}^3}\theta(d_{min} - \delta R(L_b)) \right]. \tag{3.8}$$

In place of $E_{int}$ we now have

$$\tilde{E}_{int} = \int_0^{L_b} d\tau \big[ \varepsilon_{int}(\eta(\tau), R(\tau), \Delta\Phi(\tau), g_{av})\theta(\delta R(\tau) - d_{min})\theta(d_{max} - \delta R(\tau))$$
$$+ \varepsilon_{int}(\eta(\tau), R_0 + d_{max}, \Delta\Phi(\tau), g_{av})\theta(\delta R(\tau) - d_{max}) \tag{3.9}$$
$$+ \varepsilon_{int}(\eta(\tau), R_0 + d_{min}, \Delta\Phi(\tau), g_{av})\theta(d_{min} - \delta R(\tau))\big].$$

Furthermore as $0 < \Delta\Phi < 2\pi$ we can express Eq.(3.9) as a Fourier series (supposing that $\Delta\Phi(\tau)$ varies slowly enough compared with the characteristic (Debye) screening length [21])

$$\tilde{E}_{int} = \sum_{n=-\infty}^{\infty} \int_0^{L_b} d\tau \overline{\varepsilon}_{int}(R_0, \delta R(\tau), \eta(\tau), g_{av}, n)\exp\left(-in\Delta\Phi(\tau)\right), \tag{3.10}$$

where

$$\bar{\varepsilon}_{int}(R_0, \delta R(\tau), \eta(\tau), g_{av}, n) = \frac{1}{2\pi} \int_0^{2\pi} d\Delta\Phi \exp(in\Delta\Phi)$$

$$[\varepsilon_{int}(\eta(\tau), R(\tau), \Delta\Phi, g_{av})\theta(\delta R(\tau) - d_{min})\theta(d_{max} - \delta R(\tau))$$
$$+\varepsilon_{int}(\eta(\tau), R_0 + d_{max}, \Delta\Phi, g_{av})\theta(\delta R(\tau) - d_{max}) + \varepsilon_{int}(\eta(\tau), R_0 + d_{min}, \Delta\Phi, g_{av})\theta(d_{min} - \delta R(\tau))].$$

(3.11)

We are now ready to consider the variational approximations.

## 4. Variational Approximation for strong helix specific forces

We now have the following form for the full partition function

$$Z_T = \int DR(\tau) \int D\eta(\tau) \int D\Delta\Phi(\tau) \exp\left(-\frac{\tilde{E}_T[R(\tau), \eta(\tau), \Delta\Phi(\tau)]}{k_B T}\right). \quad (4.1)$$

Note that the total free energy is given by $F_T = -k_B T \ln Z_T$. In the variational approximation we start by constructing the trial functional

$$\frac{E_0^T[\delta\eta(\tau), \delta R(\tau), \delta\Phi(\tau)]}{k_B T} = \int_0^{L_b} d\tau \left(\frac{l_p}{4}\left(\frac{d^2\delta R(\tau)}{d\tau^2}\right)^2 + \frac{\beta_R}{2}\left(\frac{d\delta R(\tau)}{d\tau}\right)^2 + \frac{\alpha_R}{2}\delta R(\tau)^2\right)$$
$$+ \int_0^{L_b} d\tau \left(\frac{l_p}{4}\left(\frac{d\delta\eta(\tau)}{d\tau}\right)^2 + \frac{\alpha_\eta}{2}\delta\eta(\tau)^2\right) + \int_0^{L_b} d\tau \left(\frac{l_c}{4}\left(\frac{d\delta\Phi(\tau)}{d\tau}\right)^2 + \frac{\alpha_\Phi}{2}\delta\Phi(\tau)^2\right), \quad (4.2)$$

where now we have that $\Delta\Phi(\tau) = \Delta\Phi_0(\tau) + \delta\Phi(\tau)$ with $\langle\delta\Phi(\tau)\rangle = 0$. The $\delta\Phi(\tau)$-fluctuations are now limited through the variational parameter $\alpha_\Phi$. Instead of $\alpha_H$, to take account of ranged interactions, we now have variational parameters $\beta_R$ and $\alpha_R$ that determine the values of both $d_R^2 = \langle\delta R(\tau)^2\rangle$ and $\theta_R^2 = \langle\delta R'(\tau)^2\rangle$. Thus we can write down the variational free energy

$$F_T \approx F_T^T = -k_B T \ln Z_T^T + \langle \tilde{E}_T[R(\tau), \eta(\tau), \Delta\Phi(\tau)] - E_0^T[\delta\eta(\tau), \delta R(\tau), \delta\Phi(\tau)]\rangle_0, \quad (4.3)$$

where

$$Z_T^T = \int DR(\tau) \int D\eta(\tau) \int D\Delta\Phi(\tau) \exp\left(-\frac{E_0^T[R(\tau), \eta(\tau), \Delta\Phi(\tau)]}{k_B T}\right), \quad (4.4)$$

and for the average of an arbitrary functional $U[R(\tau), \eta(\tau), \Delta\Phi(\tau)]$

$$\langle U[R(\tau), \eta(\tau), \Delta\Phi(\tau)]\rangle_0 = \frac{1}{Z_T^T} \int DR(\tau) \int D\eta(\tau) \int D\Delta\Phi(\tau) A[R(\tau), \eta(\tau), \Delta\Phi(\tau)] \exp\left(-\frac{E_0^T[R(\tau), \eta(\tau), \Delta\Phi(\tau)]}{k_B T}\right)$$

(4.5)

The essence of the variational approximation is Gibbs-Bogoliubov inequality that states that $F_T \leq F_T^T$. We choose the variational parameters $\Delta\Phi_0(\tau)$, $R_0$, $\eta_0$, $\alpha_\Phi$, $\alpha_R$, $\beta_R$, and $\alpha_\eta$ to minimize $F_T^T$ (described by Eq. (4.3)) to get it closest to $F_T$.

Let us consider the averages. In the averaging we will assume that the length of the braided section $L_b$ is much larger than the correlation lengths of the various fluctuations. First of all we have

$$\frac{\langle \tilde{E}_{braid,B}[\eta_0+\delta\eta(\tau), R_0+\delta R(\tau)] + \tilde{E}_{St}[\delta R(\tau)] - E_T^T[\delta\eta(\tau), \delta R(\tau)] \rangle_0}{k_B T L_b} =$$

$$-\frac{\theta_R^2 l_p}{R_0^2} \tilde{f}_1(R_0, d_R, d_{max}, d_{min}) \sin^2\left(\frac{\eta_0}{2}\right) - \frac{\alpha_\eta d_\eta^2}{2} + \frac{(\alpha_H - \alpha_R)}{2} d_R^2 - \frac{\beta_R}{2}\theta_R^2 - \frac{\alpha_\Phi d_\Phi^2}{2} - \frac{l_c}{4}\left\langle \left(\frac{d\delta\Phi(\tau)}{d\tau}\right)^2 \right\rangle_0$$

$$+ 4l_p \frac{\tilde{f}_1(R_0, d_R, d_{max}, d_{min})}{R_0^2}\left[\sin^4\left(\frac{\eta_0}{2}\right) + \frac{d_\eta^2}{2}\left(3\cos^2\left(\frac{\eta_0}{2}\right)\sin^2\left(\frac{\eta_0}{2}\right) - \sin^4\left(\frac{\eta_0}{2}\right)\right)\right].$$

(4.6)

The function $\tilde{f}_1(R_0, d_R, d_{max}, d_{min})$ is defined as

$$\tilde{f}_1(R_0, d_R, d_{max}, d_{min}) = \frac{R_0^2}{d_R \sqrt{2\pi}} \int_{d_{min}}^{d_{max}} \frac{dx}{(R_0+x)^2} \exp\left(-\frac{x^2}{2d_R^2}\right)$$

$$+ \frac{1}{2}\left(\frac{R_0^2}{(R_0+d_{min})^2}\left(1-\text{erf}\left(-\frac{d_{min}}{d_R\sqrt{2}}\right)\right) + \frac{R_0^2}{(R_0+d_{max})^2}\left(1-\text{erf}\left(\frac{d_{max}}{d_R\sqrt{2}}\right)\right)\right).$$

(4.7)

We also have the relationships between parameters

$$\theta_R^2 = \frac{1}{2\pi} \int_{-\infty}^{\infty} \frac{k^2 dk}{\frac{l_p}{2}k^4 + \beta_R k^2 + \alpha_R} = \frac{1}{2\pi\alpha_R}\left(\frac{2\alpha_R}{l_p}\right)^{3/4} \int_{-\infty}^{\infty} \frac{k^2 dk}{k^4 + \gamma k^2 + 1}, \quad \text{where } \gamma = \beta_R \left(\frac{2}{\alpha_R l_p}\right)^{1/2},$$

(4.8)

$$d_R^2 = \frac{1}{2\pi} \int_{-\infty}^{\infty} \frac{dk}{\frac{l_p}{2}k^4 + \beta_R k^2 + \alpha_R} = \frac{1}{2\pi\alpha_R}\left(\frac{2\alpha_R}{l_p}\right)^{1/4} \int_{-\infty}^{\infty} \frac{dk}{k^4 + \gamma k^2 + 1}, \tag{4.9}$$

$$d_\eta^2 = \langle \delta R(\tau)^2 \rangle = \frac{1}{2\pi} \int_{-\infty}^{\infty} \frac{dk}{\frac{l_p}{2}k^2 + \alpha_\eta} = \frac{1}{(2l_p\alpha_\eta)^{1/2}}, \tag{4.10}$$

as well as

$$d_\Phi^2 = \langle \delta\Phi(\tau)^2 \rangle = \frac{1}{2\pi} \int_{-\infty}^{\infty} dk \frac{1}{\frac{l_c}{2}k^2 + \alpha_\Phi} = \frac{2}{l_c}\frac{1}{2\pi} \int_{-\infty}^{\infty} dk \frac{1}{k^2 + \frac{1}{\lambda_h^2}} = \frac{\lambda_h}{l_c}, \quad \text{where } \lambda_h = \left(\frac{l_c}{2\alpha_\Phi}\right)^{1/2}.$$

(4.11)

In Ref. [19] we found that

$$\theta_R^2 = \frac{1}{2\alpha_R}\left(\frac{2\alpha_R}{l_p}\right)^{3/4} \frac{1}{\sqrt{\gamma+2}} \quad \text{and} \quad d_R^2 = \frac{1}{2\alpha_R}\left(\frac{2\alpha_R}{l_p}\right)^{1/4} \frac{1}{\sqrt{\gamma+2}}.$$

(4.12)

Making $\alpha_R$ and $\beta_R$ the subjects of the expressions in Eq. (4.12), we also have that [19]

$$\alpha_R = \frac{l_p}{2}\left(\frac{\theta_R}{d_R}\right)^4 \quad \text{and} \quad \beta_R = \frac{l_p}{2}\left(\frac{1}{l_p^2\theta_R^4} - 2\left(\frac{\theta_R}{d_R}\right)^2\right).$$

(4.13)

Next, we compute the average interaction energy, starting with

$$\frac{\langle \tilde{E}_{\text{int}} \rangle_0}{k_B T} = \sum_{n=-\infty}^{\infty} \int_0^{L_b} d\tau \langle \bar{\varepsilon}_{\text{int}}(R_0, \delta R(\tau), \eta(\tau), g_{av}, n) \exp(-in\Delta\Phi(\tau)) \rangle_0$$

$$= \frac{1}{(2\pi)^{1/2} d_\Phi} \sum_{n=-\infty}^{\infty} \int_{-\infty}^{\infty} d\Phi \int_0^{L_b} d\tau \langle \bar{\varepsilon}_{\text{int}}(R_0, \delta R(\tau), \eta_0 + \delta\eta(\tau), g_{av}, n) \rangle_0 \exp(-in(\Delta\Phi_0(\tau) + \Phi)) \exp\left(-\frac{\Phi^2}{2d_\Phi^2}\right)$$

$$= \sum_{n=-\infty}^{\infty} \int_0^{L_b} d\tau \langle \bar{\varepsilon}_{\text{int}}(R_0, \delta R(\tau), \eta_0 + \delta\eta(\tau), g_{av}, n) \rangle_0 \exp(-in\Delta\Phi_0(\tau)) \exp\left(-\frac{n^2 d_\Phi^2}{2}\right).$$

(4.14)

The second step comes from the Gaussian averaging formulas given in Ref. [19] or the supplemental material of Ref. [17].

Now, we come to the most crucial step in our analysis. We will now assume that $L_b$ the length of the braided section is much larger than the lengths $\lambda_R$ and $\lambda_\eta$, the correlation lengths respectively of the $\delta R$ and $\delta \eta$ fluctuations, such that

$$\frac{\langle \tilde{E}_{\text{int}} \rangle_0}{k_B T} = \frac{1}{(2\pi)d_R d_\eta} \sum_{n=-\infty}^{\infty} \int_{-\infty}^{\infty} d\eta \int_{-\infty}^{\infty} dr \int_0^{L_b} d\tau \bar{\varepsilon}_{\text{int}}(R_0, r, \eta + \eta_0, \langle g_{av,} \rangle_0, n)$$

$$\exp(-in\Delta\Phi_0(\tau)) \exp\left(-\frac{n^2 d_\Phi^2}{2}\right) \exp\left(-\frac{\eta^2}{2d_\eta^2}\right) \exp\left(-\frac{r^2}{2d_r^2}\right),$$

(4.15)

where

$$\langle g_{av,}\rangle_0 = \frac{\pi}{L}\left(Lk - \langle Wr\rangle_0\right). \tag{4.16}$$

The correlation lengths $\lambda_R$ and $\lambda_\eta$ are roughly the decay lengths with respect to $|\tau - \tau'|$ of the correlation functions $\langle \delta R(\tau)\delta R(\tau')\rangle$ and $\langle \delta\eta(\tau)\delta\eta(\tau')\rangle$, respectively. This approximation relies on expressing $\bar{\varepsilon}_{\text{int}}(R_0, \delta R(\tau), \delta\eta(\tau) + \eta_0, g_{av}, n)$ as a power series in $Wr$, as well as on the decomposition,

$$\langle \bar{\varepsilon}_{\text{int}}^n(R_0, \delta R(\tau), \delta\eta(\tau) + \eta_0, \pi Lk/L, n) Wr^n\rangle_0 \approx \langle \bar{\varepsilon}_{\text{int}}^n(R_0, \delta R(\tau), \delta\eta(\tau) + \eta_0, \pi Lk/L, n)\rangle_0 \langle Wr\rangle_0^n, \tag{4.17}$$

where $\quad \bar{\varepsilon}_{\text{int}}^n(R_0, \delta R(\tau), \delta\eta(\tau) + \eta_0, \pi Lk/L, n) = \dfrac{d^n \bar{\varepsilon}_{\text{int}}(R_0, \delta R(\tau), \delta\eta(\tau) + \eta_0, g_{av}, n)}{dg_{av}^n}\bigg|_{Wr=0}$, (4.18)

and resuming the series. To understand why these decomposition formulas (Eqs. (4.17) and (4.18)) should hold for $L_b \gg \lambda_R, \lambda_\eta$, let us examine the averages $\langle \bar{\varepsilon}_{\text{int}}^n(R_0, \delta R(\tau), \delta\eta(\tau) + \eta_0, \pi Lk/L, n) Wr\rangle_0$ and $\langle Wr^2\rangle_0$. We can write both of these averages as

$$\langle Wr^2\rangle_0 = \langle Wr\rangle_0^2 + \langle Wr^2\rangle_{0,c}, \tag{4.19}$$

$$\langle \bar{\varepsilon}_{\text{int}}^n(R_0, \delta R(\tau), \delta\eta + \eta_0, \pi Lk/L, n) Wr\rangle_0 \approx \langle \bar{\varepsilon}_{\text{int}}^n(R_0, \delta R(\tau), \delta\eta(\tau) + \eta_0, \pi Lk/L, n)\rangle_0 \langle Wr\rangle_0$$
$$+ \langle \bar{\varepsilon}_{\text{int}}^n(R_0, \delta R(\tau), \delta\eta(\tau) + \eta_0, \pi Lk/L, n) Wr\rangle_{0,c}. \tag{4.20}$$

The averaging bracket $\langle AB\rangle_{0,c}$ ( here for arbitrary quantities $A$ and $B$ ) means mean that they are correlated, when considering the average; what we call a correlated (or connected) average. This division for correlation function into connected and disconnected parts can be found in any standard statistical field theory text book [22]. Now, in general when we consider the double integral in Eq. (1.30), we can write these correlated averages as

$$\langle \bar{\varepsilon}_{\text{int}}^n(R_0, \delta R(\tau), \delta\eta(\tau) + \eta_0, \pi Lk/L, n) Wr\rangle_{0,c} = \int_0^{L_b} d\tau_1 \int_0^{L_b} d\tau_2 Y^n(\tau_2 - \tau_1, \tau_1 - \tau) \approx \int_{-\infty}^{\infty} \delta d\tau_1 \int_{-\infty}^{\infty} \delta d\tau_2 Y^n(\delta\tau_2, \delta\tau_1), \tag{4.21}$$

$$\langle Wr^2\rangle_{0,c} = \int_0^{L_b} d\tau_1 \int_0^{L_b} d\tau_2 \int_0^{L_b} d\tau_3 \int_0^{L_b} d\tau_4 X(\tau_1 - \tau_2, \tau_2 - \tau_3, \tau_3 - \tau_4) \approx L_b \int_{-\infty}^{\infty} d\delta\tau_1 \int_{-\infty}^{\infty} d\delta\tau_2 \int_{-\infty}^{\infty} d\delta\tau_3 X(\delta\tau_1, \delta\tau_2, \delta\tau_3).$$

(4.22)

In Eqs. (4.21) and (4.22) dependence on $\tau - \tau_1$ and $\tau_2 - \tau_3$, respectively, is due to correlations between the two values considered in the averaging brackets; the other dependences come from writhe integrands (see Eq. (1.30)). Due to such correlations, which have a finite range, the remaining integrals on the RHS of Eq. (4.21) and (4.22) are finite in the limit $L_b \to \infty$; both averages are expected to decay exponentially with $\tau - \tau_1$ and $\tau_2 - \tau_3$. Thus, we argue that $\langle Wr^2 \rangle_{0,c}$ and $\langle \bar{\varepsilon}_{\text{int}}^1(R_0, \delta R(\tau), \eta_0 + \delta\eta(\tau), \pi Lk/L, n) Wr \rangle_{0,c}$ scale with $L_b$ as $\langle Wr^2 \rangle_{0,c} \sim L_b$, $\langle \bar{\varepsilon}_{\text{int}}^1(R_0, \delta R(\tau), \eta_0 + \delta\eta(\tau), \pi Lk/L, n) Wr \rangle_{0,c} \sim 1$, whereas $\langle \bar{\varepsilon}_{\text{int}}^1(R_0, \delta R(\tau), \eta_0 + \delta\eta(\tau), \pi Lk/L, n) \rangle_0 \langle Wr \rangle_0 \sim L_b$ and $\langle Wr \rangle_0^2 \sim L_b^2$. As $\lambda_R$ and $\lambda_\eta$, the correlation lengths, are the scales that determine how quickly $Y(\tau_2 - \tau_1, \tau_1 - \tau)$ and $X(\tau_1 - \tau_2, \tau_2 - \tau_3, \tau_3 - \tau_4)$ go to zero as $\tau_1 - \tau \to \infty$ and $\tau_2 - \tau_3 \to \infty$, respectively, we can reasonably expect that the criteria for neglecting $\langle Wr^2 \rangle_{0,c}$ should indeed be $L_b \gg \lambda_R, \lambda_\eta$. Similar arguments can be used for other powers of $Wr$, in the expansion, to justify the use of Eq. (4.17). Indeed, an expansion of $\langle \tilde{E}_{\text{int}} \rangle_0$ in powers of $1/L_b$ could be performed by considering the correlated averages $\langle \bar{\varepsilon}_{\text{int}}^n(R_0, \delta R(\tau), \delta\eta(\tau) + \eta_0, n) Wr^n \rangle_{0,c}$, though other finite $L_b$ corrections for the braided section would need also to be considered.

Furthermore, we can assume the $\eta$ fluctuations to be small, such that $d_\eta^2 \ll 1$. Thus, we can approximate Eq. (4.13) as

$$\frac{\langle \tilde{E}_{\text{int}} \rangle_0}{k_B T} \approx \frac{1}{(2\pi)^{1/2} d_R} \sum_{n=-\infty}^{\infty} \int_{-\infty}^{\infty} dr \int_0^{L_b} d\tau \left( \bar{\varepsilon}_{\text{int},0}(R_0, r, \eta_0, \langle g_{av} \rangle_0, n) + \frac{1}{2(2l_p \alpha_\eta)^{1/2}} \bar{\varepsilon}_{\text{int},\eta\eta}(R_0, r, \eta_0, \langle g_{av} \rangle_0, n) \right)$$
$$\exp(-in\Delta\Phi_0(\tau)) \exp\left(-\frac{n^2 d_\Phi^2}{2}\right) \exp\left(-\frac{r^2}{2d_r^2}\right),$$

(4.23)

where

$$\bar{\varepsilon}_{\text{int},\eta\eta}(R_0, r, \eta_0, \langle g_{av} \rangle_0, n) = \left. \frac{d^2 \bar{\varepsilon}_{\text{int}}(R_0, r, \eta + \eta_0, \langle g_{av} \rangle_0, n)}{d\eta^2} \right|_{\eta=0}.$$
(4.24)

Now, to evaluate $\ln Z_T^T$, we consider the derivatives [19]

$$-\frac{\partial \ln Z_T^T}{\partial \alpha_\eta} = \frac{L_b}{2} \langle \delta\eta(\tau)^2 \rangle_0 = \frac{L_b}{2^{3/2} \alpha_\eta^{1/2} l_p^{1/2}}, \quad -\left(\frac{\partial \ln Z_T^T}{\partial \alpha_R}\right)_{\beta_R} = \frac{L_b}{2} \langle \delta R(\tau)^2 \rangle_0 = \frac{d_R^2 L_b}{2} \quad (4.25)$$

$$-\left(\frac{\partial \ln Z_T^T}{\partial \beta_R}\right)_{\alpha_R} = \frac{L_b}{2}\left\langle\left(\frac{d\delta R(\tau)}{d\tau}\right)^2\right\rangle_0 = \frac{\theta_R^2 L_b}{2}, \quad -\frac{\partial \ln Z_T^T}{\partial \alpha_\Phi} = \frac{L_b}{2}\left\langle\delta\Phi(\tau)^2\right\rangle_0 = \frac{L_b}{2^{3/2}\alpha_\Phi^{1/2} l_c^{1/2}}.$$

Solving these equations for $\ln Z_T^T$ and using Eqs. (4.12) and (4.13) allows us to write (see Ref. [19])

$$-\ln Z_T^T = \frac{\alpha_\eta^{1/2} L_b}{2^{1/2} l_p^{1/2}} + \frac{L_b}{2^{3/4}} \frac{\alpha_R^{1/4}}{l_p^{1/4}}\sqrt{\gamma+2} + \frac{\alpha_\Phi^{1/2} L_b}{2^{1/2} l_c^{1/2}} = \frac{\alpha_\eta^{1/2} L_b}{2^{1/2} l_p^{1/2}} + \frac{L_b}{2 l_p \theta_R^2} + \frac{L_b}{2\lambda_h}. \qquad (4.26)$$

Then, let us consider the averages of the loop elastic energies

$$\left\langle\frac{\tilde{E}_{loop1,B}}{k_B T}\right\rangle_0 = \left\langle\frac{\tilde{E}_{loop2,B}}{k_B T}\right\rangle_0 = \frac{l_p}{L_{loop}}\left(2\pi - \frac{\pi^2 R_0}{2L_{loop}}\right)^2 + \frac{l_p \pi^4}{4L_{loop}^3 d_R \sqrt{2\pi}} \int_{d_{min}}^{d_{max}} r^2 \exp\left(-\frac{r^2}{2d_R^2}\right) dr$$

$$+ \frac{l_p \pi^4 d_{max}^2}{4L_{loop}^3 d_R \sqrt{2\pi}} \int_{d_{max}}^{\infty} \exp\left(-\frac{r^2}{2d_R^2}\right) dr + \frac{l_p \pi^4 d_{min}^2}{4L_{loop}^3 d_R \sqrt{2\pi}} \int_{-\infty}^{d_{min}} \exp\left(-\frac{r^2}{2d_R^2}\right) dr$$

$$= \frac{l_p}{L_{loop}}\left(2\pi - \frac{\pi^2 R_0}{2L_{loop}}\right)^2 + \frac{l_p \pi^4}{4L_{loop}^3}\left(\frac{d_{min} d_R}{\sqrt{2\pi}}\exp\left(-\frac{d_{min}^2}{2d_R^2}\right) - \frac{d_{max} d_R}{\sqrt{2\pi}}\exp\left(-\frac{d_{max}^2}{2d_R^2}\right)\right.$$

$$+ \frac{d_R^2 - d_{min}^2}{2}\text{erf}\left(-\frac{d_{min}}{d_R \sqrt{2}}\right) + \frac{d_R^2 - d_{max}^2}{2}\text{erf}\left(\frac{d_{max}}{d_R \sqrt{2}}\right) + \frac{d_{min}^2 + d_{max}^2}{2}. \qquad (4.27)$$

Lastly, the average twisting energies are given by the expressions

$$\frac{\left\langle E_{Tw,braid,1}\right\rangle}{k_B T} = \frac{l_c}{4}\int_0^{L_b} d\tau\left[\left(\frac{d\Delta\Phi_0(\tau)}{d\tau} - \Delta g^0(\tau)\right)^2 + \left\langle\left(\frac{d\delta\Phi(\tau)}{d\tau}\right)^2\right\rangle_{\delta\Phi}\right], \qquad (4.28)$$

and

$$l_{tw} L \left\langle\left(g_{av} - \bar{g}_0\right)^2\right\rangle_0 \approx l_{tw} L \left(\left\langle g_{av}\right\rangle_0 - \bar{g}_0\right)^2, \qquad (4.29)$$

where in Eq. (4.29) we have made the same large $L_b$ approximation, as in Eq. (4.15), with the same arguments applying. Adding up Eqs.(4.6), (4.26), (4.27), (4.28) and (4.29), then yields for the free energy the expression

$$\frac{F_T^T}{k_B T} = f_{braid}(L - L_{loop}) + \frac{2l_p}{L_{loop}}\left(2\pi - \frac{\pi^2 R_0}{2L_{loop}}\right)^2 + \frac{2l_p \pi^4}{4L_{loop}^3}\left(\frac{d_{min} d_R}{\sqrt{2\pi}}\exp\left(-\frac{d_{min}^2}{2d_R^2}\right) - \frac{d_{max} d_R}{\sqrt{2\pi}}\exp\left(-\frac{d_{max}^2}{2d_R^2}\right)\right.$$

$$+ \frac{d_R^2 - d_{min}^2}{2}\text{erf}\left(-\frac{d_{min}}{d_R \sqrt{2}}\right) + \frac{d_R^2 - d_{max}^2}{2}\text{erf}\left(\frac{d_{max}}{d_R \sqrt{2}}\right) + \frac{d_{min}^2 + d_{max}^2}{2}\right) + l_{tw} L\left(\left\langle g_{av}\right\rangle_0 - \bar{g}_0\right)^2,$$

$$(4.30)$$

where

$$f_{braid} = \frac{\alpha_\eta^{1/2}}{2^{3/2} l_p^{1/2}} + \frac{1}{4 l_p \theta_R^2} + \frac{1}{4\lambda_h} + \frac{l_p \theta_R^4}{4 d_R^2} + \frac{d_R^2 \alpha_H}{2}$$
$$- \frac{\theta_R^2 l_p}{R_0^2} \tilde{f}_1(R_0, d_R, d_{max}, d_{min}) \sin^2\left(\frac{\eta_0}{2}\right) + \frac{l_c}{4 L_b} \int_0^{L_b} d\tau \left(\frac{d\Delta\Phi_0(\tau)}{d\tau} - \Delta g^0(\tau)\right)^2$$
$$+ \frac{4 l_p \tilde{f}_1(R_0, d_R, d_{max}, d_{min})}{R_0^2} \sin^4\left(\frac{\eta_0}{2}\right) + \frac{1}{k_B T L_b} \langle \tilde{E}_{int} \rangle_0$$
$$+ \frac{1}{2^{3/2} \alpha_\eta^{1/2} l_p^{1/2}} \frac{4 l_p \tilde{f}_1(R_0, d_R, d_{max}, d_{min})}{R_0^2} \left(3 \cos^2\left(\frac{\eta_0}{2}\right) \sin^2\left(\frac{\eta_0}{2}\right) - \sin^4\left(\frac{\eta_0}{2}\right)\right)$$

(4.31)

Here, we anticipate that $f_{braid}$ will not depend on $L_b$, after the final averaging over all realizations of base pair sequence. Now, at this stage, it is useful to perform a mathematical trick and introduce a Lagrange multiplier $2\pi \Upsilon_R (L \langle g_{av} \rangle_0 / \pi + \langle Wr \rangle_0)$ so that we can treat $\langle g_{av} \rangle_0$ as an independent variable for minimization, whilst enforcing the constraint on the averages, obtained from the Fuller white formula, that $Lk = \langle Tw \rangle_0 + \langle Wr \rangle_0$. The parameter $\Upsilon_R$ has some physical meaning; if we were to break the closed loop, $\Upsilon_R$ would be the twisting torque, which would have to apply to the ends, to maintain the same degree of supercoiling as the linking number in the closed loop considered. Thus, we can then write a new Gibbs like free energy $G_T = F_T^T - 2\pi \Upsilon_R (L \langle g_{av} \rangle_0 / \pi + \langle Wr \rangle_0)$ where now

$$\frac{G_T}{k_B T} = g_{braid}(L - L_{loop}) + \frac{2 l_p}{L_{loop}} \left(2\pi - \frac{\pi^2 R_0}{2 L_{loop}}\right)^2 + \frac{2 l_p \pi^4}{4 L_{loop}^3}\left(\frac{d_{min} d_R}{\sqrt{2\pi}} \exp\left(-\frac{d_{min}^2}{2 d_R^2}\right) - \frac{d_{max} d_R}{\sqrt{2\pi}} \exp\left(-\frac{d_{max}^2}{2 d_R^2}\right)\right)$$
$$+ \frac{d_R^2 - d_{min}^2}{2} \mathrm{erf}\left(-\frac{d_{min}}{d_R \sqrt{2}}\right) + \frac{d_R^2 - d_{max}^2}{2} \mathrm{erf}\left(\frac{d_{max}}{d_R \sqrt{2}}\right) + \frac{d_{min}^2 + d_{max}^2}{2}\right) - 2 L \Upsilon_R \langle g_{av} \rangle_0 + l_{tw} L \left(\langle g_{av} \rangle_0 - \overline{g}_0\right)^2,$$

(4.32)

and

$$g_{braid} = f_{braid} - \frac{2\pi \Upsilon_R \langle Wr \rangle_0}{L_b}.$$ (4.33)

We next ensemble average over realizations of $\Delta g^0(\tau)$ of the free energy, namely $\langle G_T \rangle_{\Delta g}$. To manage this we utilize the trial function [17,19]

$$\Delta \Phi_0(\tau) \approx \Delta \overline{\Phi} + \frac{1}{2} \int_{-\infty}^{\infty} \frac{(\tau - \tau')}{|\tau - \tau'|} \Delta g^0(\tau') \exp\left(-\frac{|\tau - \tau'|}{\tilde{\lambda}_h}\right) ds'.$$ (4.34)

Then, we can use Eq. (4.34) to compute the averages

$$\int_0^{L_b} d\tau \left\langle \left( \frac{d\Delta\Phi_0(\tau)}{d\tau} - \Delta g^0(\tau) \right)^2 \right\rangle_{\Delta g} = \int_0^{L_b} d\tau \frac{\int D\Delta g^0(\tau) \left( \frac{d\Delta\Phi_0(\tau)}{d\tau} - \Delta g^0(\tau) \right)^2 \exp\left( -\frac{\lambda_c^0}{4} \int_{-\infty}^{\infty} \Delta g^0(\tau'')^2 d\tau''' \right)}{\int D\Delta g^0(\tau) \exp\left( -\frac{\lambda_c^0}{4} \int_{-\infty}^{\infty} \Delta g^0(\tau'')^2 d\tau''' \right)}$$

$$= \frac{1}{4\tilde{\lambda}_h^2} \int_0^{L_b} d\tau \int_{-\infty}^{\infty} d\tau' \int_{-\infty}^{\infty} d\tau'' \frac{(\tau-\tau')}{|\tau-\tau'|} \frac{(\tau-\tau'')}{|\tau-\tau''|} \exp\left(-\frac{|\tau-\tau'|}{\tilde{\lambda}_h}\right) \exp\left(-\frac{|\tau-\tau''|}{\tilde{\lambda}_h}\right) \left\langle \Delta g^0(\tau') \Delta g^0(\tau'') \right\rangle_{\Delta g}$$

$$= \frac{1}{2\tilde{\lambda}_h^2 \lambda_c^0} \int_0^{L_b} d\tau \int_{-\infty}^{\infty} d\tau' \exp\left(-\frac{2|\tau-\tau'|}{\tilde{\lambda}_h}\right) \approx \frac{L_b}{2\tilde{\lambda}_h \lambda_c^0},$$

(4.35)

and

$$\left\langle \exp(-in\Delta\Phi_0(\tau)) \right\rangle_{\Delta g} = \frac{\int D\Delta g^0(\tau) \exp(-in\Delta\Phi_0(\tau)) \exp\left( -\frac{\lambda_c^0}{4} \int_{-\infty}^{\infty} \Delta g^0(\tau'')^2 d\tau''' \right)}{\int D\Delta g^0(\tau) \exp\left( -\frac{\lambda_c^0}{4} \int_{-\infty}^{\infty} \Delta g^0(\tau'')^2 d\tau''' \right)}$$

(4.36)

$$= \exp(-in\Delta\bar{\Phi}) \exp\left(-\frac{n^2 \tilde{\lambda}_h}{4\lambda_c^{(0)}}\right).$$

Thus, from Eqs. (4.31), (4.35), (4.36), we can write

$$\left\langle g_{braid} \right\rangle_{\Delta g} = \frac{\alpha_\eta^{1/2}}{2^{3/2} l_p^{1/2}} + \frac{1}{4 l_p \theta_R^2} + \frac{1}{4\lambda_h} + \frac{l_p \theta_R^4}{4 d_R^2} + \frac{d_R^2 \alpha_H}{2} + \frac{l_c}{8\tilde{\lambda}_h \lambda_c^0}$$

$$- \frac{\theta_R^2 l_p}{R_0^2} \tilde{f}_1(R_0, d_R, d_{\max}, d_{\min}) \sin^2\left(\frac{\eta_0}{2}\right) + \frac{4 l_p \tilde{f}_1(R_0, d_R, d_{\max}, d_{\min})}{R_0^2} \sin^4\left(\frac{\eta_0}{2}\right) - \frac{2\pi \Upsilon_R}{L_b} \left\langle Wr \right\rangle_0$$

$$+ \sum_{n=-\infty}^{\infty} \exp(-in\Delta\bar{\Phi}) \exp\left(-n^2 \left(\frac{\lambda_h}{2 l_c} + \frac{\tilde{\lambda}_h}{4\lambda_c^{(0)}}\right)\right) \tilde{f}_{\text{int},0}\left(\eta_0, R_0, d_R, \left\langle g_{av}\right\rangle_0, n, d_{\max}, d_{\min}\right)$$

(4.37)

$$+ \frac{1}{2^{3/2} \alpha_\eta^{1/2} l_p^{1/2}} \left[ \frac{4 l_p \tilde{f}_1(R_0, d_R, n, d_{\max}, d_{\min})}{R_0^2} \left( 3\cos^2\left(\frac{\eta_0}{2}\right) \sin^2\left(\frac{\eta_0}{2}\right) - \sin^4\left(\frac{\eta_0}{2}\right) \right) \right.$$

$$\left. + \sum_{n=-\infty}^{\infty} \exp(-in\Delta\bar{\Phi}) \exp\left(-n^2 \left(\frac{\lambda_h}{2 l_c} + \frac{\tilde{\lambda}_h}{4\lambda_c^{(0)}}\right)\right) \tilde{f}_{\text{int},\eta\eta}\left(\eta_0, R_0, d_R, \left\langle g_{av}\right\rangle_0, n, d_{\max}, d_{\min}\right) \right],$$

where

$$\tilde{f}_{\text{int},Y}\left(\eta_0, R_0, d_R, \langle g_{av}\rangle_0, n, d_{\max}, d_{\min}\right) = \frac{1}{d_R\sqrt{2\pi}} \int_{-\infty}^{\infty} dr\, \overline{\varepsilon}_{\text{int},Y}\left(R_0, r, \eta_0, \langle g_{av}\rangle_0, n\right) \exp\left(-\frac{r^2}{2d_R^2}\right).$$

(4.38)

The subscript $Y$ may stand either for $0$ or $\eta\eta$, depending on which interaction energy term we are dealing with. It is convenient to change to the variables [11]

$$\frac{\lambda_h^*}{2\lambda_c} = \frac{\lambda_h}{2l_c} + \frac{\tilde{\lambda}_h}{4\lambda_c^{(0)}}, \qquad \frac{\tilde{\lambda}_h^*}{2\lambda_c} = \frac{\lambda_h}{2l_c} - \frac{\tilde{\lambda}_h}{4\lambda_c^{(0)}}, \qquad (4.39)$$

and define the combined persistence length

$$\frac{1}{\lambda_c} = \frac{1}{\lambda_c^{(0)}} + \frac{1}{l_c}. \qquad (4.40)$$

From minimization with respect to $\tilde{\lambda}_h^*$ (see Refs. [11,19]) one is able to show that $\tilde{\lambda}_h = \lambda_h$ and that

$$\tilde{\lambda}_h^* = \lambda_h^* \left(\frac{2\lambda_c^{(0)} - l_c}{2\lambda_c^{(0)} + l_c}\right). \qquad (4.41)$$

Thus, we can rewrite Eq. (4.37) as

$$\begin{aligned}
\langle g_{braid}\rangle_{\Delta g} &= \frac{\alpha_\eta^{1/2}}{2^{3/2} l_p^{1/2}} + \frac{(l_c + \lambda_c)^2}{16\lambda_h^* \lambda_c l_c} + \frac{1}{4 l_p \theta_R^2} + \frac{l_p \theta_R^4}{4 d_R^2} + \frac{d_R^2 \alpha_H}{2} - \frac{2\pi \Upsilon_R}{L_b} \langle Wr\rangle_0 \\
&\quad - \frac{\theta_R^2 l_p}{R_0^2} \tilde{f}_1(R_0, d_R, d_{\max}, d_{\min}) \sin^2\left(\frac{\eta_0}{2}\right) + \frac{4 l_p \tilde{f}_1(R_0, d_R, d_{\max}, d_{\min})}{R_0^2} \sin^4\left(\frac{\eta_0}{2}\right) \\
&\quad + \sum_{n=-\infty}^{\infty} \exp(-in\Delta\overline{\Phi}) \exp\left(-\frac{n^2 \lambda_h^*}{2\lambda_c}\right) \tilde{f}_{\text{int},0}\left(\eta_0, R_0, d_R, \langle g_{av}\rangle_0, n, d_{\max}, d_{\min}\right) \\
&\quad + \frac{1}{2^{3/2} \alpha_\eta^{1/2} l_p^{1/2}} \left[\frac{4 l_p \tilde{f}_1(R_0, d_R, d_{\max}, d_{\min})}{R_0^2}\left(3\cos^2\left(\frac{\eta_0}{2}\right)\sin^2\left(\frac{\eta_0}{2}\right) - \sin^4\left(\frac{\eta_0}{2}\right)\right)\right. \\
&\quad \left. + \sum_{n=-\infty}^{\infty} \exp(-in\Delta\overline{\Phi}) \exp\left(-\frac{n^2 \lambda_h^*}{2\lambda_c}\right) \tilde{f}_{\text{int},\eta\eta}\left(\eta_0, R_0, d_R, \langle g_{av}\rangle_0, n, d_{\max}, d_{\min}\right)\right].
\end{aligned}$$

(4.42)

We now first minimize over $\overline{L}_{loop}$ which is unconstrained, which is related to $L_{loop}$, constrained to take the minimum value $\pi R_0 / 2$. through Eq. (2.7). The minimum value is found to obey the equation

$$\frac{-2\langle g_{braid}\rangle_{\Delta g} \bar{L}_{loop}^4}{3\pi^4 l_p R_0^2 A(R_0,d_R,d_{min},d_{max})} - \frac{16\bar{L}_{loop}^2}{3\pi^2 R_0^2 A(R_0,d_R,d_{min},d_{max})} + \frac{16\bar{L}_{loop}}{3\pi R_0 A(R_0,d_R,d_{min},d_{max})} - 1 = 0,$$

(4.43)

where

$$A(R_0,d_R,d_{min},d_{max}) = 1 + \frac{1}{R_0^2}\left(\frac{d_{min}d_R}{\sqrt{2\pi}}\exp\left(-\frac{d_{min}^2}{2d_R^2}\right) - \frac{d_{max}d_R}{\sqrt{2\pi}}\exp\left(-\frac{d_{max}^2}{2d_R^2}\right)\right.$$
$$\left. + \frac{d_R^2 - d_{min}^2}{2}\mathrm{erf}\left(-\frac{d_{min}}{d_R\sqrt{2}}\right) + \frac{d_R^2 - d_{max}^2}{2}\mathrm{erf}\left(\frac{d_{max}}{d_R\sqrt{2}}\right) + \frac{d_{min}^2 + d_{max}^2}{2}\right).$$

(4.44)

By introducing the rescaling $\bar{L}_{loop} = (3\pi A(R_0,d_R,d_{min},d_{max})/16) R_0 \tilde{L}_{loop}$ we can write

$$\frac{-2\langle g_{braid}\rangle_{\Delta g} R_0^2 A(R_0,d_R,d_{min},d_{max})^3}{3 l_p}\left(\frac{3}{16}\right)^4 \tilde{L}_{loop}^4 - \frac{3A(R_0,d_R,d_{min},d_{max})}{16}\tilde{L}_{loop}^2 + \tilde{L}_{loop} - 1 = 0. \quad (4.45)$$

We can express $\tilde{L}_{loop}(R_0, -g_{braid}, d_R, d_{min}, d_{max})$ as a two variable function

$$\tilde{L}_{loop}(R_0, -\langle g_{braid}\rangle_{\Delta g}, d_R, d_{min}, d_{max}) = \tilde{L}\left(-R_0^2\langle g_{braid}\rangle_{\Delta g}/l_p, A(R_0,d_R,d_{min},d_{max})\right),$$

(4.46)

where $\tilde{L}(x,y)$ is determined by the equation

$$\frac{2}{3}\left(\frac{3}{16}\right)^4 xy^3 \tilde{L}^4 - \frac{3}{16} y\tilde{L}^2 + \tilde{L} - 1 = 0.$$

(4.47)

Eq. (4.47) can be solved using the Quartic formula, though due to its cumbersome nature it is probably easier to build an interpolation function $\tilde{L}$ from the numerical solution to Eq. (4.47). We have for the minimization of the free energy with respect to $R_0$ the following equations

$$0 = \frac{\partial\langle g_{braid}\rangle_{\Delta g}}{\partial R_0}(L - \bar{L}_{loop}) - \frac{\pi^2 l_p}{\bar{L}_{loop}^2}\left(2\pi - \frac{\pi^2 R_0}{2\bar{L}_{loop}}\right) + \frac{l_p \pi^4}{2\bar{L}_{loop}^3}\left(\frac{\partial d_{max}}{\partial R_0}B(d_{max},d_R) - \frac{\partial d_{min}}{\partial R_0}B(-d_{min},d_R)\right),$$

when $\bar{L}_{loop} > \dfrac{\pi R_0}{2}$, (4.48)

$$0 = \frac{\partial \langle g_{braid} \rangle_{\Delta g}}{\partial R_0}(L - \frac{\pi R_0}{2}) - \frac{\pi}{2} \langle g_{braid} \rangle_{\Delta g} - \frac{4l_p \pi}{R_0^2}(3A(R_0, d_R, d_{min}, d_{max}) - 2) + \frac{4l_p \pi}{R_0^3}\left(\frac{\partial d_{max}}{\partial R_0} B(d_{max}, d_R) - \frac{\partial d_{min}}{dR_0} B(-d_{min}, d_R)\right),$$

$$\text{when } \bar{L}_{loop} < \frac{\pi R_0}{2}, \quad (4.49)$$

and for $d_R$ we obtain

$$0 = \frac{\partial \langle g_{braid} \rangle_{\Delta g}}{\partial d_R}(L - L_{loop}) + \frac{l_p \pi^4}{2L_{loop}^3}\left(C(d_{max}, d_R) + C(-d_{min}, d_R)\right), \quad (4.50)$$

where

$$B(d_{max}, d_R) = \left(\left(\frac{d_{max}}{d_R}\right)^2 - 1\right)\frac{d_R}{\sqrt{2\pi}}\exp\left(-\frac{d_{max}^2}{2d_R^2}\right) + d_{max}\left(1 - \text{erf}\left(\frac{d_{max}}{\sqrt{2}d_R}\right)\right) + \frac{(d_R^2 - d_{max}^2)}{d_R\sqrt{2\pi}}\exp\left(-\frac{d_{max}^2}{2d_R^2}\right),$$

(4.51)

and

$$C(d_{max}, d_R) = -\frac{d_{max}}{\sqrt{2\pi}}\left(1 + \frac{d_{max}^2}{d_R^2}\right)\exp\left(-\frac{d_{max}^2}{2d_R^2}\right) + d_R \text{erf}\left(\frac{d_{max}}{d_R\sqrt{2}}\right) - \frac{d_{max}}{d_R^2}\frac{(d_R^2 - d_{max}^2)}{\sqrt{2\pi}}\exp\left(-\frac{d_{max}^2}{2d_R^2}\right).$$

(4.52)

The derivatives $\dfrac{\partial \langle g_{braid} \rangle_{\Delta g}}{\partial R_0}$ and $\dfrac{\partial \langle g_{braid} \rangle_{\Delta g}}{\partial d_R}$ are given by

$$\frac{\partial \langle g_{braid} \rangle_{\Delta g}}{\partial R_0} = \frac{d_R^2}{2}\frac{\partial \alpha_H}{\partial R_0} - \frac{2\pi \Upsilon_R}{L_b}\frac{\partial \langle Wr \rangle_0}{\partial R_0}$$
$$+ \frac{\theta_R^2 l_p}{R_0^3}\tilde{h}_1(R_0, d_R, d_{max}, d_{min})\sin^2\left(\frac{\eta_0}{2}\right) - \frac{4l_p \tilde{h}_1(R_0, d_R, d_{max}, d_{min})}{R_0^3}\sin^4\left(\frac{\eta_0}{2}\right)$$
$$+ \sum_{n=-\infty}^{\infty}\frac{\partial \tilde{f}_{int,0}(\eta_0, R_0, d_R, \langle g_{av} \rangle_0, n, d_{max}, d_{min})}{\partial R_0}\exp(-in\Delta\bar{\Phi})\exp\left(-\frac{n^2 \lambda_h^*}{2\lambda_c}\right) \quad (4.53)$$
$$+ \frac{1}{2^{3/2}\alpha_\eta^{1/2} l_p^{1/2}}\left[-\frac{4l_p \tilde{h}_1(R_0, d_R, d_{max}, d_{min})}{R_0^3}\left(3\cos^2\left(\frac{\eta_0}{2}\right)\sin^2\left(\frac{\eta_0}{2}\right) - \sin^4\left(\frac{\eta_0}{2}\right)\right)\right.$$
$$\left.+ \sum_{n=-\infty}^{\infty}\exp(-in\Delta\bar{\Phi})\exp\left(-\frac{n^2 \lambda_h^*}{2\lambda_c}\right)\frac{\partial \tilde{f}_{int,\eta\eta}(\eta_0, R_0, d_R, \langle g_{av} \rangle_0, n, d_{max}, d_{min})}{\partial R_0}\right],$$

and

$$\frac{\partial \langle g_{braid}\rangle_{\Delta g}}{\partial d_R} = -\frac{l_p \theta_R^2}{d_R R_0^2}\tilde{l}_1(R_0,d_R,d_{max},d_{min})\sin^2\left(\frac{\eta_0}{2}\right) + d_R \alpha_H + \frac{4l_p}{d_R R_0^2}\tilde{l}_1(R_0,d_R,d_{max},d_{min})\sin^4\left(\frac{\eta_0}{2}\right)$$

$$+ \sum_{n=-\infty}^{\infty} \frac{\partial \tilde{f}_{int,0}(\eta_0,R_0,d_R,\langle g_{av}\rangle_0,n,d_{max},d_{min})}{\partial d_R}\exp(-in\Delta\bar{\Phi})\exp\left(-\frac{n^2 \lambda_h^*}{2\lambda_c}\right) - \frac{2\pi \Upsilon_R}{L_b}\frac{\partial \langle Wr\rangle_0}{\partial d_R} - \frac{l_p \theta_R^4}{2d_R^3}$$

$$+ \frac{1}{2^{3/2}\alpha_\eta^{1/2}l_p^{1/2}}\left[\frac{4l_p \tilde{l}_1(R_0,d_R,d_{max},d_{min})}{d_R R_0^2}\left(3\cos^2\left(\frac{\eta_0}{2}\right)\sin^2\left(\frac{\eta_0}{2}\right) - \sin^4\left(\frac{\eta_0}{2}\right)\right)\right.$$

$$\left. + \sum_{n=-\infty}^{\infty}\exp(-in\Delta\bar{\Phi})\exp\left(-\frac{n^2\lambda_h^*}{2\lambda_c}\right)\frac{\partial \tilde{f}_{int,\eta\eta}(\eta_0,R_0,d_R,\langle g_{av}\rangle_0,n,d_{max},d_{min})}{\partial d_R}\right].$$

(4.54)

In general $\tilde{h}_1$ is defined through the relation

$$\tilde{h}_1(R_0,d_R,d_{max},d_{min}) = 2\tilde{f}_1(R_0,d_R,d_{max},d_{min}) - R_0 \frac{\partial \tilde{f}_1(R_0,d_R,d_{max},d_{min})}{\partial R_0}. \qquad (4.55)$$

When $d_{max} = -d_{min} = (R_0 - 2a)$, it takes the following form

$$\tilde{h}_1(R_0,d_R,R_0-2a,2a-R_0) = -\frac{R_0^3}{\sqrt{2\pi}d_R}\int_{-1}^{1}dx\frac{1}{(x(R_0-2a)+R_0)^2}\exp\left(-\frac{(R_0-2a)^2 x^2}{2d_R^2}\right)$$

$$+ \frac{2(R_0-2a)R_0^3}{\sqrt{2\pi}d_R}\int_{-1}^{1}dx\frac{(x+1)}{(x(R_0-2a)+R_0)^3}\exp\left(-\frac{(R_0-2a)^2 x^2}{2d_R^2}\right)$$

$$+ \frac{(R_0-2a)^2 R_0^3}{\sqrt{2\pi}d_R^3}\int_{-1}^{1}dx\frac{x^2}{(x(R_0-2a)+R_0)^2}\exp\left(-\frac{(R_0-2a)^2 x^2}{2d_R^2}\right)$$

$$+ \frac{2R_0^3}{(2R_0-2a)^3}\left(1-\mathrm{erf}\left(\frac{R_0-2a}{d_R\sqrt{2}}\right)\right) + \frac{R_0^3}{\sqrt{2\pi}d_R}\left[\frac{1}{(2a)^2}+\frac{1}{(2R_0-2a)^2}\right]\exp\left(-\frac{(R_0-2a)^2}{2d_R^2}\right),$$

(4.56)

See Ref. [19] for another example where we choose $d_{max} = \infty$. The functions $\tilde{l}_1$ is given by

$$\tilde{l}_1(R_0, d_R, d_{max}, d_{min}) = \frac{R_0^2}{\sqrt{2\pi}} \int_{\frac{d_{min}}{d_R}}^{\frac{d_{max}}{d_R}} dx \frac{(x^2-1)}{(R_0 + d_R x)^2} \exp\left(-\frac{x^2}{2}\right)$$

$$+ \frac{R_0^2}{\sqrt{2\pi}} \left( \frac{d_{max}}{d_R} \frac{\exp\left(-\frac{1}{2}\left(\frac{d_{max}}{d_R}\right)^2\right)}{(R_0 + d_{max})^2} - \frac{d_{min}}{d_R} \frac{\exp\left(-\frac{1}{2}\left(\frac{d_{min}}{d_R}\right)^2\right)}{(R_0 + d_{min})^2} \right), \quad (4.57)$$

With minimization over $\theta_R, \eta_0, \alpha_\eta, \Delta\bar{\Phi},$ and $\lambda_h^*$ we simply need to satisfy the conditions

$$0 = \frac{\partial \langle g_{braid} \rangle_{\Delta g}}{\partial \theta_R}, \quad 0 = \frac{\partial \langle g_{braid} \rangle_{\Delta g}}{\partial \eta_0}, \quad 0 = \frac{\partial \langle g_{braid} \rangle_{\Delta g}}{\partial \alpha_\eta}, \quad 0 = \frac{\partial \langle g_{braid} \rangle_{\Delta g}}{\partial \Delta\Phi}, \quad 0 = \frac{\partial \langle g_{braid} \rangle_{\Delta g}}{\partial \lambda_h^*}.$$

(4.58)

Applying these conditions yield the Equations

$$0 = \frac{l_p \theta_R^3}{d_R^2} - \frac{1}{2l_p \theta_R^3} - \frac{2\pi \Upsilon_R}{L_b} \frac{\partial \langle Wr \rangle_0}{\partial \theta_R} - \frac{2\theta_R l_p}{R_0^2} \tilde{f}_1(R_0, d_R, d_{max}, d_{min}) \sin^2\left(\frac{\eta_0}{2}\right), \quad (4.59)$$

$$0 = \frac{d_R^2}{2} \frac{\partial \alpha_H}{\partial \eta_0} - \frac{2\pi \Upsilon_R}{L_b} \frac{\partial \langle Wr \rangle_0}{\partial \eta_0}$$

$$- \frac{\theta_R^2 l_p}{R_0^2} \tilde{f}_1(R_0, d_R, d_{max}, d_{min}) \sin\left(\frac{\eta_0}{2}\right) \cos\left(\frac{\eta_0}{2}\right) + \frac{8 l_p L \tilde{f}_1(R_0, d_R, d_{max}, d_{min})}{R_0^2} \cos\left(\frac{\eta_0}{2}\right) \sin^3\left(\frac{\eta_0}{2}\right)$$

$$+ \sum_{n=-\infty}^{\infty} \frac{\partial \tilde{f}_{int.0}(\eta_0, R_0, d_R, \langle g_{av} \rangle_0, n, d_{max}, d_{min})}{\partial \eta_0} \exp\left(-\frac{n^2 \lambda_h^*}{2\lambda_c}\right) \exp(-in\Delta\bar{\Phi}) \quad (4.60)$$

$$+ \frac{1}{2^{3/2} \alpha_\eta^{1/2} l_p^{1/2}} \left[ \frac{4 l_p \tilde{f}_1(R_0, d_R, d_{max}, d_{min})}{R_0^2} \left( 3\cos^3\left(\frac{\eta_0}{2}\right)\sin\left(\frac{\eta_0}{2}\right) - 5\sin^3\left(\frac{\eta_0}{2}\right)\cos\left(\frac{\eta_0}{2}\right) \right) \right.$$

$$+ \left. \sum_{n=-\infty}^{\infty} \frac{\partial \tilde{f}_{int.\eta\eta}(\eta_0, R_0, d_R, \langle g_{av} \rangle_0, n, d_{max}, d_{min})}{\partial \eta_0} \exp\left(-\frac{n^2 \lambda_h^*}{2\lambda_c}\right) \exp(-in\Delta\bar{\Phi}) \right],$$

$$0 = \frac{1}{2^{5/2} l_p^{1/2} \alpha_\eta^{1/2}} - \frac{2\pi \Upsilon_R}{L_b} \frac{\partial \langle Wr \rangle_0}{\partial \alpha_\eta}$$

$$- \frac{1}{2^{5/2} \alpha_\eta^{3/2} l_p^{1/2}} \left[ \frac{4 l_p \tilde{f}_1(R_0, d_R, d_{max}, d_{min})}{R_0^2} \left( 3\cos^2\left(\frac{\eta_0}{2}\right) \sin^2\left(\frac{\eta_0}{2}\right) - \sin^4\left(\frac{\eta_0}{2}\right) \right) \right. \quad (4.61)$$

$$+ \left. \sum_{n=-\infty}^{\infty} \exp(-in\Delta\bar{\Phi}) \exp\left(-\frac{n^2 \lambda_h^*}{2\lambda_c}\right) \tilde{f}_{int,\eta\eta}(\eta_0, R_0, d_R, \langle g_{av} \rangle_0, n, d_{max}, d_{min}) \right],$$

$$0 = \sum_{n=-\infty}^{\infty} in \exp(-in\Delta\bar{\Phi}) \exp\left(-\frac{n^2 \lambda_h^*}{2\lambda_c}\right) \tilde{f}_{int,0}(\eta_0, R_0, d_R, \langle g_{av} \rangle_0, n, d_{max}, d_{min})$$
$$+ \frac{1}{2^{3/2} \alpha_\eta^{1/2} l_p^{1/2}} \sum_{n=-\infty}^{\infty} in \exp(-in\Delta\bar{\Phi}) \exp\left(-\frac{n^2 \lambda_h^*}{2\lambda_c}\right) \tilde{f}_{int,\eta\eta}(\eta_0, R_0, d_R, \langle g_{av} \rangle_0, n, d_{max}, d_{min}),$$
(4.62)

$$0 = -\frac{(l_c + \lambda_c)^2}{16(\lambda_h^*)^2 \lambda_c l_c}$$
$$- \frac{1}{2\lambda_c} \sum_{n=-\infty}^{\infty} n^2 \exp(-in\Delta\bar{\Phi}) \exp\left(-\frac{n^2 \lambda_h^*}{2\lambda_c}\right) \tilde{f}_{int,0}(\eta_0, R_0, d_R, \langle g_{av} \rangle_0, n, d_{max}, d_{min})$$
$$- \frac{1}{2^{5/2} \alpha_\eta^{1/2} l_p^{1/2} \lambda_c} \sum_{n=-\infty}^{\infty} n^2 \exp(-in\Delta\bar{\Phi}) \exp\left(-\frac{n^2 \lambda_h^*}{2\lambda_c}\right) \tilde{f}_{int,\eta\eta}(\eta_0, R_0, d_R, \langle g_{av} \rangle_0, n, d_{max}, d_{min}).$$
(4.63)

Last of all, minimization over $\langle g_{av} \rangle_0$ yields

$$0 = \frac{\partial \langle g_{braid} \rangle_{\Delta g}}{\partial \langle g_{av} \rangle_0} (L - L_{loop}) - 2L\Upsilon_R + 2l_{tw} L (\langle g_{av} \rangle_0 - \bar{g}_0),$$
(4.64)

where

$$\frac{\partial \langle g_{braid} \rangle_{\Delta g}}{\partial \langle g_{av} \rangle_0} = \sum_{n=-\infty}^{\infty} \exp(-in\Delta\bar{\Phi}) \exp\left(-\frac{n^2 \lambda_h^*}{2\lambda_c}\right) \frac{\partial \tilde{f}_{int,0}(\eta_0, R_0, d_R, \langle g_{av} \rangle_0, n, d_{max}, d_{min})}{\partial \langle g_{av} \rangle_0}$$
$$+ \frac{1}{2^{3/2} \alpha_\eta^{1/2} l_p^{1/2}} \sum_{n=-\infty}^{\infty} \exp(-in\Delta\bar{\Phi}) \exp\left(-\frac{n^2 \lambda_h^*}{2\lambda_c}\right) \frac{\partial \tilde{f}_{int,\eta\eta}(\eta_0, R_0, d_R, \langle g_{av} \rangle_0, n, d_{max}, d_{min})}{\partial \langle g_{av} \rangle_0}.$$
(4.65)

For a given torque $\Upsilon_R$, the linking number can then be computed through calculating the average writhe $\langle Wr \rangle_0$ and applying the Fuller-White formula. What is left to evaluate expressions for $\langle Wr \rangle_0$ and its derivatives, which we will do in Section 6. Then Eq. (4.42), (4.47)-(4.65) will specify a full set of equations determining the braid geometry, fluctuation parameters, linking number and optimised braid free energy. In the next section we'll consider an approximation when the helix specific forces are weak.

## 5. Variational Approximation for weak helix specific forces

Here the interactions that depend on $\Delta\Phi$ are handled in a different way. Throughout this section, we will assume that $-d_{min} = d_{max} = R_0 - 2a$. We start by wring

$$\tilde{E}_T[R(\tau), \eta(\tau), \Delta\Phi(\tau)] = \tilde{E}_T^0[R(\tau), \eta(\tau), \Delta\Phi(\tau)] + \delta\tilde{E}_{int}[R(\tau), \eta(\tau), \Delta\Phi(\tau)],$$
(5.1)

$$\frac{\delta\tilde{E}_{int}[R(\tau), \eta(\tau), \Delta\Phi(\tau)]}{k_B T} = \sum_{n=-\infty}^{\infty} \int_0^{L_p} d\tau \bar{\varepsilon}_{int,0}(R_0, \delta R(\tau), \eta(\tau), g_{av}, n)(1 - \delta_{n,0}) \exp(in\Delta\Phi(\tau)),$$
(5.2)

$$\tilde{E}_T^0[R(\tau),\eta(\tau),\Delta\Phi(\tau)] = G_T^0[R(\tau),\eta(\tau)] + E_{Tw,braid,1}[\Delta\Phi(\tau)], \tag{5.3}$$

where

$$G_T^0[R(\tau),\eta(\tau)] = \tilde{E}_B[R(\tau),\eta(\tau)] + l_c L\left(g_{av} - \bar{g}_0\right)^2 + \tilde{E}_{int}^0[R(\tau),\eta(\tau)] + \tilde{E}_{St}[R(\tau),\eta(\tau)], \tag{5.4}$$

and

$$\tilde{E}_{int}^0[R(\tau),\eta(\tau)] = \int_0^{L_b} d\tau\, \bar{\varepsilon}_{int}(R_0, \delta R(\tau), \eta(\tau), 0, 0). \tag{5.5}$$

The term described by Eq. (5.5) is the helix independent, uniform cylinder mode that does not depend on $g_{av}$, so we set it to zero in that expression.

We then expand the partition function in powers of $\delta \tilde{E}_{int}[R(\tau),\eta(\tau),\Delta\Phi(\tau)]$ thus obtaining the series

$$Z_T = Z_0 + Z_1 + Z_2 + ..., \tag{5.6}$$

where

$$Z_0 = \int DR(\tau) \int D\eta(\tau) \int D\Delta\Phi(\tau) \exp\left(-\frac{\tilde{E}_T^0[R(\tau),\eta(\tau),\Delta\Phi(\tau)]}{k_B T}\right), \tag{5.7}$$

$$Z_1 = -\int DR(\tau) \int D\eta(\tau) \int D\Delta\Phi(\tau) \frac{\delta \tilde{E}_{int}[R(\tau),\eta(\tau),\Delta\Phi(\tau)]}{k_B T} \exp\left(-\frac{\tilde{E}_T^0[R(\tau),\eta(\tau),\Delta\Phi(\tau)]}{k_B T}\right),$$

$$\tag{5.8}$$

$$Z_2 = \frac{1}{2}\int DR(\tau) \int D\eta(\tau) \int D\Delta\Phi(\tau) \left(\frac{\delta \tilde{E}_{int}[R(\tau),\eta(\tau),\Delta\Phi(\tau)]}{k_B T}\right)^2 \exp\left(-\frac{\tilde{E}_T^0[R(\tau),\eta(\tau),\Delta\Phi(\tau)]}{k_B T}\right).$$

$$\tag{5.9}$$

Thus, we expand out the free energy in terms of $Z_0, Z_1$ and $Z_2$ yielding

$$F_T = -k_B T \ln Z_T \approx -k_B T \left( \ln(Z_0) + \frac{Z_1 + Z_2}{Z_0} - \frac{1}{2}\left(\frac{Z_1}{Z_0}\right)^2 \right)$$

$$\approx -k_B T \left( \ln(Z_0) - \left\langle \frac{\delta \tilde{E}_{int}[R(\tau), \eta(\tau), \Delta\Phi(\tau)]}{k_B T} \right\rangle + \frac{1}{2}\left\langle \left(\frac{\delta \tilde{E}_{int}[R(\tau), \eta(\tau), \Delta\Phi(\tau)]}{k_B T}\right)^2 \right\rangle \right. \quad (5.10)$$

$$\left. - \frac{1}{2}\left\langle \frac{\delta \tilde{E}_{int}[R(\tau), \eta(\tau), \Delta\Phi(\tau)]}{k_B T} \right\rangle^2 \right).$$

Let us first consider the averages of the form $\left\langle \delta \tilde{E}_{int}[R(\tau), \eta(\tau), \Delta\Phi(\tau)]/k_B T \right\rangle$ for these we need to consider first the thermal average

$$\left\langle \exp(in\Delta\Phi(\tau)) \right\rangle_{\Delta\Phi} = \frac{1}{Z_{\Delta\Phi}} \int D\Delta\Phi(\tau) \exp(in\Delta\Phi(\tau)) \exp\left(-\frac{E_{Tw,braid,1}[\Delta\Phi(\tau)]}{k_B T}\right), \quad (5.11)$$

where the partition function $Z_{\Delta\Phi}$ is given by

$$Z_{\Delta\Phi} = \int D\Delta\Phi(\tau) \exp\left(-\frac{E_{Tw,braid,1}[\Delta\Phi(\tau)]}{k_B T}\right). \quad (5.12)$$

We first separate $\Delta\Phi(\tau) = \Delta\Phi_0(\tau) + \delta\Phi(\tau)$ where $\left\langle \delta\Phi(\tau) \right\rangle_{\Delta g} = 0$. This allows us to write

$$\frac{E_{Tw,braid,1}[\Delta\Phi(\tau)]}{k_B T} = \frac{l_c}{4} \int_0^{L_b} d\tau \left[ \left(\frac{d\Delta\Phi_0(\tau)}{d\tau} - \Delta g^0(\tau)\right)^2 + \left(\frac{d\delta\Phi(\tau)}{d\tau}\right)^2 \right]. \quad (5.13)$$

Therefore, in principle we can use the averaging formulas developed in Refs. [17] and [19] to evaluate the average described by Eq. (5.11). This yields the following expression

$$\left\langle \exp(in\Delta\Phi(\tau)) \right\rangle_{\Delta\Phi} = \exp(in\Delta\Phi_0(\tau)) \int_{-\infty}^{\infty} d\Phi \exp(in\Phi) \exp\left(-\frac{\Phi^2}{2d_\Phi^2}\right). \quad (5.14)$$

However, since in this case (due to the form of Eq. (5.13)) we have that

$$d_\Phi^2 = \frac{1}{\pi l_c} \int_{-\infty}^{\infty} dk \frac{1}{k^2} = \infty, \quad (5.15)$$

It is required that (from Eqs. (5.14) and (5.15) combined)

$$\left\langle \exp(in\Delta\Phi(\tau)) \right\rangle_{\Delta\Phi} = 0, \qquad \text{for } n \neq 0, \quad (5.16)$$

and thus we obtain

$$\left\langle \frac{\delta \tilde{E}_{\text{int}}[R(\tau), \eta(\tau), \Delta\Phi(\tau)]}{k_B T} \right\rangle = 0. \tag{5.17}$$

Now let us consider the other type of term in Eq. (5.10), namely

$$\left( \frac{\delta \tilde{E}_{\text{int}}[R(\tau), \eta(\tau), \Delta\Phi(\tau)]}{k_B T} \right)^2 = \sum_{m=\infty}^{\infty} \sum_{n=\infty}^{\infty} \int_0^{L_b} d\tau \int_0^{L_b} d\tau' \exp(in\Delta\Phi(\tau) + im\Delta\Phi(\tau'))$$
$$(1-\delta_{n,0})(1-\delta_{m,0}) \bar{\varepsilon}_{\text{int}}(R_0, \delta R(\tau), \eta(\tau), g_{av}, n) \bar{\varepsilon}_{\text{int}}(R_0, \delta R(\tau'), \eta(\tau'), g_{av}, m). \tag{5.18}$$

We first look at the average $\langle \exp(in\Delta\Phi(\tau) + im\Delta\Phi(\tau')) \rangle$ which can be averaged using the averaging formulas developed in Appendix A. This yields the result

$$\langle \exp(in\Delta\Phi(\tau) + im\Delta\Phi(\tau')) \rangle = \exp(in\Delta\Phi_0(\tau) + im\Delta\Phi_0(\tau')) \int_{-\infty}^{\infty} d\Phi \int_{-\infty}^{\infty} d\Phi' \exp(in\Phi + im\Phi') \Gamma_{\delta\Phi}(\Phi, \Phi'; \tau - \tau'),$$
$$\tag{5.19}$$

where

$$\Gamma_{\delta\Phi}(\Phi, \Phi'; \tau - \tau') = \frac{1}{(2\pi)} \lim_{\varepsilon \to 0} \frac{1}{\sqrt{d_\Phi^4 - G_\Phi(\tau-\tau')^2}} \exp\left( -\frac{(\Phi^2 + \Phi'^2)d_\Phi^2}{2(d_\Phi^4 - G_\Phi(\tau-\tau')^2)} \right) \exp\left( -\frac{\Phi\Phi' G_X(\tau-\tau')}{(d_\Phi^4 - G_\Phi(\tau-\tau')^2)} \right), \tag{5.20}$$

and

$$d_\Phi^2 = \frac{1}{\pi l_c} \int_{-\infty}^{\infty} dk \frac{1}{k^2 + \varepsilon^2}, \qquad G_\Phi(\tau - \tau') = \frac{1}{\pi l_c} \int_{-\infty}^{\infty} dk \frac{\exp(-ik(\tau - \tau'))}{k^2 + \varepsilon^2}. \tag{5.21}$$

Integration over $\Phi$ and $\Phi'$ yields

$$\langle \exp(in\Delta\Phi(\tau) + im\Delta\Phi(\tau')) \rangle = \exp(in\Delta\Phi_0(\tau) + im\Delta\Phi_0(\tau'))$$
$$\lim_{\varepsilon \to 0} \left[ \exp\left( -\frac{d_\Phi^2}{2}(m^2 + n^2) \right) \exp(G_\Phi(\tau - \tau')mn) \right]. \tag{5.22}$$

Using the fact that we can write

$$\frac{1}{\pi l_c} \int_{-\infty}^{\infty} dk \frac{1 - \exp(-ik(\tau - \tau'))}{k^2 + \varepsilon^2} = \frac{1}{l_c}\left(\frac{1}{\varepsilon}\right)\left(1 - \exp(-\varepsilon|\tau - \tau'|)\right) \approx \frac{|\tau - \tau'|}{l_c}, \tag{5.23}$$

we find that

$$\lim_{\varepsilon \to 0} \left[ \exp\left( -\frac{d_\Phi^2}{2}(m^2 + n^2) \right) \exp(G_\Phi(\tau - \tau')mn) \right] = \delta_{n,-m} \exp\left( -\frac{n^2|\tau - \tau'|}{l_c} \right). \tag{5.24}$$

Thus, we obtain:

$$\left\langle \left( \frac{\delta \tilde{E}_{int}[R(\tau),\eta(\tau),\Delta\Phi(\tau)]}{k_B T} \right)^2 \right\rangle_{\delta\Phi} = \sum_{n=-\infty}^{\infty} \int_0^{L_b} d\tau \int_0^{L_b} d\tau' \bar{\varepsilon}_{int,0}(R(\tau),\eta(\tau),g_{av},n)(1-\delta_{n,0})$$

$$\bar{\varepsilon}_{int,0}(\eta(\tau'),R(\tau'),g_{av},-n)\exp\left(-\frac{n^2|\tau-\tau'|}{l_c}\right)\exp(in(\Delta\Phi_0(\tau)-\Delta\Phi_0(\tau'))).$$ (5.25)

Here the $\delta\Phi$ subscript on the averaging bracket corresponds averaging over only the $\delta\Phi$ fluctuations using Eq. (5.3). Note that the minimization with respect to $\Delta\Phi_0(\tau)$ in the energy functional (Eq. (5.13)) now yields

$$\exp(in(\Delta\Phi_0(\tau)-\Delta\Phi_0(\tau'))) = \exp\left(in \int_{-\tau'}^{\tau} \Delta g^0(\tau'')\right).$$ (5.26)

The ensemble average of Eq. (5.26) over all base pair sequence realizations yields

$$\langle \exp(in(\Delta\Phi_0(\tau)-\Delta\Phi_0(\tau'))) \rangle_{\Delta g}$$
$$= \frac{\int D\Delta g^0(\tau) \exp\left(in\int_{\tau'}^{\tau}\Delta g^0(\tau'')d\tau'''\right) \exp\left(-\frac{\lambda_c^0}{4}\int_{-\infty}^{\infty} \Delta g^0(\tau'')^2 d\tau'''\right)}{\int D\Delta g^0(\tau) \exp\left(-\frac{\lambda_c^0}{4}\int_{-\infty}^{\infty} \Delta g^0(\tau'')^2 d\tau'''\right)} = \exp\left(-\frac{n^2}{\lambda_c^{(0)}}|\tau-\tau'|\right).$$ (5.27)

This gives us the result on ensemble averaging:

$$\left\langle \left\langle \left( \frac{\delta \tilde{E}_{int}[R(\tau),\eta(\tau),\Delta\Phi(\tau)]}{k_B T} \right)^2 \right\rangle_{\delta\Phi} \right\rangle_{\Delta g} = \sum_{n=-\infty}^{\infty} \int_0^{L_b} d\tau \int_0^{L_b} d\tau' \bar{\varepsilon}_{int,0}(R_0,\delta R(\tau),\eta(\tau),g_{av},n)$$

$$\bar{\varepsilon}_{int}(R_0,\delta R(\tau'),\eta(\tau'),g_{av},-n)\exp\left(-\frac{n^2|\tau-\tau'|}{\lambda_c}\right)(1-\delta_{n,0}),$$ (5.28)

where the combined persistence length is defined through Eq. (4.40). Next, to handle $\ln(Z_0)$, with $Z_0$ given through Eq. (5.7), we construct the variational trial function

$$\frac{G_T^T[\delta\eta(\tau),\delta R(\tau)]}{k_B T} = \int_0^{L_b} d\tau \left( \frac{l_p}{4}\left(\frac{d^2\delta R(\tau)}{d\tau^2}\right)^2 + \frac{\beta_R}{2}\left(\frac{d\delta R(\tau)}{d\tau}\right)^2 + \frac{\alpha_R}{2}\delta R(\tau)^2 \right)$$
$$+ \int_0^{L_b} d\tau \left( \frac{l_p}{4}\left(\frac{d\delta\eta(\tau)}{d\tau}\right)^2 + \frac{\alpha_\eta}{2}\delta\eta(\tau)^2 \right),$$ (5.29)

and the partition function

$$\tilde{Z}_T^T = \int D\delta R(\tau) \int D\delta\eta(\tau) \exp\left(-\frac{G_T^T[\delta\eta(\tau),\delta R(\tau)]}{k_B T}\right).$$ (5.30)

We can then approximate $F_T$ with $\tilde{F}_T^T$ where for the ensemble average we have that

$$\left\langle \tilde{F}_T^T \right\rangle_{\Delta g} = k_B T \left( -\ln\left(\tilde{Z}_T^T\right) + \left\langle \left( \frac{G_T^0[R(\tau),\eta(\tau)] - G_T^T[\delta R(\tau),\delta\eta(\tau)]}{k_B T} \right) \right\rangle_0 - \frac{1}{2} \left\langle \left\langle \left( \frac{\delta\tilde{E}_{\text{int}}[R(\tau),\eta(\tau),\Delta\Phi(\tau)]}{k_B T} \right)^2 \right\rangle_0 \right\rangle_{\Delta g} \right)$$

(5.31)

Here, the zero subscript on the averaging bracket means averaging with $G_T^T[\delta R(\tau),\delta\eta(\tau)]$ in the Boltzmann weight of the functional integration. We use the averaging formulas developed in appendix A, where we have used the approximation for replacing $g_{av}$ with $\langle g_{av}\rangle_0$ (discussed in the previous section) valid for large $L$. We also suppose that the interaction is symmetric in $\Delta\Phi(\tau)$ such that $\bar{\varepsilon}_{\text{int},0}(R_0,r_2,\eta_0+\eta_2,\langle g_{av}\rangle_0,-n) = \bar{\varepsilon}_{\text{int},0}(R_0,r_2,\eta_0+\eta_2,\langle g_{av}\rangle_0,n)$. This yields the expression

$$\left\langle \left\langle \left( \frac{\delta\tilde{E}_{\text{int}}[R(\tau),\eta(\tau),\Delta\Phi(\tau)]}{k_B T} \right)^2 \right\rangle_0 \right\rangle_{\Delta g} \approx$$

$$2L_b \sum_{n=1}^{\infty} \int_{-\infty}^{\infty} d\eta_1 \int_{-\infty}^{\infty} d\eta_2 \int_{-\infty}^{\infty} dr_1 \int_{-\infty}^{\infty} dr_2 \int_{-\infty}^{\infty} dx \Gamma_R(r_1,r_2;x) \Gamma_\eta(\eta_1,\eta_2;x) \exp\left(-\frac{n^2|x|}{\lambda_c}\right)$$ (5.32)

$$\bar{\varepsilon}_{\text{int},0}(R_0,r_1,\eta_0+\eta_1,\langle g_{av}\rangle_0,n) \bar{\varepsilon}_{\text{int},0}(R_0,r_2,\eta_0+\eta_2,\langle g_{av}\rangle_0,n).$$

If we suppose that $l_c \gg \lambda_R, \lambda_\eta$, we may approximate

$$\left\langle \left( \frac{\delta\tilde{E}_{\text{int}}[R(\tau),\eta(\tau),\Delta\Phi(\tau)]}{k_B T} \right)^2 \right\rangle_0 \approx F_{corr,0} + F_{corr,1,1} + F_{corr,1,2},$$ (5.33)

with

$$F_{corr,0} = 4L_b \lambda_c \sum_{n=1}^{\infty} \frac{1}{n^2} \Delta_1\left(R_0,\eta_0,\langle g_{av}\rangle_0,d_R,d_\eta,n\right)^2,$$ (5.34)

$$F_{corr,1,1} = 2l_p L_b \sum_{n=1}^{\infty} \Delta_2\left(R_0,\eta_0,\langle g_{av}\rangle_0,d_R,d_\eta,n\right)^2 \Omega_{1,\eta}\left(\frac{l_p n^2}{\lambda_c},l_p \alpha_\eta\right),$$ (5.35)

$$F_{corr,1,2} = 2l_p^3 L_b \sum_{n=1}^{\infty} \Delta_3\left(R_0,\eta_0,\langle g_{av}\rangle_0,d_R,d_\eta,n\right)^2 \Omega_{1,R}\left(\frac{n^2 l_p}{\lambda_c},\alpha_R l_p^3,\beta_R l_p\right),$$ (5.36)

and

$$\Delta_1\left(R_0,\eta_0,\langle g_{av}\rangle_0,d_R,d_\eta,n\right) = \frac{1}{2\pi d_R d_\eta} \int_{-\infty}^{\infty} d\eta \int_{-\infty}^{\infty} dr \exp\left(-\frac{r^2}{2d_R^2}\right) \exp\left(-\frac{\eta^2}{2d_\eta^2}\right)$$ (5.37)

$$\bar{\varepsilon}_{\text{int},0}(R_0,r,\eta_0+\eta,\langle g_{av}\rangle_0,n),$$

$$\Delta_2\left(R_0,\eta_0,\langle g_{av}\rangle_0,d_R,d_\eta,n\right) = \frac{1}{2\pi d_R d_\eta^3} \int_{-\infty}^{\infty} d\eta \int_{-\infty}^{\infty} dr\eta \exp\left(-\frac{r^2}{2d_R^2}\right)\exp\left(-\frac{\eta^2}{2d_\eta^2}\right) \quad (5.38)$$
$$\bar{\varepsilon}_{\text{int},0}(R_0,r,\eta_0+\eta,\langle g_{av}\rangle_0,n),$$

$$\Delta_3\left(R_0,\eta_0,\langle g_{av}\rangle_0,d_R,d_\eta,n\right) = \frac{1}{2\pi d_R^3 d_\eta} \int_{-\infty}^{\infty} d\eta \int_{-\infty}^{\infty} drr \exp\left(-\frac{r^2}{2d_R^2}\right)\exp\left(-\frac{\eta^2}{2d_\eta^2}\right) \quad (5.39)$$
$$\bar{\varepsilon}_{\text{int},0}(R_0,r,\eta_0+\eta,\langle g_{av}\rangle_0,n),$$

$$\Omega_{1,\eta}\left(\frac{l_p n^2}{\lambda_c},\alpha_\eta l_p\right) = \frac{1}{l_p}\int_{-\infty}^{\infty} dx\, G_\eta(x)\exp\left(-\frac{n^2|x|}{\lambda_c}\right), \quad (5.40)$$

$$\Omega_{1,R}\left(\frac{l_p n^2}{\lambda_c},\alpha_R l_p^3,\beta_R l_p\right) = \frac{1}{l_p^3}\int_{-\infty}^{\infty} dx\, G_R(x)\exp\left(-\frac{n^2|x|}{\lambda_c}\right), \quad (5.41)$$

where $G_\eta(x) = \langle\delta\eta(0)\delta\eta(x)\rangle_0$ and $G_R(x) = \langle\delta R(0)\delta R(x)\rangle_0$. The functions $\Omega_{1,\eta}\left(\frac{l_p n^2}{\lambda_c},\alpha_\eta l_p\right)$ and $\Omega_{1,R}\left(\frac{n^2 l_p}{\lambda_c},\alpha_R l_p^3,\beta_R l_p\right)$ are evaluated to be (see Appendix C)

$$\Omega_{1,\eta}\left(\frac{l_p n^2}{\lambda_c},\alpha_\eta l_p\right) = 2\left(\frac{1}{2\alpha_\eta l_p}\right)^{1/2}\frac{1}{\left(\frac{l_p n^2}{\lambda_c}\right)+(2\alpha_\eta l_p)^{1/2}}, \quad (5.42)$$

$$\Omega_{1,R}\left(\frac{n^2 l_p}{\lambda_c},\alpha_R l_p^3,\beta_R l_p\right) \equiv \frac{4n^2 l_p}{\lambda_c}\left(2\alpha_R l_p^3\right)^{-5/4} I_1\left(2\beta_R l_p\left(\frac{1}{2\alpha_R l_p^3}\right)^{1/2},\left(\frac{n^2 l_p}{\lambda_c}\right)^2\left(\frac{1}{2\alpha_R l_p^3}\right)^{1/2}\right), \quad (5.43)$$

where

$$I_1(\gamma,\delta) = \frac{1}{\delta^2-\gamma\delta+1}\left(\frac{1}{2\sqrt{\delta}}-\frac{1}{4}\sqrt{2+\gamma}+\frac{(\delta-\gamma/2)}{2\sqrt{2+\gamma}}\right). \quad (5.44)$$

Treating $d_\eta$ small yields

$$F_{corr,0} \approx 4L_b\lambda_c \sum_{n=1}^{\infty}\frac{1}{n^2}\tilde{f}_{\text{int},0}(\eta_0,R_0,d_R,\langle g_{av}\rangle_0,n,d_{\max},d_{\min})$$
$$\left(\tilde{f}_{\text{int},0}(\eta_0,R_0,d_R,\langle g_{av}\rangle_0,n,d_{\max},d_{\min})+d_\eta^2\tilde{f}_{\text{int},\eta\eta}(\eta_0,R_0,d_R,\langle g_{av}\rangle_0,n,d_{\max},d_{\min})\right), \quad (5.45)$$

$$F_{corr,1,1} \approx 2l_p L_b \sum_{n=1}^{\infty} f_{int,\eta}\left(R_0, \eta_0, d_R, \langle g_{av}\rangle_0, n, d_{max}, d_{min}\right)^2 \Omega_{1,\eta}\left(\frac{l_p n^2}{\lambda_c}, \alpha_\eta l_p\right), \tag{5.46}$$

$$F_{corr,1,2} = 2l_p^3 L_b \sum_{n=1}^{\infty} f_{int,1}\left(R_0, \eta_0, d_R, \langle g_{av}\rangle_0, n, d_{max}, d_{min}\right)^2 \Omega_{1,R}\left(\frac{n^2 l_p}{\lambda_c}, \alpha_R l_p^3, \beta_R l_p\right), \tag{5.47}$$

where

$$f_{int,\eta}\left(R_0, \eta_0, d_R, \langle g_{av}\rangle_0, n, d_{max}, d_{min}\right) = \frac{1}{d_R\sqrt{2\pi}} \int_{-\infty}^{\infty} dr \exp\left(-\frac{r^2}{2d_R}\right) \bar{\varepsilon}_{int,\eta}(R_0, r, \eta_0, \langle g_{av}\rangle_0, n),$$

$$\tag{5.48}$$

$$f_{int,1}\left(R_0, \eta_0, d_R, \langle g_{av}\rangle_0, n, d_{max}, d_{min}\right) = \frac{1}{d_R^3\sqrt{2\pi}} \int_{-\infty}^{\infty} dr\, r \exp\left(-\frac{r^2}{2d_R}\right) \bar{\varepsilon}_{int,0}(R_0, r, \eta_0, \langle g_{av}\rangle_0, n, ),$$

$$\tag{5.49}$$

and

$$\bar{\varepsilon}_{int,\eta}\left(R_0, r, \eta_0, \langle g_{av}\rangle_0, n\right) = \left.\frac{d\bar{\varepsilon}_{int}(R_0, r, \eta_0 + \eta, \langle g_{av}\rangle_0, n)}{d\eta}\right|_{\eta=0}. \tag{5.50}$$

Now, we can write

$$-\ln\left(\tilde{Z}_T^T\right) + \left\langle\left(\frac{G_T^0[R(\tau), \eta(\tau)] - G_T^T[\delta R(\tau), \delta\eta(\tau)]}{k_B T}\right)\right\rangle_0 =$$

$$\left(L - L_{loop}\right) f_{braid}^0 + \frac{2l_p}{L_{loop}}\left(2\pi - \frac{\pi^2 R_0}{2L_{loop}}\right)^2 + \frac{2l_p \pi^4}{4L_{loop}^3}\left(\frac{d_{min} d_R}{\sqrt{2\pi}}\exp\left(-\frac{d_{min}^2}{2d_R^2}\right) - \frac{d_{max} d_R}{\sqrt{2\pi}}\exp\left(-\frac{d_{max}^2}{2d_R^2}\right)\right.$$

$$\left. + \frac{d_R^2 - d_{min}^2}{2}\operatorname{erf}\left(-\frac{d_{min}}{d_R\sqrt{2}}\right) + \frac{d_R^2 - d_{max}^2}{2}\operatorname{erf}\left(\frac{d_{max}}{d_R\sqrt{2}}\right) + \frac{d_{min}^2 + d_{max}^2}{2}\right) + l_{tw} L\left(\langle g_{av}\rangle_0 - \bar{g}_0\right)^2,$$

$$\tag{5.51}$$

where

$$f_{braid}^0 = \frac{\alpha_\eta^{1/2}}{2^{3/2} l_p^{1/2}} + \frac{1}{4 l_p \theta_R^2} + \frac{l_p \theta_R^4}{4 d_R^2} + \frac{d_R^2 \alpha_H}{2}$$

$$- \frac{\theta_R^2 l_p}{R_0^2} \tilde{f}_1(R_0, d_R, d_{max}, d_{min}) \sin^2\left(\frac{\eta_0}{2}\right) + \frac{4 l_p \tilde{f}_1(R_0, d_R, d_{max}, d_{min})}{R_0^2} \sin^4\left(\frac{\eta_0}{2}\right) \quad (5.52)$$

$$+ \frac{1}{2^{3/2} \alpha_\eta^{1/2} l_p^{1/2}} \left[ \frac{4 l_p \tilde{f}_1(R_0, d_R, d_{max}, d_{min})}{R_0^2} \left( 3\cos^2\left(\frac{\eta_0}{2}\right) \sin^2\left(\frac{\eta_0}{2}\right) - \sin^4\left(\frac{\eta_0}{2}\right) \right) \right.$$

$$\left. + \tilde{f}_{int,\eta\eta}(\eta_0, R_0, d_R, 0, d_{max}, d_{min}) \right] + \tilde{f}_{int,0}(\eta_0, R_0, d_R, 0, d_{max}, d_{min}).$$

Eqs. (5.51) and (5.52) are simply got from Eqs. (4.30) and (4.31) by taking the $\lambda_h^* \to \infty$ limit, i.e. neglecting all Fourier components for which $n \neq 0$. We again express things in terms of a total Gibbs like free energy, $\tilde{G}_T$ so that we can minimize independently over $\langle g_{av} \rangle_0$ by introducing torque $\Upsilon_R$. Thus we express $\tilde{G}_T = \langle \tilde{F}_T^T \rangle_{\Delta g} - 2\pi \Upsilon_R (L \langle g_{av} \rangle_0 / \pi + \langle Wr \rangle_0)$, which explicitly reads as

$$\langle G_T \rangle_{\Delta g} = (L - L_{loop}) \langle g_{braid} \rangle_{\Delta g} + \frac{2 l_p}{L_{loop}} \left( 2\pi - \frac{\pi^2 R_0}{2 L_{loop}} \right)^2 - 2 L \Upsilon_R \langle g_{av} \rangle_0$$

$$+ l_{tw} L \left( \langle g_{av} \rangle_0 - \overline{g}_0 \right)^2 + \frac{2 l_p \pi^4}{4 L_{loop}^3} \left( \frac{d_{min} d_R}{\sqrt{2\pi}} \exp\left( -\frac{d_{min}^2}{2 d_R^2} \right) - \frac{d_{max} d_R}{\sqrt{2\pi}} \exp\left( -\frac{d_{max}^2}{2 d_R^2} \right) \right) \quad (5.53)$$

$$+ \frac{d_R^2 - d_{min}^2}{2} \operatorname{erf}\left( -\frac{d_{min}}{d_R \sqrt{2}} \right) + \frac{d_R^2 - d_{max}^2}{2} \operatorname{erf}\left( \frac{d_{max}}{d_R \sqrt{2}} \right) + \frac{d_{min}^2 + d_{max}^2}{2},$$

with

$$\langle g_{braid} \rangle_{\Delta g} = \left( f_{braid}^0 - f_{braid}^{corr} - \frac{2\pi \Upsilon_R}{L_b} \langle Wr \rangle_0 \right). \quad (5.54)$$

Here, $f_{braid}^{corr}$ is the correction to the free energy from the helix specific forces, and can be written as

$$f_{braid}^{corr} = \frac{1}{2 L_b} \left( F_{corr,0} + F_{corr,1,1} + F_{corr,1,2} \right), \quad (5.55)$$

where expressions for $F_{corr,0}$, $F_{corr,1,1}$ and $F_{corr,1,2}$ are given by Eqs. (5.45), (5.46) and (5.47), respectively. If we treat the dependence on $\langle g_{av} \rangle_0$ in $f_{braid}^{corr}$ as a small correction, we can approximate

$$\langle g_{av} \rangle_0 \approx \frac{\Upsilon_R}{l_{tw}} + \overline{g}_0. \quad (5.56)$$

Thus the linking number is given by

$$Lk \approx \frac{2\Upsilon_R L}{\pi l_{tw}} + \frac{\overline{g}_0 L}{\pi} + \langle Wr \rangle_0. \tag{5.57}$$

When we minimize over $L_{loop}$, we obtain the same equations as Eq. (4.43)-(4.47). Minimization over $R_0$ and $d_R$ yields the same equations as Eqs. (4.48)-(4.52) with $R_0$ and $d_R$ derivatives of $\langle g_{braid} \rangle_{\Delta g}$

$$\frac{\partial \langle g_{braid} \rangle}{\partial R_0} = \left( \frac{\partial f_{braid}^0}{\partial R_0} - \frac{\partial f_{braid}^{corr}}{\partial R_0} - \frac{2\pi \Upsilon_R}{L_b} \frac{\partial \langle Wr \rangle_0}{\partial R_0} \right), \quad \frac{\partial \langle g_{braid} \rangle}{\partial d_R} = \left( \frac{\partial f_{braid}^0}{\partial d_R} - \frac{\partial f_{braid}^{corr}}{\partial d_R} - \frac{2\pi \Upsilon_R}{L_b} \frac{\partial \langle Wr \rangle_0}{\partial d_R} \right).$$

(5.58)

Minimization over $\eta_0, \theta_R$ and $d_\eta$ yield

$$0 = \left( \frac{\partial f_{braid}^0}{\partial \eta_0} + \frac{\partial f_{braid}^{corr}}{\partial \eta_0} - \frac{2\pi \Upsilon_R}{L_b} \frac{\partial \langle Wr \rangle_0}{\partial \eta_0} \right), \qquad 0 = \left( \frac{\partial f_{braid}^0}{\partial \alpha_\eta} - \frac{\partial f_{braid}^{corr}}{\partial \alpha_\eta} - \frac{2\pi \Upsilon_R}{L_b} \frac{\partial \langle Wr \rangle_0}{\partial \alpha_\eta} \right),$$

$$0 = \left( \frac{\partial f_{braid}^0}{\partial \theta_R} - \frac{\partial f_{braid}^{corr}}{\partial \theta_R} - \frac{2\pi \Upsilon_R}{L_b} \frac{\partial \langle Wr \rangle_0}{\partial \theta_R} \right). \tag{5.59}$$

The partial derivatives of $f_{braid}^0$ are given by the expressions

$$\frac{\partial f_{braid}^0}{\partial R_0} = \frac{d_R^2}{2} \frac{\partial \alpha_H}{\partial R_0} + \frac{\theta_R^2 l_p}{R_0^3} \tilde{h}_1(R_0, d_R, d_{max}, d_{min}) \sin^2\left(\frac{\eta_0}{2}\right) - \frac{4 l_p \tilde{h}_1(R_0, d_R, d_{max}, d_{min})}{R_0^3} \sin^4\left(\frac{\eta_0}{2}\right)$$

$$+ \frac{1}{2^{3/2} \alpha_\eta^{1/2} l_p^{1/2}} \left[ -\frac{4 l_p \tilde{h}_1(R_0, d_R, d_{max}, d_{min})}{R_0^3} \left( 3\cos^2\left(\frac{\eta_0}{2}\right) \sin^2\left(\frac{\eta_0}{2}\right) - \sin^4\left(\frac{\eta_0}{2}\right) \right) \right.$$

$$\left. + \frac{\partial \tilde{f}_{int,\eta\eta}(\eta_0, R_0, d_R, \langle g_{av} \rangle_0, 0, d_{max}, d_{min})}{\partial R_0} \right] + \frac{\partial \tilde{f}_{int,0}(\eta_0, R_0, d_R, \langle g_{av} \rangle_0, 0, d_{max}, d_{min})}{\partial R_0},$$

(5.60)

$$\frac{\partial f_{braid}^0}{\partial d_R} = -\frac{l_p \theta_R^2}{d_R R_0^2} \tilde{l}_1(R_0, d_R, d_{max}, d_{min}) \sin^2\left(\frac{\eta_0}{2}\right) + d_R \alpha_H - \frac{l_p \theta_R^4}{2 d_R^3} + \frac{4 l_p}{d_R R_0^2} \tilde{l}_1(R_0, d_R, d_{max}, d_{min}) \sin^4\left(\frac{\eta_0}{2}\right)$$

$$+ \frac{L}{2^{3/2} \alpha_\eta^{1/2} l_p^{1/2}} \left[ \frac{4 l_p \tilde{l}_1(R_0, d_R, d_{max}, d_{min})}{d_R R_0^2} \left( 3\cos^2\left(\frac{\eta_0}{2}\right) \sin^2\left(\frac{\eta_0}{2}\right) - \sin^4\left(\frac{\eta_0}{2}\right) \right) \right.$$

$$\left. + \frac{\partial \tilde{f}_{int,\eta\eta}(\eta_0, R_0, d_R, \langle g_{av} \rangle_0, 0, d_{max}, d_{min})}{\partial d_R} \right] + \frac{\partial \tilde{f}_{int,0}(\eta_0, R_0, d_R, \langle g_{av} \rangle_0, 0, d_{max}, d_{min})}{\partial d_R},$$

$$\frac{\partial f_{braid}^0}{\partial \eta_0} = \frac{d_R^2}{2}\frac{\partial \alpha_H}{\partial \eta_0} - \frac{\theta_R^2 l_p}{R_0^2}\tilde{f}_1(R_0,d_R,d_{max},d_{min})\sin\left(\frac{\eta_0}{2}\right)\cos\left(\frac{\eta_0}{2}\right)$$

$$+\frac{8l_p\tilde{f}_1(R_0,d_R,d_{max},d_{min})}{R_0^2}\cos\left(\frac{\eta_0}{2}\right)\sin^3\left(\frac{\eta_0}{2}\right)+\frac{\partial \tilde{f}_{int.0}(\eta_0,R_0,d_R,\langle g_{av}\rangle_0,0,d_{max},d_{min})}{\partial \eta_0}$$

$$+\frac{1}{2^{3/2}\alpha_\eta^{1/2}l_p^{1/2}}\left[\frac{4l_p\tilde{f}_1(R_0,d_R,d_{max},d_{min})}{R_0^2}\left(3\cos^3\left(\frac{\eta_0}{2}\right)\sin\left(\frac{\eta_0}{2}\right)-5\sin^3\left(\frac{\eta_0}{2}\right)\cos\left(\frac{\eta_0}{2}\right)\right)\right.$$

$$\left.+\frac{\partial \tilde{f}_{int.\eta\eta}(\eta_0,R_0,d_R,\langle g_{av}\rangle_0,0,d_{max},d_{min})}{\partial \eta_0}\right],$$

(5.62)

$$\frac{\partial f_{braid}^0}{\partial \alpha_\eta} = \frac{1}{2^{5/2}l_p^{1/2}\alpha_\eta^{1/2}} - \frac{1}{2^{5/2}\alpha_\eta^{3/2}l_p^{1/2}}\left[\frac{4l_p\tilde{f}_1(R_0,d_R,d_{max},d_{min})}{R_0^2}\left(3\cos^2\left(\frac{\eta_0}{2}\right)\sin^2\left(\frac{\eta_0}{2}\right)-\sin^4\left(\frac{\eta_0}{2}\right)\right)\right.$$

$$\left.+\tilde{f}_{int,\eta\eta}(\eta_0,R_0,d_R,\langle g_{av}\rangle_0,0,d_{max},d_{min})\right],$$

(5.63)

$$\frac{\partial f_{braid}^0}{\partial \theta_R} = \frac{l_p\theta_R^3}{d_R^2} - \frac{1}{2l_p\theta_R^3} - \frac{2\theta_R l_p}{R_0^2}\tilde{f}_1(R_0,d_R,d_{max},d_{min})\sin^2\left(\frac{\eta_0}{2}\right),$$

(5.64)

and the partial derivatives of $f_{braid}^{corr}$

$$\frac{\partial f_{braid}^{corr}}{\partial R_0} = 2\lambda_c\sum_{n=1}^{\infty}\frac{1}{n^2}\left[2\frac{\partial \tilde{f}_{int,0}(\eta_0,R_0,d_R,\langle g_{av}\rangle_0,n,d_{max},d_{min})}{\partial R_0}\tilde{f}_{int,0}(\eta_0,R_0,d_R,\langle g_{av}\rangle_0,n,d_{max},d_{min})\right.$$

$$+\frac{1}{(2l_p\alpha_\eta)^{1/2}}\left(\tilde{f}_{int,0}(\eta_0,R_0,d_R,\langle g_{av}\rangle_0,n,d_{max},d_{min})\frac{\partial \tilde{f}_{int,\eta\eta}(\eta_0,R_0,d_R,\langle g_{av}\rangle_0,n,d_{max},d_{min})}{\partial R_0}\right.$$

$$\left.\left.+\tilde{f}_{int,\eta\eta}(\eta_0,R_0,d_R,\langle g_{av}\rangle_0,n,d_{max},d_{min})\frac{\partial \tilde{f}_{int,0}(\eta_0,R_0,d_R,\langle g_{av}\rangle_0,n,d_{max},d_{min})}{\partial R_0}\right)\right]$$

$$+2l_p\sum_{n=1}^{\infty}\frac{\partial \tilde{f}_{int,\eta}\left(R_0,\eta_0,d_R,\langle g_{av}\rangle_0,n,d_{max},d_{min}\right)}{\partial R_0}\tilde{f}_{int,\eta}\left(R_0,\eta_0,d_R,\langle g_{av}\rangle_0,n,d_{max},d_{min}\right)\Omega_{1,\eta}\left(\frac{l_pn^2}{\lambda_c},\alpha_\eta l_p\right)$$

$$+2l_p^3\sum_{n=1}^{\infty}\frac{\partial \tilde{f}_{int,1}\left(R_0,\eta_0,d_R,\langle g_{av}\rangle_0,n,d_{max},d_{min}\right)}{\partial R_0}\tilde{f}_{int,1}\left(R_0,\eta_0,d_R,\langle g_{av}\rangle_0,n,d_{max},d_{min}\right)$$

$$\Omega_{1,R}\left(\frac{n^2l_p}{\lambda_c},\frac{l_p^4}{2}\left(\frac{\theta_R}{d_R}\right)^4,\frac{1}{2\theta_R^4}-\left(\frac{l_p\theta_R}{d_R}\right)^2\right),$$

(5.65)

$$\frac{\partial f_{braid}^{corr}}{\partial \eta_0} = 2\lambda_c \sum_{n=1}^{\infty} \frac{1}{n^2} \left[ 2 \frac{\partial \tilde{f}_{int,0}(\eta_0, R_0, d_R, \langle g_{av} \rangle_0, n, d_{max}, d_{min})}{\partial \eta_0} \tilde{f}_{int,0}(\eta_0, R_0, d_R, \langle g_{av} \rangle_0, n, d_{max}, d_{min}) \right.$$

$$+ \frac{1}{(2l_p \alpha_\eta)^{1/2}} \left( \tilde{f}_{int,0}(\eta_0, R_0, d_R, \langle g_{av} \rangle_0, n, d_{max}, d_{min}) \frac{\partial \tilde{f}_{int,\eta\eta}(\eta_0, R_0, d_R, \langle g_{av} \rangle_0, n, d_{max}, d_{min})}{\partial \eta_0} \right.$$

$$\left. \left. + \tilde{f}_{int,\eta\eta}(\eta_0, R_0, d_R, \langle g_{av} \rangle_0, n, d_{max}, d_{min}) \frac{\partial \tilde{f}_{int,0}(\eta_0, R_0, d_R, \langle g_{av} \rangle_0, n, d_{max}, d_{min})}{\partial \eta_0} \right) \right]$$

$$+ 2l_p \sum_{n=1}^{\infty} \frac{\partial \tilde{f}_{int,\eta}(R_0, \eta_0, d_R, \langle g_{av} \rangle_0, n, d_{max}, d_{min})}{\partial \eta_0} \tilde{f}_{int,\eta}(R_0, \eta_0, d_R, \langle g_{av} \rangle_0, n, d_{max}, d_{min}) \Omega_{1,\eta}\left(\frac{l_p n^2}{\lambda_c}, \alpha_\eta l_p\right)$$

$$+ 2l_p^3 \sum_{n=1}^{\infty} \frac{\partial \tilde{f}_{int,1}(R_0, \eta_0, d_R, \langle g_{av} \rangle_0, n, d_{max}, d_{min})}{\partial \eta_0} \tilde{f}_{int,1}(R_0, \eta_0, d_R, \langle g_{av} \rangle_0, n, d_{max}, d_{min})$$

$$\Omega_{1,R}\left(\frac{n^2 l_p}{\lambda_c}, \frac{l_p^4}{2}\left(\frac{\theta_R}{d_R}\right)^4, \frac{1}{2\theta_R^4} - \left(\frac{l_p \theta_R}{d_R}\right)^2\right),$$

(5.66)

$$\frac{\partial f_{braid}^{corr}}{\partial \alpha_\eta} = -\frac{\lambda_c \sqrt{2}}{2 l_p^{1/2} \alpha_\eta^{3/2}} \sum_{n=1}^{\infty} \frac{1}{n^2} \tilde{f}_{int,0}(\eta_0, R_0, d_R, \langle g_{av} \rangle_0, n, d_{max}, d_{min}) \tilde{f}_{int,\eta\eta}(\eta_0, R_0, d_R, \langle g_{av} \rangle_0, n, d_{max}, d_{min})$$

(5.67)

$$+ l_p \sum_{n=1}^{\infty} \tilde{f}_{int,\eta}(R_0, \eta_0, d_R, \langle g_{av} \rangle_0, n, d_{max}, d_{min})^2 \frac{\partial \Omega_{1,\eta}\left(\frac{l_p n^2}{\lambda_c}, \alpha_\eta l_p\right)}{\partial \alpha_\eta},$$

$$\frac{\partial f_{braid}^{corr}}{\partial d_R} = 2\lambda_c \sum_{n=1}^{\infty} \frac{1}{n^2} \left[ 2 \frac{\partial \tilde{f}_{int,0}(\eta_0, R_0, d_R, \langle g_{av} \rangle_0, n, d_{max}, d_{min})}{\partial d_R} \tilde{f}_{int,0}(\eta_0, R_0, d_R, \langle g_{av} \rangle_0, n, d_{max}, d_{min}) \right.$$

$$+ \frac{1}{(2l_p \alpha_\eta)^{1/2}} \left( \tilde{f}_{int,0}(\eta_0, R_0, d_R, \langle g_{av} \rangle_0, n, d_{max}, d_{min}) \frac{\partial \tilde{f}_{int,\eta\eta}(\eta_0, R_0, d_R, \langle g_{av} \rangle_0, n, d_{max}, d_{min})}{\partial d_R} \right.$$

$$\left. \left. + \tilde{f}_{int,\eta\eta}(\eta_0, R_0, d_R, \langle g_{av} \rangle_0, n, d_{max}, d_{min}) \frac{\partial \tilde{f}_{int,0}(\eta_0, R_0, d_R, \langle g_{av} \rangle_0, n, d_{max}, d_{min})}{\partial d_R} \right) \right]$$

$$+ 2l_p \sum_{n=1}^{\infty} \frac{\partial \tilde{f}_{int,\eta}(R_0, \eta_0, d_R, \langle g_{av} \rangle_0, n, d_{max}, d_{min})}{\partial d_R} \tilde{f}_{int,\eta}(R_0, \eta_0, d_R, \langle g_{av} \rangle_0, n, d_{max}, d_{min}) \Omega_{1,\eta}\left(\frac{l_p n^2}{\lambda_c}, \alpha_\eta l_p\right)$$

$$+ 2l_p^3 \sum_{n=1}^{\infty} \frac{\partial \tilde{f}_{int,1}(R_0, \eta_0, d_R, \langle g_{av} \rangle_0, n, d_{max}, d_{min})}{\partial d_R} \tilde{f}_{int,1}(R_0, \eta_0, d_R, \langle g_{av} \rangle_0, n, d_{max}, d_{min})$$

$$\Omega_{1,R}\left(\frac{n^2 l_p}{\lambda_c}, \frac{l_p^4}{2}\left(\frac{\theta_R}{d_R}\right)^4, \frac{1}{2\theta_R^4} - \left(\frac{l_p \theta_R}{d_R}\right)^2\right)$$

$$+l_p^3 \sum_{n=1}^{\infty} \tilde{f}_{\text{int},1}\left(R_0,\eta_0,d_R,\langle g_{av}\rangle_0,n,d_{\max},d_{\min}\right)^2 \frac{\partial \Omega_{1,R}\left(\frac{n^2 l_p}{\lambda_c},\frac{l_p^4}{2}\left(\frac{\theta_R}{d_R}\right)^4,\frac{1}{2\theta_R^4}-\left(\frac{l_p \theta_R}{d_R}\right)^2\right)}{\partial d_R}, \tag{5.68}$$

$$\frac{\partial f_{\text{braid}}^{\text{corr}}}{\partial \theta_R} = l_p^3 \sum_{n=1}^{\infty} \tilde{f}_{\text{int},1}\left(R_0,\eta_0,d_R,\langle g_{av}\rangle_0,n,d_{\max},d_{\min}\right)^2 \frac{\partial}{\partial \theta_R} \Omega_{1,R}\left(\frac{n^2 l_p}{\lambda_c},\frac{l_p^4}{2}\left(\frac{\theta_R}{d_R}\right)^4,\frac{1}{2\theta_R^4}-\left(\frac{l_p \theta_R}{d_R}\right)^2\right).$$

(5.69)

Expressions for the various derivatives of $\tilde{f}_{\text{int},0}$, $\tilde{f}_{\text{int},\eta}$, $\tilde{f}_{\text{int},\eta\eta}$, $\tilde{f}_{\text{int},1}$, $\Omega_{1,\eta}$ and $\Omega_{1R}$ can be found in Appendix D. The explicit $\theta_R$ and $d_R$ dependence written in $\Omega_{1R}$ can be seen from Eq. (4.13), which relates these to $\alpha_R$ and $\beta_R$. All that remains is the evaluation of $\langle Wr \rangle_0$ and its derivatives, we perform this is the next section.

## 6. Evaluation of $\langle Wr \rangle_0$

The key point in the analysis is the evaluation of $\langle Wr \rangle_0$; it is the coupling of this term to $\tau$ in the free energy that drives the value of $\eta_0$ not to be zero, generating a braided part to the plectoneme. First of all we may write

$$4\pi Wr \approx Wr_{1,1} + Wr_{2,2} - 2Wr_{1,2}, \tag{6.1}$$

where

$$Wr_{1,1} = \int_0^{L_p} d\tau \int_0^{L_p} d\tau' \frac{(\mathbf{r}_1(\tau)-\mathbf{r}_1(\tau')).\hat{\mathbf{t}}_1(\tau)\times\hat{\mathbf{t}}_1(\tau')}{|\mathbf{r}_1(\tau)-\mathbf{r}_1(\tau')|^{3/2}}, \tag{6.2}$$

$$Wr_{2,2} = \int_0^{L_p} d\tau \int_0^{L_p} d\tau' \frac{(\mathbf{r}_2(\tau)-\mathbf{r}_2(\tau')).\hat{\mathbf{t}}_2(\tau)\times\hat{\mathbf{t}}_2(\tau')}{|\mathbf{r}_2(\tau)-\mathbf{r}_2(\tau')|^{3/2}}, \tag{6.3}$$

$$Wr_{1,2} = \int_0^{L_p} d\tau \int_0^{L_p} d\tau' \frac{(\mathbf{r}_1(\tau)-\mathbf{r}_2(\tau')).\hat{\mathbf{t}}_1(\tau)\times\hat{\mathbf{t}}_2(\tau')}{|\mathbf{r}_1(\tau)-\mathbf{r}_2(\tau')|^{3/2}}. \tag{6.4}$$

Substitution of Eqs. (1.8), (1.9), (1.13) and (1.14) into Eqs. (6.2), (6.3) and (6.4) allows us to write

$$Wr_{1,1} = Wr_{2,2} = \int_0^{L_p} d\tau \int_0^{L_p} d\tau' \frac{\omega_{1,1}(\tau,\tau')}{\left(\frac{R(\tau)^2}{4}+\frac{R(\tau')^2}{4}-\frac{R(\tau)R(\tau')}{2}\cos(\theta(\tau)-\theta(\tau'))+(Z(\tau)-Z(\tau'))^2\right)^{3/2}},$$

(6.5)

$$Wr_{1,2} = \int_0^{L_b} d\tau \int_0^{L_b} d\tau' \frac{\omega_{1,2}(\tau,\tau')}{\left(\frac{R(\tau)^2}{4} + \frac{R(\tau')^2}{4} + \frac{R(\tau)R(\tau')}{2}\cos(\theta(\tau)-\theta(\tau')) + (Z(\tau)-Z(\tau'))^2\right)^{3/2}}, \quad (6.6)$$

where

$$\omega_{1,j}(\tau,\tau') = -\frac{R(\tau)}{2}\cos\left(\frac{\eta(\tau')}{2}\right)\sqrt{\sin^2\left(\frac{\eta(\tau)}{2}\right) - \frac{1}{4}\left(\frac{dR(\tau)}{d\tau}\right)^2}$$

$$-\frac{R(\tau')}{2}\cos\left(\frac{\eta(\tau)}{2}\right)\sqrt{\sin^2\left(\frac{\eta(\tau')}{2}\right) - \frac{1}{4}\left(\frac{dR(\tau)}{d\tau}\right)^2}$$

$$-(-1)^j \sin\left(\frac{\eta(\tau)}{2}\right)\cos\left(\frac{\eta(\tau')}{2}\right)\frac{R(\tau')}{2}\cos(\gamma(\tau)+\theta(\tau)-\theta(\tau')) \quad (6.7)$$

$$-(-1)^j \sin\left(\frac{\eta(\tau')}{2}\right)\cos\left(\frac{\eta(\tau)}{2}\right)\frac{R(\tau)}{2}\cos(\gamma(\tau')+\theta(\tau')-\theta(\tau))$$

$$+(-1)^j (Z(\tau)-Z(\tau'))\sin\left(\frac{\eta(\tau)}{2}\right)\sin\left(\frac{\eta(\tau')}{2}\right)\sin(\gamma(\tau)-\gamma(\tau')+\theta(\tau)-\theta(\tau')),$$

As a first approximation, we start by writing from Eq. (1.15)

$$\theta(\tau) = \theta_0 - \langle Q \rangle_0 \tau + \delta\theta(\tau), \quad (6.8)$$

where

$$\langle Q \rangle_0 = \left\langle \frac{1}{R(\tau)}\sqrt{4\sin^2\left(\frac{\eta(\tau)}{2}\right) - \left(\frac{dR(\tau)}{d\tau}\right)^2} \right\rangle_0, \quad (6.9)$$

and
$$Z(\tau) = \langle \cos(\eta(\tau)/2) \rangle_0 \tau + \delta Z(\tau). \quad (6.10)$$

We assume that the corrections $\delta\theta(\tau)$ and $\delta Z(\tau)$ are small. Physically, this supposes that the spatial averaging, through integration, in evaluating both $\theta(\tau)$ and $Z(\tau)$ is roughly equivalent to thermal averaging. This approximation should be valid when the distance $\tau - \tau'$ is larger than the correlation lengths for both $\eta(\tau)$ and $R(\tau)$. If the pitch of the braided section is much larger than the correlation length, these length scales should dominate in the evaluation of the integrals in Eqs. (6.5) and (6.6). We make this assumption, thus keeping only the first terms in a power series expansion in $\delta\theta(\tau)$ and $\delta Z(\tau)$. Thus, we approximate

$$Wr_j \approx \int_0^{L_b} d\tau \int_0^{L_b} d\tau' \frac{\tilde{\omega}_j(\tau-\tau';R(\tau),R(\tau'),R'(\tau),R'(\tau'),\eta(\tau),\eta(\tau'))}{\left(\frac{R(\tau)^2}{4}+\frac{R(\tau')^2}{4}+(-1)^j\frac{R(\tau)R(\tau')}{2}\cos\left(\langle Q\rangle_0(\tau-\tau')\right)+\left\langle\cos\left(\frac{\eta(\tau)}{2}\right)\right\rangle_0^2(\tau-\tau')^2\right)^{3/2}},$$

(6.11)

where

$$\tilde{\omega}_j(\tau-\tau';R(\tau),R(\tau'),R'(\tau),R'(\tau'),\eta(\tau),\eta(\tau')) =$$
$$-\frac{R(\tau)}{2}\cos\left(\frac{\eta(\tau')}{2}\right)\sqrt{\sin^2\left(\frac{\eta(\tau)}{2}\right)-\frac{1}{4}\left(\frac{dR(\tau)}{d\tau}\right)^2}$$
$$-\frac{R(\tau')}{2}\cos\left(\frac{\eta(\tau)}{2}\right)\sqrt{\sin^2\left(\frac{\eta(\tau')}{2}\right)-\frac{1}{4}\left(\frac{dR(\tau')}{d\tau'}\right)^2}$$
$$-(-1)^j\sin\left(\frac{\eta(\tau)}{2}\right)\cos\left(\frac{\eta(\tau')}{2}\right)\frac{R(\tau')}{2}\cos\left(\gamma(\tau)-\langle Q\rangle_0(\tau-\tau')\right)$$
$$-(-1)^j\sin\left(\frac{\eta(\tau')}{2}\right)\cos\left(\frac{\eta(\tau)}{2}\right)\frac{R(\tau)}{2}\cos\left(\gamma(\tau')-\langle Q\rangle_0(\tau'-\tau)\right)$$
$$+(-1)^j(\tau-\tau')\left\langle\cos\left(\frac{\eta(\tau)}{2}\right)\right\rangle_0\sin\left(\frac{\eta(\tau)}{2}\right)\sin\left(\frac{\eta(\tau')}{2}\right)\sin\left(\gamma(\tau)-\gamma(\tau')-\langle Q\rangle_0(\tau-\tau')\right),$$

(6.12)

and $j=1,2$. Using the Gaussian averaging formulas developed in Appendix A, we can then re-express Eq. (6.11) as

$$\langle Wr_j\rangle_0 \approx \int_{-\infty}^{\infty}d\eta_1\int_{-\infty}^{\infty}d\eta_2\int_{-\infty}^{\infty}dr_1'\int_{-\infty}^{\infty}dr_2'\int_{-\infty}^{\infty}dr_1\int_{-\infty}^{\infty}dr_2\int_0^{L_b}d\tau\int_0^{L_b}d\tau'\Gamma_\eta(\eta_1,\eta_2;\tau-\tau')\Xi_R(r_1,r_2,r_1',r_2';\tau-\tau')$$
$$\tilde{\omega}_j(\tau-\tau';R_0+r_1,R_0+r_2,r_1',r_2',\eta_0+\eta_1,\eta_0+\eta_2)\left(\frac{R_0^2}{2}\left(1+(-1)^j\cos\left(\langle Q\rangle_0(\tau-\tau')\right)\right)+\frac{r_1^2}{4}+\frac{r_2^2}{4}\right.$$
$$\left.+\frac{(r_1+r_2)R_0}{2}\left(1+(-1)^j\cos\left(\langle Q\rangle_0(\tau-\tau')\right)\right)+(-1)^j\frac{r_1r_2}{2}\cos\left(\langle Q\rangle_0(\tau-\tau')\right)+\left\langle\cos\left(\frac{\eta(\tau)}{2}\right)\right\rangle_0^2(\tau-\tau')^2\right)^{-3/2},$$

(6.13)

General expressions for $\Gamma_X(x_1,x_2;\tau-\tau')$ and $\Xi_X(x_1,x_2,y_1,y_2;\tau-\tau')$ (for which, in this case, $X=R,\eta$) are given by Eqs. (A.15) and (A.25) of Appendix A. To approximate further we suppose, again, that $\lambda_R\langle Q\rangle_0 \ll 1$ and $\lambda_\eta\langle Q\rangle_0 \ll 1$. This means that can expand out both $\Gamma_X(x_1,x_2;\tau-\tau')$ and $\Xi_X(x_1,x_2,y_1,y_2;\tau-\tau')$, which is done in Appendix B. This allows us to write (with appropriate rescaling of integrals and assuming that the braid is sufficiently long)

$$\langle Wr_j\rangle_0 \approx L_b\left(W_{j,0}+W_{j,1}+W_{j,2}+W_{j,3}+W_{j,4}\right), \tag{6.14}$$

and

$$W_{j,0} = \frac{1}{\langle Q \rangle_0 R_0} \int_{-\infty}^{\infty} d\eta_1 \int_{-\infty}^{\infty} d\eta_2 \int_{-\infty}^{\infty} dr_1' \int_{-\infty}^{\infty} dr_2' \int_{-\infty}^{\infty} dr_1 \int_{-\infty}^{\infty} dr_2 \int_{-\infty}^{\infty} dx \Gamma_{\eta,0}(\eta_1, \eta_2)$$

$$\tilde{\omega}_j(x \langle Q \rangle_0^{-1}; R_0(r_1+1), R_0(r_2+1), r_1', r_2', \eta_0+\eta_1, \eta_0+\eta_2) \Xi_{R,0,0}(R_0 r_1, R_0 r_2, r_1', r_2') \quad (6.15)$$

$$\left( \frac{1}{2}(1+(-1)^j \cos x) + \frac{r_1^2}{4} + \frac{r_2^2}{4} + \frac{(r_1+r_2)}{2}(1+(-1)^j \cos x) + (-1)^j \frac{r_1 r_2}{2} \cos x + \tilde{P}^2 x^2 \right)^{-3/2},$$

$$W_{j,1} = \frac{1}{\langle Q \rangle_0 R_0} \int_{-\infty}^{\infty} d\eta_1 \int_{-\infty}^{\infty} d\eta_2 \int_{-\infty}^{\infty} dr_1' \int_{-\infty}^{\infty} dr_2' \int_{-\infty}^{\infty} dr_1 \int_{-\infty}^{\infty} dr_2 \int_{-\infty}^{\infty} dx G_\eta\left(x \langle Q \rangle_0^{-1}\right) \Gamma_{\eta,1}(\eta_1, \eta_2)$$

$$\tilde{\omega}_j(x \langle Q \rangle_0^{-1}; R_0(r_1+1), R_0(r_2+1), r_1', r_2', \eta_0+\eta_1, \eta_0+\eta_2) \Xi_{R,0,0}(R_0 r_1, R_0 r_2, r_1', r_2') \quad (6.16)$$

$$\left( \frac{1}{2}(1+(-1)^j \cos x) + \frac{r_1^2}{4} + \frac{r_2^2}{4} + \frac{(r_1+r_2)}{2}(1+(-1)^j \cos x) + (-1)^j \frac{r_1 r_2}{2} \cos x + \tilde{P}^2 x^2 \right)^{-3/2},$$

$$W_{j,2} = \frac{1}{\langle Q \rangle_0 R_0} \int_{-\infty}^{\infty} d\eta_1 \int_{-\infty}^{\infty} d\eta_2 \int_{-\infty}^{\infty} dr_1' \int_{-\infty}^{\infty} dr_2' \int_{-\infty}^{\infty} dr_1 \int_{-\infty}^{\infty} dr_2 \int_{-\infty}^{\infty} dx G_R\left(x \langle Q \rangle_0^{-1}\right) \Gamma_{\eta,0}(\eta_1, \eta_2)$$

$$\tilde{\omega}_j(x \langle Q \rangle_0^{-1}; R_0(r_1+1), R_0(r_2+1), r_1', r_2', \eta_0+\eta_1, \eta_0+\eta_2) \Xi_{R,1,1}(R_0 r_1, R_0 r_2, r_1', r_2') \quad (6.17)$$

$$\left( \frac{1}{2}(1+(-1)^j \cos x) + \frac{r_1^2}{4} + \frac{r_2^2}{4} + \frac{(r_1+r_2)}{2}(1+(-1)^j \cos x) + (-1)^j \frac{r_1 r_2}{2} \cos x + \tilde{P}^2 x^2 \right)^{-3/2},$$

$$W_{j,3} = \frac{1}{\langle Q \rangle_0 R_0} \int_{-\infty}^{\infty} d\eta_1 \int_{-\infty}^{\infty} d\eta_2 \int_{-\infty}^{\infty} dr_1' \int_{-\infty}^{\infty} dr_2' \int_{-\infty}^{\infty} dr_1 \int_{-\infty}^{\infty} dr_2 \int_{-\infty}^{\infty} dx D_R\left(x \langle Q \rangle_0^{-1}\right) \Gamma_{\eta,0}(\eta_1, \eta_2)$$

$$\tilde{\omega}_j(x \langle Q \rangle_0^{-1}; R_0(r_1+1), R_0(r_2+1), r_1', r_2', \eta_0+\eta_1, \eta_0+\eta_2) \Xi_{R,1,2}(R_0 r_1, R_0 r_2, r_1', r_2') \quad (6.18)$$

$$\left( \frac{1}{2}(1+(-1)^j \cos x) + \frac{r_1^2}{4} + \frac{r_2^2}{4} + \frac{(r_1+r_2)}{2}(1+(-1)^j \cos x) + (-1)^j \frac{r_1 r_2}{2} \cos x + \tilde{P}^2 x^2 \right)^{-3/2},$$

$$W_{j,4} = \frac{1}{\langle Q \rangle_0 R_0} \int_{-\infty}^{\infty} d\eta_1 \int_{-\infty}^{\infty} d\eta_2 \int_{-\infty}^{\infty} dr_1' \int_{-\infty}^{\infty} dr_2' \int_{-\infty}^{\infty} dr_1 \int_{-\infty}^{\infty} dr_2 \int_{-\infty}^{\infty} dx C_R\left(x \langle Q \rangle_0^{-1}\right) \Gamma_{\eta,0}(\eta_1, \eta_2)$$

$$\tilde{\omega}_{1,j}(x \langle Q \rangle_0^{-1}; R_0(r_1+1), R_0(r_2+1), r_1', r_2', \eta_0+\eta_1, \eta_0+\eta_2) \Xi_{R,1,3}(R_0 r_1, R_0 r_2, r_1', r_2') \quad (6.19)$$

$$\left( \frac{1}{2}(1+(-1)^j \cos x) + \frac{r_1^2}{4} + \frac{r_2^2}{4} + \frac{(r_1+r_2)}{2}(1+(-1)^j \cos x) + (-1)^j \frac{r_1 r_2}{2} \cos x + \tilde{P}^2 x^2 \right)^{-3/2}.$$

Here, we have that

$$\tilde{P} = \frac{\left\langle \cos\left(\frac{\eta(\tau)}{2}\right) \right\rangle_0}{\langle Q \rangle_0 R_0}. \quad (6.20)$$

Next, we perform an expansion in $r_1'$ and $r_2'$ assuming $\theta_R$ to be small

$$\tilde{\omega}_j(x\langle Q\rangle_0^{-1}; R_0(r_1+1), R_0(r_2+1), r_1', r_2', \eta_0+\eta_1, \eta_0+\eta_2) \approx R_0\tilde{\omega}_{j,0,0}(x;r_1,r_2,\eta_1,\eta_2,\tilde{P},\eta_0)$$
$$+R_0\tilde{\omega}_{j,1,0}(x;r_1,r_2,\eta_1,\eta_2,\tilde{P},\eta_0)r_1' + R_0\tilde{\omega}_{j,0,1}(x;r_1,r_2,\eta_1,\eta_2,\tilde{P},\eta_0)r_2'$$
$$+R_0\tilde{\omega}_{j,2,0}(x;r_1,r_2,\eta_1,\eta_2,\tilde{P},\eta_0)r_1'^2 + R_0\tilde{\omega}_{j,0,2}(x;r_1,r_2,\eta_1,\eta_2,\tilde{P},\eta_0)r_2'^2$$
$$+R_0\tilde{\omega}_{j,1,1}(x;r_1,r_2,\eta_1,\eta_2,\tilde{P},\eta_0)r_1'r_2',$$
(6.21)

where

$$\tilde{\omega}_{j,0,0}(x;r_1,r_2,\eta_1,\eta_2,\tilde{P},\eta_0) = -\left(\frac{(1+r_1)}{2}\sin\left(\frac{\eta_1+\eta_0}{2}\right)\cos\left(\frac{\eta_2+\eta_0}{2}\right) + \frac{(1+r_2)}{2}\cos\left(\frac{\eta_1+\eta_0}{2}\right)\sin\left(\frac{\eta_2+\eta_0}{2}\right)\right)$$
$$-(-1)^j\left(\frac{(1+r_1)}{2}\sin\left(\frac{\eta_2+\eta_0}{2}\right)\cos\left(\frac{\eta_1+\eta_0}{2}\right) + \frac{(1+r_2)}{2}\cos\left(\frac{\eta_2+\eta_0}{2}\right)\sin\left(\frac{\eta_1+\eta_0}{2}\right)\right)\cos x$$
$$-(-1)^j\tilde{P}x\sin x\sin\left(\frac{\eta_1+\eta_0}{2}\right)\sin\left(\frac{\eta_2+\eta_0}{2}\right),$$
(6.22)

$$\tilde{\omega}_{j,1,0}(x;r_1,r_2,\eta_1,\eta_2,\tilde{P},\eta_0) = -(-1)^j\frac{(1+r_2)}{4}\cos\left(\frac{\eta_2+\eta_0}{2}\right)\sin x + (-1)^j\frac{\tilde{P}x}{2}\sin\left(\frac{\eta_2+\eta_0}{2}\right)\cos x,$$
(6.23)

$$\tilde{\omega}_{j,0,1}(x;r_1,r_2,\eta_1,\eta_2,\tilde{P},\eta_0) = (-1)^j\frac{(1+r_1)}{4}\cos\left(\frac{\eta_1+\eta_0}{2}\right)\sin x - (-1)^j\frac{\tilde{P}x}{2}\sin\left(\frac{\eta_1+\eta_0}{2}\right)\cos x$$
(6.24)

$$\tilde{\omega}_{j,2,0}(x;r_1,r_2,\eta_1,\eta_2,\tilde{P},\eta_0) = \frac{\cos\left(\frac{\eta_2+\eta_0}{2}\right)}{\sin\left(\frac{\eta_1+\eta_0}{2}\right)}\left[\frac{(1+r_1)}{16} + (-1)^j\frac{(1+r_2)}{16}\cos x\right]$$
$$+(-1)^j\frac{\tilde{P}x\sin x}{8}\frac{\sin\left(\frac{\eta_2+\eta_0}{2}\right)}{\sin\left(\frac{\eta_1+\eta_0}{2}\right)},$$
(6.25)

$$\tilde{\omega}_{j,0,2}(x;r_1,r_2,\eta_1,\eta_2,\tilde{P},\eta_0) = \frac{\cos\left(\frac{\eta_1+\eta_0}{2}\right)}{\sin\left(\frac{\eta_2+\eta_0}{2}\right)}\left[\frac{(1+r_2)}{16} + (-1)^j\frac{(1+r_1)}{16}\cos x\right]$$
$$+(-1)^j\frac{\tilde{P}x\sin x}{8}\frac{\sin\left(\frac{\eta_1+\eta_0}{2}\right)}{\sin\left(\frac{\eta_2+\eta_0}{2}\right)},$$
(6.26)

$$\tilde{\omega}_{j,1,1}(x;r_1,r_2,\eta_1,\eta_2,\tilde{P},\eta_0) = -(-1)^j \frac{\tilde{P}x\sin x}{4}. \tag{6.27}$$

Then, we can then perform the $r'$ integrations, approximating

$$W_{j,0} \approx \frac{1}{\langle Q \rangle_0} \int_{-\infty}^{\infty} d\eta_1 \int_{-\infty}^{\infty} d\eta_2 \int_{-\infty}^{\infty} dr_1 \int_{-\infty}^{\infty} dr_2 \int_{-\infty}^{\infty} dx \, \Gamma_{\eta,0}(\eta_1,\eta_2) \Gamma_{R,0}(R_0 r_1, R_0 r_2)$$
$$\left( \tilde{\omega}_{j,0,0}(x;r_1,r_2,\eta_1,\eta_2,\tilde{P},\eta_0) + \theta_R^2 \left( \tilde{\omega}_{j,2,0}(x;r_1,r_2,\eta_1,\eta_2,\tilde{P},\eta_0) + \tilde{\omega}_{j,0,2}(x;r_1,r_2,\eta_1,\eta_2,\tilde{P},\eta_0) \right) \right)$$
$$\left( \frac{1}{2}\left(1+(-1)^j \cos x\right) + \frac{r_1^2}{4} + \frac{r_2^2}{4} + \frac{(r_1+r_2)}{2}\left(1+(-1)^j \cos x\right) + (-1)^j \frac{r_1 r_2}{2}\cos x + \tilde{P}^2 x^2 \right)^{-3/2},$$

$$\tag{6.28}$$

$$W_{j,1} \approx \frac{1}{\langle Q \rangle_0} \int_{-\infty}^{\infty} d\eta_1 \int_{-\infty}^{\infty} d\eta_2 \int_{-\infty}^{\infty} dr_1 \int_{-\infty}^{\infty} dr_2 \int_{-\infty}^{\infty} dx \, \Gamma_{\eta,1}(\eta_1,\eta_2) \Gamma_{R,0}(R_0 r_1, R_0 r_2) G_\eta\left(x \langle Q \rangle_0^{-1}\right)$$
$$\left( \tilde{\omega}_{j,0,0}(x;r_1,r_2,\eta_1,\eta_2,\tilde{P},\eta_0) + \theta_R^2 \left( \tilde{\omega}_{j,2,0}(x;r_1,r_2,\eta_1,\eta_2,\tilde{P},\eta_0) + \tilde{\omega}_{j,0,2}(x;r_1,r_2,\eta_1,\eta_2,\tilde{P},\eta_0) \right) \right)$$
$$\left( \frac{1}{2}\left(1+(-1)^j \cos x\right) + \frac{r_1^2}{4} + \frac{r_2^2}{4} + \frac{(r_1+r_2)}{2}\left(1+(-1)^j \cos x\right) + (-1)^j \frac{r_1 r_2}{2}\cos x + \tilde{P}^2 x^2 \right)^{-3/2},$$

$$\tag{6.29}$$

$$W_{j,2} \approx \frac{1}{\langle Q \rangle_0} \int_{-\infty}^{\infty} d\eta_1 \int_{-\infty}^{\infty} d\eta_2 \int_{-\infty}^{\infty} dr_1 \int_{-\infty}^{\infty} dr_2 \int_{-\infty}^{\infty} dx \, G_R\left(x \langle Q \rangle_0^{-1}\right) \Gamma_{\eta,0}(\eta_1,\eta_2) \Gamma_{R,1}(R_0 r_1, R_0 r_2)$$
$$\left( \tilde{\omega}_{j,0,0}(x;r_1,r_2,\eta_1,\eta_2,\tilde{P},\eta_0) + \theta_R^2 \left( \tilde{\omega}_{j,2,0}(x;r_1,r_2,\eta_1,\eta_2,\tilde{P},\eta_0) + \tilde{\omega}_{j,0,2}(x;r_1,r_2,\eta_1,\eta_2,\tilde{P},\eta_0) \right) \right)$$
$$\left( \frac{1}{2}\left(1+(-1)^j \cos x\right) + \frac{r_1^2}{4} + \frac{r_2^2}{4} + \frac{(r_1+r_2)}{2}\left(1+(-1)^j \cos x\right) + (-1)^j \frac{r_1 r_2}{2}\cos x + \tilde{P}^2 x^2 \right)^{-3/2},$$

$$\tag{6.30}$$

$$W_{j,3} \approx \frac{1}{\langle Q \rangle_0} \int_{-\infty}^{\infty} d\eta_1 \int_{-\infty}^{\infty} d\eta_2 \int_{-\infty}^{\infty} dr_1 \int_{-\infty}^{\infty} dr_2 \int_{-\infty}^{\infty} dx \, D_R\left(x \langle Q \rangle_0^{-1}\right) \Gamma_{\eta,0}(\eta_1,\eta_2)$$
$$\tilde{\omega}_{j,1,1}(x;r_1,r_2,\eta_1,\eta_2,\tilde{P},\eta_0) \Gamma_{R,0}(R_0 r_1, R_0 r_2) \tag{6.31}$$
$$\left( \frac{1}{2}\left(1+(-1)^j \cos x\right) + \frac{r_1^2}{4} + \frac{r_2^2}{4} + \frac{(r_1+r_2)}{2}\left(1+(-1)^j \cos x\right) + (-1)^j \frac{r_1 r_2}{2}\cos x + \tilde{P}^2 x^2 \right)^{-3/2},$$

$$W_{j,4} \approx \frac{R_0}{\langle Q \rangle_0 d_R^2} \int_{-\infty}^{\infty} d\eta_1 \int_{-\infty}^{\infty} d\eta_2 \int_{-\infty}^{\infty} dr_1 \int_{-\infty}^{\infty} dr_2 \int_{-\infty}^{\infty} dx\, C_R\left(x\langle Q\rangle_0^{-1}\right) \Gamma_{\eta,0}(\eta_1,\eta_2) \Gamma_{R,0}(R_0 r_1, R_0 r_2)$$

$$\left(r_1 \tilde{\omega}_{j,0,1}(x;r_1,r_2,\eta_1,\eta_2,\tilde{P},\eta_0) - r_2 \tilde{\omega}_{j,1,0}(x;r_1,r_2,\eta_1,\eta_2,\tilde{P},\eta_0)\right) \quad (6.32)$$

$$\left(\frac{1}{2}(1+(-1)^j \cos x) + \frac{r_1^2}{4} + \frac{r_2^2}{4} + \frac{(r_1+r_2)}{2}(1+(-1)^j \cos x) + (-1)^j \frac{r_1 r_2}{2}\cos x + \tilde{P}^2 x^2\right)^{-3/2}.$$

We further approximate by expanding out for small $\eta_1$ and $\eta_2$. Also, with $W_{j,1}$, $W_{j,2}$, $W_{j,3}$ and $W_{j,4}$ we expand everything in both numerator and denominator in powers of $x$ expect for the correlation functions $G_\eta\left(x\langle Q\rangle_0^{-1}\right)$, $G_R\left(x\langle Q\rangle_0^{-1}\right)$, $D_R\left(x\langle Q\rangle_0^{-1}\right)$ and $C_R\left(x\langle Q\rangle_0^{-1}\right)$. This approximation is again valid provided that $\lambda_R \langle Q\rangle_0 \ll 1$ and $\lambda_\eta \langle Q\rangle_0 \ll 1$. We have neglected the $\theta_R^2$ correction in $W_{j,1}$ and $W_{j,2}$. Thus, we obtain (where we have changed the $r_1$ and $r_2$ integrations into polar coordinates)

$$W_{j,0} \approx -\frac{1}{2\pi d_R^2 \langle Q\rangle_0} \int_0^{2\pi} d\phi_r \int_0^{\infty} r\, dr \int_{-\infty}^{\infty} dx \exp\left(-\frac{R_0^2 r^2}{2 d_R^2}\right)$$

$$\left[\left(\sin\left(\frac{\eta_0}{2}\right)\cos\left(\frac{\eta_0}{2}\right)(1+(-1)^j \cos x)(1+r\sin\phi_r) + (-1)^j \tilde{P} \sin^2\left(\frac{\eta_0}{2}\right) x \sin x\right)\left(1 - \frac{d_\eta^2}{4}\right)\right.$$

$$\left. + \theta_R^2\left(-\frac{1}{8}\cot\left(\frac{\eta_0}{2}\right)(1+(-1)^j \cos x)(1+r\sin\phi_r) - \frac{(-1)^j}{4}\tilde{P} x \sin x\right)\right]$$

$$\left(\frac{1}{2}(1+(-1)^j \cos(x)) + \frac{r^2}{4} + \frac{r(\cos\phi_r + \sin\phi_r)}{2}(1+(-1)^j \cos(x)) + (-1)^j \frac{r^2 \sin\phi_r \cos\phi_r}{2}\cos(x) + \tilde{P}^2 x^2\right)^{-3/2}$$

(6.33)

$$W_{j,1} \approx \frac{1}{2\pi d_R^2 \langle Q\rangle_0} \int_0^{2\pi} d\phi_r \int_0^{\infty} r\, dr \int_{-\infty}^{\infty} dx\, G_\eta\left(x\langle Q\rangle_0^{-1}\right) \exp\left(-\frac{R_0^2 r^2}{2 d_R^2}\right)$$

$$\left(\frac{1}{4}\cos\left(\frac{\eta_0}{2}\right)\sin\left(\frac{\eta_0}{2}\right)\left(2\delta_{j,2} - (-1)^j \left(\frac{x^2}{2}\right)\right)(1+r\sin\phi_r) - \frac{\tilde{P} x^2}{4}(-1)^j \cos^2\left(\frac{\eta_0}{2}\right)\right) \quad (6.34)$$

$$\left(\delta_{j,2} + \frac{r^2}{4} + r(\cos\phi_r + \sin\phi_r)\delta_{j,2} + (-1)^j \frac{r^2}{2}\cos\phi_r \sin\phi_r\right.$$

$$\left. + x^2\left(\tilde{P}^2 - \frac{(-1)^j}{4} - \frac{r(\cos\phi_r + \sin\phi_r)(-1)^j}{4} - \frac{(-1)^j r^2 \cos\phi_r \sin\phi_r}{4}\right)\right)^{-3/2},$$

$$W_{j,2} \approx -\frac{R_0^2}{2\pi d_R^6 \langle Q\rangle_0} \int_0^{2\pi} d\phi_r \int_0^\infty r\, dr \int_{-\infty}^\infty dx \exp\left(-\frac{R_0^2 r^2}{2d_R^2}\right) G_R\left(x\langle Q\rangle_0^{-1}\right)$$

$$\left[\left(\sin\left(\frac{\eta_0}{2}\right)\cos\left(\frac{\eta_0}{2}\right)\left(2\delta_{j,2}-(-1)^j\left(\frac{x^2}{2}\right)\right)+(-1)^j \tilde{P}\sin^2\left(\frac{\eta_0}{2}\right)x^2\right)r^2 \sin\phi_r \cos\phi_r\right.$$

$$\left.+\left(\sin\left(\frac{\eta_0}{2}\right)\cos\left(\frac{\eta_0}{2}\right)\left(2\delta_{j,2}-(-1)^j\left(\frac{x^2}{2}\right)\right)\right)r^3 \sin^2\phi_r \cos\phi_r\right]$$

$$\left(\delta_{j,2}+\frac{r^2}{4}+r(\cos\phi_r+\sin\phi_r)\delta_{j,2}+(-1)^j\frac{r^2}{2}\cos\phi_r\sin\phi_r\right.$$

$$\left.+x^2\left(\tilde{P}^2-\frac{(-1)^j}{4}-\frac{r(\cos\phi_r+\sin\phi_r)(-1)^j}{4}-\frac{(-1)^j r^2 \cos\phi_r \sin\phi_r}{4}\right)\right)^{-3/2},$$

(6.35)

$$W_{j,3} \approx -\frac{1}{8\pi d_R^2 \langle Q\rangle_0} \int_0^{2\pi} d\phi_r \int_0^\infty r\, dr \int_{-\infty}^\infty dx \exp\left(-\frac{R_0^2 r^2}{2d_R^2}\right) D_R\left(x\langle Q\rangle_0^{-1}\right)(-1)^j \tilde{P} x^2$$

$$\left(\delta_{j,2}+\frac{r^2}{4}+r(\cos\phi_r+\sin\phi_r)\delta_{j,2}+(-1)^j\frac{r^2}{2}\cos\phi_r\sin\phi_r\right.$$

$$\left.+x^2\left(\tilde{P}^2-\frac{(-1)^j}{4}-\frac{r(\cos\phi_r+\sin\phi_r)(-1)^j}{4}-\frac{(-1)^j r^2 \cos\phi_r \sin\phi_r}{4}\right)\right)^{-3/2},$$

(6.36)

$$W_{j,4} \approx \frac{R_0}{4\pi \langle Q\rangle_0 d_R^4} \int_0^{2\pi} d\phi_r \int_0^\infty r\, dr \int_{-\infty}^\infty dx\, C_R\left(x\langle Q\rangle_0^{-1}\right)\exp\left(-\frac{R_0^2 r^2}{2d_R^2}\right)(-1)^j$$

$$\left[r\sin\phi_r\left[\cos\left(\frac{\eta_0}{2}\right)x-2\tilde{P}\sin\left(\frac{\eta_0}{2}\right)x\right]+r^2\sin^2\phi_r \cos\left(\frac{\eta_0}{2}\right)x\right]$$

$$\left(\delta_{j,2}+\frac{r^2}{4}+r(\cos\phi_r+\sin\phi_r)\delta_{j,2}+(-1)^j\frac{r^2}{2}\cos\phi_r\sin\phi_r\right.$$

$$\left.+x^2\left(\tilde{P}^2-\frac{(-1)^j}{4}-\frac{r(\cos\phi_r+\sin\phi_r)(-1)^j}{4}-\frac{(-1)^j r^2 \cos\phi_r \sin\phi_r}{4}\right)\right)^{-3/2}.$$

(6.37)

Now, we suppose that $d_R/R_0 \ll 1$. Then, in Eqs. (6.34)-(6.37), we can approximate these expressions further by expanding out the $r$ dependent terms that multiply $x^2$ in the denominator, and so we can rescale the $x$ integrations by $\left(\tilde{P}^2-(-1)^j/4\right)^{-1/2}$. Thus, this expansion represents a power series expansion in $\left(\tilde{P}^2-(-1)^j/4\right)^{-1}$, for which we retain only terms up to 1$^{st}$ order in $\left(\tilde{P}^2-(-1)^j/4\right)^{-1}$. We can also represent the correlation functions as

$$G_\eta\left(x\langle Q\rangle_0^{-1}\left(\tilde{P}^2-\frac{(-1)^j}{4}\right)^{-1/2}\right)=\frac{2\hat{G}_\eta\left(x,\tilde{\alpha}_{\eta,j}\right)}{l_p \langle Q\rangle_0 \left(\tilde{P}^2-\frac{(-1)^j}{4}\right)^{1/2}},$$

(6.38)

$$G_R\left(x\langle Q\rangle_0^{-1}\left(\tilde{P}^2-\frac{(-1)^j}{4}\right)^{-1/2}\right)=\frac{2\hat{G}_R(x,\tilde{\alpha}_{R,j},\tilde{\gamma})}{l_p\langle Q\rangle_0^3\left(\tilde{P}^2-\frac{(-1)^j}{4}\right)^{3/2}}, \tag{6.39}$$

$$C_R\left(x\langle Q\rangle_0^{-1}\left(\tilde{P}^2-\frac{(-1)^j}{4}\right)^{-1/2}\right)=\frac{2\hat{C}_R(x,\tilde{\alpha}_{R,j},\tilde{\gamma})}{l_p\langle Q\rangle_0^2\left(\tilde{P}^2-\frac{(-1)^j}{4}\right)}, \tag{6.40}$$

$$D_R\left(x\langle Q\rangle_0^{-1}\left(\tilde{P}^2-\frac{(-1)^j}{4}\right)^{-1/2}\right)=\frac{2\bar{D}_R(x,\tilde{\alpha}_{R,j},\tilde{\gamma})}{l_p\langle Q\rangle_0\left(\tilde{P}^2-\frac{(-1)^j}{4}\right)^{1/2}}, \tag{6.41}$$

where we have defined the rescaled variables

$$\tilde{\alpha}_{\eta,j}=\frac{2\langle Q\rangle_0^{-2}}{l_p}\left(\tilde{P}^2-\frac{(-1)^j}{4}\right)^{-1}\alpha_\eta=\frac{2R_0^2}{l_p\cos^2\left(\frac{\eta_0}{2}\right)}\left(1-\frac{(-1)^j}{4\tilde{P}^2}\right)^{-1}\left(\alpha_\eta+\frac{1}{4}\left(\frac{\alpha_\eta}{2l_p}\right)^{1/2}\right), \tag{6.42}$$

$$\tilde{\gamma}=\sqrt{2}\beta_R l_p^{-1/2}\alpha_R^{-1/2}=d_R^2 l_p^{-2}\theta_R^{-6}-2, \tag{6.43}$$

$$\tilde{\alpha}_{R,j}=\frac{2\langle Q\rangle_0^{-4}}{l_p}\left(\tilde{P}^2-\frac{(-1)^j}{4}\right)^{-2}\alpha_R=\frac{2R_0^4\alpha_R}{l_p\cos^4\left(\frac{\eta_0}{2}\right)}\left(1-\frac{(-1)^j}{4\tilde{P}^2}\right)^{-2}\left(1+\frac{1}{2}\left(\frac{1}{2l_p\alpha_\eta}\right)^{1/2}\right). \tag{6.44}$$

Thus, we are able to rewrite Eqs. (6.34)-(6.37) as

$$W_{j,1}(\tilde{\alpha}_{\eta,j},\tilde{R}_0,\tilde{Q}_j,\tilde{P},\eta_0)\approx\frac{1}{4\pi\tilde{Q}_j^2 l_p}\int_0^\infty rdr\int_{-\infty}^\infty dx\exp\left(-\frac{\tilde{R}_0^2 r^2}{2}\right)\hat{G}(x,\tilde{\alpha}_{\eta,j})\left[\cos\left(\frac{\eta_0}{2}\right)\sin\left(\frac{\eta_0}{2}\right)\right.$$
$$\left(2\delta_{j,2}-(-1)^j\left(\tilde{P}^2-\frac{(-1)^j}{4}\right)^{-1}\left(\frac{x^2}{2}\right)\right)(J_{j,0,0}(x,r)+J_{j,1,0}(x,r))$$
$$\left.-\tilde{P}x^2\left(\tilde{P}^2-\frac{(-1)^j}{4}\right)^{-1}(-1)^j\cos^2\left(\frac{\eta_0}{2}\right)J_{j,0,0}(x,r)\right]+\delta_{j,2}\delta\bar{W}_1(\tilde{\alpha}_{\eta,j},\tilde{R}_0,\tilde{Q}_j,\tilde{P},\eta_0),$$

$$\tag{6.45}$$

$$W_{j,2}(\tilde{\alpha}_{R,j},\tilde{\gamma},\tilde{R}_0,\tilde{Q}_j,\tilde{P},\eta_0) \approx -\frac{\tilde{R}_0^2}{\pi \tilde{Q}_j^4 l_p} \int_0^\infty rdr \int_{-\infty}^\infty dx \exp\left(-\frac{\tilde{R}_0^2 r^2}{2}\right) \hat{G}_R(x,\tilde{\alpha}_{R,j},\tilde{\gamma})$$

$$\left[\left(\sin\left(\frac{\eta_0}{2}\right)\cos\left(\frac{\eta_0}{2}\right)\left(2\delta_{j,2}-(-1)^j\left(\tilde{P}^2-\frac{(-1)^j}{4}\right)^{-1}\left(\frac{x^2}{2}\right)\right)+(-1)^j \tilde{P}\left(\tilde{P}^2-\frac{(-1)^j}{4}\right)^{-1}\sin^2\left(\frac{\eta_0}{2}\right)x^2\right)J_{j,1,1}(x,r)\right.$$

$$\left.+\left(\sin\left(\frac{\eta_0}{2}\right)\cos\left(\frac{\eta_0}{2}\right)\left(2\delta_{j,2}-(-1)^j\left(\tilde{P}^2-\frac{(-1)^j}{4}\right)^{-1}\left(\frac{x^2}{2}\right)\right)\right)J_{j,2,1}(x,r)\right] + \delta_{k,2}\delta\overline{W}_2(\tilde{\alpha}_{R,j},\tilde{\gamma},\tilde{R}_0,\tilde{Q}_j,\tilde{P},\eta_0),$$

(6.46)

$$W_{j,3}(\tilde{\alpha}_{R,j},\tilde{\gamma},\tilde{R}_0,\tilde{Q}_j,\tilde{P},\eta_0) \approx$$
$$-\frac{1}{4\pi\tilde{Q}_j^2 l_p}\int_0^\infty rdr \int_{-\infty}^\infty dx \exp\left(-\frac{\tilde{R}_0^2 r^2}{2}\right)\hat{D}_R(x,\tilde{\alpha}_{R,j},\tilde{\gamma})(-1)^j \tilde{P}\left(\tilde{P}^2-\frac{(-1)^j}{4}\right)^{-1} x^2 J_{j,0,0}(x,r),$$

(6.47)

$$W_{j,4}(\tilde{\alpha}_{R,j},\tilde{\gamma},\tilde{R}_0,\tilde{Q}_j,\tilde{P},\eta_0) \approx \frac{\tilde{R}_0}{2\pi\tilde{Q}_j^3 l_p}\left(\tilde{P}^2-\frac{(-1)^j}{4}\right)^{-1/2}\int_0^\infty rdr \int_{-\infty}^\infty dx \hat{C}_R(x,\tilde{\alpha}_{R,j},\tilde{\gamma})\exp\left(-\frac{\tilde{R}_0^2 r^2}{2}\right)$$

$$(-1)^j x\left[\left[\cos\left(\frac{\eta_0}{2}\right)-2\tilde{P}\sin\left(\frac{\eta_0}{2}\right)\right]J_{j,1,0}(x,r)+\cos\left(\frac{\eta_0}{2}\right)J_{j,2,0}(x,r)\right],$$

(6.48)

$$\delta\overline{W}_1(\tilde{\alpha}_{\eta,j},\tilde{R}_0,\tilde{Q}_j,\tilde{P},\eta_0) = \frac{3\cos\left(\frac{\eta_0}{2}\right)\sin\left(\frac{\eta_0}{2}\right)}{16\pi\tilde{Q}_2^2 l_p}\left(\tilde{P}^2-\frac{1}{4}\right)^{-1}\int_0^\infty rdr \int_{-\infty}^\infty x^2 dx \exp\left(-\frac{\tilde{R}_0^2 r^2}{2}\right)\hat{G}_\eta(x,\tilde{\alpha}_{\eta,j})$$

$$\left(H_{2,1}(x,r)+H_{2,0}(x,r)+2H_{1,1}(x,r)+H_{0,1}(x,r)+H_{1,0}(x,r)\right),$$

(6.49)

$$\delta\overline{W}_2(\tilde{\alpha}_{R,j},\tilde{\gamma},\tilde{R}_0,\tilde{Q}_j,\tilde{P},\eta_0) = -\frac{3\cos\left(\frac{\eta_0}{2}\right)\sin\left(\frac{\eta_0}{2}\right)\tilde{R}_0^2}{4\pi\tilde{Q}_2^4 l_p}\left(\tilde{P}^2-\frac{1}{4}\right)^{-1}\int_0^\infty rdr \int_{-\infty}^\infty x^2 dx \exp\left(-\frac{\tilde{R}_0^2 r^2}{2}\right)\hat{G}_R(x,\tilde{\alpha}_{R,j},\tilde{\gamma})$$

$$\left(H_{1,2}(x,r)+H_{2,1}(x,r)+2H_{2,2}(x,r)+H_{3,2}(x,r)+H_{3,1}(x,r)\right),$$

(6.50)

where $\tilde{R}_0 = R_0/d_R$

$$\tilde{Q}_j = d_R \langle Q \rangle_0 \left( \tilde{P}^2 - \frac{(-1)^j}{4} \right)^{1/2} \approx \frac{1}{\tilde{R}_0} \cos\left(\frac{\eta_0}{2}\right) \left( 1 - \frac{1}{8(2\alpha_\eta l_p)^{1/2}} \right) \left( 1 - \frac{(-1)^j}{4\tilde{P}^2} \right)^{1/2} \tag{6.51}$$

and the functions $J_{j,k,l}(x,r)$ and $H_{k,l}(x,r)$ are defined as

$$J_{j,k,l}(x,r) = r^{k+l} \int_0^{2\pi} d\phi_R \sin^k \phi_r \cos^l \phi_r \left( \delta_{j,2} + \frac{r^2}{4}\left(1 + (-1)^j \sin 2\phi_r\right) + r(\cos\phi_r + \sin\phi_r)\delta_{j,2} + x^2 \right)^{-3/2}. \tag{6.52}$$

$$H_{k,l}(x,r) = r^{k+l} \int_0^{2\pi} d\phi_R \sin^k \phi_r \cos^l \phi_r \left( 1 + \frac{r^2}{4}\left(1 + \sin 2\phi_r\right) + r(\cos\phi_r + \sin\phi_r) + x^2 \right)^{-5/2}. \tag{6.53}$$

As always $r<1$, we need to perform a power series expansion in $r$ for the functions $J_{2,k,l}(x,r)$ and $H_{k,l}(x,r)$ to prevent unphysical singularities arising from $r>1$ that cause the r and x-integrals to diverge. These unphysical divergences arise from not putting the cut-off $d_{\min}$ on the r-integrals, as was discussed for the other terms in Section 4. However, in the interests of simplicity, if we can indeed assume $\tilde{R}_0 \gg 1$, we can simply perform the expansion in $r$. Physically, what should insure this is the presence of sufficiently strong interactions between the braided segments in the plectoneme. Expressions for terms in the expansion of $J_{2,k,l}(x,r)$ and $H_{k,l}(x,r)$ can be found in Appendix D. The integrations $r$ over can then easily be performed. Thus we obtain

$$W_{2,1}(\tilde{\alpha}_{\eta,2}, \tilde{R}_0, \tilde{Q}_2, \tilde{P}, \eta_0) \approx \frac{1}{\tilde{Q}_2^2 l_p} \cos\left(\frac{\eta_0}{2}\right)\sin\left(\frac{\eta_0}{2}\right) \left[ \frac{2A_{3,0}(\tilde{\alpha}_{\eta,j})}{\tilde{R}_0^2} - \frac{9A_{5,0}(\tilde{\alpha}_{\eta,j})}{2\tilde{R}_0^4} \right.$$
$$+ A_{7,0}(\tilde{\alpha}_{\eta,2})\left[ \frac{15}{2\tilde{R}_0^4} + \frac{225}{16\tilde{R}_0^6} \right] - \frac{525 A_{9,0}(\tilde{\alpha}_{\eta,2})}{8\tilde{R}_0^6} + \frac{945 A_{11,0}(\tilde{\alpha}_{\eta,2})}{16\tilde{R}_0^6} \right]$$
$$- \frac{1}{\tilde{Q}_2^2 l_p}\left( \tilde{P}^2 - \frac{1}{4} \right)^{-1} \left[ \frac{A_{3,2}(\tilde{\alpha}_{\eta,j})}{\tilde{R}_0^2} \left( \frac{1}{2}\cos\left(\frac{\eta_0}{2}\right)\sin\left(\frac{\eta_0}{2}\right) + \tilde{P}\cos^2\left(\frac{\eta_0}{2}\right) \right) \right.$$
$$+ \frac{A_{5,2}(\tilde{\alpha}_{\eta,2})}{\tilde{R}_0^4}\left( \frac{15}{8}\cos\left(\frac{\eta_0}{2}\right)\sin\left(\frac{\eta_0}{2}\right) + \frac{3\tilde{P}}{4}\cos^2\left(\frac{\eta_0}{2}\right) \right)$$
$$+ A_{7,2}(\tilde{\alpha}_{\eta,2})\left( \tilde{P}\cos^2\left(\frac{\eta_0}{2}\right)\left( \frac{45}{32\tilde{R}_0^6} + \frac{15}{4\tilde{R}_0^4} \right) + \cos\left(\frac{\eta_0}{2}\right)\sin\left(\frac{\eta_0}{2}\right)\left( \frac{45}{8\tilde{R}_0^4} + \frac{585}{64\tilde{R}_0^6} \right) \right)$$
$$- \frac{A_{9,2}(\tilde{\alpha}_{\eta,2})}{\tilde{R}_0^6}\left( \frac{1995}{32}\cos\left(\frac{\eta_0}{2}\right)\sin\left(\frac{\eta_0}{2}\right) + \frac{315}{16}\tilde{P}\cos^2\left(\frac{\eta_0}{2}\right) \right)$$
$$+ \frac{945 A_{11,2}(\tilde{\alpha}_{\eta,2})}{32\tilde{R}_0^6}\left( \frac{5}{2}\cos\left(\frac{\eta_0}{2}\right)\sin\left(\frac{\eta_0}{2}\right) + \tilde{P}\cos^2\left(\frac{\eta_0}{2}\right) \right) \right], \tag{6.54}$$

$$W_{2,2}(\tilde{\alpha}_{R,2},\tilde{\gamma},\tilde{R}_0,\tilde{Q}_2,\tilde{P},\eta_0) \approx -\frac{1}{\tilde{Q}_2^4 l_p}\sin\left(\frac{\eta_0}{2}\right)\cos\left(\frac{\eta_0}{2}\right)\left[-\frac{9B_{5,0}(\tilde{\alpha}_{R,2},\tilde{\gamma})}{\tilde{R}_0^4}\right.$$

$$+B_{7,0}(\tilde{\alpha}_{R,2},\tilde{\gamma})\left(\frac{15}{\tilde{R}_0^4}+\frac{225}{4\tilde{R}_0^6}\right)-\frac{525 B_{9,0}(\tilde{\alpha}_{R,2},\tilde{\gamma})}{2\tilde{R}_0^6}+\frac{945 B_{11,0}(\tilde{\alpha}_{R,2},\tilde{\gamma})}{4\tilde{R}_0^6}\right]$$

$$-\frac{\left(\tilde{P}^2-\frac{1}{4}\right)^{-1}}{\tilde{Q}_2^2 l_p}\left[\frac{B_{5,2}(\tilde{\alpha}_{R,2},\tilde{\gamma})}{\tilde{R}_0^4}\left(\frac{21}{4}\sin\left(\frac{\eta_0}{2}\right)\cos\left(\frac{\eta_0}{2}\right)-\frac{3\tilde{P}}{2}\sin^2\left(\frac{\eta_0}{2}\right)\right)\right.$$

$$+B_{7,2}(\tilde{\alpha}_{R,2},\tilde{\gamma})\left[\tilde{P}\sin^2\left(\frac{\eta_0}{2}\right)\left(\frac{15}{2\tilde{R}_0^4}+\frac{45}{8\tilde{R}_0^6}\right)-\left(\frac{715}{16\tilde{R}_0^6}+\frac{45}{4\tilde{R}_0^4}\right)\sin\left(\frac{\eta_0}{2}\right)\cos\left(\frac{\eta_0}{2}\right)\right],$$

$$-\frac{B_{9,2}(\tilde{\alpha}_{R,2},\tilde{\gamma})}{\tilde{R}_0^6}\left(\frac{315\tilde{P}}{8}\sin^2\left(\frac{\eta_0}{2}\right)-\frac{525}{2}\sin\left(\frac{\eta_0}{2}\right)\cos\left(\frac{\eta_0}{2}\right)\right)$$

$$+\frac{B_{11,2}(\tilde{\alpha}_{R,2},\tilde{\gamma})}{\tilde{R}_0^6}\left(\frac{945}{8}\tilde{P}\sin^2\left(\frac{\eta_0}{2}\right)-\frac{4725}{16}\sin\left(\frac{\eta_0}{2}\right)\cos\left(\frac{\eta_0}{2}\right)\right)\right]$$

(6.55)

$$W_{2,3}(\tilde{\alpha}_{R,2},\tilde{\gamma},\tilde{R}_0,\tilde{Q}_2,\tilde{P},\eta_0) \approx -\frac{\tilde{P}}{\tilde{Q}_2^2 l_p}\left(\tilde{P}^2-\frac{1}{4}\right)^{-1}\left[\frac{C_{3,2}(\tilde{\alpha}_{R,2},\tilde{\gamma})}{\tilde{R}_0^2}-\frac{3C_{5,2}(\tilde{\alpha}_{R,2},\tilde{\gamma})}{4\tilde{R}_0^4}\right.$$

$$+C_{7,2}(\tilde{\alpha}_{R,2},\tilde{\gamma})\left[\frac{45}{32\tilde{R}_0^6}+\frac{15}{4\tilde{R}_0^4}\right]-\frac{315 C_{9,2}(\tilde{\alpha}_{R,2},\tilde{\gamma})}{16\tilde{R}_0^6}+\frac{945 C_{11,2}(\tilde{\alpha}_{R,2},\tilde{\gamma})}{32\tilde{R}_0^6}\right],$$

(6.56)

$$W_{2,4}(\tilde{\alpha}_{R,2},\tilde{\gamma},\tilde{R}_0,\tilde{Q}_2,\tilde{P},\eta_0) \approx \frac{1}{\tilde{Q}_2^3 l_p}\left(\tilde{P}^2-\frac{1}{4}\right)^{-1/2}\left[\frac{2D_{3,1}(\tilde{\alpha}_{R,2},\tilde{\gamma})}{\tilde{R}_0^3}\cos\left(\frac{\eta_0}{2}\right)\right.$$

$$+D_{5,1}(\tilde{\alpha}_{R,2},\tilde{\gamma})\left(\frac{3}{\tilde{R}_0^3}\left(2\tilde{P}\sin\left(\frac{\eta_0}{2}\right)-\cos\left(\frac{\eta_0}{2}\right)\right)-\frac{3}{\tilde{R}_0^5}\cos\left(\frac{\eta_0}{2}\right)\right)$$

$$+\frac{D_{7,1}(\tilde{\alpha}_{R,2},\tilde{\gamma})}{\tilde{R}_0^5}\left(\frac{105}{4}\cos\left(\frac{\eta_0}{2}\right)-\frac{45}{2}\tilde{P}\sin\left(\frac{\eta_0}{2}\right)\right)-\frac{D_{9,1}(\tilde{\alpha}_{R,2},\tilde{\gamma})}{\tilde{R}_0^5}\left(\frac{105}{4}\cos\left(\frac{\eta_0}{2}\right)-\frac{105}{2}\tilde{P}\sin\left(\frac{\eta_0}{2}\right)\right)\right],$$

(6.57)

Where we define the functions

$$\int_0^\infty dx \frac{x^m \hat{G}_\eta(x,\tilde{\alpha}_\eta)}{(1+x^2)^{n/2}} = A_{n,m}(\tilde{\alpha}_\eta), \qquad 2\int_0^\infty dx \frac{x^m \hat{G}_R(x,\tilde{\alpha}_\eta,\tilde{\gamma})}{(1+x^2)^{n/2}} = B_{n,m}(\tilde{\alpha}_\eta,\tilde{\gamma}),$$

(6.58)

$$\int_0^\infty dx \frac{x^m \hat{D}_R(x,\tilde{\alpha}_\eta,\tilde{\gamma})}{(1+x^2)^{n/2}} = C_{n,m}(\tilde{\alpha}_\eta,\tilde{\gamma}), \qquad \int_0^\infty dx \frac{x^m \hat{C}_R(x,\tilde{\alpha}_\eta,\tilde{\gamma})}{(1+x^2)^{n/2}} = D_{n,m}(\tilde{\alpha}_\eta,\tilde{\gamma}).$$

(6.59)

However, for $J_{1,k,l}(x,r)$ the power series in $r$ is not convergent over the x-integration, and cannot be performed, however there are no problems arising from $r>1$. The unphysical contribution is negligible, again provided that $\tilde{R}_0 \gg 1$. Thus, we leave the $J_{1,k,l}(x,r)$ untouched, and for these functions the angular integrals can be performed analytically. This yields

$$J_{1,0,0}(x,r) = \frac{4}{x^2} \frac{1}{\sqrt{x^2 + \frac{r^2}{2}}} \left[ E\left( \arcsin\left( \sqrt{\frac{r^2 + 2x^2}{r^2 + 4x^2}} \right), \frac{r^2}{r^2 + 2x^2} \right) + E\left( \frac{\pi}{4}, \frac{r^2}{r^2 + 2x^2} \right) \right]$$

$$- \frac{r^2}{x^2} \frac{1}{\left(x^2 + \frac{r^2}{2}\right)\left(x^2 + \frac{r^2}{4}\right)^{1/2}},$$

(6.60)

$$J_{1,1,0}(x,r) = 0, \tag{6.61}$$

$$J_{1,2,0}(x,r) = \frac{r^2}{2} J_{1,0,0}(x,r), \tag{6.62}$$

$$J_{1,1,1}(x,r) = \frac{1}{2}\left(r^2 + 4x^2\right) J_{1,0,0}(x,r) - \frac{8}{\sqrt{x^2 + \frac{r^2}{2}}} \left[ F\left( \arcsin\left( \sqrt{\frac{r^2 + 2x^2}{r^2 + 4x^2}} \right), \frac{r^2}{r^2 + 2x^2} \right) + F\left( \frac{\pi}{4}, \frac{r^2}{r^2 + 2x^2} \right) \right],$$

(6.63)

$$J_{1,2,1}(x,r) = 0. \tag{6.64}$$

The x-integrals and r-integrals can further be rescaled to yield

$$W_{1,1}(\bar{\alpha}_\eta, \tilde{R}_0, \tilde{Q}_1, \tilde{P}, \eta_0) \approx -\frac{1}{4\pi \tilde{Q}_1^2 l_p \tilde{R}_0^3} \left( \tilde{P}^2 + \frac{1}{4} \right)^{-1} \left( -\frac{1}{2} \cos\left(\frac{\eta_0}{2}\right) \sin\left(\frac{\eta_0}{2}\right) - \tilde{P} \cos^2\left(\frac{\eta_0}{2}\right) \right) \tilde{A}(\bar{\alpha}_\eta),$$

(6.65)

$$W_{1,2}(\bar{\alpha}_R, \tilde{\gamma}, \tilde{R}_0, \tilde{Q}_1, \tilde{P}, \eta_0) \approx -\frac{1}{\pi \tilde{Q}_1^4 l_p \tilde{R}_0^5} \left( \tilde{P}^2 + \frac{1}{4} \right)^{-1} \left( \frac{1}{2} \sin\left(\frac{\eta_0}{2}\right) \cos\left(\frac{\eta_0}{2}\right) - \tilde{P} \sin^2\left(\frac{\eta_0}{2}\right) \right) \tilde{B}(\bar{\alpha}_R, \tilde{\gamma}),$$

(6.66)

$$W_{1,3}(\bar{\alpha}_R, \tilde{\gamma}, \tilde{R}_0, \tilde{Q}_1, \tilde{P}, \eta_0) \approx \frac{1}{4\pi \tilde{Q}_1^2 l_p \tilde{R}_0^3} \tilde{P} \left( \tilde{P}^2 + \frac{1}{4} \right)^{-1} \tilde{C}(\bar{\alpha}_R, \tilde{\gamma}), \tag{6.67}$$

$$W_{1,4}(\bar{\alpha}_R,\tilde{\gamma},\tilde{R}_0,\tilde{Q}_j,\eta_0,\tilde{P}) \approx -\frac{1}{2\pi\tilde{Q}_1^3 l_p \tilde{R}_0^4}\cos\left(\frac{\eta_0}{2}\right)\left(\tilde{P}^2+\frac{1}{4}\right)^{-1/2}\tilde{D}(\bar{\alpha}_R,\tilde{\gamma}), \tag{6.68}$$

with the functions

$$\tilde{A}(\bar{\alpha}_\eta) = \int_{-\infty}^{\infty} x^2 dx \int_0^{\infty} r dr \exp\left(-\frac{r^2}{2}\right) J_{1,0,0}(x,r)\hat{G}_\eta(x,\bar{\alpha}_\eta), \tag{6.69}$$

$$\tilde{B}(\bar{\alpha}_R,\tilde{\gamma}) = \int_{-\infty}^{\infty} x^2 dx \int_0^{\infty} r dr \exp\left(-\frac{r^2}{2}\right) J_{1,1,1}(x,r)\hat{G}_R(x,\bar{\alpha}_R,\tilde{\gamma}), \tag{6.70}$$

$$\tilde{C}(\bar{\alpha}_R,\tilde{\gamma}) = \int_{-\infty}^{\infty} x^2 dx \int_0^{\infty} r dr \exp\left(-\frac{r^2}{2}\right) J_{1,0,0}(x,r)\hat{D}_R(x,\bar{\alpha}_R,\tilde{\gamma}), \tag{6.71}$$

$$\tilde{D}(\bar{\alpha}_R,\tilde{\gamma}) = \int_{-\infty}^{\infty} x dx \int_0^{\infty} r dr \exp\left(-\frac{r^2}{2}\right) J_{1,2,0}(x,r)\hat{C}_R(x,\bar{\alpha}_R,\tilde{\gamma}), \tag{6.72}$$

where we have introduced the rescaled variables $\bar{\alpha}_\eta = \tilde{R}_0^{-2}\alpha_{\eta,1}$ and $\bar{\alpha}_R = \tilde{R}_0^{-4}\alpha_{R,1}$.

By considering the form of the trial functional (either Eq. (4.2) or Eq. (5.29) depending on the strength of helix-helix correlations) and the results of Appendix A we are able to write the integrals for the rescaled correlation functions

$$\hat{G}_\eta(x,\tilde{\alpha}_\eta) = \frac{1}{2\pi}\int_{-\infty}^{\infty} dk \frac{\exp(-ikx)}{k^2 + \tilde{\alpha}_\eta}, \tag{6.73}$$

$$\hat{G}_R(x,\tilde{\alpha}_R,\tilde{\gamma}) = \frac{1}{2\pi}\int_{-\infty}^{\infty} dk \frac{\exp(-ikx)}{k^4 + \tilde{\alpha}_R^{1/2}\tilde{\gamma}k^2 + \tilde{\alpha}_R}, \tag{6.74}$$

$$\hat{C}_R(x,\tilde{\alpha}_R,\tilde{\gamma}) = \frac{1}{2\pi}\int_{-\infty}^{\infty} dk \frac{ik\exp(-ikx)}{k^4 + \tilde{\alpha}_R^{1/2}\tilde{\gamma}k^2 + \tilde{\alpha}_R}, \tag{6.75}$$

$$\hat{D}_R(x,\tilde{\alpha}_R,\tilde{\gamma}) = \frac{1}{2\pi}\int_{-\infty}^{\infty} dk \frac{k^2\exp(-ikx)}{k^4 + \tilde{\alpha}_R^{1/2}\tilde{\gamma}k^2 + \tilde{\alpha}_R}. \tag{6.76}$$

In Appendix F we evaluate Eqs. (6.73)-(6.76), and so obtain

$$\hat{G}_\eta(x,\tilde{\alpha}_\eta) = \frac{1}{2\tilde{\alpha}_\eta^{1/2}}\exp\left(-\tilde{\alpha}_\eta^{1/2}|x|\right), \tag{6.77}$$

$$\hat{G}_R(x,\tilde{\alpha}_R,\tilde{\gamma}) = \frac{\exp\left(-\frac{\tilde{\alpha}_R^{1/4}|x|}{\sqrt{2}}\sqrt{1+\frac{\tilde{\gamma}}{2}}\right)}{\sqrt{2}\tilde{\alpha}_R^{3/4}\sqrt{4-\tilde{\gamma}^2}}\left(\sqrt{1-\frac{\tilde{\gamma}}{2}}\cos\left(\frac{\tilde{\alpha}_R^{1/4}|x|}{\sqrt{2}}\sqrt{1-\frac{\tilde{\gamma}}{2}}\right)+\sqrt{1+\frac{\tilde{\gamma}}{2}}\sin\left(\frac{\tilde{\alpha}_R^{1/4}|x|}{\sqrt{2}}\sqrt{1-\frac{\tilde{\gamma}}{2}}\right)\right)\theta(2-\tilde{\gamma})$$

$$-\frac{1}{2\sqrt{2}\tilde{\alpha}_R^{3/4}\sqrt{\tilde{\gamma}^2-4}}\left[\left(\tilde{\gamma}-\sqrt{\tilde{\gamma}^2-4}\right)^{1/2}\exp\left(-\frac{\tilde{\alpha}_R^{1/4}|x|}{\sqrt{2}}\left(\tilde{\gamma}+\sqrt{\tilde{\gamma}^2-4}\right)^{1/2}\right)\right.$$

$$\left.-\left(\tilde{\gamma}+\sqrt{\tilde{\gamma}^2-4}\right)^{1/2}\exp\left(-\frac{\tilde{\alpha}_R^{1/4}|x|}{\sqrt{2}}\left(\tilde{\gamma}-\sqrt{\tilde{\gamma}^2-4}\right)^{1/2}\right)\right]\theta(\tilde{\gamma}-2),$$

(6.78)

$$\hat{D}_R(x,\tilde{\alpha}_R,\tilde{\gamma}) = \frac{\exp\left(-\frac{\tilde{\alpha}_R^{1/4}|x|}{\sqrt{2}}\sqrt{1+\frac{\tilde{\gamma}}{2}}\right)}{\tilde{\alpha}_R^{1/4}\sqrt{2}\sqrt{4-\tilde{\gamma}^2}}\left[\sqrt{1-\frac{\tilde{\gamma}}{2}}\cos\left(\frac{\tilde{\alpha}_R^{1/4}|x|}{\sqrt{2}}\sqrt{1-\frac{\tilde{\gamma}}{2}}\right)-\sqrt{1+\frac{\tilde{\gamma}}{2}}\sin\left(\frac{\tilde{\alpha}_R^{1/4}|x|}{\sqrt{2}}\sqrt{1-\frac{\tilde{\gamma}}{2}}\right)\right]\theta(2-\tilde{\gamma})$$

$$+\frac{1}{2\sqrt{2}\tilde{\alpha}_R^{1/4}\sqrt{\tilde{\gamma}^2-4}}\left(\left(\tilde{\gamma}+\sqrt{\tilde{\gamma}^2-4}\right)^{1/2}\exp\left(-\frac{\tilde{\alpha}_R^{1/4}|x|}{\sqrt{2}}\left(\tilde{\gamma}+\sqrt{\tilde{\gamma}^2-4}\right)^{1/2}\right)\right.$$

$$\left.-\left(\tilde{\gamma}-\sqrt{\tilde{\gamma}^2-4}\right)^{1/2}\exp\left(-\frac{\tilde{\alpha}_R^{1/4}|x|}{\sqrt{2}}\left(\tilde{\gamma}-\sqrt{\tilde{\gamma}^2-4}\right)^{1/2}\right)\right)\theta(\tilde{\gamma}-2),$$

(6.79)

$$\hat{C}_R(x,\tilde{\alpha}_R,\tilde{\gamma}) = \frac{\mathrm{sgn}(x)\exp\left(-\frac{\tilde{\alpha}_R^{1/4}|x|}{\sqrt{2}}\sqrt{1+\frac{\tilde{\gamma}}{2}}\right)}{\tilde{\alpha}_R^{1/2}\sqrt{4-\tilde{\gamma}^2}}\sin\left(\frac{\tilde{\alpha}_R^{1/4}|x|}{\sqrt{2}}\sqrt{1-\frac{\tilde{\gamma}}{2}}\right)\theta(2-\tilde{\gamma})$$

$$-\frac{\mathrm{sgn}(x)}{2\tilde{\alpha}_R^{1/2}\sqrt{\tilde{\gamma}^2-4}}\left(\exp\left(-\frac{\tilde{\alpha}_R^{1/4}|x|}{\sqrt{2}}\left(\tilde{\gamma}+\sqrt{\tilde{\gamma}^2-4}\right)^{1/2}\right)-\exp\left(-\frac{\tilde{\alpha}_R^{1/4}|x|}{\sqrt{2}}\left(\tilde{\gamma}-\sqrt{\tilde{\gamma}^2-4}\right)^{1/2}\right)\right)\theta(\tilde{\gamma}-2),$$

(6.80)

(always $\tilde{\gamma}>-2$).

We treat $W_{j,0}$ differently; we express it as a function of the variables $\tilde{R}_0, \tilde{P}$, $\alpha_\eta, \theta_R$ and $\eta_0$

$$d_R W_{j,0} \equiv \overline{W}_j(\tilde{R}_0.\tilde{P},\alpha_\eta,\theta_R,\eta_0) \approx -\frac{\tilde{P}\tilde{R}_0}{2\pi \cos\left(\frac{\eta_0}{2}\right)} \int_0^\infty rdr \exp\left(-\frac{\tilde{R}_0^2 r^2}{2}\right)$$

$$\left\{\left[\left(\sin\left(\frac{\eta_0}{2}\right)\cos\left(\frac{\eta_0}{2}\right)K_{j,1,0}(r,\tilde{P})+(-1)^j K_{j,2,0}(r,\tilde{P})\sin^2\left(\frac{\eta_0}{2}\right)\right)\left(1-\frac{1}{8(2l_p\alpha_\eta)^{1/2}}\right)\right.\right.$$

$$\left.+\theta_R^2\left(-\frac{1}{8}\cot\left(\frac{\eta_0}{2}\right)K_{j,1,0}(r,\tilde{P})-\frac{(-1)^j}{4}K_{j,2,0}(r,\tilde{P})\right)\right]+\left[\sin\left(\frac{\eta_0}{2}\right)\cos\left(\frac{\eta_0}{2}\right)\left(1-\frac{1}{8(2l_p\alpha_\eta)^{1/2}}\right)\right.$$

$$\left.\left.-\frac{\theta_R^2}{8}\cot\left(\frac{\eta_0}{2}\right)\right]K_{j,1,1}(r,\tilde{P})\right\},$$ (6.81)

where the functions $K_{j,i,k}(r,\tilde{P})$ are given by the double integrals

$$K_{j,1,k}(r,\tilde{P}) = r^k \int_0^{2\pi} d\phi_r \int_{-\infty}^\infty dx \left(1+(-1)^j \cos x\right)\sin^k \phi_r \left(\frac{1}{2}\left(1+(-1)^j \cos x\right)+\frac{r^2}{4}\right.$$

$$\left.+\frac{r(\cos\phi_r+\sin\phi_r)}{2}\left(1+(-1)^j \cos x\right)+(-1)^j \frac{r^2 \sin 2\phi_r}{4}\cos x+\tilde{P}^2 x^2\right)^{-3/2},$$ (6.82)

$$K_{j,2,k}(r,\tilde{P}) = r^k \int_0^{2\pi} d\phi_r \int_{-\infty}^\infty dx \tilde{P}x \sin x \sin^k \phi_r \left(\frac{1}{2}\left(1+(-1)^j \cos(x)\right)+\frac{r^2}{4}\right.$$

$$\left.+\frac{r(\cos\phi_r+\sin\phi_r)}{2}\left(1+(-1)^j \cos x\right)+(-1)^j \frac{r^2 \sin 2\phi_r}{4}\cos x+\tilde{P}^2 x^2\right)^{-3/2},$$ (6.83)

Again we expand out for $K_{2,1,k}(r,\tilde{P})$ and $K_{2,2,k}(r,\tilde{P})$ for small $r$, but keep $K_{1,1,k}(r,\tilde{P})$ and $K_{1,2,k}(r,\tilde{P})$ unexpanded, as these power series again leads to divergent x-integrals. Thus we can write

$$K_{2,1,k}(r,\tilde{P}) \approx r^k \left(K_{2,1,k}^0(\tilde{P}) + K_{2,1,k}^1(\tilde{P})r + K_{2,1,k}^2(\tilde{P})r^2 + K_{2,1,k}^3(\tilde{P})r^3 + K_{2,1,k}^4(\tilde{P})r^4 + \ldots\right)$$ (6.84)

$$K_{2,2,k}(r,\tilde{P}) \approx r^k \left(K_{2,2,k}^0(\tilde{P}) + K_{2,2,k}^1(\tilde{P})r + K_{2,2,k}^2(\tilde{P})r^2 + K_{2,2,k}^3(\tilde{P})r^3 + K_{2,2,k}^4(\tilde{P})r^4 + \ldots\right)$$ (6.85)

Expressions for the various $K_{2,1,k}^p(\tilde{P})$ and $K_{2,2,k}^p(\tilde{P})$ are given in Appendix G, note that only these functions multiplied by an even power of $r$ are not zero. Using Eqs. (6.84) and (6.85) we can easily perform the r-integrations giving us

$$\bar{W}_2(\tilde{R}_0, \tilde{P}, \alpha_\eta, \theta_R, \eta_0) \approx -\frac{\tilde{P}}{2\pi \cos\left(\frac{\eta_0}{2}\right)} \left\{ \left[ \sin\left(\frac{\eta_0}{2}\right) \cos\left(\frac{\eta_0}{2}\right) \left(1 - \frac{1}{8(2l_p \alpha_\eta)^{1/2}}\right) - \frac{\theta_R^2}{8} \cot\left(\frac{\eta_0}{2}\right) \right] \right.$$

$$\left( \frac{K_{2,1,0}^0(\tilde{P})}{\tilde{R}_0} + \frac{2\left(K_{2,1,0}^2(\tilde{P}) + K_{2,1,1}^1(\tilde{P})\right)}{\tilde{R}_0^3} + \frac{8\left(K_{2,1,0}^4(\tilde{P}) + K_{2,1,1}^3(\tilde{P})\right)}{\tilde{R}_0^5} \right)$$

$$+ \left[ \sin^2\left(\frac{\eta_0}{2}\right) \left(1 - \frac{1}{8(2l_p \alpha_\eta)^{1/2}}\right) - \frac{\theta_R^2}{4} \right] \left( \frac{K_{2,2,0}^0(\tilde{P})}{\tilde{R}_0} + \frac{2K_{2,2,0}^2(\tilde{P})}{\tilde{R}_0^3} + \frac{8K_{2,2,0}^4(\tilde{P})}{\tilde{R}_0^5} \right) \right\}.$$

(6.86)

Finally we have $\tilde{P}$ evaluating the averages for small $\eta_0$ and $R'(\tau)$

$$\tilde{P} \approx \frac{1}{2\tan\left(\frac{\eta_0}{2}\right)} f(\tilde{R}_0) \frac{\left(1 - \frac{d_\eta^2}{8}\right)}{\left(1 - \frac{d_\eta^2}{8} - \frac{\theta_R^2}{8\sin^2\left(\frac{\eta_0}{2}\right)}\right)} \approx \frac{1}{2\tan\left(\frac{\eta_0}{2}\right)} f(\tilde{R}_0) \left(1 + \frac{\theta_R^2}{8\sin^2\left(\frac{\eta_0}{2}\right)}\right),$$

(6.87)

where

$$f(\tilde{R}_0) \approx 1 - \frac{1}{\tilde{R}_0^2} - \frac{2}{\tilde{R}_0^4}.$$ (6.88)

The function $f(\tilde{R}_0)$ represents the average $\langle R_0 / R(\tau) \rangle_0^{-1}$, again we have expanded out for large values $\tilde{R}_0$ to prevent, again, unphysical divergences in the averaging integral, which comes from not including the cuttoff $d_{\min}$. The final result for the average writhe can be written as

$$\frac{2\pi \langle Wr \rangle_0}{L_b} = \frac{1}{d_R} \left( \bar{W}_1(\tilde{R}_0, \tilde{P}, \alpha_\eta, \theta_R, \eta_0) - \bar{W}_2(\tilde{R}_0, \tilde{P}, \alpha_\eta, \theta_R, \eta_0) \right)$$
$$+ \left[ \left( W_{1,1}(\bar{\alpha}_\eta, \tilde{R}_0, \tilde{Q}_j, \tilde{P}, \eta_0) - W_{2,1}(\tilde{\alpha}_{\eta,2}, \tilde{R}_0, \tilde{Q}_j, \tilde{P}, \eta_0) \right) + W_T(\tilde{\alpha}_{R,1}, \tilde{\alpha}_{R,2}, \tilde{\gamma}, \tilde{R}_0, \tilde{Q}_1, \tilde{Q}_2, \tilde{P}, \eta_0) \right],$$

(6.89)

where

$$W_T(\tilde{\alpha}_{R,1},\tilde{\alpha}_{R,2},\tilde{\gamma},\tilde{R}_0,\tilde{Q}_1,\tilde{Q}_2,\tilde{P},\eta_0) = W_{1,2}(\bar{\alpha}_R,\tilde{\gamma},\tilde{R}_0,\tilde{Q}_1,\tilde{P},\eta_0) - W_{2,2}(\tilde{\alpha}_{R,2},\tilde{\gamma},\tilde{R}_0,\tilde{Q}_2,\tilde{P},\eta_0)$$
$$+W_{1,3}(\bar{\alpha}_R,\tilde{\gamma},\tilde{R}_0,\tilde{Q}_1,\tilde{P},\eta_0) - W_{2,3}(\tilde{\alpha}_{R,2},\tilde{\gamma},\tilde{R}_0,\tilde{Q}_2,\tilde{P},\eta_0) + W_{1,4}(\bar{\alpha}_R,\tilde{\gamma},\tilde{R}_0,\tilde{Q}_1,\tilde{P},\eta_0)$$
$$-W_{2,4}(\tilde{\alpha}_{R,2},\tilde{\gamma},\tilde{R}_0,\tilde{Q}_2,\tilde{P},\eta_0).$$

(6.90)

In the next section we derive expressions for the derivatives of $\langle Wr \rangle_0$ appearing in the systems of Eqs. (4.48)-(4.61) and Eqs. (5.58)-(5.59).

## 7. Derivatives of $\langle Wr \rangle_0$

To compute the derivatives $\langle Wr \rangle_0$ it is useful to consider the derivatives of first $\bar{W}_j(\tilde{R}_0,\tilde{P},d_\eta,\theta_R,\eta_0)$

$$\bar{W}_j^R(\tilde{R}_0,\tilde{P},\alpha_\eta,\theta_R,\eta_0) = \frac{\partial}{\partial \tilde{R}_0}\bar{W}_j(\tilde{R}_0,\tilde{P},\alpha_\eta,\theta_R,\eta_0) \approx -\frac{\tilde{P}}{2\pi\cos\left(\frac{\eta_0}{2}\right)}\int_0^\infty rdr\left(1-r^2\tilde{R}_0^2\right)\exp\left(-\frac{\tilde{R}_0^2 r^2}{2}\right)$$

$$\left\{\left[\left(\sin\left(\frac{\eta_0}{2}\right)\cos\left(\frac{\eta_0}{2}\right)K_{j,1,0}(r,\tilde{P})+(-1)^j K_{j,2,0}(r,\tilde{P})\sin^2\left(\frac{\eta_0}{2}\right)\right)\left(1-\frac{1}{8(2l_p\alpha_\eta)^{1/2}}\right)\right.\right.$$

$$\left.-\theta_R^2\left(\frac{1}{8}\cot\left(\frac{\eta_0}{2}\right)K_{j,1,0}(r,\tilde{P})+\frac{(-1)^j}{4}K_{j,2,0}(r,\tilde{P})\right)\right]+\left[\sin\left(\frac{\eta_0}{2}\right)\cos\left(\frac{\eta_0}{2}\right)\left(1-\frac{1}{8(2l_p\alpha_\eta)^{1/2}}\right)\right.$$

$$\left.\left.-\frac{\theta_R^2}{8}\cot\left(\frac{\eta_0}{2}\right)\right]K_{j,1,1}(r,\tilde{P})\right\},$$

(7.1)

$$\bar{W}_j^P(\tilde{R}_0,\tilde{P},\alpha_\eta,\theta_R,\eta_0) = \frac{\partial}{\partial \tilde{P}}\bar{W}_j(\tilde{R}_0,\tilde{P},\alpha_\eta,\theta_R,\eta_0) \approx -\frac{\tilde{R}_0}{2\pi\cos\left(\frac{\eta_0}{2}\right)}\int_0^\infty rdr\exp\left(-\frac{\tilde{R}_0^2 r^2}{2}\right)$$

$$\left\{\left[\left(\sin\left(\frac{\eta_0}{2}\right)\cos\left(\frac{\eta_0}{2}\right)\left(K_{j,1,0}(r,\tilde{P})+\tilde{P}K_{j,1,0}^P(r,\tilde{P})\right)+(-1)^j\left(K_{j,2,0}(r,\tilde{P})+\tilde{P}K_{j,2,0}^P(r,\tilde{P})\right)\sin^2\left(\frac{\eta_0}{2}\right)\right)\right.\right.$$

$$\left.\left(1-\frac{1}{8(2l_p\alpha_\eta)^{1/2}}\right)+\theta_R^2\left(-\frac{1}{8}\cot\left(\frac{\eta_0}{2}\right)\left(K_{j,1,0}(r,\tilde{P})+\tilde{P}K_{j,1,0}^P(r,\tilde{P})\right)+\frac{(-1)^j}{4}\left(K_{j,2,0}(r,\tilde{P})+\tilde{P}K_{j,2,0}^P(r,\tilde{P})\right)\right)\right]$$

$$\left.+\left[\sin\left(\frac{\eta_0}{2}\right)\cos\left(\frac{\eta_0}{2}\right)\left(1-\frac{1}{8(2l_p\alpha_\eta)^{1/2}}\right)-\frac{\theta_R^2}{8}\cot\left(\frac{\eta_0}{2}\right)\right]\left(K_{j,1,1}(r,\tilde{P})+\tilde{P}K_{j,1,1}^P(r,\tilde{P})\right)\right\},$$

(7.2)

$$\overline{W}_j^{\alpha_\eta}(\tilde{R}_0,\tilde{P},\alpha_\eta,\theta_R,\eta_0) = \frac{\partial}{\partial \alpha_\eta}\overline{W}_j(\tilde{R}_0,\tilde{P},\alpha_\eta,\theta_R,\eta_0) \approx -\frac{\tilde{R}_0\tilde{P}}{2\pi\cos\left(\frac{\eta_0}{2}\right)}\left(\frac{1}{16\alpha_\eta^{3/2}(2l_p)^{1/2}}\right)\int_0^\infty rdr\exp\left(-\frac{\tilde{R}_0^2 r^2}{2}\right)$$

$$\left[\left(\sin\left(\frac{\eta_0}{2}\right)\cos\left(\frac{\eta_0}{2}\right)K_{j,1,0}(r,\tilde{P})+(-1)^j K_{j,2,0}(r,\tilde{P})\sin^2\left(\frac{\eta_0}{2}\right)\right)+\left(\sin\left(\frac{\eta_0}{2}\right)\cos\left(\frac{\eta_0}{2}\right)K_{j,1,1}(r,\tilde{P})\right)\right],$$

(7.3)

$$\overline{W}_j^{\theta_R}(\tilde{R}_0,\tilde{P},\alpha_\eta,\theta_R,\eta_0) = \frac{d}{d\theta_R}\overline{W}_j(\tilde{R}_0,\tilde{P},\alpha_\eta,\theta_R,\eta_0) \approx \frac{\tilde{R}_0\tilde{P}\theta_R}{2\pi\cos\left(\frac{\eta_0}{2}\right)}\int_0^\infty rdr\exp\left(-\frac{\tilde{R}_0^2 r^2}{2}\right)$$

$$\left(\frac{1}{4}\cot\left(\frac{\eta_0}{2}\right)(K_{j,1,0}(r,\tilde{P})+K_{j,1,1}(r,\tilde{P}))+\frac{(-1)^j}{2}K_{j,2,0}(r,\tilde{P})\right),$$

(7.4)

$$\overline{W}_{j,0}^{\eta}(\tilde{R}_0,\tilde{P},\alpha_\eta,\theta_R,\eta_0) = \frac{d}{d\eta_0}\overline{W}_{j,0}(\tilde{R}_0,\tilde{P},\alpha_\eta,\theta_R,\eta_0) \approx -\frac{\tilde{P}\tilde{R}_0}{4\pi}\int_0^\infty rdr\exp\left(-\frac{\tilde{R}_0^2 r^2}{2}\right)$$

$$\left\{\left[\left(\cos\left(\frac{\eta_0}{2}\right)K_{j,1,0}(r,\tilde{P})+(-1)^j K_{j,2,0}(r,\tilde{P})\left(\sin\left(\frac{\eta_0}{2}\right)+\sec\left(\frac{\eta_0}{2}\right)\tan\left(\frac{\eta_0}{2}\right)\right)\right)\left(1-\frac{1}{8(2l_p\alpha_\eta)^{1/2}}\right)\right.\right.$$

$$\left.+\theta_R^2\left[\frac{1}{8}\cot\left(\frac{\eta_0}{2}\right)\mathrm{cosec}\left(\frac{\eta_0}{2}\right)K_{j,1,0}(r,\tilde{P})-\frac{(-1)^j}{4}\tan\left(\frac{\eta_0}{2}\right)\sec\left(\frac{\eta_0}{2}\right)K_{j,2,0}(r,\tilde{P})\right]\right]$$

$$\left.+\left[\cos\left(\frac{\eta_0}{2}\right)\left(1-\frac{1}{8(2l_p\alpha_\eta)^{1/2}}\right)+\theta_R^2\frac{1}{8}\cot\left(\frac{\eta_0}{2}\right)\mathrm{cosec}\left(\frac{\eta_0}{2}\right)\right]K_{j,1,1}(r,\tilde{P})\right\}.$$

(7.5)

Here the functions $K_{j,k,l}^P(r,\tilde{P})$ are given by

$$K_{j,1,k}^P(r,\tilde{P}) = -3\tilde{P}r^k\int_0^{2\pi}d\phi_r\int_{-\infty}^{\infty}dx\, x^2\left(1+(-1)^j\cos x\right)\sin^k\phi_r\left(\frac{1}{2}(1+(-1)^j\cos x)+\frac{r^2}{4}\right.$$

$$\left.+\frac{r(\cos\phi_r+\sin\phi_r)}{2}(1+(-1)^j\cos x)+(-1)^j\frac{r^2\sin 2\phi_r}{4}\cos x+\tilde{P}^2 x^2\right)^{-5/2},$$

(7.6)

$$K_{j,2,k}^P(r,\tilde{P}) = K_{j,2,k}(r,\tilde{P})/\tilde{P}-3\tilde{P}^2 r^k\int_0^{2\pi}d\phi_r\int_{-\infty}^{\infty}dx\, x^3\sin x\sin^l\phi_r\left(\frac{1}{2}(1+(-1)^j\cos x)+\frac{r^2}{4}\right.$$

$$\left.+\frac{r(\cos\phi_r+\sin\phi_r)}{2}(1+(-1)^j\cos x)+(-1)^j\frac{r^2\sin 2\phi}{4}\cos x+\tilde{P}^2 x^2\right)^{-5/2},$$

(7.7)

Expanding out $K_{2,1,k}^P(r,\tilde{P})$ and $K_{2,2,k}^P(r,\tilde{P})$, we can write down expressions for the derivatives of $\overline{W}_2(\tilde{R}_0,\tilde{P},d_\eta,\theta_R,\eta_0)$, which are given in Appendix H.

For $W_{1,1}$, it is useful to define functions in terms of its derivatives

$$W_{1,1}^{\alpha_\eta}(\bar{\alpha}_\eta, \tilde{R}_0, \tilde{Q}_1, \tilde{P}, \eta_0) = \frac{\partial W_{1,1}(\tilde{R}_0^{-2}\tilde{\alpha}_{\eta,1}, \tilde{R}_0, \tilde{Q}_1, \tilde{P}, \eta_0)}{\partial \tilde{\alpha}_{\eta,1}}, \quad W_{1,1}^R(\bar{\alpha}_\eta, \tilde{R}_0, \tilde{Q}_1, \tilde{P}, \eta_0) = \frac{\partial W_{1,1}(\tilde{R}_0^{-2}\tilde{\alpha}_{\eta,1}, \tilde{R}_0, \tilde{Q}_1, \tilde{P}, \eta_0)}{\partial \tilde{R}_0},$$

$$W_{1,1}^Q(\bar{\alpha}_\eta, \tilde{R}_0, \tilde{Q}_1, \tilde{P}, \eta_0) = \frac{\partial W_{1,1}^{\alpha_\eta}(\bar{\alpha}_\eta, \tilde{R}_0, \tilde{Q}_1, \tilde{P}, \eta_0)}{\partial \tilde{Q}_1}, \quad W_{1,1}^P(\bar{\alpha}_\eta, \tilde{R}_0, \tilde{Q}_1, \tilde{P}, \eta_0) = \frac{\partial W_{1,1}(\bar{\alpha}_\eta, \tilde{R}_0, \tilde{Q}_1, \tilde{P}, \eta_0)}{\partial \tilde{P}},$$

$$W_{1,1}^\eta(\bar{\alpha}_\eta, \tilde{R}_0, \tilde{Q}_1, \tilde{P}, \eta_0) = \frac{\partial W_{1,1}(\bar{\alpha}_\eta, \tilde{R}_0, \tilde{Q}_1, \tilde{P}, \eta_0)}{\partial \eta_0}, \tag{7.8}$$

and for $W_{2,1}$

$$W_{2,1}^{\alpha_\eta}(\tilde{\alpha}_{\eta,2}, \tilde{R}_0, \tilde{Q}_1, \tilde{P}, \eta_0) = \frac{\partial W_{2,1}(\tilde{\alpha}_{\eta,2}, \tilde{R}_0, \tilde{Q}_1, \tilde{P}, \eta_0)}{\partial \tilde{\alpha}_\eta}, \quad W_{2,1}^R(\tilde{\alpha}_{\eta,2}, \tilde{R}_0, \tilde{Q}_2, \tilde{P}, \eta_0) = \frac{\partial W_{2,1}(\tilde{\alpha}_{\eta,2}, \tilde{R}_0, \tilde{Q}_2, \tilde{P}, \eta_0)}{\partial \tilde{R}_0},$$

$$W_{2,1}^Q(\tilde{\alpha}_{\eta,2}, \tilde{R}_0, \tilde{Q}_2, \tilde{P}, \eta_0) = \frac{\partial W_{2,1}(\tilde{\alpha}_{\eta,2}, \tilde{R}_0, \tilde{Q}_2, \tilde{P}, \eta_0)}{\partial \tilde{Q}_2}, \quad W_{2,1}^P(\tilde{\alpha}_{\eta,2}, \tilde{R}_0, \tilde{Q}_2, \tilde{P}, \eta_0) = \frac{\partial W_{2,1}(\tilde{\alpha}_{\eta,2}, \tilde{R}_0, \tilde{Q}_2, \tilde{P}, \eta_0)}{\partial \tilde{P}},$$

$$W_{2,1}^\eta(\tilde{\alpha}_{\eta,2}, \tilde{R}_0, \tilde{Q}_2, \tilde{P}, \eta_0) = \frac{\partial W_{2,1}(\tilde{\alpha}_{\eta,2}, \tilde{R}_0, \tilde{Q}_2, \tilde{P}, \eta_0)}{\partial \eta_0}. \tag{7.9}$$

It is straight forward to obtain expressions for the functions defined by both Eqs.(7.8) and (7.9). These are given in Appendix H.

Then for $W_{1,2}, W_{1,3}$ and $W_{1,4}$ we define the functions ($k = 2,3,4$)

$$W_{1,k}^{\alpha_R}(\bar{\alpha}_R, \tilde{\gamma}, \tilde{R}_0, \tilde{Q}_1, \tilde{P}, \eta_0) = \frac{\partial W_{1,k}(\tilde{R}_0^{-4}\tilde{\alpha}_R, \tilde{\gamma}, \tilde{R}_0, \tilde{Q}_1, \tilde{P}, \eta_0)}{\partial \tilde{\alpha}_R}, \quad W_{1,k}^\gamma(\bar{\alpha}_R, \tilde{\gamma}, \tilde{R}_0, \tilde{Q}_1, \tilde{P}, \eta_0) = \frac{\partial W_{1,k}(\bar{\alpha}_R, \tilde{\gamma}, \tilde{R}_0, \tilde{Q}_1, \tilde{P}, \eta_0)}{\partial \tilde{\gamma}},$$

$$W_{1,k}^R(\bar{\alpha}_R, \tilde{\gamma}, \tilde{R}_0, \tilde{Q}_1, \tilde{P}, \eta_0) = \frac{\partial W_{1,k}(\tilde{R}_0^{-4}\tilde{\alpha}_{R,1}, \tilde{\gamma}, \tilde{R}_0, \tilde{Q}_1, \tilde{P}, \eta_0)}{\partial \tilde{R}_0}, \quad W_{1,k}^Q(\bar{\alpha}_R, \tilde{\gamma}, \tilde{R}_0, \tilde{Q}_1, \tilde{P}, \eta_0) = \frac{\partial W_{1,k}(\bar{\alpha}_R, \tilde{\gamma}, \tilde{R}_0, \tilde{Q}_1, \tilde{P}, \eta_0)}{\partial \tilde{Q}_1},$$

$$W_{1,k}^P(\bar{\alpha}_R, \tilde{\gamma}, \tilde{R}_0, \tilde{Q}_1, \tilde{P}, \eta_0) = \frac{\partial W_{1,k}(\bar{\alpha}_R, \tilde{\gamma}, \tilde{R}_0, \tilde{Q}_1, \tilde{P}, \eta_0)}{\partial \tilde{P}}, \quad W_{1,k}^\eta(\bar{\alpha}_R, \tilde{\gamma}, \tilde{R}_0, \tilde{Q}_1, \tilde{P}, \eta_0) = \frac{\partial W_{1,k}(\bar{\alpha}_R, \tilde{\gamma}, \tilde{R}_0, \tilde{Q}_1, \tilde{P}, \eta_0)}{\partial \eta_0}.$$

$$\tag{7.10}$$

Expressions for these are again given in Appendix H.

Next we consider the derivatives of $W_{2,2}, W_{2,3}$ and $W_{2,4}$, which we can define the following functions for ($k = 2,3,4$)

$$W_{2,k}^{\alpha_R}(\tilde{\alpha}_{R,1},\tilde{\gamma},\tilde{R}_0,\tilde{Q}_2,\tilde{P},\eta_0) = \frac{\partial W_{2,k}(\tilde{\alpha}_{R,1},\tilde{\gamma},\tilde{R}_0,\tilde{Q}_2,\tilde{P},\eta_0)}{\partial \tilde{\alpha}_R},$$

$$W_{2,k}^{\gamma}(\tilde{\alpha}_{R,2},\tilde{\gamma},\tilde{R}_0,\tilde{Q}_2,\tilde{P},\eta_0) = \frac{\partial W_{2,k}(\tilde{\alpha}_{R,2},\tilde{\gamma},\tilde{R}_0,\tilde{Q}_2,\tilde{P},\eta_0)}{\partial \tilde{\gamma}},$$

$$W_{2,k}^{R}(\tilde{\alpha}_{R,2},\tilde{\gamma},\tilde{R}_0,\tilde{Q}_2,\tilde{P},\eta_0) = \frac{\partial W_{2,k}(\tilde{\alpha}_{R,2},\tilde{\gamma},\tilde{R}_0,\tilde{Q}_2,\tilde{P},\eta_0)}{\partial \tilde{R}_0}, \quad W_{2,k}^{Q}(\tilde{\alpha}_{R,2},\tilde{\gamma},\tilde{R}_0,\tilde{Q}_2,\tilde{P},\eta_0) = \frac{\partial W_{2,k}(\tilde{\alpha}_{R,2},\tilde{\gamma},\tilde{R}_0,\tilde{Q}_2,\tilde{P},\eta_0)}{\partial \tilde{Q}_2},$$

$$W_{2,k}^{P}(\tilde{\alpha}_{R,2},\tilde{\gamma},\tilde{R}_0,\tilde{Q}_2,\tilde{P},\eta_0) = \frac{\partial W_{2,k}(\tilde{\alpha}_{R,2},\tilde{\gamma},\tilde{R}_0,\tilde{Q}_2,\tilde{P},\eta_0)}{\partial \tilde{P}}, \quad W_{2,k}^{\eta}(\tilde{\alpha}_{R,2},\tilde{\gamma},\tilde{R}_0,\tilde{Q}_2,\tilde{P},\eta_0) = \frac{\partial W_{2,k}(\tilde{\alpha}_{R,2},\tilde{\gamma},\tilde{R}_0,\tilde{Q}_2,\tilde{P},\eta_0)}{\partial \eta_0}.$$

(7.11)

Again, expressions for these are given in Appendix H.

Then, we construct the following functions

$$W_T^X(\tilde{\alpha}_{R,1},\tilde{\alpha}_{R,2},\tilde{\gamma},\tilde{R}_0,\tilde{Q}_1,\tilde{Q}_2,\tilde{P},\eta_0) = W_{1,2}^X(\bar{\alpha}_R,\tilde{\gamma},\tilde{R}_0,\tilde{Q}_1,\tilde{P},\eta_0) - W_{2,2}^X(\tilde{\alpha}_{R,2},\tilde{\gamma},\tilde{R}_0,\tilde{Q}_2,\tilde{P},\eta_0)$$
$$+ W_{1,3}^X(\bar{\alpha}_R,\tilde{\gamma},\tilde{R}_0,\tilde{Q}_1,\tilde{P},\eta_0) - W_{2,3}^X(\tilde{\alpha}_{R,2},\tilde{\gamma},\tilde{R}_0,\tilde{Q}_2,\tilde{P},\eta_0) + W_{1,4}^X(\bar{\alpha}_R,\tilde{\gamma},\tilde{R}_0,\tilde{Q}_1,\tilde{P},\eta_0)$$
$$- W_{2,4}^X(\tilde{\alpha}_{R,2},\tilde{\gamma},\tilde{R}_0,\tilde{Q}_2,\tilde{P},\eta_0),$$

$$\text{where} \quad X = \gamma, R, P, \eta. \quad (7.12)$$

and (using Eqs. (H.37), (H.43), (H.49), (H.67), (H.73) and (H.79) of Appendix H)

$$W_{T,j}^Q(\tilde{\alpha}_{R,1},\tilde{\alpha}_{R,2},\tilde{\gamma},\tilde{R}_0,\tilde{Q}_1,\tilde{Q}_2,\tilde{P},\eta_0) = \delta_{j,1}(4W_{1,2}(\bar{\alpha}_R,\tilde{\gamma},\tilde{R}_0,\tilde{Q}_1,\tilde{P},\eta_0) + 2W_{1,3}(\bar{\alpha}_R,\tilde{\gamma},\tilde{R}_0,\tilde{Q}_1,\tilde{P},\eta_0)$$
$$+ 3W_{1,4}(\bar{\alpha}_R,\tilde{\gamma},\tilde{R}_0,\tilde{Q}_1,\tilde{P},\eta_0)) - \delta_{j,2}(4W_{2,2}(\tilde{\alpha}_R,\tilde{\gamma},\tilde{R}_0,\tilde{Q}_2,\tilde{P},\eta_0) - 2W_{2,3}(\tilde{\alpha}_R,\tilde{\gamma},\tilde{R}_0,\tilde{Q}_2,\tilde{P},\eta_0)$$
$$- 3W_{2,4}(\tilde{\alpha}_R,\tilde{\gamma},\tilde{R}_0,\tilde{Q}_2,\tilde{P},\eta_0)),$$

(7.13)

$$W_{T,j}^{\alpha_R}(\tilde{\alpha}_{R,1},\tilde{\alpha}_{R,2},\tilde{\gamma},\tilde{R}_0,\tilde{Q}_1,\tilde{Q}_2,\tilde{P},\eta_0) = \delta_{j,1}(W_{1,2}^{\alpha_R}(\bar{\alpha}_R,\tilde{\gamma},\tilde{R}_0,\tilde{Q}_1,\tilde{P},\eta_0) + W_{1,3}^{\alpha_R}(\bar{\alpha}_R,\tilde{\gamma},\tilde{R}_0,\tilde{Q}_1,\tilde{P},\eta_0)$$
$$+ W_{1,4}^{\alpha_R}(\bar{\alpha}_R,\tilde{\gamma},\tilde{R}_0,\tilde{Q}_1,\tilde{P},\eta_0)) - \delta_{j,2}(W_{2,2}^{\alpha_R}(\tilde{\alpha}_R,\tilde{\gamma},\tilde{R}_0,\tilde{Q}_2,\tilde{P},\eta_0) + W_{2,3}^{\alpha_R}(\tilde{\alpha}_R,\tilde{\gamma},\tilde{R}_0,\tilde{Q}_2,\tilde{P},\eta_0) \quad (7.14)$$
$$+ W_{2,4}^{\alpha_R}(\tilde{\alpha}_R,\tilde{\gamma},\tilde{R}_0,\tilde{Q}_2,\tilde{P},\eta_0)),$$

where the various functions are defined through Eq. (7.10)-(7.11).

Next, we note that $\tilde{P} = \tilde{P}(\eta_0,\tilde{R}_0,\theta_R)$, $\tilde{\alpha}_\eta = \tilde{\alpha}_\eta(\eta_0,R_0,d_R,\theta_R,\alpha_\eta)$, $\tilde{\alpha}_R = \tilde{\alpha}_R(\eta_0,R_0,\theta_R,d_R,\alpha_\eta)$, $\tilde{\gamma} = \tilde{\gamma}(\theta_R,d_R)$, and $\tilde{Q}_j = \tilde{Q}_j(\eta_0,\tilde{R}_0,\theta_R,\alpha_\eta)$. Thus, using, the chain rule we can obtain expressions for the derivatives of the average writhe

$$2\pi \frac{\partial \langle Wr \rangle_0}{\partial R_0} = \frac{1}{d_R^2} \left( \overline{W}_1^R(\tilde{R}_0, \tilde{P}, \alpha_\eta, \theta_R, \eta_0) - \overline{W}_2^R(\tilde{R}_0, \tilde{P}, \alpha_\eta, \theta_R, \eta_0) \right)$$

$$+ \frac{1}{d_R} \Big[ \left( W_{1,1}^R(\bar{\alpha}_\eta, \tilde{R}_0, \tilde{Q}_1, \tilde{P}, \eta_0) - W_{2,1}^R(\tilde{\alpha}_{\eta,2}, \tilde{R}_0, \tilde{Q}_2, \tilde{P}, \eta_0) \right)$$

$$+ W_T^R(\tilde{\alpha}_{R,1}, \tilde{\alpha}_{R,2}, \tilde{\gamma}, \tilde{R}_0, \tilde{Q}_1, \tilde{Q}_2, \tilde{P}, \eta_0) \Big] + \frac{1}{d_R} \frac{\partial \tilde{P}}{\partial \tilde{R}_0} \Bigg[ \frac{1}{d_R} \left( \overline{W}_1^P(\tilde{R}_0, \tilde{P}, \alpha_\eta, \theta_R, \eta_0) - \overline{W}_2^P(\tilde{R}_0, \tilde{P}, \alpha_\eta, \theta_R, \eta_0) \right)$$

$$+ \left( W_{1,1}^P(\bar{\alpha}_\eta, \tilde{R}_0, \tilde{Q}_1, \tilde{P}, \eta_0) - W_{2,1}^P(\tilde{\alpha}_\eta, \tilde{R}_0, \tilde{Q}_2, \tilde{P}, \eta_0) \right) + W_T^P(\tilde{\alpha}_{R,1}, \tilde{\alpha}_{R,2}, \tilde{\gamma}, \tilde{R}_0, \tilde{Q}_1, \tilde{Q}_2, \tilde{P}, \eta_0) \Bigg]$$

$$- \sum_{j=1}^{2} \Bigg\{ \frac{1}{d_R \tilde{Q}_j} \frac{\partial \tilde{Q}_j}{\partial \tilde{R}_0} \Big[ 2 \left( \delta_{j,1} W_{1,1}(\bar{\alpha}_\eta, \tilde{R}_0, \tilde{Q}_1, \tilde{P}, \eta_0) - \delta_{j,2} W_{2,1}(\tilde{\alpha}_{\eta,2}, \tilde{R}_0, \tilde{Q}_2, \tilde{P}, \eta_0) \right)$$

$$- W_{T,j}^Q(\tilde{\alpha}_{R,1}, \tilde{\alpha}_{R,2}, \tilde{\gamma}, \tilde{R}_0, \tilde{Q}_1, \tilde{Q}_2, \tilde{P}, \eta_0) \Big] - \frac{\partial \tilde{\alpha}_{\eta,j}}{\partial R_0} \left( \delta_{j,1} W_{1,1}^{\alpha_\eta}(\tilde{\alpha}_\eta, \tilde{R}_0, \tilde{Q}_1, \tilde{P}, \eta_0) - \delta_{j,2} W_{2,1}^{\alpha_\eta}(\tilde{\alpha}_\eta, \tilde{R}_0, \tilde{Q}_2, \tilde{P}, \eta_0) \right)$$

$$- \frac{\partial \tilde{\alpha}_{R,j}}{\partial R_0} W_{T,j}^{\alpha_R}(\tilde{\alpha}_{R,1}, \tilde{\alpha}_{R,2}, \tilde{\gamma}, \tilde{R}_0, \tilde{Q}_1, \tilde{Q}_2, \tilde{P}, \eta_0) \Bigg\},$$

(7.15)

$$2\pi \frac{\partial \langle Wr \rangle_0}{\partial d_R} = \frac{2 d_R}{l_p^2 \theta_R^6} W_T^\gamma(\tilde{\alpha}_{R,1}, \tilde{\alpha}_{R,2}, \tilde{\gamma}, \tilde{R}_0, \tilde{Q}_1, \tilde{Q}_2, \tilde{P}, \eta_0)$$

$$- \frac{1}{d_R^2} \left( \overline{W}_1(\tilde{R}_0, \tilde{P}, \alpha_\eta, \theta_R, \eta_0) - \overline{W}_2(\tilde{R}_0, \tilde{P}, \alpha_\eta, \theta_R, \eta_0) + \tilde{R}_0 \overline{W}_1^R(\tilde{R}_0, \tilde{P}, \alpha_\eta, \theta_R, \eta_0) - \tilde{R}_0 \overline{W}_2^R(\tilde{R}_0, \tilde{P}, \alpha_\eta, \theta_R, \eta_0) \right)$$

$$- \frac{\tilde{R}_0}{d_R} \left( W_{1,1}^R(\bar{\alpha}_\eta, \tilde{R}_0, \tilde{Q}_1, \tilde{P}, \eta_0) - W_{2,1}^R(\tilde{\alpha}_{\eta,2}, \tilde{R}_0, \tilde{Q}_2, \tilde{P}, \eta_0) + W_T^R(\tilde{\alpha}_{R,1}, \tilde{\alpha}_{R,2}, \tilde{\gamma}, \tilde{R}_0, \tilde{Q}_1, \tilde{Q}_2, \tilde{P}, \eta_0) \right)$$

$$- \frac{\tilde{R}_0}{d_R^2} \frac{\partial \tilde{P}}{\partial \tilde{R}_0} \Bigg[ \left( \overline{W}_1^P(\tilde{R}_0, \tilde{P}, \alpha_\eta, \theta_R, \eta_0) - \overline{W}_2^P(\tilde{R}_0, \tilde{P}, \alpha_\eta, \theta_R, \eta_0) \right) + d_R W_T^P(\tilde{\alpha}_{R,1}, \tilde{\alpha}_{R,2}, \tilde{\gamma}, \tilde{R}_0, \tilde{Q}_1, \tilde{Q}_2, \tilde{P}, \eta_0)$$

$$+ d_R \left( W_{1,1}^P(\bar{\alpha}_\eta, \tilde{R}_0, \tilde{Q}_1, \tilde{P}, \eta_0) - W_{2,1}^P(\tilde{\alpha}_{\eta,2}, \tilde{R}_0, \tilde{Q}_2, \tilde{P}, \eta_0) \right) \Bigg]$$

$$+ \sum_{j=1}^{2} \Bigg\{ \frac{\tilde{R}_0}{d_R \tilde{Q}_j} \frac{\partial \tilde{Q}_j}{\partial \tilde{R}_0} \Big[ 2 \left( \delta_{j,1} W_{1,1}(\bar{\alpha}_\eta, \tilde{R}_0, \tilde{Q}_1, \tilde{P}, \eta_0) - \delta_{j,2} W_{2,1}(\tilde{\alpha}_{\eta,2}, \tilde{R}_0, \tilde{Q}_2, \tilde{P}, \eta_0) \right)$$

$$+ W_{T,j}^Q(\tilde{\alpha}_{R,1}, \tilde{\alpha}_{R,2}, \tilde{\gamma}, \tilde{R}_0, \tilde{Q}_1, \tilde{Q}_2, \tilde{P}, \eta_0) \Big] + \frac{\partial \tilde{\alpha}_{R,j}}{\partial d_R} W_{T,j}^{\alpha_R}(\tilde{\alpha}_{R,1}, \tilde{\alpha}_{R,2}, \tilde{\gamma}, \tilde{R}_0, \tilde{Q}_1, \tilde{Q}_2, \tilde{P}, \eta_0)$$

$$+ \frac{\partial \tilde{\alpha}_{\eta,j}}{\partial d_R} \left( \delta_{j,1} W_{1,1}^{\alpha_\eta}(\tilde{\alpha}_\eta, \tilde{R}_0, \tilde{Q}_1, \tilde{P}, \eta_0) - \delta_{j,2} W_{2,1}^{\alpha_\eta}(\tilde{\alpha}_\eta, \tilde{R}_0, \tilde{Q}_2, \tilde{P}, \eta_0) \right) \Bigg\},$$

(7.16)

$$2\pi \frac{\partial \langle Wr \rangle_0}{\partial \theta_R} = \frac{1}{d_R}\left(\bar{W}_1^\theta(\tilde{R}_0, \tilde{P}, \alpha_\eta, \theta_R, \eta_0) - \bar{W}_2^\theta(\tilde{R}_0, \tilde{P}, \alpha_\eta, \theta_R, \eta_0)\right)$$

$$+ \frac{\partial \tilde{P}}{\partial \theta_R}\left[\frac{1}{d_R}\left(\bar{W}_1^P(\tilde{R}_0, \tilde{P}, \alpha_\eta, \theta_R, \eta_0) - \bar{W}_2^P(\tilde{R}_0, \tilde{P}, \alpha_\eta, \theta_R, \eta_0)\right) + W_{1,1}^P(\bar{\alpha}_\eta, \tilde{R}_0, \tilde{Q}_1, \tilde{P}, \eta_0)\right.$$

$$\left. - W_{2,1}^P(\tilde{\alpha}_{\eta,2}, \tilde{R}_0, \tilde{Q}_2, \tilde{P}, \eta_0) + W_T^P(\tilde{\alpha}_{R,1}, \tilde{\alpha}_{R,2}, \tilde{\gamma}, \tilde{R}_0, \tilde{Q}_1, \tilde{Q}_2, \tilde{P}, \eta_0)\right] - \frac{6 d_R^2}{l_p^2 \theta_R^7} W_T^\gamma(\tilde{\alpha}_{R,1}, \tilde{\alpha}_{R,2}, \tilde{\gamma}, \tilde{R}_0, \tilde{Q}_1, \tilde{Q}_2, \tilde{P}, \eta_0)$$

$$- \sum_{j=1}^2 \left\{ \frac{1}{\tilde{Q}_j} \frac{\partial \tilde{Q}_j}{\partial \theta_R}\left[2\left(\delta_{j,1} W_{1,1}(\bar{\alpha}_\eta, \tilde{R}_0, \tilde{Q}_1, \tilde{P}, \eta_0) - \delta_{j,2} W_{2,1}(\tilde{\alpha}_{\eta,2}, \tilde{R}_0, \tilde{Q}_2, \tilde{P}, \eta_0)\right)\right.\right.$$

$$\left. + W_{T,j}^Q(\tilde{\alpha}_{R,1}, \tilde{\alpha}_{R,2}, \tilde{\gamma}, \tilde{R}_0, \tilde{Q}_1, \tilde{Q}_2, \tilde{P}, \eta_0)\right] - \frac{\partial \tilde{\alpha}_{R,j}}{\partial \theta_R} W_{T,j}^{\alpha_R}(\tilde{\alpha}_{R,1}, \tilde{\alpha}_{R,2}, \tilde{\gamma}, \tilde{R}_0, \tilde{Q}_1, \tilde{Q}_2, \tilde{P}, \eta_0)$$

$$\left. - \frac{\partial \tilde{\alpha}_{\eta,j}}{\partial \theta_R}\left(\delta_{j,1} W_{1,1}^{\alpha_\eta}(\bar{\alpha}_\eta, \tilde{R}_0, \tilde{Q}_1, \tilde{P}, \eta_0) - \delta_{j,2} W_{2,1}^{\alpha_\eta}(\tilde{\alpha}_{\eta,2}, \tilde{R}_0, \tilde{Q}_2, \tilde{P}, \eta_0)\right)\right\},$$

(7.17)

$$2\pi \frac{\partial \langle Wr \rangle_0}{\partial \eta_0} = \frac{1}{d_R}\left(\bar{W}_1^\eta(\tilde{R}_0, \tilde{P}, \alpha_\eta, \theta_R, \eta_0) - \bar{W}_2^\eta(\tilde{R}_0, \tilde{P}, \alpha_\eta, \theta_R, \eta_0)\right)$$

$$+ W_{1,1}^\eta(\bar{\alpha}_\eta, \tilde{R}_0, \tilde{Q}_1, \tilde{P}, \eta_0) - W_{2,1}^\eta(\tilde{\alpha}_{\eta,2}, \tilde{R}_0, \tilde{Q}_2, \tilde{P}, \eta_0) + W_T^\eta(\tilde{\alpha}_{R,1}, \tilde{\alpha}_{R,2}, \tilde{\gamma}, \tilde{R}_0, \tilde{Q}_1, \tilde{Q}_2, \tilde{P}, \eta_0)$$

$$+ \frac{\partial \tilde{P}}{\partial \eta_0}\left[\frac{1}{d_R}\left(\bar{W}_1^P(\tilde{R}_0, \tilde{P}, \alpha_\eta, \theta_R, \eta_0) - \bar{W}_2^P(\tilde{R}_0, \tilde{P}, \alpha_\eta, \theta_R, \eta_0)\right)\right.$$

$$\left. + W_{1,1}^P(\bar{\alpha}_\eta, \tilde{R}_0, \tilde{Q}_1, \tilde{P}, \eta_0) - W_{2,1}^P(\tilde{\alpha}_{\eta,2}, \tilde{R}_0, \tilde{Q}_2, \tilde{P}, \eta_0) + W_T^P(\tilde{\alpha}_{R,1}, \tilde{\alpha}_{R,2}, \tilde{\gamma}, \tilde{R}_0, \tilde{Q}_1, \tilde{Q}_2, \tilde{P}, \eta_0)\right]$$

$$- \sum_{j=1}^2 \left\{ \frac{1}{\tilde{Q}_j} \frac{\partial \tilde{Q}_j}{\partial \eta_0}\left[2\left(\delta_{j,1} W_{1,1}^Q(\bar{\alpha}_\eta, \tilde{R}_0, \tilde{Q}_1, \tilde{P}, \eta_0) - \delta_{j,2} W_{2,1}^Q(\tilde{\alpha}_{\eta,2}, \tilde{R}_0, \tilde{Q}_2, \tilde{P}, \eta_0)\right)\right.\right.$$

$$\left. + W_{T,j}^Q(\tilde{\alpha}_{R,1}, \tilde{\alpha}_{R,2}, \tilde{\gamma}, \tilde{R}_0, \tilde{Q}_1, \tilde{Q}_2, \tilde{P}, \eta_0)\right] - \frac{\partial \tilde{\alpha}_{R,j}}{\partial \eta_0} W_{T,j}^{\alpha_R}(\tilde{\alpha}_{R,1}, \tilde{\alpha}_{R,2}, \tilde{\gamma}, \tilde{R}_0, \tilde{Q}_1, \tilde{Q}_2, \tilde{P}, \eta_0)$$

$$\left. - \frac{\partial \tilde{\alpha}_{\eta,j}}{\partial \eta_0}\left(\delta_{j,1} W_{1,1}^{\alpha_\eta}(\bar{\alpha}_\eta, \tilde{R}_0, \tilde{Q}_1, \tilde{P}, \eta_0) - \delta_{j,2} W_{2,1}^{\alpha_\eta}(\tilde{\alpha}_{\eta,2}, \tilde{R}_0, \tilde{Q}_2, \tilde{P}, \eta_0)\right)\right\},$$

(7.18)

$$2\pi \frac{\partial \langle Wr \rangle_0}{\partial \alpha_\eta} = \frac{1}{d_R}\left(\bar{W}_1^{\alpha_\eta}(\tilde{R}_0, \tilde{P}, \alpha_\eta, \theta_R, \eta_0) - \bar{W}_2^{\alpha_\eta}(\tilde{R}_0, \tilde{P}, \alpha_\eta, \theta_R, \eta_0)\right)$$

$$- \sum_{j=1}^2 \left\{ \frac{1}{\tilde{Q}_j} \frac{\partial \tilde{Q}_j}{\partial \alpha_\eta}\left[2\left(\delta_{j,1} W_{1,1}(\bar{\alpha}_\eta, \tilde{R}_0, \tilde{Q}_1, \tilde{P}, \eta_0) - \delta_{j,2} W_{2,1}(\tilde{\alpha}_{\eta,2}, \tilde{R}_0, \tilde{Q}_2, \tilde{P}, \eta_0)\right)\right.\right.$$

$$\left. + W_{T,j}^Q(\tilde{\alpha}_{R,1}, \tilde{\alpha}_{R,2}, \tilde{\gamma}, \tilde{R}_0, \tilde{Q}_1, \tilde{Q}_2, \tilde{P}, \eta_0)\right] - \frac{\partial \tilde{\alpha}_{R,j}}{\partial \alpha_\eta} W_{T,j}^{\alpha_R}(\tilde{\alpha}_{R,1}, \tilde{\alpha}_{R,2}, \tilde{\gamma}, \tilde{R}_0, \tilde{Q}_1, \tilde{Q}_2, \tilde{P}, \eta_0)$$

$$\left. - \frac{\partial \tilde{\alpha}_{\eta,j}}{\partial \alpha_\eta}\left(\delta_{j,1} W_{1,1}^{\alpha_\eta}(\bar{\alpha}_\eta, \tilde{R}_0, \tilde{Q}_1, \tilde{P}, \eta_0) - \delta_{j,2} W_{2,1}^{\alpha_\eta}(\tilde{\alpha}_{\eta,2}, \tilde{R}_0, \tilde{Q}_2, \tilde{P}, \eta_0)\right)\right\}.$$

(7.19)

The partial derivatives of $\tilde{P}$ are given by

$$\frac{\partial \tilde{P}}{\partial \tilde{R}_0} \approx \frac{1}{2\tan\left(\frac{\eta_0}{2}\right)} g(\tilde{R}_0)\left[1+\frac{\theta_R^2}{8\sin^2\left(\frac{\eta_0}{2}\right)}\right], \qquad \frac{\partial \tilde{P}}{\partial \theta_R} \approx \frac{\theta_R f(\tilde{R}_0)}{8} \frac{\cos\left(\frac{\eta_0}{2}\right)}{\sin^3\left(\frac{\eta_0}{2}\right)}, \qquad (7.20)$$

$$\frac{\partial \tilde{P}}{\partial \eta_0} \approx -\frac{f(\tilde{R}_0)}{\sin^2\left(\frac{\eta_0}{2}\right)}\left[\frac{1}{4}+\frac{\theta_R^2}{32}\frac{\left(2\cos^2\left(\frac{\eta_0}{2}\right)+1\right)}{\sin^2\left(\frac{\eta_0}{2}\right)}\right], \qquad (7.21)$$

where

$$g(\tilde{R}_0) \approx \frac{2}{\tilde{R}_0^3}+\frac{8}{\tilde{R}_0^5}. \qquad (7.22)$$

The partial derivatives of $\tilde{Q}_j$, $\tilde{\alpha}_\eta$, $\tilde{\alpha}_R$ and $\tilde{\gamma}$ are given in Appendix I.

Substitution of the results of this section into either into the system of equations Eqs. (4.48)-(4.61) or Eqs. (5.58)-(5.59) gives explicit relations defining $R_0$, $\theta_R$, $d_R$, $\eta_0$ and $\alpha_\eta$. Thus, we have a full set of explicitly defined equations for both approximations; weak and strong helix dependent forces.

## 7. The ground state limit of the average writhe.

Let us retrieve a limiting expression for the writhe in the absence of thermal fluctuations. For small thermal fluctuations the significant term in Eq. (6.14) is $W_{j,0}$. Expressions for $W_{j,0}$ simplify to

$$W_{1,0} \approx -\frac{\tilde{P}\tilde{R}_0^2}{2\pi R_0}\sin\left(\frac{\eta_0}{2}\right)\int_0^\infty r\,dr\,\exp\left(-\frac{\tilde{R}_0^2 r^2}{2}\right)\left((K_{1,1,0}(r,\tilde{P})+K_{1,1,1}(r,\tilde{P}))-K_{1,2,0}(r,\tilde{P})\tan\left(\frac{\eta_0}{2}\right)\right),$$

(8.1)

$$W_{2,0} \approx -\frac{\tilde{P}}{2\pi R_0}\sin\left(\frac{\eta_0}{2}\right)\left[K_{2,1,0}^0(\tilde{P})+\tan\left(\frac{\eta_0}{2}\right)K_{2,2,0}^0(\tilde{P})\right], \qquad (8.2)$$

and the expression for $\tilde{P}$

$$\tilde{P} \approx \frac{1}{2}\cot\left(\frac{\eta_0}{2}\right). \qquad (8.3)$$

Furthermore, in the absence of thermal fluctuations we can replace $\tilde{R}_0^2 r \exp(-\tilde{R}_0^2 r^2 / 2)$ in Eq. (8.1) with the delta function $\delta(r)$. Thus, we obtain, using Eqs. (6.82), (6.83) and (8.3),

$$W_{1,0} \approx -\frac{\tilde{P}}{2\pi R_0} \sin\left(\frac{\eta_0}{2}\right) \left( K_{1,1,0}(0, \tilde{P}) - \frac{1}{2\tilde{P}} K_{1,2,0}(0, \tilde{P}) \right) \equiv -\frac{2\tilde{P}}{R_0} \sin\left(\frac{\eta_0}{2}\right) I_1(\tilde{P}), \quad (8.4)$$

$$W_{2,0} \approx -\frac{2\tilde{P}}{R_0} \sin\left(\frac{\eta_0}{2}\right) I_2(\tilde{P}), \quad (8.5)$$

where

$$I_j(\tilde{P}) = \int_0^\infty dx \, \frac{1 + (-1)^j \cos x + (-1)^j \frac{x \sin x}{2}}{\left( \frac{1}{2}(1 + (-1)^j \cos x) + \tilde{P}^2 x^2 \right)^{3/2}}. \quad (8.6)$$

The total writhe, in the absence of thermal fluctuations, can then be written as

$$Wr = \frac{L_b \tilde{P}}{\pi R_0} \sin\left(\frac{\eta_0}{2}\right) \left( I_2(\tilde{P}) - I_1(\tilde{P}) \right). \quad (8.7)$$

Let us consider the case when $\tilde{P}$ is large corresponding to $\eta_0$ being small. This allows us to approximate

$$Wr \approx \frac{L_b}{\pi R_0} \sin\left(\frac{\eta_0}{2}\right) \left( 2 - 4 \tan^2\left(\frac{\eta_0}{2}\right) I_{1,0} \right) \approx \frac{L_b}{\pi R_0} \eta_0, \quad (8.8)$$

where

$$I_{1,0} = \int_0^\infty dx \, \frac{1 - \cos x - \frac{x \sin x}{2}}{x^3} = \frac{1}{4}. \quad (8.9)$$

This is exactly the result used in Ref. [4] with $\eta_0 = 2\alpha$.

## 9. More accurate treatment of end loops and the steric interaction.

Here, we discuss the use of the exact steric interaction potential in the exact partition function for the braided section, and a better treatment of the end loops. We start by considering a partition function for the braided section for given configurations at the ends of the braided section, the set of values $\{R_1, \eta_1\}$ and $\{R_2, \eta_2\}$, which are attached to the end loops. These end values then are summed over in total partition function, presented below. If the braided section is sufficiently long enough, and tight enough, we can still approximate both bending and interaction energies with Eqs. (2.4) and (2.14). Thus, we can write for the partition function of the braid for a particular end configuration

$$Z_{braid}\left(R_1, R_2, \eta_1, \eta_2, L_{loop}\right) = \int_{R(L_b)=R_1}^{R(L_b)=R_2} dR(s) \int_{\eta(L_b)=\eta_1}^{\eta(L_b)=\eta_2} d\eta(s) \int D\Delta\Phi(\tau) \Omega_{st}[R(\tau)] \exp\left(-\frac{\bar{E}_T[R(\tau), \eta(\tau), \Delta\Phi(\tau)]}{k_B T}\right).$$

(9.1)

The limits on the functional integrals represent the end boundary condition constraints on both $R(\tau)$ and $\eta(\tau)$. The function $\Delta\Phi(\tau)$ is considered unconstrained (the change in the its end values represents sliding of the molecule relative to the position of the braided section in the plectoneme). The energy functional is given by

$$\bar{E}_T[R(\tau), \eta(\tau), \Delta\Phi(\tau)] = E_{R,B}[R(\tau), \eta(\tau)] + E_{Tw,braid,1}[\Delta\Phi(\tau)] + l_{tw} L\left(g_{av}[R(\tau), \eta(\tau)] - \bar{g}_0\right)^2 \\ + E_{int}[R(\tau), \eta(\tau), \Delta\Phi(\tau)],$$

(9.2)

where $E_B[R(\tau), \eta(\tau)]$, $E_{Tw,braid,1}[\Delta\Phi(\tau)]$, and $E_{int}[R(\tau), \eta(\tau), \Delta\Phi(\tau)]$ are given by Eqs. (2.3), (2.4) ,(2.10) and (2.14). We also have that

$$g_{av}[R(\tau), \eta(\tau)] = \pi\left(Lk - Wr[R(\tau), \eta(\tau)]\right)/L,$$

(9.3)

and $Wr[R(\tau), \eta(\tau)]$ is given by Eq. (6.1) and (6.5), (6.6), and (6.7). Here, we still are assuming that $L_b$ is sufficiently long to neglect the writhe contribution from the end loops; however, in principle, their writhe contribution could also be calculated. The exclusion functional $\Omega_{st}[R(\tau)]$ is for the steric interactions. It is $\Omega_{st}[R(\tau)] = 0$ at any points along the braid (values of $\tau$) where $R(\tau) < 2a$ otherwise its value is $\Omega_{st}[R(\tau)] = 1$.

The end pieces can be dealt in a more accurate way than what was presented before. One approach is not to worry too much about thermal fluctuations and use a more elaborate trial function to describe the trajectories of the end loops, as in Ref. [18]. A second approach is to take the thermal fluctuations into account and write the partition functions

$$Z_{end,1}\left(R_2, \eta_2, L_{loop}\right) = \int_{\hat{\mathbf{t}}(0)=\hat{\mathbf{t}}_1(L_b)}^{\hat{\mathbf{t}}(L_{loop})=-\hat{\mathbf{t}}_2(L_b)} D\hat{\mathbf{t}}(s) \Gamma[\hat{\mathbf{t}}(s), R(L_b), \eta(L_b)] \exp\left(-\frac{l_p}{2} \int_0^{L_{loop}} ds \left(\frac{d\hat{\mathbf{t}}(s)}{ds}\right)^2\right),$$

(9.4)

$$Z_{end,2}\left(R_1, \eta_1, L_{loop}\right) = \int_{\hat{\mathbf{t}}(0)=-\hat{\mathbf{t}}_2(0)}^{\hat{\mathbf{t}}(L_{loop})=\hat{\mathbf{t}}_1(0)} D\hat{\mathbf{t}}(s) \Gamma[\hat{\mathbf{t}}(s), R(0), \eta(0)] \exp\left(-\frac{l_p}{2} \int_0^{L_{loop}} ds \left(\frac{d\hat{\mathbf{t}}(s)}{ds}\right)^2\right),$$

(9.5)

where expressions for the tangent vectors $\hat{\mathbf{t}}_1(L_b)$, $\hat{\mathbf{t}}_2(L_b)$, $\hat{\mathbf{t}}_1(0)$, and $\hat{\mathbf{t}}_2(0)$, can be found from Eqs. (1.13) and (1.14), where we set $R_1 = R(0)$, $R_2 = R(L_b)$, $\eta_1 = \eta(0)$, and $\eta_2 = \eta(L_b)$. Strictly speaking, we have constraint functionals $\Gamma[\hat{\mathbf{t}}(s), R(L_b), \eta(L_b)]$ and $\Gamma[\hat{\mathbf{t}}(s), R(0), \eta(0)]$ (these can be represented as functional delta functions) that only allow for non-zero contributions from configurations in the functional integrals such that

$$\int_0^{L_{loop}} ds\hat{\mathbf{t}}(s) = R(L_b)\left(\cos\theta(L_b)\hat{\mathbf{i}} + \sin\theta(L_b)\hat{\mathbf{j}}\right), \quad \int_0^{L_{loop}} ds\hat{\mathbf{t}}(s) = R(0)\left(\cos\theta(0)\hat{\mathbf{i}} + \sin\theta(0)\hat{\mathbf{j}}\right) \quad (9.6)$$

respectively. Now, the tangent vectors in the loop can be parameterized as

$$\hat{\mathbf{t}}(s) = \cos(\theta(s)+\theta_0)\sin\psi(s)\hat{\mathbf{i}} + \sin(\theta(s)+\theta_0)\sin\psi(s)\hat{\mathbf{j}} + \cos\psi(s)\hat{\mathbf{k}}, \quad (9.7)$$

and $\quad\quad\quad\quad \theta(0) = 0 \quad$ and $\quad \theta(L_{loop}) = \pi, \quad\quad\quad\quad (9.8)$

where $\theta_0$ is a constant rotation chosen, for each end loop, so to satisfy the conditions on vector orientations (the values of either $\theta(L_b)$ or $\theta(0)$), given in Eq. (9.6).

To obtain $\theta_0$ is rather tricky as the overall orientation of the end loop depends on the fluctuations in the braided section (see Eq. (1.15) for an expression for $\theta$ for the braid in terms of other parameters). However, for the energy the actual value of $\theta_0$ is unimportant, and so we need only consider $\theta_0 = 0$. Also, we simply require that

$$\int_0^{L_{loop}} ds\hat{\mathbf{t}}(s) = R(L_b)\hat{\mathbf{i}} = R_2\hat{\mathbf{i}}, \quad \text{or} \quad \int_0^{L_{loop}} ds\hat{\mathbf{t}}(s) = R(0)\hat{\mathbf{i}} = R_1\hat{\mathbf{i}}. \quad (9.9)$$

The tangent matching conditions $\hat{\mathbf{t}}(L_{loop}) = -\hat{\mathbf{t}}_2(L_b)$ and $\hat{\mathbf{t}}(0) = \hat{\mathbf{t}}_1(L_b)$ can be reconsidered as

$$\psi(0) = -\psi(L_{loop}) = -\eta(L_b) = \eta_2, \quad (9.10)$$

while the other matching conditions, for the other end loop $\hat{\mathbf{t}}(L_{loop}) = \hat{\mathbf{t}}_1(L_b)$ and $\hat{\mathbf{t}}(0) = -\hat{\mathbf{t}}_2(L_b)$ we simply require

$$-\psi(0) = \psi(L_{loop}) = -\eta(0) = -\eta_1. \quad (9.11)$$

Thus with the constraints (Eqs. (9.8), (9.9), (9.10) and (9.11)) we can in principle compute the partition functions $Z_{end,1}(R_2, \eta_2, L_{loop})$ and $Z_{end,2}(R_1, \eta_1, L_{loop})$ (Eqs. (9.4) and (9.5)).

The total partition function then can be written as

$$Z_{total} = \int_0^L dL_{loop} \int_{2a}^\infty dR_1 \int_{2a}^\infty dR_2 \int_{-\infty}^\infty d\eta_1 \int_{-\infty}^\infty d\eta_2 Z_{braid}(R_1, R_2, \eta_1, \eta_2, L_{loop}) Z_{end,1}(R_2, \eta_2, L_{loop}) Z_{end,2}(R_1, \eta_1, L_{loop}).$$
$$(9.12)$$

The total free energy should be in principle be computed by MC simulations for each specific partition function given in Eq. (9.12). For the loop, the geometrical constraints (Eqs. (9.8)-(9.11)) can probably be enforced by starting off with an allowed configuration known to satisfy the constraints for a given $L_{loop}$ and for fixed values of $R_1, R_2, \eta_1$ and $\eta_2$ (i.e. a ground state solution). Then, one considers MC moves where at one point, on the end loop, the tangent vector changes by

$$\delta\hat{\mathbf{t}} = \left(\cos(\theta(s))\cos\psi(s)\hat{\mathbf{i}} + \sin(\theta(s))\cos\psi(s)\hat{\mathbf{j}} - \sin\psi(s)\hat{\mathbf{k}}\right)\delta\psi \qquad (9.13)$$
$$+ \left(-\sin(\theta(s))\sin\psi(s)\hat{\mathbf{i}} + \cos(\theta(s))\sin\psi(s)\hat{\mathbf{j}}\right)\delta\theta,$$

followed by the choice of **another point**, at another random location, where the tangent vector changes by $-\delta\hat{\mathbf{t}}$. Both $\delta\psi$ and $\delta\theta$ are randomly chosen. This would be followed by a rescaling of the length tangent vectors to insure that $|\hat{\mathbf{t}}| = 1$ along the loop.

Another way might be considering changing the values of $R_1, R_2, \eta_1$ and $\eta_2$ also as additional MC moves and so to consider the total energy of both loops and braided section in constructing the free energy. Here, for each move that changes either the values of $R_1$ or $R_2$, the separation of each discrete element (of the discretization of $s$) in the end loop has to be changed by a fixed amount, and so length of the end loop $L_{loop}$ is appropriately changed, so that Eq. (9.9) is indeed satisfied.

## Acknowledgements

D.J. Lee would like to acknowledge useful discussions with R. Cortini, A. Korte, A. A. Korynshev, E. L. Starostin and G.H.M. van der Heijden, and. This work was initially inspired by joint work that has been supported by the United Kingdom Engineering and Physical Sciences Research Council (grant EP/H004319/1). He would also like to acknowledge the support of the Human Frontiers Science Program (grant RGP0049/2010-C102).

## Appendix A: Gaussian averaging formulas

Here, we will consider two types of general Gaussian average. The simplest is

$$\langle f(X(\tau), X(\tau'))\rangle_X = \frac{1}{Z}\int DX(\tau) f(X(\tau), X(\tau'))\exp\left(-E_X[X(\tau)]\right), \qquad (A.1)$$

where

$$Z_X = \int DX(\tau)\exp\left(-E_X[X(\tau)]\right), \qquad (A.2)$$

and for any Gaussian functional (in a translational invariant system) we have that

$$E_X[X(\tau)] = \frac{1}{2}\int_{-\infty}^{\infty} d\tau \int_{-\infty}^{\infty} d\tau' X(\tau) G_X^{-1}(\tau - \tau') X(\tau'). \qquad (A.3)$$

The general field $X(\tau)$ may correspond to the fields $\delta R(\tau)$, $\delta\eta(\tau)$ or $\delta\Phi(\tau)$ considered in the main text, and the variational trial functionals for each of the fields (see Eqs.(4.2), (5.13) and (5.29)) can be written in the form Eq. (A.3). As a particular example lets us consider the energy functional

$$E_\eta[\delta\eta(\tau)] = \int_0^{L_b} d\tau \left( \frac{l_p}{4}\left(\frac{d\delta\eta(\tau)}{d\tau}\right)^2 + \frac{\alpha_\eta}{2}\delta\eta(\tau)^2 \right) \approx \int_{-\infty}^{\infty} d\tau \left( \frac{l_p}{4}\left(\frac{d\delta\eta(\tau)}{d\tau}\right)^2 + \frac{\alpha_\eta}{2}\delta\eta(\tau)^2 \right) \quad (A.4)$$

that appears in the trial functional Eq. (4.2). The approximation we write on the right hand side of Eq. (A.4) is valid when $L_b \gg \lambda_\eta$, where $\lambda_\eta$ is the correlation length of the $\eta$-fluctuations. We may write this as Eq. (A.3), namely

$$E_\eta[\delta\eta(\tau)] = \int_{-\infty}^{\infty} d\tau \int_{-\infty}^{\infty} d\tau' \delta\eta(\tau') \left( -\frac{l_p}{4}\frac{d^2}{d\tau^2}\delta(\tau-\tau') + \frac{\alpha_\eta}{2}{}^2 \delta(\tau-\tau') \right) \delta\eta(\tau) \quad (A.5)$$

The more complicated average is

$$\langle g(X(\tau), X(\tau'), X'(\tau), X'(\tau'))\rangle_X = \frac{1}{Z_X} \int DX(\tau) g(X(\tau), X(\tau'), X'(\tau), X'(\tau')) \exp(-E_X[X(\tau)]).$$

(A.6)

Here $X'(\tau)$ refers to the differential of the field $X(\tau)$ with respect to its argument. Let us first focus on Eq.(A.1), we can first rewrite Eq. (A.1) as

$$\langle f(X(\tau), X(\tau'))\rangle_X = \frac{1}{Z_X} \int_{-\infty}^{\infty} dx \int_{-\infty}^{\infty} dx' f(x,x') \int DX(\tau) \delta(x - X(\tau))\delta(x' - X(\tau'))\exp(-E_X[X(\tau)])$$

$$\equiv \frac{1}{Z_X} \int_{-\infty}^{\infty} dx \int_{-\infty}^{\infty} dx' f(x,x') \Gamma_X(x,x';\tau-\tau'),$$

(A.7)

where we can write

$$\Gamma_X(x,x';\tau-\tau') = \frac{1}{Z_X (2\pi)^2} \int_{-\infty}^{\infty} dp \int_{-\infty}^{\infty} dp' \int DX(\tau) \exp(ip(x-X(\tau)))\exp(ip'(x'-X(\tau')))\exp(-E_X[X(\tau)]).$$

(A.8)

We can first rewrite Eq. (A.8) using Fourier transforms in terms for the transformed field $\tilde{X}(k)$. This reads as

$$\Gamma_X(x,x';\tau-\tau') = \frac{1}{\tilde{Z}_X(2\pi)^2} \int_{-\infty}^{\infty} dp \int_{-\infty}^{\infty} dp' \int D\tilde{X}(k) \exp\left(-\frac{i}{4\pi} \int_{-\infty}^{\infty} X(k)\left(p\exp(ik\tau)+p'\exp(ik\tau')\right)\right)$$

$$\exp\left(-\frac{i}{4\pi} \int_{-\infty}^{\infty} X(-k)\left(p\exp(-ik\tau)+p'\exp(-ik\tau')\right)\right) \exp\left(ip'x'+ipx\right) \exp\left(-E_X[X(k)]\right),$$

(A.9)

where

$$\tilde{Z}_X = \int D\tilde{X}(k) \exp\left(-E_X[\tilde{X}(k)]\right),$$

(A.10)

and

$$E_X[\tilde{X}(k)] = \frac{1}{2\pi} \int_{-\infty}^{\infty} dk \frac{\tilde{X}(k)\tilde{G}_X^{-1}(k)\tilde{X}(-k)}{2}.$$

(A.11)

Therefore, we can perform the Gaussian integration over $\tilde{X}(k)$. This yields

$$\Gamma_X(x,x';\tau-\tau') = \frac{1}{(2\pi)^2} \int_{-\infty}^{\infty} dp \int_{-\infty}^{\infty} dp' \exp\left(-\frac{d_X^2}{2}\left(p^2+p'^2\right)\right) \exp\left(-G_X(\tau-\tau')pp'\right)$$

$$\exp\left(ip'x'+ipx\right).$$

(A.12)

where

$$d_X^2 = \frac{1}{2\pi} \int_{-\infty}^{\infty} dk\tilde{G}_X(k), \quad G_X(\tau-\tau') = \frac{1}{2\pi} \int_{-\infty}^{\infty} dk\tilde{G}_X(k)\exp\left(-ik(\tau-\tau')\right),$$

(A.13)

and

$$G_X^{-1}(\tau-\tau') = \frac{1}{2\pi} \int_{-\infty}^{\infty} dk \frac{1}{\tilde{G}_X(k)} \exp\left(-ik(\tau-\tau')\right).$$

(A.14)

Integration over $p$ and $p'$ in Eq. (A.12) is straightforward yielding

$$\Gamma_X(x,x';\tau-\tau') = \frac{1}{(2\pi)} \frac{1}{\sqrt{d_X^4 - G_X(\tau-\tau')^2}} \exp\left(-\frac{(x^2+x'^2)d_X^2}{2(d_X^4 - G_X(\tau-\tau')^2)}\right) \exp\left(-\frac{xx'G_X(\tau-\tau')}{(d_X^4 - G_X(\tau-\tau')^2)}\right),$$

(A.15)

Now, let us focus on the average given by Eq. (A.6). This can be rewritten as

$$\langle g(X(\tau), X(\tau'), X'(\tau), X'(\tau'))\rangle_X = \int_{-\infty}^{\infty} dx \int_{\infty}^{\infty} dx' \int_{-\infty}^{\infty} dy \int_{\infty}^{\infty} dy' g(x,x',y,y') \Xi_X(x,x',y,y';\tau-\tau'),$$

(A.16)

where

$$\Xi_X(x,x',y,y';\tau-\tau') = \frac{1}{Z_X}\int DX(\tau)\delta(x-X(\tau))\delta(x'-X(\tau'))\delta(y-X'(\tau))\delta(y'-X'(\tau'))\exp(-E_X[X(\tau)]).$$

(A.17)

We can rewrite Eq. (A.17) as

$$\Xi_X(x,x',y,y';\tau-\tau') = \frac{1}{(2\pi)^4 Z_X}\int_{-\infty}^{\infty}dp\int_{-\infty}^{\infty}dp'\int_{-\infty}^{\infty}dq\int_{-\infty}^{\infty}dq'\int DX(\tau)\exp(ip(x-X(\tau)))\exp(ip'(x'-X(\tau')))$$
$$\exp(iq(y-X'(\tau)))\exp(iq'(y'-X'(\tau')))\exp(-E_X[X(\tau)]).$$

(A.18)

We can further write

$$\Xi_X(x,x',y,y';\tau-\tau') = \frac{1}{\tilde{Z}_X(2\pi)^4}\int_{-\infty}^{\infty}dp\int_{-\infty}^{\infty}dp'\int_{-\infty}^{\infty}dq\int_{-\infty}^{\infty}dq'\int D\tilde{X}(k)\exp\left(-\frac{i}{4\pi}\int_{-\infty}^{\infty}\tilde{X}(k)\big(p\exp(ik\tau)+p'\exp(ik\tau')\big)\right)$$
$$\exp\left(-\frac{i}{4\pi}\int_{-\infty}^{\infty}\tilde{X}(k)\big(iqk\exp(ik\tau)+iq'k\exp(ik\tau')\big)\right)\exp\left(-\frac{i}{4\pi}\int_{-\infty}^{\infty}\tilde{X}(-k)\big(p\exp(-ik\tau)+p'\exp(-ik\tau')\big)\right)$$
$$\exp\left(-\frac{i}{4\pi}\int_{-\infty}^{\infty}\tilde{X}(-k)\big(-iqk\exp(-ik\tau)-iq'k\exp(-ik\tau')\big)\right)\exp(ip'x'+ipx+iqy+iq'y')\exp(-E_X[\tilde{X}(k)]).$$

(A.19)

We can, again, perform the integration over $\tilde{X}(k)$, which leaves us with the form

$$\Xi_X(x,x',y,y';\tau-\tau') = \frac{1}{(2\pi)^4}\int d\mathbf{P}\exp\left(-\frac{\mathbf{P}^T\mathbf{M}(\tau-\tau')\mathbf{P}}{2}\right)\exp\left(i\frac{\mathbf{X}^T\mathbf{P}+\mathbf{P}^T\mathbf{X}}{2}\right),$$

(A.20)

where

$$\mathbf{P} = \begin{pmatrix} p \\ p' \\ q \\ q' \end{pmatrix}, \quad \mathbf{X} = \begin{pmatrix} x \\ x' \\ y \\ y' \end{pmatrix}, \text{ and } \mathbf{M}(\tau-\tau') = \begin{pmatrix} d_X^2 & G_X(\tau-\tau') & 0 & C_X(\tau-\tau') \\ G_X(\tau-\tau') & d_X^2 & -C_X(\tau-\tau') & 0 \\ 0 & -C_X(\tau-\tau') & \theta_X^2 & D_X(\tau-\tau') \\ C_X(\tau-\tau') & 0 & D_X(\tau-\tau') & \theta_X^2 \end{pmatrix},$$

(A.21)

and

$$\theta_X^2 = \frac{1}{2\pi}\int_{-\infty}^{\infty}k^2\tilde{G}_X(k)dk,$$

(A.22)

$$D_X(\tau-\tau') = \frac{1}{2\pi}\int_{-\infty}^{\infty}k^2\tilde{G}_X(k)\exp(-ik(\tau-\tau'))dk,$$

(A.23)

$$C_X(\tau-\tau') = \frac{1}{2\pi}\int_{-\infty}^{\infty} ik\tilde{G}_X(k)\exp(-ik(\tau-\tau'))dk. \qquad (A.24)$$

Here, $\mathbf{X}^T$ and $\mathbf{P}^T$ are the transposes of the column vectors $\mathbf{X}$ and $\mathbf{P}$, and $d\mathbf{P} = dpdp'dqdq'$ respectively. The integration over $\mathbf{P}$ can be performed, yielding

$$\Xi_X(x,x',y,y';\tau-\tau') = \frac{1}{(2\pi)^2}\sqrt{\frac{1}{\det(\mathbf{M}(\tau-\tau'))}}\exp\left(-\frac{\mathbf{X}^T\mathbf{M}^{-1}(\tau-\tau')\mathbf{X}}{2}\right), \qquad (A.25)$$

where

$$\det(\mathbf{M}(\tau-\tau')) = d_X^4\theta_X^4 - G_X(\tau-\tau')^2\theta_X^4 - D_X(\tau-\tau')^2 d_R^4 + D_X(\tau-\tau')^2 G_X(\tau-\tau')^2 \\ -2d_X^2\theta_X^2 C_X(\tau-\tau')^2 + 2C_X(\tau-\tau')^2 G_X(\tau-\tau')D_X(\tau-\tau') + C_X(\tau-\tau')^4, \qquad (A.26)$$

and the inverse matrix is given by

$$\mathbf{M}^{-1}(\tau-\tau') = \frac{1}{\det(\mathbf{M}(\tau-\tau'))}\mathbf{N}(\tau-\tau'), \qquad (A.27)$$

with

$$\mathbf{N}(\tau-\tau') = \begin{pmatrix} N_1(\tau-\tau') & N_2(\tau-\tau') & N_3(\tau-\tau') & N_4(\tau-\tau') \\ N_2(\tau-\tau') & N_1(\tau-\tau') & -N_4(\tau-\tau') & -N_3(\tau-\tau') \\ N_3(\tau-\tau') & -N_4(\tau-\tau') & N_5(\tau-\tau') & N_6(\tau-\tau') \\ N_4(\tau-\tau') & -N_3(\tau-\tau') & N_6(\tau-\tau') & N_5(\tau-\tau') \end{pmatrix}, \qquad (A.28)$$

and

$$N_1(\tau-\tau') = d_X^2\theta_X^4 - d_X^2 D_X(\tau-\tau')^2 - \theta_X^2 C_X(\tau-\tau')^2, \qquad (A.29)$$

$$N_2(\tau-\tau') = -G_X(\tau-\tau')\theta_R^4 + G_X(\tau-\tau')D_X(\tau-\tau')^2 + D_X(\tau-\tau')C_X(\tau-\tau')^2, \qquad (A.30)$$

$$N_3(\tau-\tau') = -G_X(\tau-\tau')C_X(\tau-\tau')\theta_X^2 + d_X^2 D_X(\tau-\tau')C_X(\tau-\tau'), \qquad (A.31)$$

$$N_4(\tau-\tau') = -d_X^2\theta_X^2 C_X(\tau-\tau') + G_X(\tau-\tau')D_X(\tau-\tau')C_X(\tau-\tau') + C_X(\tau-\tau')^3, \qquad (A.32)$$

$$N_5(\tau-\tau') = d_X^4\theta_X^2 - G_X(\tau-\tau')^2\theta_X^2 - C_X(\tau-\tau')^2 d_X^2, \qquad (A.33)$$

$$N_6(\tau-\tau') = -d_X^4 D_X(\tau-\tau') + G_X(\tau-\tau')^2 D_X(\tau-\tau') + G_X(\tau-\tau')C_X(\tau-\tau')^2. \qquad (A.34)$$

## Appendix B: Large $\tau-\tau'$ approximation of averaging formulas

Here we will assume that $\tau - \tau'$ can be taken to be large. Thus we can write (easiest to obtain from Eq.(A.12))

$$\Gamma_X(x,x';\tau-\tau') \approx \Gamma_{X,0}(x,x') + \Gamma_{X,1}(x,x')G_X(\tau-\tau'), \tag{B.1}$$

where

$$\Gamma_{X,0}(x,x') = \frac{1}{\tilde{Z}_X(2\pi)^2} \int_{-\infty}^{\infty} dp \int_{-\infty}^{\infty} dp' \exp\left(-\frac{d_X^2}{2}(p^2+p'^2)\right)\exp(ip'x'+ipx)$$
$$= \frac{1}{2\pi d_X^2} \exp\left(-\frac{x^2+x'^2}{2d_X^2}\right), \tag{B.2}$$

$$\Gamma_{X,1}(x,x') = -\frac{1}{\tilde{Z}_X(2\pi)^2} \int_{-\infty}^{\infty} dp \int_{-\infty}^{\infty} dp' pp' \exp\left(-\frac{d_X^2}{2}(p^2+p'^2)\right)\exp(ip'x'+ipx)$$
$$= \frac{xx'}{2\pi d_X^6} \exp\left(-\frac{x^2+x'^2}{2d_X^2}\right). \tag{B.3}$$

We can also write (these expressions are obtained easiest from Eq. (A.20))

$$\Xi_X(x,x',y,y';\tau-\tau') \approx \Xi_{X,0}(x,x',y,y') + \Xi_{X,1,1}(x,x',y,y')G_X(\tau-\tau') + \Xi_{X,1,2}(x,x',y,y')D_X(\tau-\tau')$$
$$+\Xi_{X,1,3}(x,x',y,y')C_X(\tau-\tau'), \tag{B.4}$$

where

$$\Xi_{X,0}(x,x',y,y') = \frac{1}{\tilde{Z}_X(2\pi)^4} \int_{-\infty}^{\infty} dq \int_{-\infty}^{\infty} dq' \int_{-\infty}^{\infty} dp \int_{-\infty}^{\infty} dp' \exp\left(-\frac{d_X^2}{2}(p^2+p'^2)\right)\exp\left(-\frac{\theta_X^2}{2}(q^2+q'^2)\right)$$
$$\exp(ip'x'+ipx)\exp(iqy+iq'y')$$
$$= \frac{1}{(2\pi)^2 d_X^2 \theta_X^2} \exp\left(-\frac{x^2+x'^2}{2d_X^2}\right)\exp\left(-\frac{y^2+y'^2}{2\theta_X^2}\right),$$

$$\tag{B.5}$$

$$\Xi_{X,1,1}(x,x',y,y') = -\frac{1}{\tilde{Z}_X(2\pi)^4} \int_{-\infty}^{\infty} dq \int_{-\infty}^{\infty} dq' \int_{-\infty}^{\infty} dp \int_{-\infty}^{\infty} dp' pp' \exp\left(-\frac{d_X^2}{2}(p^2+p'^2)\right)\exp\left(-\frac{\theta_X^2}{2}(q^2+q'^2)\right)$$
$$\exp(ip'x'+ipx)\exp(iqy+iq'y')$$
$$= \frac{xx'}{(2\pi)^2 d_X^6 \theta_X^2} \exp\left(-\frac{x^2+x'^2}{2d_X^2}\right)\exp\left(-\frac{y^2+y'^2}{2\theta_X^2}\right),$$

$$\tag{B.6}$$

$$\Xi_{X,1,2}(x,x',y,y') = -\frac{1}{\tilde{Z}_X (2\pi)^4} \int_{-\infty}^{\infty} dq \int_{-\infty}^{\infty} dq' \int_{-\infty}^{\infty} dp \int_{-\infty}^{\infty} dp' qq' \exp\left(-\frac{d_X^2}{2}(p^2+p'^2)\right)\exp\left(-\frac{\theta_X^2}{2}(q^2+q'^2)\right)$$
$$\exp(ip'x'+ipx)\exp(iqy+iq'y')$$
$$= \frac{yy'}{(2\pi)^2 d_2^2 \theta_X^6} \exp\left(-\frac{x^2+x'^2}{2d_X^2}\right)\exp\left(-\frac{y^2+y'^2}{2\theta_X^2}\right),$$

(B.7)

$$\Xi_{X,1,3}(x,x',y,y') = -\frac{1}{\tilde{Z}_X (2\pi)^4} \int_{-\infty}^{\infty} dq \int_{-\infty}^{\infty} dq' \int_{-\infty}^{\infty} dp \int_{-\infty}^{\infty} dp' \exp\left(-\frac{d_X^2}{2}(p^2+p'^2)\right)\exp\left(-\frac{\theta_X^2}{2}(q^2+q'^2)\right)$$
$$(pq'-qp')\exp(ip'x'+ipx)\exp(iqy+iq'y')$$
$$= \frac{(xy'-yx')}{(2\pi)^2 d_X^4 \theta_X^4} \exp\left(-\frac{x^2+x'^2}{2d_X^2}\right)\exp\left(-\frac{y^2+y'^2}{2\theta_X^2}\right).$$

(B.8)

## Appendix C Useful expressions for section 5

Let's consider the general case:

$$\Omega_{1,X} = \frac{1}{l_p} \int_{-\infty}^{\infty} dx G_X(x) \exp\left(-\frac{n^2|x|}{\lambda_c}\right) = \frac{1}{2\pi l_p} \int_0^{\infty} dx \int_{-\infty}^{\infty} dk \tilde{G}_X(k) \exp(-ikx) \exp\left(-\frac{n^2 x}{\lambda_c}\right)$$
$$+ \frac{1}{2\pi l_p} \int_{-\infty}^{0} dx \int_{-\infty}^{\infty} dk \tilde{G}_X(k) \exp(-ikx) \exp\left(\frac{n^2 x}{\lambda_c}\right) = \frac{1}{2\pi l_p} \int_{-\infty}^{\infty} dk \tilde{G}_X(k) \left(\frac{1}{ik+n^2/\lambda_c} + \frac{1}{n^2/\lambda_c - ik}\right)$$
$$= \frac{n^2}{\pi \lambda_c l_p} \int_{-\infty}^{\infty} dk \frac{\tilde{G}_X(k)}{k^2 + (n^2/\lambda_c)^2}.$$

(C.1)

Thus, we may write

$$\Omega_{1,\eta} = \frac{1}{l_p} \int_{-\infty}^{\infty} dx G_\eta(x) \exp\left(-\frac{n^2|x|}{\lambda_c}\right) = \frac{2n^2}{\pi \lambda_c l_p^2} \int_{-\infty}^{\infty} dk \frac{1}{k^2 + \frac{2\alpha_\eta}{l_p}} \frac{1}{k^2 + \left(\frac{n^2}{\lambda_c}\right)^2}$$
$$= \frac{2n^2}{\lambda_c l_p^2} \left(\left(\frac{l_p}{2\alpha_\eta}\right)^{1/2} - \frac{\lambda_c}{n^2}\right) \frac{1}{\left(\frac{n^2}{\lambda_c}\right)^2 - \left(\frac{2\alpha_\eta}{l_p}\right)} = 2\left(\frac{1}{2\alpha_\eta l_p}\right)^{1/2} \frac{1}{\left(\frac{n^2 l_p}{\lambda_c}\right) + (2\alpha_\eta l_p)^{1/2}}.$$

(C.2)

The last line gives Eq. (5.42) of the main text. We may also write

$$\Omega_{1,R} = \frac{1}{l_p^3} \int_{-\infty}^{\infty} dx G_R(x) \exp\left(-\frac{n^2|x|}{\lambda_c}\right) = \frac{2n^2 l_p}{\pi \lambda_c} \int_{-\infty}^{\infty} dk \frac{1}{k^4 + 2\beta_R l_p k^2 + 2\alpha_R l_p^3} \frac{1}{k^2 + \left(\frac{n^2 l_p}{\lambda_c}\right)^2} \quad \text{(C.3)}$$

$$= \frac{2n^2 l_p}{\pi \lambda_c} (2\alpha_R l_p^3)^{-5/4} \int_{-\infty}^{\infty} dk \frac{1}{k^4 + \gamma k^2 + 1} \frac{1}{k^2 + \delta} \equiv \frac{4n^2 l_p}{\lambda_c} (2\alpha_R l_p^3)^{-5/4} I_1,$$

where

$$\tilde{\gamma} = 2^{1/2} l_p^{-1/2} \beta_R \alpha_R^{-1/2}, \qquad \delta = \left(\frac{n^2 l_p}{\lambda_c}\right)^2 \left(\frac{1}{2\alpha_R l_p^3}\right)^{1/2}. \quad \text{(C.4)}$$

Following the analysis of Ref. [19] we can express

$$I_1 = \frac{1}{2\pi} \int_{-\infty}^{\infty} dk \frac{1}{(k^2 - K^+)(k^2 - K^-)} \frac{1}{(k^2 + \delta)}, \quad \text{(C.5)}$$

where

$$K^\pm = \frac{-\tilde{\gamma} \pm \sqrt{\tilde{\gamma}^2 - 4}}{2}. \quad \text{(C.6)}$$

We need to deal with two separate cases: $-2 < \tilde{\gamma} < 2$, where $K^+$ and $K^+$ are complex, and the case $\tilde{\gamma} \geq 2$ where the roots are negative and real. We then complete factorization by finding the roots to the equations $k^2 = K^+$ and $k^2 = K^-$, $k_1$, $k_2$, $k_3$ and $k_4$. This allows us to write

$$I_1 = \frac{1}{2\pi} \int_{-\infty}^{\infty} dk \frac{1}{(k - k_1)(k - k_2)(k - k_3)(k - k_4)} \frac{1}{(k + i\sqrt{\delta})} \frac{1}{(k - i\sqrt{\delta})}, \quad \text{(C.7)}$$

where for $-2 < \tilde{\gamma} < 2$

$$k_1 = \frac{1}{\sqrt{2}} \left( \sqrt{1 - \frac{\tilde{\gamma}}{2}} + i\sqrt{1 + \frac{\tilde{\gamma}}{2}} \right) = -k_2 \quad \text{and} \quad k_3 = \frac{1}{\sqrt{2}} \left( -\sqrt{1 - \frac{\tilde{\gamma}}{2}} + i\sqrt{1 + \frac{\tilde{\gamma}}{2}} \right) = -k_4, \quad \text{(C.8)}$$

and for $\tilde{\gamma} \geq 2$

$$k_1 = \frac{i}{\sqrt{2}} \left(\tilde{\gamma} + \sqrt{\tilde{\gamma}^2 - 4}\right)^{1/2} = -k_2 \quad \text{and} \quad k_3 = \frac{i}{\sqrt{2}} \left(\tilde{\gamma} - \sqrt{\tilde{\gamma}^2 - 4}\right)^{1/2} = -k_4. \quad \text{(C.9)}$$

Then we can use the standard method of contour integration to evaluate the integrals

$$I_1 = \frac{i}{(k_1-k_2)(k_1-k_3)(k_1-k_4)} \frac{1}{\delta^2+k_1^2} + \frac{i}{(k_3-k_2)(k_3-k_1)(k_3-k_4)} \frac{1}{\delta^2+k_3^2}$$
$$+ \frac{1}{\delta^2-\gamma\delta+1} \frac{1}{2\sqrt{\delta}}.$$
(C.10)

Substituting in Eqs. (C.8) or (C.9) we obtain

$$I_1(\tilde{\gamma},\delta) = \frac{1}{\delta^2-\tilde{\gamma}\delta+1}\left(\frac{1}{2\sqrt{\delta}} - \frac{1}{4}\sqrt{2+\tilde{\gamma}} + \frac{(\delta-\tilde{\gamma}/2)}{2\sqrt{2+\tilde{\gamma}}}\right).$$
(C.11)

Thus, we have for $\Omega_{1,R}$ Eqs. (5.43) and (5.44) of the main text.

# Appendix D Differentials of $f_{int,0}$, $f_{int,\eta}$, $f_{int,\eta\eta}$, $f_{int,1}$, $\Omega_{1,\eta}$ and $\Omega_{1,R}$

We have

$$\frac{\partial \tilde{f}_{int,X}\left(R_0,\eta_0,d_R,\langle g_{av}\rangle_0,n,d_{max},d_{min}\right)}{\partial R_0} = \frac{1}{d_R\sqrt{2\pi}} \int_{-\infty}^{\infty} dr \exp\left(-\frac{r^2}{2d_R}\right) \frac{\partial \overline{\varepsilon}_{int,X}(R_0,r,\eta_0,\langle g_{av}\rangle_0,n)}{\partial R_0},$$
(D.1)

$$\frac{\partial \tilde{f}_{int,X}\left(R_0,\eta_0,d_R,\langle g_{av}\rangle_0,n,d_{max},d_{min}\right)}{\partial \eta_0} = \frac{1}{d_R\sqrt{2\pi}} \int_{-\infty}^{\infty} dr \exp\left(-\frac{r^2}{2d_R}\right) \frac{\partial \overline{\varepsilon}_{int,X}(R_0,r,\eta_0,\langle g_{av}\rangle_0,n)}{\partial \eta_0},$$
(D.2)

$$\frac{\partial \tilde{f}_{int,X}\left(R_0,\eta_0,d_R,\langle g_{av}\rangle_0,n,d_{max},d_{min}\right)}{\partial d_R} = \frac{1}{d_R^2\sqrt{2\pi}} \int_{-\infty}^{\infty} dr \left(\frac{r^2}{d_R^2}-1\right)\exp\left(-\frac{r^2}{2d_R}\right) \overline{\varepsilon}_{int,X}(R_0,r,\eta_0,\langle g_{av}\rangle_0,n),$$
(D.3)

where $X=0,\eta$ or $\eta\eta$. For the derivatives of $\tilde{f}_{int,1}$ we have

$$\frac{\partial \tilde{f}_{int,1}\left(R_0,\eta_0,\langle g_{av}\rangle_0,d_R,n,d_{max},d_{min}\right)}{\partial R_0} = \frac{1}{d_R^3\sqrt{2\pi}} \int_{-\infty}^{\infty} dr\, r \exp\left(-\frac{r^2}{2d_R}\right) \frac{\partial \overline{\varepsilon}_{int,0}(R_0,r,\eta_0,\langle g_{av}\rangle_0,n)}{\partial R_0},$$
(D.4)

$$\frac{\partial \tilde{f}_{int,1}\left(R_0,\eta_0,\langle g_{av}\rangle_0,d_R,n,d_{max},d_{min}\right)}{\partial \eta_0} = \frac{1}{d_R^3\sqrt{2\pi}} \int_{-\infty}^{\infty} dr\, r \exp\left(-\frac{r^2}{2d_R}\right) \frac{\partial \overline{\varepsilon}_{int,0}(R_0,r,\eta_0,\langle g_{av}\rangle_0,n)}{\partial \eta_0},$$
(D.5)

$$\frac{\partial \tilde{f}_{int,1}(R_0,\eta_0,\langle g_{av}\rangle_0,d_R,n,d_{max},d_{min})}{\partial d_R} = \frac{1}{d_R^4\sqrt{2\pi}}\int_{-\infty}^{\infty} dr\, r\left(\frac{r^2}{d_R^2}-3\right)\exp\left(-\frac{r^2}{2d_R}\right)\bar{\varepsilon}_{int,0}(R_0,r,\eta_0,\langle g_{av}\rangle_0,n).$$

(D.6)

Finally, for the derivatives of $\Omega_{1,\eta}$ and $\Omega_{1,R}$ we have

$$\frac{\partial \Omega_{1,\eta}\left(\frac{l_p n^2}{\lambda_c},\alpha_\eta l_p\right)}{\partial \alpha_\eta} = -2l_p\left[\frac{1}{(2\alpha_\eta l_p)^{3/2}}\frac{1}{\frac{l_p n^2}{\lambda_c}+(2\alpha_\eta l_p)^{1/2}} + \frac{1}{(2\alpha_\eta l_p)}\frac{1}{\left(\frac{l_p n^2}{\lambda_c}+(2\alpha_\eta l_p)^{1/2}\right)^2}\right]$$

(D.7)

$$\frac{\partial}{\partial d_R}\Omega_{1,R}\left(\frac{n^2 l_p}{\lambda_c},\frac{l_p^4}{2}\left(\frac{\theta_R}{d_R}\right)^4,\frac{1}{2}\left(\frac{1}{\theta_R^4}-2\left(\frac{l_p\theta_R}{d_R}\right)^2\right)\right) = \frac{20n^2}{\lambda_c\theta_R}\left(\frac{d_R}{l_p\theta_R}\right)^4 I_1\left(\frac{d_R^2}{l_p^2\theta_R^6}-2,\left(\frac{n^2 d_R}{\lambda_c\theta_R}\right)^2\right)$$
$$+\frac{8n^2}{\lambda_c\theta_R^5}\left(\frac{d_R}{l_p\theta_R}\right)^6 I_1^\gamma\left(\frac{d_R^2}{l_p^2\theta_R^6}-2,\left(\frac{n^2 d_R}{\lambda_c\theta_R}\right)^2\right) + \frac{8}{l_p\theta_R}\left(\frac{d_R}{l_p\theta_R}\right)^6\left(\frac{n^2 l_p}{\lambda_c}\right)^3 I_1^\delta\left(\frac{d_R^2}{l_p^2\theta_R^6}-2,\left(\frac{n^2 d_R}{\lambda_c\theta_R}\right)^2\right),$$

(D.8)

$$\frac{\partial}{\partial \theta_R}\Omega_{1,R}\left(\frac{n^2 l_p}{\lambda_c},\frac{l_p^4}{2}\left(\frac{\theta_R}{d_R}\right)^4,\frac{1}{2}\left(\frac{1}{\theta_R^4}-2\left(\frac{l_p\theta_R}{d_R}\right)^2\right)\right) = -\frac{20 n^2 l_p}{\lambda_c\theta_R}\left(\frac{d_R}{l_p\theta_R}\right)^5 I_1\left(\frac{d_R^2}{l_p^2\theta_R^6}-2,\left(\frac{n^2 d_R}{\lambda_c\theta_R}\right)^2\right)$$
$$-\frac{24 n^2 l_p}{\lambda_c\theta_R^5}\left(\frac{d_R}{l_p\theta_R}\right)^7 I_1^\gamma\left(\frac{d_R^2}{l_p^2\theta_R^6}-2,\left(\frac{n^2 d_R}{\lambda_c\theta_R}\right)^2\right) - \frac{8n^2 l_p}{\lambda_c\theta_R}\left(\frac{d_R}{l_p\theta_R}\right)^7\left(\frac{n^2 l_p}{\lambda_c}\right)^2 I_1^\delta\left(\frac{d_R^2}{l_p^2\theta_R^6}-2,\left(\frac{n^2 d_R}{\lambda_c\theta_R}\right)^2\right),$$

(D.9)

as well as

$$I_1^\gamma(\gamma,\delta) = \frac{\partial I_1(\gamma,\delta)}{\partial \gamma} = \frac{\delta}{(\delta^2-\gamma\delta+1)^2}\left(\frac{1}{2\sqrt{\delta}} - \frac{1}{4}\sqrt{2+\gamma} + \frac{(\delta-\gamma/2)}{2\sqrt{2+\gamma}}\right)$$
$$-\frac{1}{\delta^2-\gamma\delta+1}\left[\frac{3}{8}\frac{1}{\sqrt{2+\gamma}} + \frac{(\delta-\gamma/2)}{4(2+\gamma)^{3/2}}\right],$$

(D.10)

$$I_1^\delta(\gamma,\delta) = \frac{\partial I_1(\gamma,\delta)}{\partial \delta} = \frac{\gamma - 2\delta}{\left(\delta^2 - \gamma\delta + 1\right)^2}\left(\frac{1}{2\sqrt{\delta}} - \frac{1}{4}\sqrt{2+\gamma} + \frac{(\delta - \gamma/2)}{2\sqrt{2+\gamma}}\right)$$
$$+ \frac{1}{\delta^2 - \gamma\delta + 1}\left[\frac{1}{2\sqrt{2+\gamma}} - \frac{1}{4\delta^{3/2}}\right].$$
(D.11)

## Appendix E Results for the small $r$ expansions of $J_{2,k,l}(x,r)$ and $H_{k,l}(x,r)$

For various results in the small $r$ expansion of $J_{2,k,l}(x,r)$ have

$$J_{2,0,0}(x,r) \approx \frac{2\pi}{\left(1+x^2\right)^{3/2}} + \frac{\pi r^2}{4}\left[\frac{15}{\left(1+x^2\right)^{7/2}} - \frac{3}{\left(1+x^2\right)^{5/2}}\right]$$
$$+ \frac{45\pi r^4}{128}\left[\frac{21}{\left(1+x^2\right)^{11/2}} - \frac{14}{\left(1+x^2\right)^{9/2}} + \frac{1}{\left(1+x^2\right)^{7/2}}\right]$$
(E.1)

$$J_{2,1,0}(x,r) \approx -\frac{3\pi r^2}{2\left(1+x^2\right)^{5/2}} + \frac{3\pi r^4}{32}\left(\frac{15}{(1+x^2)^{7/2}} - \frac{35}{(1+x^2)^{9/2}}\right)$$
(E.2)

$$J_{2,2,0}(x,r) \approx \frac{\pi r^2}{\left(1+x^2\right)^{3/2}} + \frac{3\pi r^4}{8}\left[\frac{5}{\left(1+x^2\right)^{7/2}} - \frac{1}{\left(1+x^2\right)^{5/2}}\right]$$
(E.3)

$$J_{2,1,1}(x,r) \approx \frac{\pi r^4}{16}\left[\frac{15}{\left(1+x^2\right)^{7/2}} - \frac{3}{\left(1+x^2\right)^{5/2}}\right] + \frac{15\pi r^6}{128}\left[\frac{1}{\left(1+x^2\right)^{7/2}} - \frac{14}{\left(1+x^2\right)^{9/2}} + \frac{21}{\left(1+x^2\right)^{11/2}}\right]$$
(E.4)

$$J_{2,2,1}(x,r) \approx -\frac{3\pi r^4}{8}\frac{1}{\left(1+x^2\right)^{5/2}} + \frac{\pi r^6}{32}\left[\frac{15}{\left(1+x^2\right)^{7/2}} - \frac{35}{\left(1+x^2\right)^{9/2}}\right]$$
(E.5)

For the expansion of $H_{k,l}(x,r)$ we have

$$H_{0,1}(x,r) = H_{1,0}(x,r) \approx -\frac{5\pi r^2}{2}\frac{1}{(1+x^2)^{7/2}} + \frac{3\pi r^4}{32}\left[\frac{35}{\left(1+x^2\right)^{9/2}} - \frac{105}{\left(1+x^2\right)^{11/2}}\right]$$
(E.6)

$$H_{1,1}(x,r) \approx \frac{\pi r^4}{16}\left[\frac{35}{\left(1+x^2\right)^{9/2}} - \frac{5}{\left(1+x^2\right)^{7/2}}\right]$$
(E.7)

$$H_{2,0}(x,r) \approx \frac{\pi r^2}{\left(1+x^2\right)^{5/2}} + \frac{\pi r^4}{8}\left[\frac{35}{\left(1+x^2\right)^{9/2}} - \frac{5}{\left(1+x^2\right)^{7/2}}\right] \tag{E.8}$$

$$H_{2,1}(x,r) = H_{1,2}(x,r) \approx -\frac{5\pi}{8}\frac{r^4}{\left(1+x^2\right)^{7/2}} + \frac{\pi r^6}{32}\left(\frac{35}{\left(1+x^2\right)^{9/2}} - \frac{105}{\left(1+x^2\right)^{11/2}}\right) \tag{E.9}$$

$$H_{3,1}(x,r) \approx \frac{\pi r^6}{32}\left(\frac{35}{\left(1+x^2\right)^{9/2}} - \frac{5}{\left(1+x^2\right)^{7/2}}\right) \tag{E.10}$$

$$H_{2,2}(x,r) \approx \frac{\pi r^4}{4\left(1+x^2\right)^{5/2}} + \frac{\pi r^6}{32}\left(\frac{35}{\left(1+x^2\right)^{9/2}} - \frac{5}{\left(1+x^2\right)^{7/2}}\right) \tag{E.11}$$

$$H_{3,2}(x,r) \approx -\frac{5\pi r^6}{16\left(1+x^2\right)^{7/2}} \tag{E.12}$$

## Appendix F Evaluation of the rescaled correlation functions

Now, we can write the integrals in Eqs. (6.73)-(6.76) as

$$\hat{G}_\eta(x,\tilde{\alpha}_\eta) = \frac{1}{2\pi}\int_{-\infty}^{\infty} dk \, \frac{\exp(-ikx)}{k^2 + \tilde{\alpha}_\eta}, \tag{F.1}$$

$$\hat{G}_R(x,\tilde{\alpha}_R,\tilde{\gamma}) = \frac{1}{2\pi\tilde{\alpha}_R^{3/4}}\int_0^{2\pi} dk \, \frac{\exp(-ikx\tilde{\alpha}_R^{1/4})}{k^4 + \tilde{\gamma}k^2 + 1}, \tag{F.2}$$

$$\hat{C}_R(x,\tilde{\alpha}_\eta,\tilde{\gamma}) = \frac{1}{2\pi\tilde{\alpha}_R^{1/2}}\int_{-\infty}^{\infty} dk \, \frac{ik\exp(-ikx\tilde{\alpha}_R^{1/4})}{k^4 + \tilde{\gamma}k^2 + 1}, \tag{F.3}$$

$$\hat{D}_R(x,\tilde{\beta}_R,\tilde{\gamma}) = \frac{1}{2\pi\tilde{\alpha}_R^{1/4}}\int_{-\infty}^{\infty} dk \, \frac{k^2\exp(-ikx\tilde{\alpha}_R^{1/4})}{k^4 + \tilde{\gamma}k^2 + 1}. \tag{F.4}$$

Eq. (F.1) is easily evaluated in the complex plane using the residue theorem and is given by

$$\hat{G}_\eta(x,\tilde{\alpha}_\eta) = \frac{1}{2\tilde{\alpha}_\eta^{1/2}}\exp(-\tilde{\alpha}_\eta^{1/2}|x|). \tag{F.5}$$

Thus, we have Eq. (6.77) of the main text. Eqs. (F.2)-(F.4) are slightly more complicated, however we may express them as

$$\hat{G}_R(x,\tilde{\alpha}_R,\tilde{\gamma}) = \frac{1}{2\pi\tilde{\alpha}_R^{3/4}} \int_{-\infty}^{\infty} dk \frac{\exp(-ikx\tilde{\alpha}_R^{1/4})}{(k-k_1)(k-k_2)(k-k_3)(k-k_4)}, \tag{F.6}$$

$$\hat{C}_R(x,\tilde{\alpha}_R,\tilde{\gamma}) = \frac{1}{2\pi\tilde{\alpha}_R^{1/2}} \int_{-\infty}^{\infty} dk \frac{ik\exp(-ikx\tilde{\alpha}_R^{1/4})}{(k-k_1)(k-k_2)(k-k_3)(k-k_4)}, \tag{F.7}$$

$$\hat{D}_R(x,\tilde{\alpha}_R,\tilde{\gamma}) = \frac{1}{2\pi\tilde{\alpha}_R^{1/4}} \int_{-\infty}^{\infty} dk \frac{k^2\exp(-ikx\tilde{\alpha}_R^{1/4})}{(k-k_1)(k-k_2)(k-k_3)(k-k_4)}. \tag{F.8}$$

Expressions for the roots $k_1, k_2, k_3$ and $k_4$ are given by Eqs. (C.8) and (C.9). When $x < 0$ we close in the top half of the complex plane picking up roots $k_1$ and $k_3$, when $x > 0$ we close in the bottom half picking up the roots $k_2$ and $k_4$, and a factor of $-1$ for performing the contour integral in the clockwise direction. Thus we obtain

$$\hat{G}_R(x,\tilde{\beta}_R,\tilde{\gamma}) = \frac{i\theta(-x)}{\tilde{\alpha}_R^{3/4}}\left(\frac{\exp(-ik_1x\tilde{\alpha}_R^{1/4})}{(k_1-k_2)(k_1-k_3)(k_1-k_4)} + \frac{\exp(-ik_3x\tilde{\alpha}_R^{1/4})}{(k_3-k_1)(k_3-k_2)(k_3-k_4)}\right)$$
$$-\frac{i\theta(x)}{\tilde{\alpha}_R^{3/4}}\left(\frac{\exp(-ik_2x\tilde{\alpha}_R^{1/4})}{(k_2-k_1)(k_2-k_3)(k_2-k_4)} + \frac{\exp(-ik_4x\tilde{\alpha}_R^{1/4})}{(k_4-k_1)(k_4-k_2)(k_4-k_3)}\right), \tag{F.9}$$

$$\hat{C}_R(x,\tilde{\beta}_R,\tilde{\gamma}) = \frac{-\theta(-x)}{\tilde{\alpha}_R^{1/2}}\left(\frac{k_1\exp(-ik_1x\tilde{\alpha}_R^{1/4})}{(k_1-k_2)(k_1-k_3)(k_1-k_4)} + \frac{k_3\exp(-ik_3x\tilde{\alpha}_R^{1/4})}{(k_3-k_1)(k_3-k_2)(k_3-k_4)}\right)$$
$$+\frac{\theta(x)}{\tilde{\alpha}_R^{1/2}}\left(\frac{k_2\exp(-ik_2x\tilde{\alpha}_R^{1/4})}{(k_2-k_1)(k_2-k_3)(k_2-k_4)} + \frac{k_4\exp(-ik_4x\tilde{\alpha}_R^{1/4})}{(k_4-k_1)(k_4-k_2)(k_4-k_3)}\right), \tag{F.10}$$

$$\hat{D}_R(x,\tilde{\alpha}_R,\tilde{\gamma}) = \frac{i\theta(-x)}{\tilde{\alpha}_R^{1/4}}\left(\frac{k_1^2\exp(-ik_1x\tilde{\alpha}_R^{1/4})}{(k_1-k_2)(k_1-k_3)(k_1-k_4)} + \frac{k_3^2\exp(-ik_3x\tilde{\alpha}_R^{1/4})}{(k_3-k_1)(k_3-k_2)(k_3-k_4)}\right)$$
$$-\frac{i\theta(x)}{\tilde{\alpha}_R^{1/4}}\left(\frac{k_2^2\exp(-ik_2x\tilde{\alpha}_R^{1/4})}{(k_2-k_1)(k_2-k_3)(k_2-k_4)} + \frac{k_4^2\exp(-ik_4x\tilde{\alpha}_R^{1/4})}{(k_4-k_1)(k_4-k_2)(k_4-k_3)}\right). \tag{F.11}$$

First using the fact that $k_2 = -k_1$ and $k_3 = -k_4$ we may re-express Eqs. (F.9)-(F.11) as

$$\hat{G}_R(x,\tilde{\alpha}_R,\tilde{\gamma}) = \frac{i}{2\tilde{\alpha}_R^{3/4}(k_1^2-k_3^2)}\left(\frac{\exp(ik_1\tilde{\alpha}_R^{1/4}|x|)}{k_1} - \frac{\exp(ik_3\tilde{\alpha}_R^{1/4}|x|)}{k_3}\right), \tag{F.12}$$

$$\hat{C}_R(x,\tilde{\alpha}_R,\tilde{\gamma}) = \frac{\text{sgn}(x)}{2\tilde{\alpha}_R^{1/2}(k_1^2-k_3^2)}\left(\exp(-ik_1\tilde{\alpha}_R^{1/4}|x|) - \exp(-ik_3\tilde{\alpha}_R^{1/4}|x|)\right), \tag{F.13}$$

$$\hat{D}_R(x,\tilde{\alpha}_R,\tilde{\gamma}) = \frac{i}{2\tilde{\alpha}_R^{1/4}(k_1^2 - k_3^2)}\left(k_1 \exp(-ik_1\tilde{\alpha}_R^{1/4}|x|) - k_3 \exp(-ik_3\tilde{\alpha}_R^{1/4}|x|)\right). \tag{F.14}$$

Substitution of the roots for $-2 < \tilde{\gamma} < 2$ (Eq. (C.8)) into Eqs. (F.12)-(F.14) yields the expressions

$$\hat{G}_R(x,\tilde{\alpha}_R,\tilde{\gamma}) = \frac{\exp\left(-\frac{\tilde{\alpha}_R^{1/4}|x|}{\sqrt{2}}\sqrt{1+\frac{\tilde{\gamma}}{2}}\right)}{\sqrt{2}\tilde{\alpha}_R^{3/4}\sqrt{4-\tilde{\gamma}^2}}\left(\frac{\exp\left(i\frac{\tilde{\alpha}_R^{1/4}|x|}{\sqrt{2}}\sqrt{1-\frac{\tilde{\gamma}}{2}}\right)}{\left(\sqrt{1-\frac{\tilde{\gamma}}{2}}+i\sqrt{1+\frac{\tilde{\gamma}}{2}}\right)} + \frac{\exp\left(-i\frac{\tilde{\alpha}_R^{1/4}|x|}{\sqrt{2}}\sqrt{1-\frac{\tilde{\gamma}}{2}}|x|\right)}{\left(\sqrt{1-\frac{\tilde{\gamma}}{2}}-i\sqrt{1+\frac{\tilde{\gamma}}{2}}\right)}\right)$$

$$= \frac{\exp\left(-\frac{\tilde{\alpha}_R^{1/4}|x|}{\sqrt{2}}\sqrt{1+\frac{\gamma}{2}}|x|\right)}{\sqrt{2}\tilde{\alpha}_R^{3/4}\sqrt{4-\gamma^2}}\left(\sqrt{1-\frac{\tilde{\gamma}}{2}}\cos\left(\frac{\tilde{\alpha}_R^{1/4}|x|}{\sqrt{2}}\sqrt{1-\frac{\tilde{\gamma}}{2}}\right) + \sqrt{1+\frac{\tilde{\gamma}}{2}}\sin\left(\frac{\tilde{\alpha}_R^{1/4}|x|}{\sqrt{2}}\sqrt{1-\frac{\tilde{\gamma}}{2}}\right)\right),$$

(F.15)

$$\hat{C}_R(x,\tilde{\alpha}_R,\tilde{\gamma}) = -\frac{i\,\mathrm{sgn}(x)\exp\left(-\frac{\tilde{\alpha}_R^{1/4}|x|}{\sqrt{2}}\sqrt{1+\frac{\tilde{\gamma}}{2}}\right)}{2\tilde{\alpha}_R^{1/2}\sqrt{4-\tilde{\gamma}^2}}\left(\exp\left(i\frac{\tilde{\alpha}_R^{1/4}|x|}{\sqrt{2}}\sqrt{1-\frac{\tilde{\gamma}}{2}}\right) - \exp\left(-i\frac{\tilde{\alpha}_R^{1/4}|x|}{\sqrt{2}}\sqrt{1-\frac{\tilde{\gamma}}{2}}\right)\right)$$

$$= \frac{\mathrm{sgn}(x)\exp\left(-\frac{\tilde{\alpha}_R^{1/4}|x|}{\sqrt{2}}\sqrt{1+\frac{\tilde{\gamma}}{2}}\right)}{\tilde{\alpha}_R^{1/2}\sqrt{4-\tilde{\gamma}^2}}\sin\left(\frac{\tilde{\alpha}_R^{1/4}|x|}{\sqrt{2}}\sqrt{1-\frac{\tilde{\gamma}}{2}}\right),$$

(F.16)

$$\hat{D}_R(x,\tilde{\alpha}_R,\tilde{\gamma}) = \frac{\exp\left(-\frac{\tilde{\alpha}_R^{1/4}|x|}{\sqrt{2}}\sqrt{1+\frac{\tilde{\gamma}}{2}}\right)}{2\sqrt{2}\tilde{\alpha}_R^{1/4}\sqrt{4-\tilde{\gamma}^2}}\left[\left(\sqrt{1-\frac{\tilde{\gamma}}{2}}+i\sqrt{1+\frac{\tilde{\gamma}}{2}}\right)\exp\left(i\frac{\tilde{\alpha}_R^{1/4}|x|}{\sqrt{2}}\sqrt{1-\frac{\tilde{\gamma}}{2}}\right)\right.$$

$$\left. +\left(\sqrt{1-\frac{\tilde{\gamma}}{2}}-i\sqrt{1+\frac{\tilde{\gamma}}{2}}\right)\exp\left(-i\frac{\tilde{\alpha}_R^{1/4}|x|}{\sqrt{2}}\sqrt{1-\frac{\tilde{\gamma}}{2}}\right)\right]$$

(F.17)

$$= \frac{\exp\left(-\frac{\tilde{\alpha}_R^{1/4}|x|}{\sqrt{2}}\sqrt{1+\frac{\tilde{\gamma}}{2}}\right)}{\sqrt{2}\tilde{\alpha}_R^{1/4}\sqrt{4-\tilde{\gamma}^2}}\left[\sqrt{1-\frac{\tilde{\gamma}}{2}}\cos\left(\frac{\tilde{\alpha}_R^{1/4}|x|}{\sqrt{2}}\sqrt{1-\frac{\tilde{\gamma}}{2}}\right) - \sqrt{1+\frac{\tilde{\gamma}}{2}}\sin\left(\frac{\tilde{\alpha}_R^{1/4}|x|}{\sqrt{2}}\sqrt{1-\frac{\tilde{\gamma}}{2}}\right)\right].$$

For $\tilde{\gamma} \geq 2$, substituting Eq. (C.9) into Eqs. (F.12)-(F.14), we obtain

$$\hat{G}_R(x,\tilde{\alpha}_R,\tilde{\gamma})$$

$$= -\frac{1}{\sqrt{2}\tilde{\alpha}_R^{3/4}\sqrt{\tilde{\gamma}^2-4}}\left[\frac{\exp\left(-\frac{\tilde{\alpha}_R^{1/4}|x|}{\sqrt{2}}\left(\tilde{\gamma}+\sqrt{\tilde{\gamma}^2-4}\right)^{1/2}\right)}{\left(\tilde{\gamma}+\sqrt{\tilde{\gamma}^2-4}\right)^{1/2}} - \frac{\exp\left(-\frac{\tilde{\alpha}_R^{1/4}|x|}{\sqrt{2}}\left(\tilde{\gamma}-\sqrt{\tilde{\gamma}^2-4}\right)^{1/2}\right)}{\left(\tilde{\gamma}-\sqrt{\tilde{\gamma}^2-4}\right)^{1/2}}\right] \quad \text{(F.18)}$$

$$= -\frac{1}{2\sqrt{2}\tilde{\alpha}_R^{3/4}\sqrt{\tilde{\gamma}^2-4}}\left[\left(\tilde{\gamma}-\sqrt{\tilde{\gamma}^2-4}\right)^{1/2}\exp\left(-\frac{\tilde{\alpha}_R^{1/4}|x|}{\sqrt{2}}\left(\tilde{\gamma}+\sqrt{\tilde{\gamma}^2-4}\right)^{1/2}\right)\right.$$
$$\left.-\left(\tilde{\gamma}+\sqrt{\tilde{\gamma}^2-4}\right)^{1/2}\exp\left(-\frac{\tilde{\alpha}_R^{1/4}|x|}{\sqrt{2}}\left(\tilde{\gamma}-\sqrt{\tilde{\gamma}^2-4}\right)^{1/2}\right)\right],$$

$$\hat{C}_R(x,\tilde{\alpha}_R,\tilde{\gamma}) = -\frac{\text{sgn}(x)}{2\tilde{\alpha}_R^{1/2}\sqrt{\tilde{\gamma}^2-4}}\left(\exp\left(-\frac{\tilde{\alpha}_R^{1/4}|x|}{\sqrt{2}}\left(\tilde{\gamma}+\sqrt{\tilde{\gamma}^2-4}\right)^{1/2}\right)-\exp\left(-\frac{\tilde{\alpha}_R^{1/4}|x|}{\sqrt{2}}\left(\tilde{\gamma}-\sqrt{\tilde{\gamma}^2-4}\right)^{1/2}\right)\right),$$
$$\text{(F.19)}$$

$$\hat{D}_R(x,\tilde{\alpha}_R,\tilde{\gamma}) = \frac{1}{2\sqrt{2}\tilde{\alpha}_R^{1/4}\sqrt{\tilde{\gamma}^2-4}}\left(\left(\tilde{\gamma}+\sqrt{\tilde{\gamma}^2-4}\right)^{1/2}\exp\left(-\frac{\tilde{\alpha}_R^{1/4}|x|}{\sqrt{2}}\left(\tilde{\gamma}+\sqrt{\tilde{\gamma}^2-4}\right)^{1/2}\right)\right.$$
$$\left.-\left(\tilde{\gamma}-\sqrt{\tilde{\gamma}^2-4}\right)^{1/2}\exp\left(-\frac{\tilde{\alpha}_R^{1/4}|x|}{\sqrt{2}}\left(\tilde{\gamma}-\sqrt{\tilde{\gamma}^2-4}\right)^{1/2}\right)\right). \quad \text{(F.20)}$$

Using Eqs. (F.15)-(F.20) we may write Eqs. (6.78)-(6.80) of the main text.

# Appendix G Terms in the expansion of $K_{2,j,k}(\tilde{P})$

The terms given in the expansions of $K_{2,j,k}(\tilde{P})$, Eqs. (6.84) and (6.85), are given by the expressions

$$K_{2,1,0}^0(\tilde{P}) = 2\pi\int_{-\infty}^{\infty} dx \frac{(1+\cos x)}{\left(\frac{1}{2}(1+\cos x)+\tilde{P}^2 x^2\right)^{3/2}}, \quad \text{(G.1)}$$

$$K_{2,1,0}^1(\tilde{P}) = 0, \quad \text{(G.2)}$$

$$K_{2,1,0}^2(\tilde{P}) = 2\pi\int_{-\infty}^{\infty} dx\left(\frac{15(1+\cos x)^3}{32\left(\frac{1}{2}(1+\cos x)+\tilde{P}^2 x^2\right)^{7/2}} - \frac{3(1+\cos x)}{8\left(\frac{1}{2}(1+\cos x)+\tilde{P}^2 x^2\right)^{5/2}}\right), \quad \text{(G.3)}$$

$$K_{2,1,0}^3(\tilde{P}) = 0, \quad \text{(G.4)}$$

$$K_{2,1,0}^{4}(\tilde{P}) = \frac{15\pi}{128} \int_{-\infty}^{\infty} dx \left( \frac{(1+\cos x)(2+\cos^2 x)}{\left(\frac{1}{2}(1+\cos x)+\tilde{P}^2 x^2\right)^{7/2}} - \frac{7(2+\cos x)(1+\cos x)^3}{2\left(\frac{1}{2}(1+\cos x)+\tilde{P}^2 x^2\right)^{9/2}} + \frac{63(1+\cos x)^5}{16\left(\frac{1}{2}(1+\cos x)+\tilde{P}^2 x^2\right)^{11/2}} \right),$$
(G.5)

$$K_{2,1,1}^{0}(\tilde{P}) = 0,$$
(G.6)

$$K_{2,1,1}^{1}(\tilde{P}) = -\frac{3\pi}{4} \int_{-\infty}^{\infty} dx \frac{(1+\cos x)^2}{\left(\frac{1}{2}(1+\cos x)+\tilde{P}^2 x^2\right)^{5/2}},$$
(G.7)

$$K_{2,1,1}^{2}(\tilde{P}) = 0,$$
(G.8)

$$K_{2,1,1}^{3}(\tilde{P}) = \frac{\pi}{2} \int_{-\infty}^{\infty} dx \left( \frac{15(2+\cos x)(1+\cos x)^2}{32\left(\frac{1}{2}(1+\cos x)+\tilde{P}^2 x^2\right)^{7/2}} - \frac{105(1+\cos x)^4}{128\left(\frac{1}{2}(1+\cos x)+\tilde{P}^2 x^2\right)^{9/2}} \right),$$
(G.9)

$$K_{2,1,1}^{4}(\tilde{P}) = 0,$$
(G.10)

$$K_{2,2,0}^{0}(\tilde{P}) = 2\pi \int_{-\infty}^{\infty} dx \frac{\tilde{P} x \sin x}{\left(\frac{1}{2}(1+\cos x)+\tilde{P}^2 x^2\right)^{3/2}},$$
(G.11)

$$K_{2,2,0}^{1}(\tilde{P}) = 0,$$
(G.12)

$$K_{2,2,0}^{2}(\tilde{P}) = 2\pi \tilde{P} \int_{-\infty}^{\infty} dx\, x \sin x \left( \frac{15(1+\cos x)^2}{32\left(\frac{1}{2}(1+\cos x)+\tilde{P}^2 x^2\right)^{7/2}} - \frac{3}{8\left(\frac{1}{2}(1+\cos x)+\tilde{P}^2 x^2\right)^{5/2}} \right),$$
(G.13)

$$K_{2,2,0}^{3}(\tilde{P}) = 0,$$
(G.14)

and

$$K_{2,2,0}^{4}(\tilde{P}) = \frac{15\pi \tilde{P}}{128} \int_{-\infty}^{\infty} dx\, x \sin x$$

$$\left( \frac{2+\cos^2 x}{\left(\frac{1}{2}(1+\cos x)+\tilde{P}^2 x^2\right)^{7/2}} - \frac{7(1+\cos x)^2(2+\cos x)}{2\left(\frac{1}{2}(1+\cos x)+\tilde{P}^2 x^2\right)^{9/2}} + \frac{63(1+\cos x)^4}{16\left(\frac{1}{2}(1+\cos x)+\tilde{P}^2 x^2\right)^{11/2}} \right).$$
(G.15)

# Appendix H Explicit expressions for the derivatives of $\bar{W}_2$, $W_{1,j}$ and $W_{2,j}$

For the functions describing derivatives of the function $\bar{W}_2(\tilde{R}_0, \tilde{P}, d_\eta, \theta_R, \eta_0)$ (defined by Eqs. (7.1)-(7.5) of the main text) we have for the expansion in $r$ the following expressions

$$\bar{W}_2^R(\tilde{R}_0, \tilde{P}, d_\eta, \theta_R, \eta_0) \approx \frac{\tilde{P}}{2\pi \cos\left(\frac{\eta_0}{2}\right)} \left\{ \left[ \sin\left(\frac{\eta_0}{2}\right) \cos\left(\frac{\eta_0}{2}\right) \left(1 - \frac{1}{8(2l_p \alpha_\eta)^{1/2}}\right) - \frac{\theta_R^2}{8} \cot\left(\frac{\eta_0}{2}\right) \right] \right.$$

$$\left( \frac{K_{2,1,0}^0(\tilde{P})}{\tilde{R}_0^2} + \frac{6\left(K_{2,1,0}^2(\tilde{P}) + K_{2,1,1}^2(\tilde{P})\right)}{\tilde{R}_0^4} + \frac{40\left(K_{2,1,0}^4(\tilde{P}) + K_{2,1,1}^4(\tilde{P})\right)}{\tilde{R}_0^6} \right) \quad \text{(H.1)}$$

$$+ \left( \sin^2\left(\frac{\eta_0}{2}\right) \left(1 - \frac{1}{8(2l_p \alpha_\eta)^{1/2}}\right) - \frac{\theta_R^2}{4} \right) \left( \frac{K_{2,2,0}^0(\tilde{P})}{\tilde{R}_0^2} + \frac{6 K_{2,2,0}^2(\tilde{P})}{\tilde{R}_0^4} + \frac{40 K_{2,2,0}^4(\tilde{P})}{\tilde{R}_0^6} \right) \right\},$$

$$\bar{W}_2^P(\tilde{R}_0, \tilde{P}, d_\eta, \theta_R, \eta_0) = -\frac{1}{2\pi \cos\left(\frac{\eta_0}{2}\right)} \left\{ \left[ \sin\left(\frac{\eta_0}{2}\right) \cos\left(\frac{\eta_0}{2}\right) \left(1 - \frac{1}{8(2l_p \alpha_\eta)^{1/2}}\right) - \frac{\theta_R^2}{8} \cot\left(\frac{\eta_0}{2}\right) \right] \right.$$

$$\left( \frac{\left(K_{2,1,0}^0(\tilde{P}) + \tilde{P} K_{2,1,0}^{P,0}(\tilde{P})\right)}{\tilde{R}_0} + \frac{2\left(K_{2,1,0}^2(\tilde{P}) + K_{2,1,1}^2(\tilde{P}) + \tilde{P} K_{2,1,0}^{P,2}(\tilde{P}) + \tilde{P} K_{2,1,1}^{P,2}(\tilde{P})\right)}{\tilde{R}_0^3} \right.$$

$$\left. + \frac{8\left(K_{2,1,0}^4(\tilde{P}) + K_{2,1,1}^4(\tilde{P}) + \tilde{P} K_{2,1,0}^{P,4}(\tilde{P}) + \tilde{P} K_{2,1,1}^{P,4}(\tilde{P})\right)}{\tilde{R}_0^5} \right) + \left( \sin^2\left(\frac{\eta_0}{2}\right) \left(1 - \frac{1}{8(2l_p \alpha_\eta)^{1/2}}\right) - \frac{\theta_R^2}{4} \right)$$

$$\left( \frac{\left(K_{2,2,0}^0(\tilde{P}) + \tilde{P} K_{2,2,0}^{P,0}(\tilde{P})\right)}{\tilde{R}_0} + \frac{2\left(K_{2,2,0}^2(\tilde{P}) + \tilde{P} K_{2,2,0}^{P,2}(\tilde{P})\right)}{\tilde{R}_0^3} + \frac{8\left(K_{2,2,0}^4(\tilde{P}) + \tilde{P} K_{2,2,0}^{P,4}(\tilde{P})\right)}{\tilde{R}_0^5} \right) \right\},$$

(H.2)

$$\bar{W}_2^{\alpha_\eta}(\tilde{R}_0, \tilde{P}, \alpha_\eta, \theta_R, \eta_0) = -\frac{1}{16(2l_p)^{1/2} \alpha_\eta^{3/2}} \frac{\tilde{P}}{2\pi \cos\left(\frac{\eta_0}{2}\right)} \left\{ \sin\left(\frac{\eta_0}{2}\right) \cos\left(\frac{\eta_0}{2}\right) \right.$$

$$\left[ \frac{K_{2,1,0}^0(\tilde{P})}{\tilde{R}_0} + \frac{2\left(K_{2,1,0}^2(\tilde{P}) + K_{2,1,1}^2(\tilde{P})\right)}{\tilde{R}_0^3} + \frac{8\left(K_{2,1,0}^2(\tilde{P}) + K_{2,1,1}^2(\tilde{P})\right)}{\tilde{R}_0^5} \right] + \sin^2\left(\frac{\eta_0}{2}\right) \quad \text{(H.3)}$$

$$\left[ \frac{K_{2,2,0}^0(\tilde{P})}{\tilde{R}_0} + \frac{2 K_{2,2,0}^2(\tilde{P})}{\tilde{R}_0^3} + \frac{8 K_{2,2,0}^4(\tilde{P})}{\tilde{R}_0^5} \right] \right\},$$

$$\bar{W}_2^{\theta_R}(\tilde{R}_0,\tilde{P},\alpha_\eta,\theta_R,\eta_0) \approx \frac{\tilde{P}\theta_R}{4\pi \cos\left(\frac{\eta_0}{2}\right)} \left\{ \left( \frac{K_{2,2,0}^0(\tilde{P})}{\tilde{R}_0} + \frac{2K_{2,2,0}^2(\tilde{P})}{\tilde{R}_0^3} + \frac{8K_{2,2,0}^4(\tilde{P})}{\tilde{R}_0^5} \right) \right.$$

$$\left. + \frac{1}{2}\cot\left(\frac{\eta_0}{2}\right) \left( \frac{K_{2,1,0}^0(\tilde{P})}{\tilde{R}_0} + \frac{2\left(K_{2,1,0}^2(\tilde{P})+K_{2,1,1}^2(\tilde{P})\right)}{\tilde{R}_0^3} + \frac{8\left(K_{2,1,0}^4(\tilde{P})+K_{2,1,1}^4(\tilde{P})\right)}{\tilde{R}_0^5} \right) \right\},$$

(H.4)

$$\bar{W}_2^{\eta}(\tilde{R}_0,\tilde{P},\alpha_\eta,\theta_R,\eta_0) \approx -\frac{\tilde{P}}{4\pi} \left\{ \left[ \cos\left(\frac{\eta_0}{2}\right)\left(1-\frac{1}{8(2l_p\alpha_\eta)^{1/2}}\right) + \frac{\theta_R^2}{8}\operatorname{cosec}\left(\frac{\eta_0}{2}\right)\cot\left(\frac{\eta_0}{2}\right) \right] \right.$$

$$\left( \frac{K_{2,1,0}^0(\tilde{P})}{\tilde{R}_0} + \frac{2\left(K_{2,1,0}^2(\tilde{P})+K_{2,1,1}^2(\tilde{P})\right)}{\tilde{R}_0^3} + \frac{8\left(K_{2,1,0}^4(\tilde{P})+K_{2,1,1}^4(\tilde{P})\right)}{\tilde{R}_0^5} \right)$$

$$+ \left[ \sin\left(\frac{\eta_0}{2}\right)\left(1-\frac{1}{8(2l_p\alpha_\eta)^{1/2}}\right) + \sec\left(\frac{\eta_0}{2}\right)\tan\left(\frac{\eta_0}{2}\right)\left(1-\frac{1}{8(2l_p\alpha_\eta)^{1/2}} - \frac{\theta_R^2}{4}\right) \right]$$

$$\left. \left( \frac{K_{2,2,0}^0(\tilde{P})}{\tilde{R}_0} + \frac{2K_{2,2,0}^2(\tilde{P})}{\tilde{R}_0^3} + \frac{8K_{2,2,0}^4(\tilde{P})}{\tilde{R}_0^5} \right) \right\}.$$

(H.5)

The functions $K_{2,1,0}^{P,k}(\tilde{P})$, $K_{2,1,1}^{P,k}(\tilde{P})$ and $K_{2,2,0}^{P,k}(\tilde{P})$ (appearing in Eq. (H.2)) are given by:

$$K_{2,1,0}^{P,0}(\tilde{P}) = -6\pi\tilde{P}\int_{-\infty}^{\infty} dx \frac{x^2(1+\cos x)}{\left(\frac{1}{2}(1+\cos x)+\tilde{P}^2 x^2\right)^{5/2}},$$

(H.6)

$$K_{2,1,0}^{P,1}(\tilde{P}) = 0,$$

(H.7)

$$K_{2,1,0}^{P,2}(\tilde{P}) = -2\pi\tilde{P}\int_{-\infty}^{\infty} x^2 dx \left( \frac{105(1+\cos x)^3}{32\left(\frac{1}{2}(1+\cos x)+\tilde{P}^2 x^2\right)^{9/2}} - \frac{15(1+\cos x)}{8\left(\frac{1}{2}(1+\cos x)+\tilde{P}^2 x^2\right)^{7/2}} \right),$$

(H.8)

$$K_{2,1,0}^{P,3}(\tilde{P}) = 0,$$

(H.9)

$$K_{2,1,0}^{P,4}(\tilde{P}) = -\frac{15\pi\tilde{P}}{128}\int_{-\infty}^{\infty} x^2 dx \left( \frac{7(1+\cos x)(2+\cos^2 x)}{\left(\frac{1}{2}(1+\cos x)+\tilde{P}^2 x^2\right)^{9/2}} - \frac{63(2+\cos x)(1+\cos x)^3}{2\left(\frac{1}{2}(1+\cos x)+\tilde{P}^2 x^2\right)^{11/2}} + \frac{693(1+\cos x)^5}{16\left(\frac{1}{2}(1+\cos x)+\tilde{P}^2 x^2\right)^{13/2}} \right),$$

(H.10)

$$K_{2,1,1}^{P,0}(\tilde{P}) = 0,$$

(H.11)

$$K_{2,1,1}^{P,1}(\tilde{P}) = \frac{15\pi \tilde{P}}{4} \int_{-\infty}^{\infty} x^2 dx \frac{(1+\cos x)^2}{\left(\frac{1}{2}(1+\cos x)+\tilde{P}^2 x^2\right)^{7/2}}, \tag{H.12}$$

$$K_{2,1,1}^{P,2}(\tilde{P}) = 0, \tag{H.13}$$

$$K_{2,1,1}^{P,3}(\tilde{P}) = -\frac{\pi \tilde{P}}{2} \int_{-\infty}^{\infty} x^2 dx \left( \frac{105(2+\cos x)(1+\cos x)^2}{32\left(\frac{1}{2}(1+\cos x)+\tilde{P}^2 x^2\right)^{9/2}} - \frac{945(1+\cos x)^4}{128\left(\frac{1}{2}(1+\cos x)+\tilde{P}^2 x^2\right)^{11/2}} \right), \tag{H.14}$$

$$K_{2,1,1}^{P,4}(\tilde{P}) = 0, \tag{H.15}$$

$$K_{2,2,0}^{P,0}(\tilde{P}) = 2\pi \int_{-\infty}^{\infty} dx \left( \frac{x \sin x}{\left(\frac{1}{2}(1+\cos x)+\tilde{P}^2 x^2\right)^{3/2}} - \frac{3\tilde{P}^2 x^3 \sin x}{\left(\frac{1}{2}(1+\cos x)+\tilde{P}^2 x^2\right)^{5/2}} \right), \tag{H.16}$$

$$K_{2,2,0}^{P,1}(\tilde{P}) = 0, \tag{H.17}$$

$$K_{2,2,0}^{P,2}(\tilde{P}) = 2\pi \int_{-\infty}^{\infty} dx\, x \sin x \left( -\frac{105\tilde{P}^2 x^2 (1+\cos x)^2}{32\left(\frac{1}{2}(1+\cos x)+\tilde{P}^2 x^2\right)^{9/2}} - \frac{3}{8\left(\frac{1}{2}(1+\cos x)+\tilde{P}^2 x^2\right)^{5/2}} \right.$$
$$\left. + \frac{15\left((1+\cos x)^2 + 4x^2 \tilde{P}^2\right)}{32\left(\frac{1}{2}(1+\cos x)+\tilde{P}^2 x^2\right)^{7/2}} \right), \tag{H.18}$$

$$K_{2,2,0}^{P,3}(\tilde{P}) = 0, \tag{H.19}$$

and

$$K_{2,2,0}^{4}(\tilde{P}) = \frac{15\pi}{128} \int_{-\infty}^{\infty} dx\, x \sin x$$

$$\left( \frac{2+\cos^2 x}{\left(\frac{1}{2}(1+\cos x)+\tilde{P}^2 x^2\right)^{7/2}} - \frac{7\left[2\tilde{P}^2 x^2(2+\cos^2 x)+(1+\cos x)^2(2+\cos x)\right]}{2\left(\frac{1}{2}(1+\cos x)+\tilde{P}^2 x^2\right)^{9/2}} \right. \tag{H.20}$$

$$\left. + \frac{63(1+\cos x)^2\left[8\tilde{P}^2 x^2(2+\cos x)+(1+\cos x)^2\right]}{16\left(\frac{1}{2}(1+\cos x)+\tilde{P}^2 x^2\right)^{11/2}} - \frac{693\tilde{P}^2 x^2(1+\cos x)^4}{16\left(\frac{1}{2}(1+\cos x)+\tilde{P}^2 x^2\right)^{13/2}} \right),$$

The functions that are derivatives of $W_{1,1}(\bar{\alpha}_\eta, \tilde{R}_0, \tilde{Q}_1, \tilde{P}, \eta_0)$ and $W_{2,1}(\bar{\alpha}_\eta, \tilde{R}_0, \tilde{Q}_1, \tilde{P}, \eta_0)$, defined through Eqs. (7.8) and (7.9) are given by the expressions

$$W_{1,1}^{\alpha_\eta}(\bar{\alpha}_\eta, \tilde{R}_0, \tilde{Q}_1, \tilde{P}, \eta_0) \approx \frac{1}{4\pi \tilde{Q}_1^2 l_p \tilde{R}_0^5}\left(\tilde{P}^2+\frac{1}{4}\right)^{-1}\left(\frac{1}{2}\cos\left(\frac{\eta_0}{2}\right)\sin\left(\frac{\eta_0}{2}\right)+\tilde{P}\cos^2\left(\frac{\eta_0}{2}\right)\right)\tilde{A}^\alpha(\bar{\alpha}_\eta),$$

(H.21)

$$W_{1,1}^{R}(\bar{\alpha}_\eta, \tilde{R}_0, \tilde{Q}_1, \tilde{P}, \eta_0) \approx -\frac{\left(3W_{1,1}(\bar{\alpha}_\eta, \tilde{R}_0, \tilde{Q}_1, \tilde{P}, \eta_0) + 2\tilde{\alpha}_{\eta,1}W_{1,1}^{\alpha_R}(\bar{\alpha}_\eta, \tilde{R}_0, \tilde{Q}_1, \tilde{P}, \eta_0)\right)}{\tilde{R}_0}, \tag{H.22}$$

$$W_{1,1}^{Q}(\bar{\alpha}_\eta, \tilde{R}_0, \tilde{Q}_1, \tilde{P}, \eta_0) \approx -\frac{2W_{1,1}(\bar{\alpha}_\eta, \tilde{R}_0, \tilde{Q}_1, \tilde{P}, \eta_0)}{\tilde{Q}_1}, \tag{H.23}$$

$$W_{1,1}^{P}(\bar{\alpha}_\eta, \tilde{R}_0, \tilde{Q}_1, \tilde{P}, \eta_0) \approx$$
$$-\frac{\tilde{A}(\bar{\alpha}_\eta)}{4\pi \tilde{Q}_1^2 l_p \tilde{R}_0^3}\left[\left(\tilde{P}^2+\frac{1}{4}\right)^{-2}\left(\frac{1}{2}\cos\left(\frac{\eta_0}{2}\right)\sin\left(\frac{\eta_0}{2}\right)+\tilde{P}\cos^2\left(\frac{\eta_0}{2}\right)\right)-\left(\tilde{P}^2+\frac{1}{4}\right)^{-1}\cos^2\left(\frac{\eta_0}{2}\right)\right],$$

(H.24)

$$W_{1,1}^{\eta}(\bar{\alpha}_\eta, \tilde{R}_0, \tilde{Q}_1, \tilde{P}, \eta_0) \approx -\frac{\tilde{A}(\bar{\alpha}_\eta)}{4\pi \tilde{Q}_1^2 l_p \tilde{R}_0^3}\left(\tilde{P}^2+\frac{1}{4}\right)^{-1}\left[\frac{1}{2}\tilde{P}\sin\eta_0 - \frac{1}{4}\cos\eta_0\right], \tag{H.25}$$

and

$$W_{2,1}^{\alpha_\eta}(\bar{\alpha}_{\eta,1}, \tilde{R}_0, \tilde{Q}_1, \tilde{P}, \eta_0) \approx \frac{1}{\tilde{Q}_2^2 l_p} \cos\left(\frac{\eta_0}{2}\right) \sin\left(\frac{\eta_0}{2}\right) \left[ \frac{2 A_{3,0}^\alpha(\tilde{\alpha}_{\eta,1})}{\tilde{R}_0^2} - \frac{9 A_{5,0}^\alpha(\tilde{\alpha}_{\eta,1})}{2\tilde{R}_0^4} \right.$$

$$+ A_{7,0}^\alpha(\tilde{\alpha}_{\eta,1}) \left[ \frac{225}{16\tilde{R}_0^6} + \frac{15}{2\tilde{R}_0^4} \right] - \frac{525 A_{9,0}^\alpha(\tilde{\alpha}_{\eta,1})}{8\tilde{R}_0^6} + \frac{945 A_{11,0}^\alpha(\tilde{\alpha}_{\eta,1})}{16\tilde{R}_0^6} \right]$$

$$- \frac{1}{\tilde{Q}_2^2 l_p} \left( \tilde{P}^2 - \frac{1}{4} \right)^{-1} \left[ \frac{A_{3,2}^\alpha(\tilde{\alpha}_{\eta,1})}{\tilde{R}_0^2} \left( \frac{1}{2} \cos\left(\frac{\eta_0}{2}\right) \sin\left(\frac{\eta_0}{2}\right) + \tilde{P} \cos^2\left(\frac{\eta_0}{2}\right) \right) \right.$$

$$- \frac{A_{5,2}^\alpha(\tilde{\alpha}_{\eta,1})}{\tilde{R}_0^4} \left( \frac{15}{8} \cos\left(\frac{\eta_0}{2}\right) \sin\left(\frac{\eta_0}{2}\right) + \frac{3\tilde{P}}{4} \cos^2\left(\frac{\eta_0}{2}\right) \right)$$

$$+ A_{7,2}^\alpha(\tilde{\alpha}_{\eta,1}) \left( \tilde{P} \cos^2\left(\frac{\eta_0}{2}\right) \left( \frac{45}{32\tilde{R}_0^6} + \frac{15}{4\tilde{R}_0^4} \right) - \cos\left(\frac{\eta_0}{2}\right) \sin\left(\frac{\eta_0}{2}\right) \left( \frac{45}{8\tilde{R}_0^4} + \frac{585}{64\tilde{R}_0^6} \right) \right)$$

$$- \frac{A_{9,2}^\alpha(\tilde{\alpha}_{\eta,1})}{\tilde{R}_0^6} \left( \frac{1995}{32} \cos\left(\frac{\eta_0}{2}\right) \sin\left(\frac{\eta_0}{2}\right) + \frac{315}{16} \tilde{P} \cos^2\left(\frac{\eta_0}{2}\right) \right)$$

$$\left. + \frac{945 A_{11,2}^\alpha(\tilde{\alpha}_{\eta,1})}{32 \tilde{R}_0^6} \left( \frac{5}{2} \cos\left(\frac{\eta_0}{2}\right) \sin\left(\frac{\eta_0}{2}\right) + \tilde{P} \cos^2\left(\frac{\eta_0}{2}\right) \right) \right], \tag{H.26}$$

$$W_{2,1}^R(\tilde{\alpha}_{\eta,1}, \tilde{R}_0, \tilde{Q}_2, \tilde{P}, \eta_0) \approx -\frac{1}{\tilde{Q}_2^2 l_p} \cos\left(\frac{\eta_0}{2}\right) \sin\left(\frac{\eta_0}{2}\right) \left[ \frac{4 A_{3,0}(\tilde{\alpha}_{\eta,1})}{\tilde{R}_0^4} - \frac{18 A_{5,0}(\tilde{\alpha}_{\eta,1})}{\tilde{R}_0^5} \right.$$

$$+ A_{7,0}(\tilde{\alpha}_{\eta,1}) \left[ \frac{675}{8\tilde{R}_0^7} + \frac{30}{\tilde{R}_0^5} \right] - \frac{1575 A_{9,0}(\tilde{\alpha}_{\eta,1})}{4\tilde{R}_0^7} + \frac{2835 A_{11,0}(\tilde{\alpha}_{\eta,1})}{4\tilde{R}_0^7} \right]$$

$$+ \frac{1}{\tilde{Q}_j^2 l_p} \left( \tilde{P}^2 - \frac{1}{4} \right)^{-1} \left[ \frac{2 A_{3,2}(\tilde{\alpha}_{\eta,1})}{\tilde{R}_0^3} \left( \frac{1}{2} \cos\left(\frac{\eta_0}{2}\right) \sin\left(\frac{\eta_0}{2}\right) + \tilde{P} \cos^2\left(\frac{\eta_0}{2}\right) \right) \right.$$

$$+ \frac{A_{5,2}(\tilde{\alpha}_{\eta,1})}{\tilde{R}_0^5} \left( \frac{15}{2} \cos\left(\frac{\eta_0}{2}\right) \sin\left(\frac{\eta_0}{2}\right) - 3\tilde{P} \cos^2\left(\frac{\eta_0}{2}\right) \right)$$

$$+ A_{7,2}(\tilde{\alpha}_{\eta,1}) \left( \tilde{P} \cos^2\left(\frac{\eta_0}{2}\right) \left( \frac{135}{16\tilde{R}_0^7} + \frac{15}{\tilde{R}_0^5} \right) + \cos\left(\frac{\eta_0}{2}\right) \sin\left(\frac{\eta_0}{2}\right) \left( \frac{45}{2\tilde{R}_0^5} + \frac{1755}{32\tilde{R}_0^7} \right) \right)$$

$$- \frac{3 A_{9,2}(\tilde{\alpha}_{\eta,1})}{\tilde{R}_0^7} \left( \frac{1995}{16} \cos\left(\frac{\eta_0}{2}\right) \sin\left(\frac{\eta_0}{2}\right) + \frac{315}{4} \tilde{P} \cos^2\left(\frac{\eta_0}{2}\right) \right)$$

$$\left. + \frac{2835 A_{11,2}(\tilde{\alpha}_{\eta,1})}{16\tilde{R}_0^7} \left( \frac{5}{2} \cos\left(\frac{\eta_0}{2}\right) \sin\left(\frac{\eta_0}{2}\right) + \tilde{P} \cos\left(\frac{\eta_0}{2}\right) \right) \right], \tag{H.27}$$

$$W_{2,1}^Q(\tilde{\alpha}_{\eta,1}, \tilde{R}_0, \tilde{Q}_2, \tilde{P}, \eta_0) \approx -\frac{2 W_{2,1}(\tilde{R}_0^{-1} \tilde{\alpha}_{\eta,1}, \tilde{R}_0, \tilde{Q}_2, \tilde{P}, \eta_0)}{\tilde{Q}_2}, \tag{H.28}$$

$$W_{2,1}^P(\tilde{\alpha}_\eta, \tilde{R}_0, \tilde{Q}_2, \tilde{P}, \eta_0) \approx \frac{2\tilde{P}}{\tilde{Q}_2^2 l_p}\left(\tilde{P}^2 - \frac{1}{4}\right)^{-2}\left[\frac{A_{3,2}(\tilde{\alpha}_{\eta,1})}{\tilde{R}_0^2}\left(\frac{1}{2}\cos\left(\frac{\eta_0}{2}\right)\sin\left(\frac{\eta_0}{2}\right) + \tilde{P}\cos^2\left(\frac{\eta_0}{2}\right)\right)\right.$$

$$-\frac{A_{5,2}(\tilde{\alpha}_{\eta,1})}{\tilde{R}_0^4}\left(\frac{15}{8}\cos\left(\frac{\eta_0}{2}\right)\sin\left(\frac{\eta_0}{2}\right) + \frac{3\tilde{P}}{4}\cos^2\left(\frac{\eta_0}{2}\right)\right)$$

$$+A_{7,2}(\tilde{\alpha}_{\eta,1})\left(\tilde{P}\cos^2\left(\frac{\eta_0}{2}\right)\left(\frac{45}{32\tilde{R}_0^6} + \frac{15}{4\tilde{R}_0^4}\right) - \cos\left(\frac{\eta_0}{2}\right)\sin\left(\frac{\eta_0}{2}\right)\left(\frac{45}{8\tilde{R}_0^4} + \frac{585}{64\tilde{R}_0^6}\right)\right)$$

$$-\frac{A_{9,2}(\tilde{\alpha}_{\eta,1})}{\tilde{R}_0^6}\left(\frac{1995}{32}\cos\left(\frac{\eta_0}{2}\right)\sin\left(\frac{\eta_0}{2}\right) + \frac{315}{16}\tilde{P}\cos^2\left(\frac{\eta_0}{2}\right)\right)$$

$$+\frac{945 A_{11,2}(\tilde{\alpha}_{\eta,1})}{32\tilde{R}_0^6}\left(\frac{5}{2}\cos\left(\frac{\eta_0}{2}\right)\sin\left(\frac{\eta_0}{2}\right) + \tilde{P}\cos^2\left(\frac{\eta_0}{2}\right)\right)\right]$$

$$-\frac{1}{\tilde{Q}_2^2 l_p}\left(\tilde{P}^2 - \frac{1}{4}\right)^{-1}\cos^2\left(\frac{\eta_0}{2}\right)\left[\frac{A_{3,2}(\tilde{\alpha}_{\eta,1})}{\tilde{R}_0^2} - \frac{3 A_{5,2}(\tilde{\alpha}_{\eta,1})}{4\tilde{R}_0^4} + A_{7,2}(\tilde{\alpha}_{\eta,1})\left(\frac{45}{32\tilde{R}_0^6} + \frac{15}{4\tilde{R}_0^4}\right)\right.$$

$$\left.-\frac{315 A_{9,2}(\tilde{\alpha}_{\eta,1})}{16\tilde{R}_0^6} + \frac{945 A_{11,2}(\tilde{\alpha}_{\eta,1})}{32\tilde{R}_0^6}\right]$$

(H.29)

$$W_{2,1}^\eta(\tilde{\alpha}_{\eta,1}, \tilde{R}_0, \tilde{Q}_2, \tilde{P}, \eta_0)$$

$$\approx \frac{\cos\eta_0}{2\tilde{Q}_2^2 l_p}\left[\frac{2A_{3,0}(\tilde{\alpha}_{\eta,1})}{\tilde{R}_0^2} - \frac{9A_{5,0}(\tilde{\alpha}_{\eta,1})}{2\tilde{R}_0^4} - A_{7,0}(\tilde{\alpha}_{\eta,1})\left[\frac{225}{16\tilde{R}_0^6} + \frac{15}{2\tilde{R}_0^4}\right] - \frac{525 A_{9,0}(\tilde{\alpha}_{\eta,1})}{8\tilde{R}_0^6} - \frac{945 A_{11,0}(\tilde{\alpha}_{\eta,1})}{16\tilde{R}_0^6}\right]$$

$$-\frac{1}{\tilde{Q}_2^2 l_p}\left(\tilde{P}^2 - \frac{1}{4}\right)^{-1}\left[\frac{A_{3,2}(\tilde{\alpha}_{\eta,1})}{\tilde{R}_0^2}\left(\frac{1}{4}\cos\eta_0 - \frac{\tilde{P}}{2}\sin\eta_0\right) - \frac{A_{5,2}(\tilde{\alpha}_{\eta,1})}{\tilde{R}_0^4}\left(\frac{15}{16}\cos\eta_0 - \frac{3\tilde{P}}{8}\sin\eta_0\right)\right.$$

$$+A_{7,2}(\tilde{\alpha}_{\eta,1})\left(\cos\eta_0\left(\frac{45}{16\tilde{R}_0^4} + \frac{585}{128\tilde{R}_0^6}\right) - \tilde{P}\sin\eta_0\left(\frac{45}{64\tilde{R}_0^6} + \frac{15}{8\tilde{R}_0^4}\right)\right)$$

$$\left.-\frac{A_{9,2}(\tilde{\alpha}_{\eta,1})}{\tilde{R}_0^6}\left(\frac{1995}{64}\cos\eta_0 - \frac{315}{32}\tilde{P}\sin\eta_0\right) - \frac{945 A_{11,2}(\tilde{\alpha}_{\eta,1})}{64\tilde{R}_0^6}\left(\frac{5}{2}\cos\eta_0 - \tilde{P}\sin\eta_0\right)\right].$$

(H.30)

The functions $\tilde{A}^\alpha(\bar{\alpha}_\eta)$ and $A_{n,m}^\alpha(\tilde{\alpha}_\eta)$ are derivatives of $\tilde{A}(\bar{\alpha}_\eta)$ and $A_{n,m}(\tilde{\alpha}_\eta)$ with respect to argument, and are given by the expressions

$$\tilde{A}^\alpha(\bar{\alpha}_\eta) = \int_{-\infty}^{\infty} x^2 dx \int_0^\infty r dr \exp\left(-\frac{r^2}{2}\right) J_{1,0,0}(x,r)\frac{\partial \hat{G}_\eta(x,\bar{\alpha}_\eta)}{\partial \bar{\alpha}_\eta},$$

(H.31)

$$A_{n,m}^\alpha(\tilde{\alpha}_\eta) = \int_0^\infty dx \frac{x^m}{(1+x^2)^{n/2}}\frac{\partial \hat{G}_\eta(x,\tilde{\alpha}_\eta)}{\partial \tilde{\alpha}_\eta},$$

(H.32)

where

$$\frac{\partial \hat{G}_\eta(x,\tilde{\alpha}_\eta)}{\partial \tilde{\alpha}_\eta} = -\frac{1}{4}\left[\frac{|x|}{\tilde{\alpha}_\eta} + \frac{1}{\tilde{\alpha}_\eta^{3/2}}\right]\exp\left(-\tilde{\alpha}_\eta^{1/2}|x|\right). \tag{H.33}$$

Now, the functions $W_{1,j}^{\alpha_R}(\bar{\alpha}_R,\tilde{\gamma},\tilde{R}_0,\tilde{Q}_1,\tilde{P},\eta_0)$, $W_{1,j}^{\gamma}(\bar{\alpha}_R,\tilde{\gamma},\tilde{R}_0,\tilde{Q}_1,\tilde{P},\eta_0)$, $W_{1,j}^{R}(\bar{\alpha}_R,\tilde{\gamma},\tilde{R}_0,\tilde{Q}_1,\tilde{P},\eta_0)$, $W_{1,j}^{Q}(\bar{\alpha}_R,\tilde{\gamma},\tilde{R}_0,\tilde{Q}_1,\tilde{P},\eta_0)$, $W_{1,j}^{P}(\bar{\alpha}_R,\tilde{\gamma},\tilde{R}_0,\tilde{Q}_1,\tilde{P},\eta_0)$, and $W_{1,j}^{\eta}(\bar{\alpha}_R,\tilde{\gamma},\tilde{R}_0,\tilde{Q}_1,\tilde{P},\eta_0)$ are given by (for $j=2,3,4$)

$$W_{1,2}^{\alpha_R}(\bar{\alpha}_R,\tilde{\gamma},\tilde{R}_0,\tilde{Q}_1,\tilde{P},\eta_0) \approx -\frac{1}{\pi\tilde{Q}_1^4 l_p \tilde{R}_0^9}\left(\tilde{P}^2 - \frac{1}{4}\right)^{-1}\left(\frac{1}{2}\sin\left(\frac{\eta_0}{2}\right)\cos\left(\frac{\eta_0}{2}\right) - \tilde{P}\sin^2\left(\frac{\eta_0}{2}\right)\right)\tilde{B}^\alpha(\bar{\alpha}_R,\tilde{\gamma}),$$

$$\tag{H.34}$$

$$W_{1,2}^{\gamma}(\bar{\alpha}_R,\tilde{\gamma},\tilde{R}_0,\tilde{Q}_1,\tilde{P},\eta_0) \approx -\frac{1}{\pi\tilde{Q}_1^4 l_p \tilde{R}_0^5}\left(\tilde{P}^2 - \frac{1}{4}\right)^{-1}\left(\frac{1}{2}\sin\left(\frac{\eta_0}{2}\right)\cos\left(\frac{\eta_0}{2}\right) - \tilde{P}\sin^2\left(\frac{\eta_0}{2}\right)\right)\tilde{B}^\gamma(\bar{\alpha}_R,\tilde{\gamma}),$$

$$\tag{H.35}$$

$$W_{1,2}^{R}(\bar{\alpha}_R,\tilde{\gamma},\tilde{R}_0,\tilde{Q}_1,\tilde{P},\eta_0) \approx -\frac{\left(5W_{1,2}(\bar{\alpha}_R,\tilde{\gamma},\tilde{R}_0,\tilde{Q}_1,\tilde{P},\eta_0) + 4\tilde{\alpha}_{R,1}W_{1,2}^{\alpha_R}(\bar{\alpha}_R,\tilde{\gamma},\tilde{R}_0,\tilde{Q}_1,\tilde{P},\eta_0)\right)}{\tilde{R}_0},$$

$$\tag{H.36}$$

$$W_{1,2}^{Q}(\bar{\alpha}_R,\tilde{\gamma},\tilde{R}_0,\tilde{Q}_1,\tilde{P},\eta_0) \approx -\frac{4W_{1,2}(\bar{\alpha}_R,\tilde{\gamma},\tilde{R}_0,\tilde{Q}_1,\tilde{P},\eta_0)}{\tilde{Q}_1}, \tag{H.37}$$

$$W_{1,2}^{P}(\bar{\alpha}_R,\tilde{\gamma},\tilde{R}_0,\tilde{Q}_1,\tilde{P},\eta_0)$$
$$\approx \frac{\tilde{B}(\bar{\alpha}_R,\tilde{\gamma})}{\pi\tilde{Q}_1^4 l_p \tilde{R}_0^5}\left[2\tilde{P}\left(\tilde{P}^2 - \frac{1}{4}\right)^{-2}\left(\frac{1}{2}\sin\left(\frac{\eta_0}{2}\right)\cos\left(\frac{\eta_0}{2}\right) - \tilde{P}\sin^2\left(\frac{\eta_0}{2}\right)\right) + \left(\tilde{P}^2 - \frac{1}{4}\right)^{-1}\sin^2\left(\frac{\eta_0}{2}\right)\right],$$

$$\tag{H.38}$$

$$W_{1,2}^{\eta}(\bar{\alpha}_R,\tilde{\gamma},\tilde{R}_0,\tilde{Q}_1,\tilde{P},\eta_0) \approx -\frac{\tilde{B}(\bar{\alpha}_R,\tilde{\gamma})}{\pi\tilde{Q}_1^4 l_p \tilde{R}_0^5}\left(\tilde{P}^2 - \frac{1}{4}\right)^{-1}\left(\frac{1}{4}\cos\eta_0 - \frac{\tilde{P}}{2}\sin\eta_0\right), \tag{H.39}$$

$$W_{1,3}^{\alpha_R}(\bar{\alpha}_R,\tilde{\gamma},\tilde{R}_0,\tilde{Q}_1,\tilde{P},\eta_0) \approx \frac{\tilde{C}^\alpha(\bar{\alpha}_R,\tilde{\gamma})}{4\pi\tilde{Q}_1^2 l_p \tilde{R}_0^7}\tilde{P}\left(\tilde{P}^2 + \frac{1}{4}\right)^{-1}, \tag{H.40}$$

$$W_{1,3}^{\gamma}(\bar{\alpha}_R,\tilde{\gamma},\tilde{R}_0,\tilde{Q}_1,\tilde{P},\eta_0) \approx \frac{\tilde{C}^{\gamma}(\bar{\alpha}_R,\tilde{\gamma})}{4\pi\tilde{Q}_1^2 l_p \tilde{R}_0^3}\tilde{P}\left(\tilde{P}^2+\frac{1}{4}\right)^{-1}, \tag{H.41}$$

$$W_{1,3}^{R}(\bar{\alpha}_R,\tilde{\gamma},\tilde{R}_0,\tilde{Q}_1,\tilde{P},\eta_0) \approx -\frac{\left(3W_{1,3}(\bar{\alpha}_R,\tilde{\gamma},\tilde{R}_0,\tilde{Q}_1,\tilde{P},\eta_0)+4\tilde{\alpha}_{R,1}W_{1,3}^{\alpha_R}(\bar{\alpha}_R,\tilde{\gamma},\tilde{R}_0,\tilde{Q}_1,\tilde{P},\eta_0)\right)}{\tilde{R}_0}, \tag{H.42}$$

$$W_{1,3}^{Q}(\bar{\alpha}_R,\tilde{\gamma},\tilde{R}_0,\tilde{Q}_1,\tilde{P},\eta_0) \approx -\frac{2W_{1,3}(\bar{\alpha}_R,\tilde{\gamma},\tilde{R}_0,\tilde{Q}_1,\tilde{P},\eta_0)}{\tilde{Q}_1}, \tag{H.43}$$

$$W_{1,3}^{P}(\bar{\alpha}_R,\tilde{\gamma},\tilde{R}_0,\tilde{Q}_1,\tilde{P},\eta_0) = \frac{\partial W_{1,3}(\bar{\alpha}_R,\tilde{\gamma},\tilde{R}_0,\tilde{Q}_1,\tilde{P},\eta_0)}{\partial \tilde{P}} \approx \frac{\tilde{C}(\bar{\alpha}_R,\tilde{\gamma})}{4\pi\tilde{Q}_1^2 l_p \tilde{R}_0^3}\left(\left(\tilde{P}^2+\frac{1}{4}\right)^{-1}-2\tilde{P}^2\left(\tilde{P}^2+\frac{1}{4}\right)^{-2}\right), \tag{H.44}$$

$$W_{1,3}^{\eta}(\bar{\alpha}_R,\tilde{\gamma},\tilde{R}_0,\tilde{Q}_1,\tilde{P},\eta_0) \approx 0, \tag{H.45}$$

$$W_{1,4}^{\alpha_R}(\bar{\alpha}_R,\tilde{\gamma},\tilde{R}_0,\tilde{Q}_1,\tilde{P},\eta_0) \approx -\frac{\tilde{D}^{\alpha}(\bar{\alpha}_R,\tilde{\gamma})}{2\pi\tilde{Q}_1^3 l_p \tilde{R}_0^8}\cos\left(\frac{\eta_0}{2}\right)\left(\tilde{P}^2+\frac{1}{4}\right)^{-1/2}, \tag{H.46}$$

$$W_{1,4}^{\gamma}(\bar{\alpha}_R,\tilde{\gamma},\tilde{R}_0,\tilde{Q}_1,\tilde{P},\eta_0) \approx -\frac{\tilde{D}^{\gamma}(\bar{\alpha}_R,\tilde{\gamma})}{2\pi\tilde{Q}_1^3 l_p \tilde{R}_0^4}\cos\left(\frac{\eta_0}{2}\right)\left(\tilde{P}^2+\frac{1}{4}\right)^{-1/2}, \tag{H.47}$$

$$W_{1,4}^{R}(\bar{\alpha}_R,\tilde{\gamma},\tilde{R}_0,\tilde{Q}_1,\tilde{P},\eta_0) \approx -\frac{4\left(W_{1,4}(\bar{\alpha}_R,\tilde{\gamma},\tilde{R}_0,\tilde{Q}_1,\tilde{P},\eta_0)+\tilde{\alpha}_{R,1}W_{1,4}^{\alpha_R}(\bar{\alpha}_R,\tilde{\gamma},\tilde{R}_0,\tilde{Q}_1,\tilde{P},\eta_0)\right)}{\tilde{R}_0}, \tag{H.48}$$

$$W_{1,4}^{Q}(\bar{\alpha}_R,\tilde{\gamma},\tilde{R}_0,\tilde{Q}_1,\tilde{P},\eta_0) \approx -\frac{3W_{1,4}(\bar{\alpha}_R,\tilde{\gamma},\tilde{R}_0,\tilde{Q}_1,\tilde{P},\eta_0)}{\tilde{Q}_1}, \tag{H.49}$$

$$W_{1,4}^{P}(\bar{\alpha}_R,\tilde{\gamma},\tilde{R}_0,\tilde{Q}_1,\tilde{P},\eta_0) \approx -\frac{\tilde{P}}{\left(\tilde{P}^2+\frac{1}{4}\right)}W_{1,4}(\bar{\alpha}_R,\tilde{\gamma},\tilde{R}_0,\tilde{Q}_1,\tilde{P},\eta_0), \tag{H.50}$$

and

$$W_{1,4}^{\eta}(\bar{\alpha}_R,\tilde{\gamma},\tilde{R}_0,\tilde{Q}_1,\tilde{P},\eta_0) \approx \frac{\tilde{D}^{\gamma}(\bar{\alpha}_R,\tilde{\gamma})}{4\pi\tilde{Q}_j^3 l_p \tilde{R}_0^4}\sin\left(\frac{\eta_0}{2}\right)\left(\tilde{P}^2+\frac{1}{4}\right)^{-1/2}. \tag{H.51}$$

The functions $B^{\alpha}(\bar{\alpha}_R,\tilde{\gamma})$, $C^{\alpha}(\bar{\alpha}_R,\tilde{\gamma})$ and $D^{\alpha}(\bar{\alpha}_R,\tilde{\gamma})$ are the partial derivatives of $B(\bar{\alpha}_R,\tilde{\gamma})$, $C(\bar{\alpha}_R,\tilde{\gamma})$ and $D(\bar{\alpha}_R,\tilde{\gamma})$, respectively, with respect to $\bar{\alpha}_R$, whereas $B^{\gamma}(\bar{\alpha}_R,\tilde{\gamma})$, $C^{\gamma}(\bar{\alpha}_R,\tilde{\gamma})$ and $D^{\gamma}(\bar{\alpha}_R,\tilde{\gamma})$ are their partial derivatives with respect to $\tilde{\gamma}$. These are given by

$$\tilde{B}^{\alpha}(\bar{\alpha}_R,\tilde{\gamma}) = \int_{-\infty}^{\infty} x^2 dx \int_{0}^{\infty} r dr \exp\left(-\frac{r^2}{2}\right) J_{1,1,1}(x,r) \frac{\partial \hat{G}_R(x,\bar{\alpha}_R,\tilde{\gamma})}{\partial \bar{\alpha}_R}, \tag{H.52}$$

$$\tilde{B}^{\gamma}(\bar{\alpha}_R,\tilde{\gamma}) = \int_{-\infty}^{\infty} x^2 dx \int_{0}^{\infty} r dr \exp\left(-\frac{r^2}{2}\right) J_{1,1,1}(x,r) \frac{\partial \hat{G}_R(x,\bar{\alpha}_R,\tilde{\gamma})}{\partial \tilde{\gamma}}, \tag{H.53}$$

$$\tilde{C}^{\alpha}(\bar{\alpha}_R,\tilde{\gamma}) = \int_{-\infty}^{\infty} x^2 dx \int_{0}^{\infty} r dr \exp\left(-\frac{r^2}{2}\right) J_{1,0,0}(x,r) \frac{\partial \hat{D}_R(x,\bar{\alpha}_R,\tilde{\gamma})}{\partial \bar{\alpha}_R}, \tag{H.54}$$

$$\tilde{C}^{\gamma}(\bar{\alpha}_R,\tilde{\gamma}) = \int_{-\infty}^{\infty} x^2 dx \int_{0}^{\infty} r dr \exp\left(-\frac{r^2}{2}\right) J_{1,0,0}(x,r) \frac{\partial \hat{D}_R(x,\bar{\alpha}_R,\tilde{\gamma})}{\partial \tilde{\gamma}}, \tag{H.55}$$

$$\tilde{D}^{\alpha}(\bar{\alpha}_R,\tilde{\gamma}) = \int_{-\infty}^{\infty} x dx \int_{0}^{\infty} r dr \exp\left(-\frac{r^2}{2}\right) J_{1,2,0}(x,r) \frac{\partial \hat{C}_R(x,\bar{\alpha}_R,\tilde{\gamma})}{\partial \bar{\alpha}_R}, \tag{H.56}$$

and

$$\tilde{D}^{\gamma}(\bar{\alpha}_R,\tilde{\gamma}) = \int_{-\infty}^{\infty} x dx \int_{0}^{\infty} r dr \exp\left(-\frac{r^2}{2}\right) J_{1,2,0}(x,r) \frac{\partial \hat{C}_R(x,\bar{\alpha}_R,\tilde{\gamma})}{\partial \tilde{\gamma}}. \tag{H.57}$$

where the partial derivatives of the rescaled correlation functions are given by (in the case of the above $\tilde{\alpha}_R$ replaced by $\bar{\alpha}_R$)

$$\frac{\partial \hat{G}_R(x,\tilde{\alpha}_R,\tilde{\gamma})}{\partial \tilde{\alpha}_R} = -\frac{\exp\left(-\frac{\tilde{\alpha}_R^{1/4}|x|}{\sqrt{2}}\sqrt{1+\frac{\tilde{\gamma}}{2}}\right)}{4\sqrt{2}\tilde{\alpha}_R^{3/4}\sqrt{4-\tilde{\gamma}^2}} \left(\frac{3}{\tilde{\alpha}_R} + \frac{\sqrt{1+\frac{\tilde{\gamma}}{2}}|x|}{\sqrt{2}\tilde{\alpha}_R^{3/4}}\right)$$

$$\left(\sqrt{1-\frac{\tilde{\gamma}}{2}}\cos\left(\frac{\tilde{\alpha}_R^{1/4}|x|}{\sqrt{2}}\sqrt{1-\frac{\tilde{\gamma}}{2}}\right) + \sqrt{1+\frac{\tilde{\gamma}}{2}}\sin\left(\frac{\tilde{\alpha}_R^{1/4}|x|}{\sqrt{2}}\sqrt{1-\frac{\tilde{\gamma}}{2}}\right)\right)\theta(2-\tilde{\gamma})$$

$$+ \frac{|x|\exp\left(-\tilde{\alpha}_R^{1/4}\sqrt{1+\frac{\tilde{\gamma}}{2}}|x|\right)}{8\tilde{\alpha}_R^{3/2}\sqrt{4-\tilde{\gamma}^2}}\left(\left(\frac{\tilde{\gamma}}{2}-1\right)\sin\left(\frac{\tilde{\alpha}_R^{1/4}|x|}{\sqrt{2}}\sqrt{1-\frac{\tilde{\gamma}}{2}}\right) + \sqrt{1-\frac{\tilde{\gamma}^2}{4}}\cos\left(\frac{\tilde{\alpha}_R^{1/4}|x|}{\sqrt{2}}\sqrt{1-\frac{\tilde{\gamma}}{2}}\right)\right)\theta(2-\tilde{\gamma})$$

$$+ \frac{3}{8\sqrt{2}\tilde{\alpha}_R^{7/4}\sqrt{\tilde{\gamma}^2-4}}\left[\left(\tilde{\gamma}-\sqrt{\tilde{\gamma}^2-4}\right)^{1/2}\exp\left(-\frac{\tilde{\alpha}_R^{1/4}|x|}{\sqrt{2}}\left(\tilde{\gamma}+\sqrt{\tilde{\gamma}^2-4}\right)^{1/2}\right)\right.$$

$$\left. -\left(\tilde{\gamma}+\sqrt{\tilde{\gamma}^2-4}\right)^{1/2}\exp\left(-\frac{\tilde{\alpha}_R^{1/4}|x|}{\sqrt{2}}\left(\tilde{\gamma}-\sqrt{\tilde{\gamma}^2-4}\right)^{1/2}\right)\right]\theta(\tilde{\gamma}-2)$$

$$+ \frac{|x|}{8\tilde{\alpha}_R^{3/2}\sqrt{\tilde{\gamma}^2-4}}\left[\exp\left(-\frac{|x|\tilde{\alpha}_R^{1/4}}{\sqrt{2}}\left(\tilde{\gamma}+\sqrt{\tilde{\gamma}^2-4}\right)^{1/2}\right) - \exp\left(-\frac{|x|\tilde{\alpha}_R^{1/4}}{\sqrt{2}}\left(\tilde{\gamma}-\sqrt{\tilde{\gamma}^2-4}\right)^{1/2}\right)\right]\theta(\tilde{\gamma}-2),$$

$$\tag{H.58}$$

$$\frac{\partial \hat{D}_R\left(x,\tilde{\alpha}_R,\tilde{\gamma}\right)}{\partial \tilde{\alpha}_R} = -\frac{\exp\left(-\frac{\tilde{\alpha}_R^{1/4}|x|}{\sqrt{2}}\sqrt{1+\frac{\tilde{\gamma}}{2}}\right)}{4\sqrt{2}\tilde{\alpha}_R^{1/4}\sqrt{4-\tilde{\gamma}^2}}\left(\frac{1}{\tilde{\alpha}_R}+\frac{\sqrt{1+\frac{\tilde{\gamma}}{2}}|x|}{\tilde{\alpha}_R^{3/4}\sqrt{2}}\right)$$

$$\left[\sqrt{1-\frac{\tilde{\gamma}}{2}}\cos\left(\frac{\tilde{\alpha}_R^{1/4}|x|}{\sqrt{2}}\sqrt{1-\frac{\tilde{\gamma}}{2}}\right)-\sqrt{1+\frac{\tilde{\gamma}}{2}}\sin\left(\frac{\tilde{\alpha}_R^{1/4}|x|}{\sqrt{2}}\sqrt{1-\frac{\tilde{\gamma}}{2}}\right)\right]\theta(2-\tilde{\gamma})$$

$$-\frac{|x|\exp\left(-\frac{\tilde{\alpha}_R^{1/4}|x|}{\sqrt{2}}\sqrt{1+\frac{\tilde{\gamma}}{2}}\right)}{8\tilde{\alpha}_R\sqrt{4-\tilde{\gamma}^2}}\left[\left(1-\frac{\tilde{\gamma}}{2}\right)\sin\left(\frac{\tilde{\alpha}_R^{1/4}|x|}{\sqrt{2}}\sqrt{1-\frac{\tilde{\gamma}}{2}}\right)+\sqrt{1-\frac{\tilde{\gamma}^2}{4}}\cos\left(\frac{\tilde{\alpha}_R^{1/4}|x|}{\sqrt{2}}\sqrt{1-\frac{\tilde{\gamma}}{2}}\right)\right]\theta(2-\tilde{\gamma})$$

$$-\frac{1}{8\sqrt{2}\tilde{\alpha}_R^{5/4}\sqrt{\tilde{\gamma}^2-4}}\left(\left(\tilde{\gamma}+\sqrt{\tilde{\gamma}^2-4}\right)^{1/2}\exp\left(-\frac{\tilde{\alpha}_R^{1/4}|x|}{\sqrt{2}}\left(\tilde{\gamma}+\sqrt{\tilde{\gamma}^2-4}\right)^{1/2}\right)\right.$$

$$\left.-\left(\tilde{\gamma}-\sqrt{\tilde{\gamma}^2-4}\right)^{1/2}\exp\left(-\frac{\tilde{\alpha}_R^{1/4}|x|}{\sqrt{2}}\left(\tilde{\gamma}-\sqrt{\tilde{\gamma}^2-4}\right)^{1/2}\right)\right)\theta(\tilde{\gamma}-2)$$

$$-\frac{|x|}{16\tilde{\alpha}_R\sqrt{\tilde{\gamma}^2-4}}\left(\left(\tilde{\gamma}+\sqrt{\tilde{\gamma}^2-4}\right)\exp\left(-\frac{\tilde{\alpha}_R^{1/4}|x|}{\sqrt{2}}\left(\tilde{\gamma}+\sqrt{\tilde{\gamma}^2-4}\right)^{1/2}\right)\right.$$

$$\left.-\left(\tilde{\gamma}-\sqrt{\tilde{\gamma}^2-4}\right)\exp\left(-\frac{\tilde{\alpha}_R^{1/4}|x|}{\sqrt{2}}\left(\tilde{\gamma}-\sqrt{\tilde{\gamma}^2-4}\right)^{1/2}\right)\right)\theta(\tilde{\gamma}-2),\qquad\text{(H.59)}$$

$$\frac{\partial \hat{C}_R\left(x,\tilde{\alpha}_R,\tilde{\gamma}\right)}{\partial \tilde{\alpha}_R} = -\frac{\text{sgn}(x)|x|\exp\left(-\frac{\tilde{\alpha}_R^{1/4}|x|}{\sqrt{2}}\sqrt{1+\frac{\tilde{\gamma}}{2}}\right)}{4\sqrt{2}\tilde{\alpha}_R^{5/4}\sqrt{4-\tilde{\gamma}^2}}$$

$$\left[\sqrt{1+\frac{\tilde{\gamma}}{2}}\sin\left(\frac{\tilde{\alpha}_R^{1/4}|x|}{\sqrt{2}}\sqrt{1-\frac{\tilde{\gamma}}{2}}\right)-\sqrt{1-\frac{\tilde{\gamma}}{2}}\cos\left(\frac{\tilde{\alpha}_R^{1/4}|x|}{\sqrt{2}}\sqrt{1-\frac{\tilde{\gamma}}{2}}\right)\right]\theta(2-\tilde{\gamma})$$

$$-\frac{\text{sgn}(x)\exp\left(-\frac{\tilde{\alpha}_R^{1/4}|x|}{\sqrt{2}}\sqrt{1+\frac{\tilde{\gamma}}{2}}\right)}{2\tilde{\alpha}_R^{3/2}\sqrt{4-\tilde{\gamma}^2}}\sin\left(\frac{\tilde{\alpha}_R^{1/4}|x|}{\sqrt{2}}\sqrt{1-\frac{\gamma}{2}}\right)\theta(2-\tilde{\gamma})$$

$$+\left[\frac{\text{sgn}(x)|x|}{8\sqrt{2}\tilde{\alpha}_R^{5/4}\sqrt{\tilde{\gamma}^2-4}}\left(\left(\tilde{\gamma}+\sqrt{\tilde{\gamma}^2-4}\right)^{1/2}\exp\left(-\frac{\tilde{\alpha}_R^{1/4}|x|}{\sqrt{2}}\left(\tilde{\gamma}+\sqrt{\tilde{\gamma}^2-4}\right)^{1/2}\right)\right.\right.$$

$$\left.-\left(\tilde{\gamma}-\sqrt{\tilde{\gamma}^2-4}\right)^{1/2}\exp\left(-\frac{\tilde{\alpha}_R^{1/4}|x|}{\sqrt{2}}\left(\tilde{\gamma}-\sqrt{\tilde{\gamma}^2-4}\right)^{1/2}\right)\right)$$

$$\left.+\frac{\text{sgn}(x)}{4\tilde{\alpha}_R^{3/2}\sqrt{\tilde{\gamma}^2-4}}\left(\exp\left(-\frac{\tilde{\alpha}_R^{1/4}|x|}{\sqrt{2}}\left(\tilde{\gamma}+\sqrt{\tilde{\gamma}^2-4}\right)^{1/2}\right)-\exp\left(-\frac{\tilde{\alpha}_R^{1/4}|x|}{\sqrt{2}}\left(\tilde{\gamma}-\sqrt{\tilde{\gamma}^2-4}\right)^{1/2}\right)\right)\right]\theta(\tilde{\gamma}-2),$$

$$\text{(H.60)}$$

$$\frac{\partial \hat{G}_R(x,\tilde{\alpha}_R,\tilde{\gamma})}{\partial \tilde{\gamma}} = -\frac{|x|\exp\left(-\frac{\tilde{\alpha}_R^{1/4}|x|}{\sqrt{2}}\sqrt{1+\frac{\tilde{\gamma}}{2}}\right)}{8\tilde{\alpha}_R^{1/2}\sqrt{2+\tilde{\gamma}}}\left(\frac{1}{\sqrt{2+\tilde{\gamma}}}\cos\left(\frac{\tilde{\alpha}_R^{1/4}|x|}{\sqrt{2}}\sqrt{1-\frac{\tilde{\gamma}}{2}}\right)+\frac{1}{\sqrt{2-\tilde{\gamma}}}\sin\left(\frac{\tilde{\alpha}_R^{1/4}|x|}{\sqrt{2}}\sqrt{1-\frac{\tilde{\gamma}}{2}}\right)\right)\theta(2-\tilde{\gamma})$$

$$-\frac{\exp\left(-\frac{\tilde{\alpha}_R^{1/4}|x|}{\sqrt{2}}\sqrt{1+\frac{\tilde{\gamma}}{2}}\right)}{4\tilde{\alpha}_R^{3/4}}\left(\frac{1}{(2+\tilde{\gamma})^{3/2}}\cos\left(\frac{\tilde{\alpha}_R^{1/4}|x|}{\sqrt{2}}\sqrt{1-\frac{\tilde{\gamma}}{2}}\right)-\frac{1}{(2-\tilde{\gamma})^{3/2}}\sin\left(\frac{\tilde{\alpha}_R^{1/4}|x|}{\sqrt{2}}\sqrt{1-\frac{\tilde{\gamma}}{2}}\right)\right)\theta(2-\tilde{\gamma})$$

$$+\frac{|x|}{8\tilde{\alpha}_R^{1/2}\sqrt{2-\tilde{\gamma}}}\exp\left(-\frac{\tilde{\alpha}_R^{1/4}|x|}{\sqrt{2}}\sqrt{1+\frac{\tilde{\gamma}}{2}}\right)\left(\frac{1}{(2+\tilde{\gamma})^{1/2}}\sin\left(\frac{\tilde{\alpha}_R^{1/4}|x|}{\sqrt{2}}\sqrt{1-\frac{\tilde{\gamma}}{2}}\right)\right.$$

$$\left.-\frac{1}{(2-\tilde{\gamma})^{1/2}}\cos\left(\frac{\tilde{\alpha}_R^{1/4}|x|}{\sqrt{2}}\sqrt{1-\frac{\tilde{\gamma}}{2}}\right)\right)\theta(2-\tilde{\gamma})+\frac{\tilde{\gamma}}{2\sqrt{2}\tilde{\alpha}_R^{3/4}(\tilde{\gamma}^2-4)^{3/2}}\left[\left(\tilde{\gamma}-\sqrt{\tilde{\gamma}^2-4}\right)^{1/2}\exp\left(-\frac{\tilde{\alpha}_R^{1/4}|x|}{\sqrt{2}}\left(\tilde{\gamma}+\sqrt{\tilde{\gamma}^2-4}\right)^{1/2}\right)\right.$$

$$\left.-\left(\tilde{\gamma}+\sqrt{\tilde{\gamma}^2-4}\right)^{1/2}\exp\left(-\frac{\tilde{\alpha}_R^{1/4}|x|}{\sqrt{2}}\left(\tilde{\gamma}-\sqrt{\tilde{\gamma}^2-4}\right)^{1/2}\right)\right]\theta(\tilde{\gamma}-2)$$

$$-\frac{1}{4\sqrt{2}\tilde{\alpha}_R^{3/4}\left(\tilde{\gamma}-\sqrt{\tilde{\gamma}^2-4}\right)^{1/2}}\left(\frac{1}{\sqrt{\tilde{\gamma}^2-4}}-\frac{\tilde{\gamma}}{\tilde{\gamma}^2-4}\right)\left[\exp\left(-\frac{\tilde{\alpha}_R^{1/4}|x|}{\sqrt{2}}\left(\tilde{\gamma}+\sqrt{\tilde{\gamma}^2-4}\right)^{1/2}\right)\right.$$

$$\left.+\frac{|x|\tilde{\alpha}_R^{1/4}}{\sqrt{2}}\left(\tilde{\gamma}+\sqrt{\tilde{\gamma}^2-4}\right)^{1/2}\exp\left(-\frac{\tilde{\alpha}_R^{1/4}|x|}{\sqrt{2}}\left(\tilde{\gamma}-\sqrt{\tilde{\gamma}^2-4}\right)^{1/2}\right)\right]\theta(\tilde{\gamma}-2)$$

$$+\frac{1}{4\sqrt{2}\tilde{\alpha}_R^{3/4}\left(\tilde{\gamma}+\sqrt{\tilde{\gamma}^2-4}\right)^{1/2}}\left(\frac{1}{\sqrt{\tilde{\gamma}^2-4}}+\frac{\tilde{\gamma}}{\tilde{\gamma}^2-4}\right)\left[\exp\left(-\frac{\tilde{\alpha}_R^{1/4}|x|}{\sqrt{2}}\left(\tilde{\gamma}-\sqrt{\tilde{\gamma}^2-4}\right)^{1/2}\right)\right.$$

$$\left.+\frac{|x|\tilde{\alpha}_R^{1/4}}{\sqrt{2}}\left(\tilde{\gamma}-\sqrt{\tilde{\gamma}^2-4}\right)^{1/2}\exp\left(-\frac{\tilde{\alpha}_R^{1/4}|x|}{\sqrt{2}}\left(\tilde{\gamma}+\sqrt{\tilde{\gamma}^2-4}\right)^{1/2}\right)\right]\theta(\tilde{\gamma}-2),$$

$$\tag{H.61}$$

$$\frac{\partial \hat{D}_R(x,\tilde{\alpha}_R,\tilde{\gamma})}{\partial \tilde{\gamma}} = -\frac{|x|\exp\left(-\frac{\tilde{\alpha}_R^{1/4}|x|}{\sqrt{2}}\sqrt{1+\frac{\tilde{\gamma}}{2}}\right)}{8(2+\tilde{\gamma})^{1/2}}\left[\frac{1}{\sqrt{2+\tilde{\gamma}}}\cos\left(\frac{\tilde{\alpha}_R^{1/4}|x|}{\sqrt{2}}\sqrt{1-\frac{\tilde{\gamma}}{2}}\right) - \frac{1}{\sqrt{2-\tilde{\gamma}}}\sin\left(\frac{\tilde{\alpha}_R^{1/4}|x|}{\sqrt{2}}\sqrt{1-\frac{\tilde{\gamma}}{2}}\right)\right]\theta(2-\tilde{\gamma})$$

$$+\frac{|x|\exp\left(-\frac{\tilde{\alpha}_R^{1/4}|x|}{\sqrt{2}}\sqrt{1+\frac{\tilde{\gamma}}{2}}\right)}{8(2-\tilde{\gamma})^{1/2}}\left[\frac{1}{\sqrt{2+\tilde{\gamma}}}\sin\left(\frac{\tilde{\alpha}_R^{1/4}|x|}{\sqrt{2}}\sqrt{1-\frac{\tilde{\gamma}}{2}}\right) + \frac{1}{\sqrt{2-\tilde{\gamma}}}\cos\left(\frac{\tilde{\alpha}_R^{1/4}|x|}{\sqrt{2}}\sqrt{1-\frac{\tilde{\gamma}}{2}}\right)\right]\theta(2-\tilde{\gamma})$$

$$-\frac{\exp\left(-\frac{\tilde{\alpha}_R^{1/4}|x|}{\sqrt{2}}\sqrt{1+\frac{\tilde{\gamma}}{2}}\right)}{4\tilde{\alpha}_R^{1/4}}\left[\frac{1}{(2+\tilde{\gamma})^{3/2}}\cos\left(\frac{\tilde{\alpha}_R^{1/4}|x|}{\sqrt{2}}\sqrt{1-\frac{\tilde{\gamma}}{2}}\right) + \frac{1}{(2-\tilde{\gamma})^{3/2}}\sin\left(\frac{\tilde{\alpha}_R^{1/4}|x|}{\sqrt{2}}\sqrt{1-\frac{\tilde{\gamma}}{2}}\right)\right]\theta(2-\tilde{\gamma})$$

$$-\frac{\tilde{\gamma}}{2\sqrt{2}\tilde{\alpha}_R^{1/4}(\tilde{\gamma}^2-4)^{3/2}}\left(\left(\tilde{\gamma}+\sqrt{\tilde{\gamma}^2-4}\right)^{1/2}\exp\left(-\frac{\tilde{\alpha}_R^{1/4}|x|}{\sqrt{2}}\left(\tilde{\gamma}+\sqrt{\tilde{\gamma}^2-4}\right)^{1/2}\right)\right.$$

$$\left.-\left(\tilde{\gamma}-\sqrt{\tilde{\gamma}^2-4}\right)^{1/2}\exp\left(-\frac{\tilde{\alpha}_R^{1/4}|x|}{\sqrt{2}}\left(\tilde{\gamma}-\sqrt{\tilde{\gamma}^2-4}\right)^{1/2}\right)\right)\theta(\tilde{\gamma}-2)$$

$$+\frac{1}{4\tilde{\alpha}_R^{1/4}}\left(\frac{1}{\sqrt{\tilde{\gamma}^2-4}}+\frac{\tilde{\gamma}}{\tilde{\gamma}^2-4}\right)\left(\frac{1}{\sqrt{2}}\frac{1}{\left(\tilde{\gamma}+\sqrt{\tilde{\gamma}^2-4}\right)^{1/2}}-\frac{|x|\tilde{\alpha}_R^{1/4}}{2}\right)\exp\left(-\frac{\tilde{\alpha}_R^{1/4}|x|}{\sqrt{2}}\left(\tilde{\gamma}+\sqrt{\tilde{\gamma}^2-4}\right)^{1/2}\right)\theta(\tilde{\gamma}-2)$$

$$-\frac{1}{4\tilde{\alpha}_R^{1/4}}\left(\frac{1}{\sqrt{\tilde{\gamma}^2-4}}-\frac{\tilde{\gamma}}{\tilde{\gamma}^2-4}\right)\left(\frac{1}{\sqrt{2}}\frac{1}{\left(\tilde{\gamma}-\sqrt{\tilde{\gamma}^2-4}\right)^{1/2}}-\frac{|x|\tilde{\alpha}_R^{1/4}}{2}\right)\exp\left(-\frac{\tilde{\alpha}_R^{1/4}|x|}{\sqrt{2}}\left(\tilde{\gamma}-\sqrt{\tilde{\gamma}^2-4}\right)^{1/2}\right)\theta(\tilde{\gamma}-2),$$

(H.62)

$$\frac{\partial \hat{C}_R(x,\tilde{\alpha}_R,\tilde{\gamma})}{\partial \tilde{\gamma}} = \frac{\tilde{\gamma}\,\text{sgn}(x)\exp\left(-\frac{\tilde{\alpha}_R^{1/4}|x|}{\sqrt{2}}\sqrt{1+\frac{\tilde{\gamma}}{2}}\right)}{\tilde{\alpha}_R^{1/2}(4-\tilde{\gamma}^2)^{3/2}}\sin\left(\frac{\tilde{\alpha}_R^{1/4}|x|}{\sqrt{2}}\sqrt{1-\frac{\tilde{\gamma}}{2}}\right)\theta(2-\tilde{\gamma})$$

$$-\frac{|x|\text{sgn}(x)\exp\left(-\frac{\tilde{\alpha}_R^{1/4}|x|}{\sqrt{2}}\sqrt{1+\frac{\tilde{\gamma}}{2}}\right)}{4\tilde{\alpha}_R^{1/4}\sqrt{4-\tilde{\gamma}^2}}\left[\frac{1}{\sqrt{2+\tilde{\gamma}}}\sin\left(\frac{\tilde{\alpha}_R^{1/4}|x|}{\sqrt{2}}\sqrt{1-\frac{\tilde{\gamma}}{2}}\right) + \frac{1}{\sqrt{2-\tilde{\gamma}}}\cos\left(\frac{\tilde{\alpha}_R^{1/4}|x|}{\sqrt{2}}\sqrt{1-\frac{\tilde{\gamma}}{2}}\right)\right]\theta(2-\tilde{\gamma})$$

$$+\frac{\tilde{\gamma}\,\text{sgn}(x)}{2\tilde{\alpha}_R^{1/2}(\tilde{\gamma}^2-4)^{3/2}}\left(\exp\left(-\frac{\tilde{\alpha}_R^{1/4}|x|}{\sqrt{2}}\left(\tilde{\gamma}+\sqrt{\tilde{\gamma}^2-4}\right)^{1/2}\right) - \exp\left(-\frac{\tilde{\alpha}_R^{1/4}|x|}{\sqrt{2}}\left(\tilde{\gamma}-\sqrt{\tilde{\gamma}^2-4}\right)^{1/2}\right)\right)\theta(\tilde{\gamma}-2)$$

$$+\frac{\text{sgn}(x)}{4\sqrt{2}\tilde{\alpha}_R^{1/4}}\frac{|x|}{\left(\tilde{\gamma}+\sqrt{\tilde{\gamma}^2-4}\right)^{1/2}}\left(\frac{1}{\sqrt{\tilde{\gamma}^2-4}}+\frac{\tilde{\gamma}}{(\tilde{\gamma}^2-4)}\right)\exp\left(-\frac{\tilde{\alpha}_R^{1/4}|x|}{\sqrt{2}}\left(\tilde{\gamma}+\sqrt{\tilde{\gamma}^2-4}\right)^{1/2}\right)\theta(\tilde{\gamma}-2)$$

$$-\frac{\text{sgn}(x)}{4\sqrt{2}\tilde{\alpha}_R^{1/4}}\frac{|x|}{\left(\tilde{\gamma}-\sqrt{\tilde{\gamma}^2-4}\right)^{1/2}}\left(\frac{1}{\sqrt{\tilde{\gamma}^2-4}}-\frac{\tilde{\gamma}}{(\tilde{\gamma}^2-4)}\right)\exp\left(-\frac{\tilde{\alpha}_R^{1/4}|x|}{\sqrt{2}}\left(\tilde{\gamma}-\sqrt{\tilde{\gamma}^2-4}\right)^{1/2}\right)\theta(\tilde{\gamma}-2).$$

(H.63)

The functions $W_{2,j}^{\alpha_R}(\tilde{\alpha}_R,\tilde{\gamma},\tilde{R}_0,\tilde{Q}_2,\tilde{P},\eta_0)$, $W_{2,j}^{\gamma}(\tilde{\alpha}_R,\tilde{\gamma},\tilde{R}_0,\tilde{Q}_2,\tilde{P},\eta_0)$, $W_{2,j}^{R}(\tilde{\alpha}_R,\tilde{\gamma},\tilde{R}_0,\tilde{Q}_2,\tilde{P},\eta_0)$, $W_{2,j}^{Q}(\tilde{\alpha}_R,\tilde{\gamma},\tilde{R}_0,\tilde{Q}_2,\tilde{P},\eta_0)$, $W_{2,j}^{P}(\tilde{\alpha}_R,\tilde{\gamma},\tilde{R}_0,\tilde{Q}_2,\tilde{P},\eta_0)$, and $W_{2,j}^{\eta}(\tilde{\alpha}_R,\tilde{\gamma},\tilde{R}_0,\tilde{Q}_2,\tilde{P},\eta_0)$ are given by (for $j=2,3,4$)

$$W_{2,2}^{\alpha_R}(\tilde{\alpha}_R,\tilde{\gamma},\tilde{R}_0,\tilde{Q}_2,\tilde{P},\eta_0)$$
$$\approx -\frac{1}{\tilde{Q}_2^4 l_p}\sin\left(\frac{\eta_0}{2}\right)\cos\left(\frac{\eta_0}{2}\right)\left[-\frac{9B_{5,0}^{\alpha}(\tilde{\alpha}_R,\gamma)}{\tilde{R}_0^4}+B_{7,0}^{\alpha}(\tilde{\alpha}_R,\gamma)\left(\frac{15}{\tilde{R}_0^4}+\frac{225}{4\tilde{R}_0^6}\right)-\frac{525B_{9,0}^{\alpha}(\tilde{\alpha}_R,\gamma)}{2\tilde{R}_0^6}\right.$$
$$\left.+\frac{945B_{11,0}^{\alpha}(\tilde{\alpha}_R,\gamma)}{4\tilde{R}_0^6}\right]-\frac{1}{\tilde{Q}_2^4 l_p}\left(\tilde{P}^2-\frac{1}{4}\right)^{-1}\left[\frac{B_{5,2}^{\alpha}(\tilde{\alpha}_R,\gamma)}{\tilde{R}_0^4}\left(\frac{21}{4}\sin\left(\frac{\eta_0}{2}\right)\cos\left(\frac{\eta_0}{2}\right)-\frac{3\tilde{P}}{2}\sin^2\left(\frac{\eta_0}{2}\right)\right)\right.$$
$$+B_{7,2}^{\alpha}(\tilde{\alpha}_R,\gamma)\left(\tilde{P}\sin^2\left(\frac{\eta_0}{2}\right)\left(\frac{15}{2\tilde{R}_0^4}+\frac{45}{8\tilde{R}_0^6}\right)-\sin\left(\frac{\eta_0}{2}\right)\cos\left(\frac{\eta_0}{2}\right)\left(\frac{715}{16\tilde{R}_0^6}+\frac{45}{4\tilde{R}_0^4}\right)\right)$$
$$-\frac{B_{9,2}^{\alpha}(\tilde{\alpha}_R,\gamma)}{\tilde{R}_0^6}\left(\frac{315\tilde{P}}{8}\sin^2\left(\frac{\eta_0}{2}\right)-\frac{525}{2}\sin\left(\frac{\eta_0}{2}\right)\cos\left(\frac{\eta_0}{2}\right)\right)$$
$$\left.+\frac{B_{11,2}^{\alpha}(\tilde{\alpha}_R,\gamma)}{\tilde{R}_0^6}\left(\frac{945}{8}\tilde{P}\sin^2\left(\frac{\eta_0}{2}\right)+\frac{4725}{16}\sin\left(\frac{\eta_0}{2}\right)\cos\left(\frac{\eta_0}{2}\right)\right)\right],$$

(H.64)

$$W_{2,2}^{\gamma}(\tilde{\alpha}_R,\tilde{\gamma},\tilde{R}_0,\tilde{Q}_2,\tilde{P},\eta_0)$$
$$\approx -\frac{1}{\tilde{Q}_2^4 l_p}\sin\left(\frac{\eta_0}{2}\right)\cos\left(\frac{\eta_0}{2}\right)\left[-\frac{9B_{5,0}^{\gamma}(\tilde{\alpha}_R,\gamma)}{\tilde{R}_0^4}+B_{7,0}^{\gamma}(\tilde{\alpha}_R,\gamma)\left(\frac{15}{\tilde{R}_0^4}+\frac{225}{4\tilde{R}_0^6}\right)-\frac{525B_{9,0}^{\gamma}(\tilde{\alpha}_R,\gamma)}{2\tilde{R}_0^6}\right.$$
$$\left.+\frac{945B_{11,0}^{\gamma}(\tilde{\alpha}_R,\gamma)}{4\tilde{R}_0^6}\right]-\frac{1}{\tilde{Q}_2^4 l_p}\left(\tilde{P}^2-\frac{1}{4}\right)^{-1}\left[\frac{B_{5,2}^{\gamma}(\tilde{\alpha}_R,\gamma)}{\tilde{R}_0^4}\left(\frac{21}{4}\sin\left(\frac{\eta_0}{2}\right)\cos\left(\frac{\eta_0}{2}\right)-\frac{3\tilde{P}}{2}\sin^2\left(\frac{\eta_0}{2}\right)\right)\right.$$
$$+B_{7,2}^{\gamma}(\tilde{\alpha}_R,\gamma)\left(\tilde{P}\sin^2\left(\frac{\eta_0}{2}\right)\left(\frac{15}{2\tilde{R}_0^4}+\frac{45}{8\tilde{R}_0^6}\right)-\sin\left(\frac{\eta_0}{2}\right)\cos\left(\frac{\eta_0}{2}\right)\left(\frac{715}{16\tilde{R}_0^6}+\frac{45}{4\tilde{R}_0^4}\right)\right)$$
$$-\frac{B_{9,2}^{\gamma}(\tilde{\alpha}_R,\gamma)}{\tilde{R}_0^6}\left(\frac{315\tilde{P}}{8}\sin^2\left(\frac{\eta_0}{2}\right)-\frac{525}{2}\sin\left(\frac{\eta_0}{2}\right)\cos\left(\frac{\eta_0}{2}\right)\right)$$
$$\left.+\frac{B_{11,2}^{\gamma}(\tilde{\alpha}_R,\gamma)}{\tilde{R}_0^6}\left(\frac{945}{8}\tilde{P}\sin^2\left(\frac{\eta_0}{2}\right)+\frac{4725}{16}\sin\left(\frac{\eta_0}{2}\right)\cos\left(\frac{\eta_0}{2}\right)\right)\right],$$

(H.65)

$$W_{2,2}^R(\tilde{\alpha}_R, \tilde{\gamma}, \tilde{R}_0, \tilde{Q}_2, \tilde{P}, \eta_0)$$

$$\approx \frac{1}{\tilde{Q}_2^4 l_p} \sin\left(\frac{\eta_0}{2}\right)\cos\left(\frac{\eta_0}{2}\right)\left[-\frac{36 B_{5,0}(\tilde{\alpha}_R, \tilde{\gamma})}{\tilde{R}_0^5} - B_{7,0}(\tilde{\alpha}_R, \tilde{\gamma})\left(\frac{60}{\tilde{R}_0^5} + \frac{675}{2\tilde{R}_0^7}\right) - \frac{1575 B_{9,0}(\tilde{\alpha}_R, \tilde{\gamma})}{\tilde{R}_0^7}\right.$$

$$\left.+\frac{2835 B_{11,0}(\tilde{\alpha}_R, \tilde{\gamma})}{2\tilde{R}_0^7}\right] + \frac{1}{\tilde{Q}_2^4 l_p}\left(\tilde{P}^2 - \frac{1}{4}\right)^{-1}\left[\frac{B_{5,2}(\tilde{\alpha}_R, \tilde{\gamma})}{\tilde{R}_0^5}\left(21\sin\left(\frac{\eta_0}{2}\right)\cos\left(\frac{\eta_0}{2}\right) - 6\tilde{P}\sin^2\left(\frac{\eta_0}{2}\right)\right)\right.$$

$$+ B_{7,2}(\tilde{\alpha}_R, \tilde{\gamma})\left[\left(\frac{30}{\tilde{R}_0^5} + \frac{135}{4\tilde{R}_0^7}\right)\tilde{P}\sin^2\left(\frac{\eta_0}{2}\right) - \left(\frac{2145}{8\tilde{R}_0^7} + \frac{45}{\tilde{R}_0^5}\right)\sin\left(\frac{\eta_0}{2}\right)\cos\left(\frac{\eta_0}{2}\right)\right]$$

$$-\frac{3 B_{9,2}(\tilde{\alpha}_R, \tilde{\gamma})}{\tilde{R}_0^7}\left(\frac{315\tilde{P}}{4}\sin^2\left(\frac{\eta_0}{2}\right) - 525\sin\left(\frac{\eta_0}{2}\right)\cos\left(\frac{\eta_0}{2}\right)\right)$$

$$\left.+\frac{3 B_{11,2}(\tilde{\alpha}_R, \tilde{\gamma})}{\tilde{R}_0^7}\left(\frac{945}{4}\tilde{P}\sin^2\left(\frac{\eta_0}{2}\right) + \frac{4725}{8}\sin\left(\frac{\eta_0}{2}\right)\cos\left(\frac{\eta_0}{2}\right)\right)\right],$$

(H.66)

$$W_{2,2}^Q(\tilde{\alpha}_R, \tilde{\gamma}, \tilde{R}_0, \tilde{Q}_2, \tilde{P}, \eta_0) \approx -\frac{4 W_{2,2}(\tilde{\alpha}_R, \tilde{\gamma}, \tilde{R}_0, \tilde{Q}_2, \tilde{P}, \eta_0)}{\tilde{Q}_2}, \tag{H.67}$$

$$W_{2,2}^P(\tilde{\alpha}_R, \tilde{\gamma}, \tilde{R}_0, \tilde{Q}_2, \tilde{P}, \eta_0)$$

$$\approx \frac{2\tilde{P}}{\tilde{Q}_2^4 l_p}\left(\tilde{P}^2 - \frac{1}{4}\right)^{-2}\left[\frac{B_{5,2}(\tilde{\alpha}_R, \gamma)}{\tilde{R}_0^4}\left(\frac{21}{4}\sin\left(\frac{\eta_0}{2}\right)\cos\left(\frac{\eta_0}{2}\right) - \frac{3\tilde{P}}{2}\sin^2\left(\frac{\eta_0}{2}\right)\right)\right.$$

$$+ B_{7,2}(\tilde{\alpha}_R, \gamma)\left(\tilde{P}\sin^2\left(\frac{\eta_0}{2}\right)\left(\frac{15}{2\tilde{R}_0^4} + \frac{45}{8\tilde{R}_0^6}\right) - \sin\left(\frac{\eta_0}{2}\right)\cos\left(\frac{\eta_0}{2}\right)\left(\frac{715}{16\tilde{R}_0^6} + \frac{45}{4\tilde{R}_0^4}\right)\right)$$

$$-\frac{B_{9,2}(\tilde{\alpha}_R, \gamma)}{\tilde{R}_0^6}\left(\frac{315\tilde{P}}{8}\sin^2\left(\frac{\eta_0}{2}\right) - \frac{525}{2}\sin\left(\frac{\eta_0}{2}\right)\cos\left(\frac{\eta_0}{2}\right)\right)$$

$$\left.+\frac{B_{11,2}(\tilde{\alpha}_R, \gamma)}{\tilde{R}_0^6}\left(\frac{945}{8}\tilde{P}\sin^2\left(\frac{\eta_0}{2}\right) + \frac{4725}{16}\sin\left(\frac{\eta_0}{2}\right)\cos\left(\frac{\eta_0}{2}\right)\right)\right]$$

$$-\frac{1}{\tilde{Q}_2^4 l_p}\left(\tilde{P}^2 - \frac{1}{4}\right)^{-1}\sin^2\left(\frac{\eta_0}{2}\right)\left[-\frac{3 B_{5,2}(\tilde{\alpha}_R, \gamma)}{2\tilde{R}_0^4} + B_{7,2}(\tilde{\alpha}_R, \gamma)\left(\frac{15}{2\tilde{R}_0^4} + \frac{45}{8\tilde{R}_0^6}\right)\right.$$

$$\left.-\frac{315 B_{9,2}^\gamma(\tilde{\alpha}_R, \gamma)}{8\tilde{R}_0^6} + \frac{945 B_{11,2}^\gamma(\tilde{\alpha}_R, \gamma)}{8\tilde{R}_0^6}\right]$$

(H.68)

$$W_{2,2}^{\eta}(\tilde{\alpha}_R,\tilde{\gamma},\tilde{R}_0,\tilde{Q}_2,\tilde{P},\eta_0)$$
$$\approx -\frac{\cos\eta_0}{2\tilde{Q}_2^4 l_p}\left[-\frac{9B_{5,0}(\tilde{\alpha}_R,\tilde{\gamma})}{\tilde{R}_0^4}+B_{7,0}(\tilde{\alpha}_R,\tilde{\gamma})\left(\frac{15}{\tilde{R}_0^4}+\frac{225}{4\tilde{R}_0^6}\right)-\frac{525B_{9,0}(\tilde{\alpha}_R,\tilde{\gamma})}{2\tilde{R}_0^6}+\frac{945B_{11,0}(\tilde{\alpha}_R,\tilde{\gamma})}{4\tilde{R}_0^6}\right]$$
$$-\frac{1}{\tilde{Q}_2^4 l_p}\left(\tilde{P}^2-\frac{1}{4}\right)^{-1}\left[\frac{B_{5,2}(\tilde{\alpha}_R,\tilde{\gamma})}{\tilde{R}_0^4}\left(\frac{21}{8}\cos\eta_0-\frac{3\tilde{P}}{4}\sin\eta_0\right)+B_{7,2}(\tilde{\alpha}_R,\tilde{\gamma})\left(\tilde{P}\sin\eta_0\left(\frac{15}{4\tilde{R}_0^4}+\frac{45}{16\tilde{R}_0^6}\right)\right.\right.$$
$$\left.-\cos\eta_0\left(\frac{715}{32\tilde{R}_0^6}+\frac{45}{8\tilde{R}_0^4}\right)\right)-\frac{B_{9,2}(\tilde{\alpha}_R,\tilde{\gamma})}{\tilde{R}_0^6}\left(\frac{315\tilde{P}}{16}\sin\eta_0-\frac{525}{4}\cos\eta_0\right)$$
$$\left.+\frac{B_{11,2}(\tilde{\alpha}_R,\tilde{\gamma})}{\tilde{R}_0^6}\left(\frac{945}{16}\tilde{P}\sin\eta_0+\frac{4725}{32}\cos\eta_0\right)\right],$$

(H.69)

$$W_{2,3}^{\alpha_R}(\tilde{\alpha}_R,\tilde{\gamma},\tilde{R}_0,\tilde{Q}_2,\tilde{P},\eta_0) \approx -\frac{\tilde{P}}{\tilde{Q}_2^2 l_p}\left(\tilde{P}^2+\frac{1}{4}\right)^{-1}\left[\frac{C_{3,2}^{\alpha}(\tilde{\alpha}_R,\tilde{\gamma})}{\tilde{R}_0^2}\right.$$
$$\left.-\frac{3C_{5,2}^{\alpha}(\tilde{\alpha}_R,\tilde{\gamma})}{4\tilde{R}_0^4}+C_{7,2}^{\alpha}(\tilde{\alpha}_R,\tilde{\gamma})\left(\frac{45}{32\tilde{R}_0^6}+\frac{15}{4\tilde{R}_0^4}\right)-\frac{315C_{9,2}^{\alpha}(\tilde{\alpha}_R,\tilde{\gamma})}{16\tilde{R}_0^6}+\frac{945C_{9,2}^{\alpha}(\tilde{\alpha}_R,\tilde{\gamma})}{32\tilde{R}_0^6}\right],$$

(H.70)

$$W_{2,3}^{\gamma}(\tilde{\alpha}_R,\tilde{\gamma},\tilde{R}_0,\tilde{Q}_2,\tilde{P},\eta_0) \approx -\frac{\tilde{P}}{\tilde{Q}_2^2 l_p}\left(\tilde{P}^2+\frac{1}{4}\right)^{-1}\left[\frac{C_{3,2}^{\gamma}(\tilde{\alpha}_R,\tilde{\gamma})}{\tilde{R}_0^2}\right.$$
$$\left.-\frac{3C_{5,2}^{\gamma}(\tilde{\alpha}_R,\tilde{\gamma})}{4\tilde{R}_0^4}+C_{7,2}^{\gamma}(\tilde{\alpha}_R,\tilde{\gamma})\left(\frac{45}{32\tilde{R}_0^6}+\frac{15}{4\tilde{R}_0^4}\right)-\frac{315C_{9,2}^{\gamma}(\tilde{\alpha}_R,\tilde{\gamma})}{16\tilde{R}_0^6}+\frac{945C_{9,2}^{\gamma}(\tilde{\alpha}_R,\tilde{\gamma})}{32\tilde{R}_0^6}\right],$$

(H.71)

$$W_{2,3}^{R}(\tilde{\alpha}_R,\tilde{\gamma},\tilde{R}_0,\tilde{Q}_2,\tilde{P},\eta_0) \approx \frac{\tilde{P}}{\tilde{Q}_2^2 l_p}\left(\tilde{P}^2+\frac{1}{4}\right)^{-1}\left[\frac{2C_{3,2}(\tilde{\alpha}_R,\tilde{\gamma})}{\tilde{R}_0^3}\right.$$
$$\left.-\frac{3C_{5,2}(\tilde{\alpha}_R,\tilde{\gamma})}{\tilde{R}_0^5}+C_{7,2}(\tilde{\alpha}_R,\tilde{\gamma})\left(\frac{135}{16\tilde{R}_0^7}+\frac{15}{\tilde{R}_0^5}\right)-\frac{945C_{9,2}(\tilde{\alpha}_R,\tilde{\gamma})}{8\tilde{R}_0^7}+\frac{2835C_{11,2}(\tilde{\alpha}_R,\tilde{\gamma})}{16\tilde{R}_0^7}\right],$$

(H.72)

$$W_{2,3}^{Q}(\tilde{\alpha}_R,\tilde{\gamma},\tilde{R}_0,\tilde{Q}_2,\tilde{P},\eta_0) \approx -\frac{2W_{2,3}(\tilde{\alpha}_R,\tilde{\gamma},\tilde{R}_0,\tilde{Q}_2,\tilde{P},\eta_0)}{\tilde{Q}_2},$$

(H.73)

$$W_{2,3}^{P}(\tilde{\alpha}_R,\tilde{\gamma},\tilde{R}_0,\tilde{Q}_2,\tilde{P},\eta_0) \approx \left[\frac{1}{\tilde{P}}-\frac{8\tilde{P}}{(4\tilde{P}^2-1)}\right]W_{2,3}(\tilde{\alpha}_R,\tilde{\gamma},\tilde{R}_0,\tilde{Q}_2,\tilde{P},\eta_0),$$

(H.74)

$$W_{2,3}^{\eta}(\tilde{\alpha}_R,\tilde{\gamma},\tilde{R}_0,\tilde{Q}_2,\tilde{P},\eta_0) \approx 0,$$

(H.75)

$$W_{2,4}^{\alpha_R}(\tilde{\alpha}_R,\tilde{\gamma},\tilde{R}_0,\tilde{Q}_2,\tilde{P},\eta_0)$$

$$\approx \frac{1}{\tilde{Q}_2^3 l_p}\left(\tilde{P}^2 - \frac{1}{4}\right)^{-1/2}\left[\frac{2D_{3.1}^{\alpha}(\tilde{\alpha}_R,\tilde{\gamma})}{\tilde{R}_0^3}\cos\left(\frac{\eta_0}{2}\right) + D_{5.1}^{\alpha}(\tilde{\alpha}_R,\tilde{\gamma})\left(\frac{3}{\tilde{R}_0^3}\left(2\tilde{P}\sin\left(\frac{\eta_0}{2}\right) - \cos\left(\frac{\eta_0}{2}\right)\right) - \frac{3}{\tilde{R}_0^5}\cos\left(\frac{\eta_0}{2}\right)\right)\right.$$

$$\left. + \frac{D_{7.1}^{\alpha}(\tilde{\alpha}_R,\tilde{\gamma})}{\tilde{R}_0^5}\left(\frac{105}{4}\cos\left(\frac{\eta_0}{2}\right) - \frac{45\tilde{P}}{2}\sin\left(\frac{\eta_0}{2}\right)\right) + \frac{D_{9.1}^{\alpha}(\tilde{\alpha}_R,\tilde{\gamma})}{\tilde{R}_0^5}\left(\frac{105\tilde{P}}{2}\sin\left(\frac{\eta_0}{2}\right) - \frac{105}{4}\cos\left(\frac{\eta_0}{2}\right)\right)\right],$$

(H.76)

$$W_{2,4}^{\gamma}(\tilde{\alpha}_R,\tilde{\gamma},\tilde{R}_0,\tilde{Q}_2,\tilde{P},\eta_0)$$

$$\approx \frac{1}{\tilde{Q}_2^3 l_p}\left(\tilde{P}^2 - \frac{1}{4}\right)^{-1/2}\left[\frac{2D_{3.1}^{\gamma}(\tilde{\alpha}_R,\tilde{\gamma})}{\tilde{R}_0^3}\cos\left(\frac{\eta_0}{2}\right) + D_{5.1}^{\gamma}(\tilde{\alpha}_R,\tilde{\gamma})\left(\frac{3}{\tilde{R}_0^3}\left(2\tilde{P}\sin\left(\frac{\eta_0}{2}\right) - \cos\left(\frac{\eta_0}{2}\right)\right) - \frac{3}{\tilde{R}_0^5}\cos\left(\frac{\eta_0}{2}\right)\right)\right.$$

$$\left. + \frac{D_{7.1}^{\gamma}(\tilde{\alpha}_R,\tilde{\gamma})}{\tilde{R}_0^5}\left(\frac{105}{4}\cos\left(\frac{\eta_0}{2}\right) - \frac{45\tilde{P}}{2}\sin\left(\frac{\eta_0}{2}\right)\right) + \frac{D_{9.1}^{\gamma}(\tilde{\alpha}_R,\tilde{\gamma})}{\tilde{R}_0^5}\left(\frac{105\tilde{P}}{2}\sin\left(\frac{\eta_0}{2}\right) - \frac{105}{4}\cos\left(\frac{\eta_0}{2}\right)\right)\right],$$

(H.77)

$$W_{2,4}^{R}(\tilde{\alpha}_R,\tilde{\gamma},\tilde{R}_0,\tilde{Q}_2,\tilde{P},\eta_0)$$

$$\approx -\frac{1}{l_p \tilde{Q}_2^3}\left(\tilde{P}^2 - \frac{1}{4}\right)^{-1/2}\left[\frac{6D_{3,1}(\tilde{\alpha}_R,\tilde{\gamma})}{\tilde{R}_0^4}\cos\left(\frac{\eta_0}{2}\right) + D_{5,1}(\tilde{\alpha}_R,\tilde{\gamma})\left(\frac{9}{\tilde{R}_0^4}\left(2\tilde{P}\sin\left(\frac{\eta_0}{2}\right) - \cos\left(\frac{\eta_0}{2}\right)\right) - \frac{15}{\tilde{R}_0^6}\cos\left(\frac{\eta_0}{2}\right)\right)\right.$$

$$\left. + \frac{5D_{7,1}(\tilde{\alpha}_R,\tilde{\gamma})}{\tilde{R}_0^6}\left(\frac{105}{4}\cos\left(\frac{\eta_0}{2}\right) - \frac{15\tilde{P}}{2}\sin\left(\frac{\eta_0}{2}\right)\right) + \frac{5D_{9,1}(\tilde{\alpha}_R,\tilde{\gamma})}{\tilde{R}_0^6}\left(\frac{105\tilde{P}}{2}\sin\left(\frac{\eta_0}{2}\right) - \frac{105}{4}\cos\left(\frac{\eta_0}{2}\right)\right)\right],$$

(H.78)

$$W_{2,4}^{Q}(\tilde{\alpha}_R,\tilde{\gamma},\tilde{R}_0,\tilde{Q}_2,\tilde{P},\eta_0) \approx -\frac{3W_{2,4}(\tilde{\alpha}_R,\tilde{\gamma},\tilde{R}_0,\tilde{Q}_2,\tilde{P},\eta_0)}{\tilde{Q}_2},$$

(H.79)

$$W_{2,4}^{P}(\tilde{\alpha}_R,\tilde{\gamma},\tilde{R}_0,\tilde{Q}_2,\tilde{P},\eta_0) \approx -4\tilde{P}\left(4\tilde{P}^2 - 1\right)^{-1}W_{2,4}(\tilde{\alpha}_R,\tilde{\gamma},\tilde{R}_0,\tilde{Q}_2,\tilde{P},\eta_0)$$

$$+ \frac{1}{\tilde{Q}_2^3}\left(\tilde{P}^2 - 1/4\right)^{-1/2}\sin\left(\frac{\eta_0}{2}\right)\left[\frac{6}{\tilde{R}_0^3}D_{5,1}(\tilde{\alpha}_R,\tilde{\gamma}) - \frac{45}{2\tilde{R}_0^5}D_{7,1}(\tilde{\alpha}_R,\tilde{\gamma}) + \frac{105}{2\tilde{R}_0^5}D_{9,1}(\tilde{\alpha}_R,\tilde{\gamma})\right],$$

(H.80)

and

$$W_{2,4}^{\eta}(\tilde{\alpha}_R,\tilde{\gamma},\tilde{R}_0,\tilde{Q}_2,\tilde{P},\eta_0) \approx \frac{1}{2l_p \tilde{Q}_2^3}\left(\tilde{P}^2 - \frac{1}{4}\right)^{-1/2}\left[-\frac{2D_{3,1}(\tilde{\alpha}_R,\tilde{\gamma})}{\tilde{R}_0^3}\sin\left(\frac{\eta_0}{2}\right)\right.$$

$$+ D_{5,1}(\tilde{\alpha}_R,\tilde{\gamma})\left(\frac{3}{\tilde{R}_0^3}\left(2\tilde{P}\cos\left(\frac{\eta_0}{2}\right) + \sin\left(\frac{\eta_0}{2}\right)\right) + \frac{3}{\tilde{R}_0^5}\sin\left(\frac{\eta_0}{2}\right)\right)$$

$$\left. - \frac{D_{7,1}(\tilde{\alpha}_R,\tilde{\gamma})}{\tilde{R}_0^5}\left(\frac{105}{4}\sin\left(\frac{\eta_0}{2}\right) + \frac{45}{2}\tilde{P}\cos\left(\frac{\eta_0}{2}\right)\right) + \frac{D_{9,1}(\tilde{\alpha}_R,\tilde{\gamma})}{\tilde{R}_0^5}\left(\frac{105}{4}\sin\left(\frac{\eta_0}{2}\right) + \frac{105}{2}\tilde{P}\cos\left(\frac{\eta_0}{2}\right)\right)\right].$$

(H.81)

The functions $B_{n,m}^{\alpha}(\tilde{\alpha}_R, \tilde{\gamma})$, $C_{n,m}^{\alpha}(\tilde{\alpha}_R, \tilde{\gamma})$ and $D_{n,m}^{\alpha}(\tilde{\alpha}_R, \tilde{\gamma})$ are the partial derivatives of $B_{n,m}(\bar{\alpha}_R, \tilde{\gamma})$, $C_{n,m}(\bar{\alpha}_R, \tilde{\gamma})$ and $D_{n,m}(\bar{\alpha}_R, \tilde{\gamma})$, respectively, with respect to $\tilde{\alpha}_R$, whereas $B_{n,m}^{\gamma}(\tilde{\alpha}_R, \tilde{\gamma})$, $C_{n,m}^{\gamma}(\tilde{\alpha}_R, \tilde{\gamma})$ and $D_{n,m}^{\gamma}(\tilde{\alpha}_R, \tilde{\gamma})$ are their partial derivatives with respect to $\tilde{\gamma}$. These are given by

$$B_{n,m}^{\alpha}(\tilde{\alpha}_R, \tilde{\gamma}) = 2\int_0^{\infty} dx \frac{x^m}{(1+x^2)^{n/2}} \frac{\partial \hat{G}_R(x, \tilde{\alpha}_R, \tilde{\gamma})}{\partial \tilde{\alpha}_R}, \tag{H.82}$$

$$B_{n,m}^{\gamma}(\tilde{\alpha}_R, \tilde{\gamma}) = 2\int_0^{\infty} dx \frac{x^m}{(1+x^2)^{n/2}} \frac{\partial \hat{G}_R(x, \tilde{\alpha}_R, \tilde{\gamma})}{\partial \tilde{\gamma}}, \tag{H.83}$$

$$C_{n,m}^{\alpha}(\tilde{\alpha}_R, \tilde{\gamma}) = \int_0^{\infty} dx \frac{x^m}{(1+x^2)^{n/2}} \frac{\partial \hat{D}_R(x, \tilde{\alpha}_R, \tilde{\gamma})}{\partial \tilde{\alpha}_R}, \tag{H.84}$$

$$C_{n,m}^{\gamma}(\tilde{\alpha}_R, \tilde{\gamma}) = \int_0^{\infty} dx \frac{x^m}{(1+x^2)^{n/2}} \frac{\partial \hat{D}_R(x, \tilde{\alpha}_R, \tilde{\gamma})}{\partial \tilde{\gamma}}, \tag{H.85}$$

$$D_{n,m}^{\alpha}(\tilde{\alpha}_R, \tilde{\gamma}) = \int_0^{\infty} dx \frac{x^m}{(1+x^2)^{n/2}} \frac{\partial \hat{C}_R(x, \tilde{\alpha}_R, \tilde{\gamma})}{\partial \tilde{\alpha}_R}, \tag{H.86}$$

and

$$D_{n,m}^{\gamma}(\tilde{\alpha}_R, \tilde{\gamma}) = \int_0^{\infty} dx \frac{x^m}{(1+x^2)^{n/2}} \frac{\partial \hat{C}_R(x, \tilde{\alpha}_R, \tilde{\gamma})}{\partial \tilde{\gamma}}. \tag{H.87}$$

The derivatives of the correlation functions are again given by Eqs. (H.58)-(H.63).

## Appendix I. Partial derivatives of $\tilde{Q}_j, \tilde{\alpha}_\eta,$ and $\tilde{\alpha}_R$

The partial derivatives of $\tilde{Q}_j, \tilde{\alpha}_\eta,$ and $\tilde{\alpha}_R$ that appear in Eqs. (7.15)-(7.19) are given by

$$\frac{\partial \tilde{Q}_j}{\partial \tilde{R}_0} = -\frac{1}{\tilde{R}_0} \tilde{Q}_j + \frac{\partial \tilde{P}}{\partial \tilde{R}_0} \frac{1}{\tilde{R}_0} \frac{(-1)^j}{4\tilde{P}^3} \cos\left(\frac{\eta_0}{2}\right)\left(1 - \frac{1}{8(2\alpha_\eta l_p)^{1/2}}\right)\left(1 - \frac{(-1)^j}{4\tilde{P}^2}\right)^{-1/2}, \tag{I.1}$$

$$\frac{\partial \tilde{Q}_j}{\partial \eta_0} = -\frac{1}{2}\tan\left(\frac{\eta_0}{2}\right)\tilde{Q}_j + \frac{\partial \tilde{P}}{\partial \eta_0} \frac{1}{\tilde{R}_0} \frac{(-1)^j}{4\tilde{P}^3} \cos\left(\frac{\eta_0}{2}\right)\left(1 - \frac{1}{8(2\alpha_\eta l_p)^{1/2}}\right)\left(1 - \frac{(-1)^j}{4\tilde{P}^2}\right)^{-1/2}, \tag{I.2}$$

$$\frac{\partial \tilde{Q}_j}{\partial \theta_R} = \frac{\partial \tilde{P}}{\partial \theta_R} \frac{1}{\tilde{R}_0} \frac{(-1)^j}{4\tilde{P}^3} \cos\left(\frac{\eta_0}{2}\right)\left(1 - \frac{1}{8(2\alpha_\eta l_p)^{1/2}}\right)\left(1 - \frac{(-1)^j}{4\tilde{P}^2}\right)^{-1/2}, \tag{I.3}$$

$$\frac{\partial \tilde{Q}_j}{\partial \alpha_\eta} = \frac{1}{16\alpha_\eta (2l_p \alpha_\eta)^{1/2}} \frac{1}{\tilde{R}_0} \cos\left(\frac{\eta_0}{2}\right)\left(1 - \frac{(-1)^j}{4\tilde{P}^2}\right)^{1/2}, \tag{I.4}$$

$$\frac{\partial \tilde{\alpha}_\eta}{\partial R_0} = \frac{2\tilde{\alpha}_\eta}{R_0} - \frac{\partial \tilde{P}}{\partial \tilde{R}_0} \frac{(-1)^j}{\tilde{P}^3} \frac{R_0^2}{\cos^2\left(\frac{\eta_0}{2}\right) d_R l_p} \left(\alpha_\eta + \frac{1}{4}\left(\frac{\alpha_\eta}{2l_p}\right)^{1/2}\right)\left(1 - \frac{(-1)^j}{4\tilde{P}^2}\right)^{-2}, \tag{I.5}$$

$$\frac{\partial \tilde{\alpha}_\eta}{\partial d_R} = \frac{\partial \tilde{P}}{\partial \tilde{R}_0} \frac{(-1)^j}{\tilde{P}^3} \frac{R_0^3}{\cos^2\left(\frac{\eta_0}{2}\right) d_R^2 l_p} \left(\alpha_\eta + \frac{1}{4}\left(\frac{\alpha_\eta}{2l_p}\right)^{1/2}\right)\left(1 - \frac{(-1)^j}{4\tilde{P}^2}\right)^{-2}, \tag{I.6}$$

$$\frac{\partial \tilde{\alpha}_\eta}{\partial \eta_0} = \tan\left(\frac{\eta_0}{2}\right)\tilde{\alpha}_\eta - \frac{\partial \tilde{P}}{\partial \eta_0} \frac{(-1)^j}{\tilde{P}^3} \frac{R_0^2}{\cos^2\left(\frac{\eta_0}{2}\right) l_p} \left(\alpha_\eta + \frac{1}{4}\left(\frac{\alpha_\eta}{2l_p}\right)^{1/2}\right)\left(1 - \frac{(-1)^j}{4\tilde{P}^2}\right)^{-2}, \tag{I.7}$$

$$\frac{\partial \tilde{\alpha}_\eta}{\partial \theta_R} = -\frac{\partial \tilde{P}}{\partial \theta_R} \frac{(-1)^j}{\tilde{P}^3} \frac{R_0^2}{\cos^2\left(\frac{\eta_0}{2}\right) l_p} \left(\alpha_\eta + \frac{1}{4}\left(\frac{\alpha_\eta}{2l_p}\right)^{1/2}\right)\left(1 - \frac{(-1)^j}{4\tilde{P}^2}\right)^{-2}, \tag{I.8}$$

$$\frac{\partial \tilde{\alpha}_\eta}{\partial \alpha_\eta} = \frac{2R_0^2}{\cos^2\left(\frac{\eta_0}{2}\right) l_p}\left(1 + \frac{1}{8}\left(\frac{1}{2\alpha_\eta l_p}\right)^{1/2}\right)\left(1 - \frac{(-1)^j}{4\tilde{P}^2}\right)^{-1}, \tag{I.9}$$

$$\frac{\partial \tilde{\alpha}_R}{\partial R_0} = \frac{4\tilde{\alpha}_R}{R_0} - \frac{\partial \tilde{P}}{\partial \tilde{R}_0} \frac{R_0^4}{d_R \cos^4\left(\frac{\eta_0}{2}\right)} \frac{(-1)^j}{\tilde{P}^3}\left(1 + \frac{1}{2}\left(\frac{1}{2l_p \alpha_\eta}\right)^{1/2}\right)\left(1 - \frac{(-1)^j}{4\tilde{P}^2}\right)^{-3}\left(\frac{\theta_R}{d_R}\right)^4, \tag{I.10}$$

$$\frac{\partial \tilde{\alpha}_R}{\partial d_R} = \frac{\partial \tilde{P}}{\partial \tilde{R}_0} \frac{R_0^5}{d_R^2 \cos^4\left(\frac{\eta_0}{2}\right)} \frac{(-1)^j}{\tilde{P}^3}\left(1 + \frac{1}{2}\left(\frac{1}{2l_p \alpha_\eta}\right)^{1/2}\right)\left(1 - \frac{(-1)^j}{4\tilde{P}^2}\right)^{-3}\left(\frac{\theta_R}{d_R}\right)^4 - \frac{4\tilde{\alpha}_R}{d_R}, \tag{I.11}$$

$$\frac{\partial \tilde{\alpha}_R}{\partial \theta_R} = -\frac{\partial \tilde{P}}{\partial \theta_R} \frac{R_0^4}{\cos^4\left(\frac{\eta_0}{2}\right)} \frac{(-1)^j}{\tilde{P}^3}\left(1 + \frac{1}{2}\left(\frac{1}{2l_p \alpha_\eta}\right)^{1/2}\right)\left(1 - \frac{(-1)^j}{4\tilde{P}^2}\right)^{-3}\left(\frac{\theta_R}{d_R}\right)^4 + \frac{4\tilde{\alpha}_R}{\theta_R}, \tag{I.12}$$

$$\frac{\partial \tilde{\alpha}_R}{\partial \eta_0} = 2\tan\left(\frac{\eta_0}{2}\right)\tilde{\alpha}_R - \frac{\partial \tilde{P}}{\partial \eta_0} \frac{R_0^4}{\cos^4\left(\frac{\eta_0}{2}\right)} \frac{(-1)^j}{\tilde{P}^3}\left(1 + \frac{1}{2}\left(\frac{1}{2l_p \alpha_\eta}\right)^{1/2}\right)\left(1 - \frac{(-1)^j}{4\tilde{P}^2}\right)^{-3}\left(\frac{\theta_R}{d_R}\right)^4, \tag{I.13}$$

and

$$\frac{\partial \tilde{\alpha}_R}{\partial \alpha_\eta} = -\frac{R_0^4}{4\alpha_\eta^{3/2}(2l_p)^{1/2}\cos^4\left(\frac{\eta_0}{2}\right)}\left(1 - \frac{(-1)^j}{4\tilde{P}^2}\right)^{-2}\left(\frac{\theta_R}{d_R}\right)^4. \tag{I.14}$$